\documentclass[twocolumn,superscriptaddress,showpacs,nofootinbib]{revtex4}

\usepackage{bm,color}

\usepackage{graphicx}
\usepackage{amsmath}
\usepackage{amssymb}
\usepackage{amsfonts}
\usepackage{bm}
\usepackage{color}
\usepackage{lipsum}

\def\be{\begin{equation}}
\def\ee{\end{equation}}
\def\bea{\begin{eqnarray}}
\def\eea{\end{eqnarray}}

\begin{document}

\title{A new logotropic model based on a complex scalar field with a logarithmic potential}

\author{Pierre-Henri Chavanis}
\email{chavanis@irsamc.ups-tlse.fr}
\affiliation{Laboratoire de Physique Th\'eorique, Universit\'e de Toulouse,
CNRS, UPS, France}

\begin{abstract}

We introduce a new logotropic model based on a complex scalar field with a
logarithmic potential that unifies dark matter and dark energy. 
The scalar field
satisfies a nonlinear
wave  equation generalizing the Klein-Gordon equation in the relativistic regime and the
Schr\"odinger equation in the nonrelativistic regime. This
model has an intrinsically quantum nature and returns the $\Lambda$CDM
model in the classical limit $\hbar\rightarrow 0$. It involves a new
fundamental constant of physics $A/c^2=2.10\times 10^{-26}\, {\rm g}\, {\rm
m}^{-3}$  responsible for the
late accelerating expansion of the Universe and superseding the Einstein
cosmological constant $\Lambda$. The logotropic
model is almost indistinguishable from
the $\Lambda$CDM model at large (cosmological) scales but solves the CDM crisis
at small (galactic) scales. It also solves the problems of the fuzzy dark matter
model. Indeed,  it leads to
cored dark matter halos with a universal surface density
$\Sigma_0^{\rm th}=5.85\,\left ({A}/{4\pi G}\right
)^{1/2}=133\, M_{\odot}/{\rm pc}^2$. This universal surface density is 
predicted from the
logotropic model without
adjustable parameter and turns out to be close to the observed value
$\Sigma_0^{\rm
obs}=141_{-52}^{+83}\, M_{\odot}/{\rm pc}^2$. We also argue
that the quantities
$\Omega_{\rm dm,0}$ and $\Omega_{\rm de,0}$, which are usually interpreted as
the present proportion of
dark
matter and dark energy in the $\Lambda$CDM model, 
are equal to $\Omega_{\rm dm,0}^{\rm th}=\frac{1}{1+e}(1-\Omega_{\rm
b,0})=0.2559$ and  $\Omega_{\rm de,0}^{\rm th}=\frac{e}{1+e}(1-\Omega_{\rm
b,0})=0.6955$  in very good agreement
with the measured values $\Omega_{\rm dm,0}^{\rm obs}=0.2589$ and $\Omega_{\rm
de,0}^{\rm obs}=0.6911$ (their ratio $2.669$ is close to the pure number
$e=2.71828...$). We point out,
however, important difficulties with the logotropic model, similar to those
encountered by the generalized Chaplygin gas model. These problems are related
to 
the difficulty of forming  large-scale structures
due to an increasing speed of sound as the Universe expands. We discuss
potential
solutions to these problems, stressing in particular the
importance to perform a nonlinear study of structure formation.

\end{abstract}



\maketitle


\section{Introduction}

The nature of dark matter (DM) and dark energy (DE) is still unknown and remains
one of the greatest mysteries of cosmology. In a previous paper \cite{epjp}
(see also \cite{lettre,jcap,pdu}), we have introduced an exotic fluid, called
the logotropic dark fluid (LDF), that unifies DM and DE in the spirit of a
generalized Chaplygin gas model.\footnote{The original logotropic model
\cite{epjp,lettre,jcap,pdu} has
been further studied in
\cite{fa,cal1,ootsm,cal2,bal,mamon,bklmp,logogen}.}
The LDF is
characterized by the equation of state \cite{epjp,lettre,jcap,pdu} 
\begin{eqnarray}
P=A\ln \left (\frac{\rho_m}{\rho_P}\right ),
\label{intro1}
\end{eqnarray}
where $\rho_m=n m$ is the rest-mass density, $\rho_P=5.16\times
10^{99}\, {\rm g/m^3}$ is the Planck density and $A/c^2=2.10\times 10^{-26}\,
{\rm g}\, {\rm
m}^{-3}$ is a constant interpreted  as
a new fundamental constant of physics superseding the Einstein cosmological
constant $\Lambda$. For $\rho_m<\rho_P$, the
pressure of the LDF is negative.\footnote{The logotropic  model is not valid in
the primordial Universe so that, in practice, the LDF exhibits a negative
pressure, as required to explain the
acceleration of the Universe today.} At early times, the pressure is negligible
with respect to
the energy density and the LDF behaves like the pressureless CDM model.
At later times, the negative pressure of the LDF becomes efficient and explains
the
acceleration of the Universe that we observe today.  We obtained very
encouraging results \cite{epjp,lettre,jcap,pdu}. At large
(cosmological) scales, the logotropic model is almost indistinguishable from
the $\Lambda$CDM model up to the present time for what concerns the evolution
of the homogeneous background. The two models will differ in about
$25$ Gyrs when the logotropic model will start to exhibit a phantom behavior, i.e.,
the energy density will increase with the scale factor, leading to a super de
Sitter era where the scale factor increases as $a\sim e^{t^2}$.\footnote{ By
contrast, in the $\Lambda$CDM model, the energy density tends to a constant
$\epsilon_\Lambda$ leading to a de Sitter era where the scale factor increases
as $a\sim e^{t}$. Note that the increase of the
energy density $\epsilon$
with $a$ in the logotropic model is slow -- logarithmic. As a result, there is no future finite time
singularity (no ``big rip'') \cite{caldwellprl}. The energy density becomes
infinite in infinite time. This is called
``little rip'' \cite{littlerip}.} At small (galactic) scales, the logotropic
model
is able to solve
the small-scale crisis of the CDM model. Indeed, contrary to the pressureless CDM model,
the logotropic equation of state provides a pressure gradient that can balance
the gravitational attraction and prevent gravitational collapse. As a result, logotropic DM halos present a central core
rather than a cusp, in agreement with the observations. In addition, a very nice
property of the logotropic equation of state is that it generates DM halos with a
universal surface density. Its predicted value $\Sigma_0^{\rm th}=5.85\,\left ({A}/{4\pi
G}\right
)^{1/2}=133\, M_{\odot}/{\rm pc}^2$
turns out to be close to the observed value $\Sigma_0^{\rm
obs}=141_{-52}^{+83}\, M_{\odot}/{\rm pc}^2$
\cite{kormendy,spano,donato}. This is
remarkable because there is no adjustable parameter in our
model \cite{epjp,lettre,jcap,pdu}. The
logotropic model also implies that the mass of dwarf galaxies enclosed within a
sphere of fixed radius
$r_{u}=300\, {\rm pc}$ has a universal value
$M_{300}^{\rm th}=1.82\times 10^{7}\, M_{\odot}$, i.e. $\log
(M_{300}^{\rm th}/M_{\odot})=7.26$, in agreement with the
observations giving  $\log
(M_{300}^{\rm obs}/M_{\odot})=7.0^{+0.3}_{-0.4}$ \cite{strigari}.
The logotropic model also reproduces the Tully-Fisher
relation $M_b\propto v_h^4$, where $M_b$ is the baryonic mass and $v_h$ the
circular velocity at the halo radius, and predicts a value of the ratio
$(M_b/v_h^4)^{\rm th}=46.4\,
M_{\odot}{\rm km}^{-4}{\rm s}^4$ which is close to the
observed value 
$(M_b/v_h^4)^{\rm obs}=47\pm 6 \, M_{\odot}{\rm km}^{-4}{\rm s}^4$
\cite{mcgaugh}.

In the present paper, we introduce a related, but different, logotropic model.
We develop a field theory based on the Klein-Gordon (KG) equation in general
relativity  for
a complex scalar field (SF) with a logarithmic potential.\footnote{Most authors
describe DM and 
DE by a real SF. Here, we consider a complex SF based on the general formalism
developed in \cite{abrilphas,action,cspoly} for an arbitrary potential. It
could be called the CSF model.} In
the
fast
oscillation regime, which is equivalent to the Thomas-Fermi (TF) approximation
$\hbar\rightarrow 0$, this complex SF generates a logotropic equation of state
similar to Eq. (\ref{intro1}) except that the rest-mass density $\rho_m$ is
replaced by a pseudo rest-mass density $\rho$ related to the squared modulus of
the SF. This new logotropic model is similar to the previous one (conserving its
main virtues), except that it does not display a phantom behavior in the future
but rather a late de Sitter era. Indeed, the energy density always decreases as the
Universe expands and tends to a constant $\epsilon_{\rm min}$ like in the
$\Lambda$CDM model but with a slightly different value. 
Correspondingly, the speed of sound is real and always
smaller than the speed of light while,  in the original logotropic
model \cite{epjp,lettre,jcap,pdu}, the
speed of sound  was diverging at the entry of the phantom regime before becoming
imaginary. Therefore, the new logotropic model avoids some pathologies of the
original logotropic model such as a phantom behavior (violation of the
dominant-energy condition $P/\epsilon>-1$ and little rip) and a
superluminal or imaginary speed of sound. It rather asymptotically
approaches a well-behaved de Sitter era. 
Therefore, the new logotropic model interpolates a regime of dust-dominated
Universe to a
vacuum energy dominated Universe, providing an explanation for the possible
accelerating phase today: For small values of the scale factor, the LDF
exhibits the same behavior 
as a pressureless fluid; for
large values of the scale factor, it approaches the equation of state of a
cosmological constant.

The logotropic complex SF model is based on a nonlinear
KG equation involving a
logarithmic potential whose strength is measured by the logotropic constant $A$.
We argue that this constant does not correspond to a
particular characteristic of the SF (such as its mass $m$ or self-interaction
constant $\lambda$) but that it has a fundamental and universal
nature.\footnote{This term should be present in all SF theories
even though it may be negligible in certain cases. In other words, a SF with a
purely logarithmic potential is considered to be massless and noninteracting. We
can then introduce  specific attributes of the SF such as a mass term $m^2
|{\varphi}|^2$ and a self-interaction term
like a quartic potential $\lambda |{\varphi}|^4$.} In our model, this constant
is
responsible for
the accelerating expansion of the Universe and, at the same time, for the
universality of the surface density of DM halos.  In the nonrelativistic limit,
the nonlinear KG equation reduces to a nonlinear Schr\"odinger equation which
has the form of a generalized Gross-Pitaevskii (GP) equation with a logarithmic
potential.

The aim of this paper is to develop the logotropic complex SF model in detail
and to compare it
with other models such as the original logotropic model, the $\Lambda$CDM model, and
the fuzzy dark matter (FDM) model. The paper is organized as follows. In Sec.
\ref{sec_csf}, we summarize the theory
developed in \cite{abrilphas} for a
spatially homogeneous complex SF with an arbitrary potential  $V(|{\varphi}|^2)$
evolving in an expanding background. We consider in particular the fast
oscillation regime, equivalent to the TF approximation, where the equations can
be simplified and where the SF behaves as a fluid with a barotropic equation of
state $P(\rho)$ determined by the SF potential. In Sec. \ref{sec_log}, we
consider a
logarithmic potential and show that it leads to a logotropic equation of state.
In Sec. \ref{sec_ie} we determine the rest-mass density and the internal energy
of the LDF which represent DM and DE respectively.
In Sec. \ref{sec_evo}, we study the evolution  of the LDF, stressing that it
behaves as DM
in the early Universe and as DE in the late Universe. In Sec. \ref{sec_valf}, we
determine
the fundamental constant $A$ (and the equivalent  dimensionless constant $B$) of
our model from cosmological considerations. We finally argue
that the quantities
$\Omega_{\rm dm,0}$ and $\Omega_{\rm de,0}$, which are usually interpreted as
the present proportion of DM and DE in the $\Lambda$CDM model, 
are equal to $\Omega_{\rm dm,0}^{\rm th}=\frac{1}{1+e}(1-\Omega_{\rm
b,0})=0.2559$ and  $\Omega_{\rm de,0}^{\rm th}=\frac{e}{1+e}(1-\Omega_{\rm
b,0})=0.6955$  in very good agreement
with the measured values $\Omega_{\rm dm,0}^{\rm obs}=0.2589$ and $\Omega_{\rm
de,0}^{\rm obs}=0.6911$ (their ratio $2.669$ is close to the pure number
$e=2.71828...$). In Sec.
\ref{sec_dime}, we
write the equations of the
problem in
dimensionless form and study the cosmic evolution of the LDF. We show that the
$\Lambda$CDM model is recovered in the limit $B\rightarrow 0$ which corresponds
to $\hbar\rightarrow 0$.  Therefore, the $\Lambda$CDM model can
be viewed as the classical limit of the logotropic model (it corresponds to a
dark fluid with a constant pressure $P=-\epsilon_\Lambda$ or to a
complex SF with a constant potential
$V=\epsilon_\Lambda$). In Sec.
\ref{sec_effdmde}, we determine and plot the effective proportion of DM and DE
in the logotropic model. In
Secs. \ref{sec_q}
and \ref{sec_cs}, we
study the evolution of the deceleration parameter and speed of sound. We show
that the speed of sound increases from $c_s=0$ to
$c_s=c$ as the Universe expands. In Sec. \ref{sec_eu}, we study the evolution of
the scale
factor. In Sec. \ref{sec_tp}, we determine the total SF potential including the
rest-mass
term and the potential term and discuss the motion of the SF in this potential
in relation to the phenomenon of spintessence. In Sec. \ref{sec_v}, we determine
the
validity of the fast oscillation regime. We first show that it imposes the
condition $m\gg m_\Lambda$ on the mass of the SF, where
$m_\Lambda=(\Lambda\hbar^2/3c^4)^{1/2}=1.20\times 10^{-33}\, {\rm
eV/c^2}$ is the cosmon mass. We then show that the SF undergoes a stiff matter
era (in the
slow oscillation regime) prior to the DM and DE eras. We also argue that the
fast oscillation regime ceases to be valid at very late times and we determine
the dynamical phase diagram of the logotropic model. In Sec. \ref{sec_diff}, we
discuss
the analogies
and the differences between the logotropic model and the  $\Lambda$CDM model. We
show that the asymptotic energy density $\epsilon_{\rm min}$ in the logotropic
model is slightly larger than the asymptotic energy density $\epsilon_{\Lambda}$
in the $\Lambda$CDM model. We also show that the logotropic model leads to DM
halos with a universal surface density consistent with the observations while
the CDM model leads to cuspy density profiles that are not observed. In Sec.
\ref{sec_lwe},
we go beyond the TF approximation and describe DM halos in terms of the
logotropic GP equation. Their equilibrium state is determined by a quantum
Lane-Emden equation of index $n=-1$. Quantum logotropic DM halos have a quantum
core (soliton), an inner logotropic envelope where the density
decreases as $\rho\sim r^{-1}$ (responsible for the constant surface density of
DM halos) and an outer Navarro-Frenk-White (NFW)
envelope where the density decreases as $r^{-3}$ (or an outer
isothermal envelope where the density decreases as $r^{-2}$). The classical
logotropic model is recovered in the TF approximation $\hbar\rightarrow 0$. 
On the other hand, the FDM model is recovered in the limit $B\rightarrow 0$. We
mention that the inner logotropic envelope solves the problems of the FDM model
reported in our previous papers \cite{clm2,chavtotal,modeldm} (see also
\cite{burkertfdm,deng}). In Sec.
\ref{sec_fw}, we study the Jeans instability of an expanding 
logotropic
Universe
by using a nonrelativistic approach. We show that the speed of sound and the
comoving Jeans length increase as the Universe expands. As a result, the density
contrast first increases like in the $\Lambda$CDM model then undergoes damped
oscillations. This is the same behavior as in the generalized Chaplygin gas
(GCG) model and the 
inverse behavior as in the FDM model. We explain that this behavior poses
problems for the formation of structures and we discuss possible solutions that
have been invoked in the context
of the GCG model. In particular, we stress the 
importance to perform a nonlinear study of structure formation. The
Appendices provide complements to our main results.
In Appendix \ref{sec_mot}, we recall the motivations for the logotropic model
and explain that it can be regarded as the simplest generalization of the
$\Lambda$CDM model. In Appendix \ref{sec_lmt}, in line with \cite{action}, we
show that
different types of logotropic models can be introduced depending on whether the
pressure is specified in terms of the energy density, the rest-mass density, or
the pseudo rest-mass density (the present model corresponds
to a logotropic of type III in the terminology of  \cite{action}). In Appendix
\ref{sec_lin}, we extend certain
results of the Jeans instability study to the case of DM with a linear
equation of state. In Appendix \ref{sec_lcdm},  we discuss
different equivalent
versions of the $\Lambda$CDM model. In Appendix \ref{sec_gfdm}, we
discuss the main
properties of the $\Lambda$FDM model. In
Appendix \ref{sec_pldm}, we describe the structure
of logotropic
DM halos. In Appendix \ref{sec_mdm}, we determine the typical mass of the DM
particle in the quantum logotropic model.

\section{Complex SF theory}
\label{sec_csf}

In this section, 
we recall the basic
equations governing the cosmological evolution of a spatially homogeneous
complex SF with an arbitrary  self-interaction potential in a
Friedmann-Lema\^itre-Robertson-Walker (FLRW) Universe. We also recall how these
equations can be simplified in the fast oscillation regime (equivalent to the
classical or TF approximation) that will be considered in the following
sections. We refer to
our previous papers \cite{abrilphas,action,cspoly} and references therein for a
more
detailed
discussion.

\subsection{Spatially homogeneous SF}
\label{sec_hsf}

Let us consider a complex  SF $\varphi$ with a
self-interaction potential $V(|\varphi|^2)$ described by the KG
equation.  For a spatially homogeneous SF $\varphi(t)$ evolving in an
expanding background, the KG equation takes the form\footnote{See, e.g.,
Refs. \cite{shapiro,abrilph,playa,abrilphas,chavmatos,action,cspoly} and Sec.
\ref{sec_lwe} for the general expression of the KG equation valid for
possibly inhomogeneous systems.}
\begin{eqnarray}
\frac{1}{c^2}\frac{d^2\varphi}{dt^2}+\frac{3H}{c^2}\frac{d\varphi}{dt}+\frac
{m^2
c^2}{\hbar^2}\varphi
+2\frac{dV}{d|\varphi|^2}\varphi=0,
\label{hsf1}
\end{eqnarray}
where $H=\dot a/a$ is the Hubble parameter and $a(t)$ is the scale factor. The
second term in Eq. (\ref{hsf1}) is the Hubble drag. The
rest-mass term (third term) can be written as $\varphi/\lambda_C^2$, where
$\lambda_C=\hbar/mc$ is the Compton wavelength ($m$
is the mass of the SF). The total potential including the rest-mass term and
the self-interaction term reads
\begin{equation}
V_{\rm tot}(|\varphi|^2)=\frac{m^2c^2}{2\hbar^2}|\varphi|^2+V(|\varphi|^2).
\label{hsf1b}
\end{equation}
The energy density
$\epsilon(t)$ and the pressure $P(t)$ of the SF are given by
\begin{equation}
\epsilon=\frac{1}{2c^2}\left |\frac{d\varphi}{d
t}\right|^2+\frac{m^2c^2}{2\hbar^2}|\varphi|^2+V(|\varphi|^2),
\label{hsf2}
\end{equation}
\begin{equation}
P=\frac{1}{2c^2}\left |\frac{d\varphi}{d
t}\right|^2-\frac{m^2c^2}{2\hbar^2}|\varphi|^2-V(|\varphi|^2).
\label{hsf3}
\end{equation}
The equation of state parameter is defined by $w=P/\epsilon$.

The Friedmann equations determining the evolution of the  homogeneous
background are
\begin{equation}
\frac{d\epsilon}{dt}+3H(\epsilon+P)=0
\label{hsf4}
\end{equation}
and 
\begin{equation}
\label{fe1}
H^2=\frac{8\pi
G}{3c^2}\epsilon-\frac{kc^2}{a^2}+\frac{\Lambda}{3},
\end{equation}
where $\Lambda$ is the cosmological constant
and $k$ determines
the curvature of space. The Universe may be closed ($k>0$), flat ($k=0$), or
open ($k<0$). In this paper, we consider a flat Universe ($k=0$) in agreement
with the inflation paradigm \cite{guthinflation} and the observations of the
cosmic
microwave background (CMB) \cite{planck2014,planck2016}. On the other
hand, we set $\Lambda=0$ because the acceleration of the expansion of
the Universe will be taken into account in the potential of the
SF (see below). The Friedmann
equation (\ref{fe1}) then reduces to the form
\begin{eqnarray}
H^2=\frac{8\pi G}{3c^2}\epsilon.
\label{hsf5}
\end{eqnarray}
The Friedmann equations can be derived from the Einstein field equations by
using the FLRW metric \cite{weinbergbook}. The  energy conservation 
equation (\ref{hsf4}) can
also be obtained from the KG equation (\ref{hsf1}) by using Eqs. (\ref{hsf2})
and (\ref{hsf3}) (see Appendix G of \cite{action}).\footnote{Inversely,
the KG equation (\ref{hsf1}) can be obtained from the energy conservation 
equation (\ref{hsf4}).} Once the SF
potential
is given, the Klein-Gordon-Friedmann (KGF) equations
provide a complete set of
equations that can in principle be solved to obtain the evolution of the
Universe assuming that the energy density is entirely due to the SF.
To complete the description one can introduce radiation and
baryonic matter as independent species but, for simplicity, we shall not
consider their effect
here.

\subsection{Charge of the SF}
\label{sec_ext}

Writing the complex SF as 
\begin{eqnarray}
\varphi=|\varphi|e^{i\theta},
\label{ext2}
\end{eqnarray}
where $|\varphi|$ is the modulus of the SF and $\theta$ its argument (angle),
inserting this decomposition into the KG equation (\ref{hsf1}), and separating
the
real and imaginary parts, we obtain the following pair of equations
\begin{eqnarray}
\frac{1}{c^2}\left (2\frac{d|\varphi|}{dt}\frac{d\theta}{dt}+|\varphi|
\frac{d^2\theta}{dt^2}\right
)+\frac{3H}{c^2}|\varphi|\frac{d\theta}{dt}=0,
\label{ext4}
\end{eqnarray}
\begin{eqnarray}
\frac{1}{c^2}\left\lbrack \frac{d^2|\varphi|}{dt^2}-|\varphi| \left
(\frac{d\theta}{dt}\right
)^2\right\rbrack+\frac{3H}{c^2}\frac{d|\varphi|}{dt}\nonumber\\
+ \frac{m^2
c^2}{\hbar^2}|\varphi|+2\frac{dV}{d|\varphi|^2}|\varphi|=0.
\label{ext3}
\end{eqnarray}
Equation (\ref{ext4}) can be rewritten as a conservation equation
\begin{eqnarray}
\frac{d}{dt}\left (a^3|\varphi|^2\frac{d\theta}{dt}\right )=0.
\label{ext6}
\end{eqnarray}
Introducing the pulsation $\omega=-\dot\theta$, we get 
\begin{eqnarray}
\omega=\frac{Q\hbar c^2}{a^3|\varphi|^2},
\label{ext7}
\end{eqnarray}
where $Q$ is a
constant of integration which represents the conserved charge of the complex
SF \cite{arbeycosmo,gh,shapiro,abrilph,abrilphas,action,cspoly} (see
Sec. \ref{sec_rmd}).\footnote{The conservation of
the charge results from the global $U(1)$ symmetry of the Lagrangian of a
complex SF. There is no such conservation law for a real SF.} Note that this
equation is exact. On the other hand, in the fast oscillation regime
$\omega=d\theta/dt\gg H=\dot a/a$ where the pulsation is high with respect to
the Hubble expansion rate, Eq.
(\ref{ext3})
reduces to
\begin{eqnarray}
\omega^2=\frac{m^2c^4}{\hbar^2}+2c^2\frac{dV}{d|\varphi|^2}.
\label{ext5}
\end{eqnarray}
For a free field ($V=0$), the pulsation $\omega$ is proportional
to the mass of the SF ($\omega=mc^2/\hbar$) and the fast oscillation
condition reduces to $mc^2/\hbar\gg H$. Combining Eqs. (\ref{ext7}) and
(\ref{ext5}), we obtain
\begin{equation}
\frac{Q^2\hbar^2c^4}{a^6|\varphi|^4}=\frac{m^2
c^4}{\hbar^2}
+2c^2\frac{dV}{d|\varphi|^2}.
\label{suna2}
\end{equation}
This equation relates the modulus $|\varphi|$  of the SF to the scale
factor $a$ in the fast oscillation regime. The pulsation $\omega$ of the SF is
then given by Eq. (\ref{ext7}) or (\ref{ext5}).

\subsection{Spintessence}
\label{sec_spin}

According to Eqs. (\ref{ext3}) and (\ref{ext7}) we have 
\begin{equation}
\frac{d^2|\varphi|}{dt^2}+3H\frac{d|\varphi|}{dt}
+ \frac{m^2
c^4}{\hbar^2}|\varphi|
+2c^2\frac{dV}{d|\varphi|^2}|\varphi|-\frac{
Q^2\hbar^2c^4}{a^6|\varphi|^3}=0,
\label{suna1}
\end{equation}
where $H$ is given by Eq. (\ref{hsf5}). This equation is exact. It determines the
evolution of the modulus of
the complex SF. It
differs from the KG equation of a real SF by the presence of the
last term and the fact that $\varphi$ is replaced by $|\varphi|$. 
The energy density and the pressure are given by
\begin{equation}
\epsilon=\frac{1}{2c^2}\left (\frac{d|\varphi|}{d
t}\right )^2+\left (\frac{\omega^2}{2c^2}+\frac{m^2c^2}{2\hbar^2}\right )|\varphi|^2+V(|\varphi|^2),
\label{hsf2b}
\end{equation}
\begin{equation}
P=\frac{1}{2c^2}\left (\frac{d|\varphi|}{d
t}\right )^2+\left (\frac{\omega^2}{2c^2}-\frac{m^2c^2}{2\hbar^2}\right )|\varphi|^2-V(|\varphi|^2).
\label{hsf3b}
\end{equation}
Eq. (\ref{suna1}) can be written as
\begin{equation}
\frac{d^2 R}{dt^2}+3H\frac{dR}{dt}=-c^2\frac{dV_{\rm
tot}}{dR}+\omega^2 R,
\label{suna1b}
\end{equation}
where $R=|\varphi|$ and $\omega=Q\hbar c^2/(a^3R^2)$. This equation is similar
to the equation of motion of a damped particle of position $R(t)$ moving in a
potential $c^2 V_{\rm tot}(R)-(1/2)\omega^2 R^2$. The last term
coming from the ``angular motion'' of the complex SF can be interpreted as a
``centrifugal force'' whose strength depends on the
charge of the complex SF
\cite{gh}. The presence
of the centrifugal force for a complex SF is a crucial difference with respect
to a real SF (that is not charged) because the fast oscillation approximation
leading to Eq. (\ref{ext5}) or (\ref{suna2}) corresponds to the
equilibrium $c^2V'_{\rm tot}(R)=\omega^2 R$  between the centrifugal force
and the force associated with the total SF potential $V_{\rm tot}$ (see Sec.
V.A. of \cite{abrilphas}). When this condition is satisfied, the phase of the
SF rotates rapidly while its modulus remains approximately constant. 
This is what Boyle {\it et al.} \cite{spintessence} call
``spintessence''. There is no relation such as  Eq. (\ref{ext5}) or
(\ref{suna2})  for a real
SF.

{\it Remark:} For a complex SF in the fast oscillation regime, only the phase
$\theta$ of the SF oscillates (spintessence). The modulus $|\varphi|$ of the SF
evolves slowly (adiabatically)
without oscillating. By contrast, for a real SF in the fast oscillation
regime, $\varphi(t)$ oscillates rapidly by taking positive and negative values.
In this connection, we note that Arbey {\it et al.} \cite{arbeycosmo} study
a complex SF
but consider a fast oscillation regime different from spintessence where the
complex SF behaves as a real SF. In the present paper, when considering the fast
oscillation regime of a complex SF, 
we shall implicitly assume that it corresponds to the spintessence regime.

\subsection{Equation of state in the fast oscillation regime}
\label{sec_eosfo}

To establish the equation of state of the SF in the fast oscillation regime, we
can proceed as follows \cite{turner,ford,pv,mul,shapiro,abrilphas}.
Multiplying the KG
equation (\ref{hsf1}) by
$\varphi^*$ and averaging over a
time interval  that is much longer than the field oscillation period
$\omega^{-1}$, but much shorter than the Hubble time $H^{-1}$,  we obtain 
\begin{eqnarray}
\frac{1}{c^2}\left\langle \left |\frac{d\varphi}{dt}\right
|^2\right\rangle=\frac{m^2c^2}{\hbar^2}\langle |\varphi|^2\rangle+2\left\langle
\frac{dV}{d|\varphi|^2}|\varphi|^2\right\rangle.
\label{ext10}
\end{eqnarray}
This relation constitutes a sort of virial theorem. On the other hand, for a
spatially
homogeneous SF, the energy density and the pressure are given
by Eqs. (\ref{hsf2}) and (\ref{hsf3}). Taking the average value of the energy
density and pressure, using Eq.
(\ref{ext10}), and making the approximation
\begin{eqnarray}
\left\langle \frac{dV}{d|\varphi|^2}|\varphi|^2\right\rangle\simeq V'(\langle
|\varphi|^2\rangle)\langle |\varphi|^2\rangle,
\label{ext11}
\end{eqnarray} 
we get
\begin{eqnarray}
\langle\epsilon\rangle=\frac{m^2c^2}{\hbar^2}\langle
|\varphi|^2\rangle+V'(\langle
|\varphi|^2\rangle)\langle |\varphi|^2\rangle+V(\langle
|\varphi|^2\rangle),\quad
\label{ext12}
\end{eqnarray} 
\begin{eqnarray}
\langle P \rangle=V'(\langle
|\varphi|^2\rangle)\langle |\varphi|^2\rangle-V(\langle |\varphi|^2\rangle).
\label{ext13}
\end{eqnarray} 
The equation of state parameter is then given by
\begin{eqnarray}
w=\frac{P}{\epsilon}=\frac{V'(\langle
|\varphi|^2\rangle)\langle |\varphi|^2\rangle-V(\langle
|\varphi|^2\rangle)}{\frac{m^2c^2}{\hbar^2}\langle |\varphi|^2\rangle+V'(\langle
|\varphi|^2\rangle)\langle |\varphi|^2\rangle+V(\langle
|\varphi|^2\rangle)}.\nonumber\\
\label{ext14}
\end{eqnarray}
We note that the averages are not strictly necessary in Eqs. (\ref{ext12})-(\ref{ext14})
since the modulus of the
SF changes slowly with time.

{\it Remark:} Eqs. (\ref{ext12}) and (\ref{ext13}) can also be obtained from
Eqs. (\ref{hsf2b}) and (\ref{hsf3b}) by using Eq. (\ref{ext5}) and neglecting
the term $(d|\varphi|/dt)^2$.

\subsection{Equation of state in the slow oscillation regime: kination and stiff matter era}
\label{sec_eosso}

In the so-called ``kination regime'' \cite{kination} where the 
kinetic term dominates
the potential term in Eqs. (\ref{hsf2}) and (\ref{hsf3}), we obtain the stiff
equation of state $P=\epsilon$ where the speed of sound
$c_s=(P'(\epsilon))^{1/2}c$ equals the speed of light. This equation of state
applies in particular to
a free massless SF ($m=V=0$) or when $H\sim \dot{|\varphi|}/|\varphi|$ is large.
The stiff matter era associated with the
kination
regime may take place in the very early
Universe before other eras associated
with the fast oscillation regime ($\omega\gg H$) occur.
The stiff matter era usually corresponds to a slow
oscillation regime ($\omega\ll H$). In that case, the
SF cannot even complete one cycle of spin
within one Hubble time so that it just rolls down the potential well, without
oscillating. Therefore, the comparison of $\omega$ and $H$ determines whether
the SF oscillates or rolls (see Sec. \ref{sec_v}). For the stiff equation of
state  $P=\epsilon$, using the
Friedmann
equations  (\ref{hsf4}) and (\ref{hsf5}), one easily gets $\epsilon\propto
a^{-6}$,
$a\propto t^{1/3}$, and $\epsilon\sim c^2/24\pi Gt^2$. One can also show 
 that $|\varphi|\sim
(3c^4/4\pi G)^{1/2}(-\ln a)$ \cite{abrilph}.

\subsection{Hydrodynamic variables and TF approximation}
\label{sec_ge}

Instead of working with the SF $\varphi(t)$, we can use hydrodynamic variables
(see our previous works  \cite{abrilph,playa,abrilphas,chavmatos,action,cspoly}
for a 
general
description valid for possibly inhomogeneous systems). We define
the pseudo rest-mass density by 
\begin{eqnarray}
\rho=\frac{m^2}{\hbar^2}|\varphi|^2.
\label{ge1}
\end{eqnarray}
We stress that it is only in the nonrelativistic limit $c\rightarrow +\infty$
that $\rho$
has the interpretation of a rest-mass density (in this limit, we also have
$\epsilon\sim \rho c^2$). In the relativistic regime,
$\rho$ does not have a clear physical interpretation but it can
always be defined as a convenient notation. We note that the total potential
(\ref{hsf1b}) can be written as
\begin{equation}
V_{\rm tot}(\rho)=\frac{1}{2}\rho c^2+V(\rho).
\label{hsf1brho}
\end{equation}
We now write the SF in the de Broglie form
\begin{eqnarray}
\varphi(t)=\frac{\hbar}{m}\sqrt{\rho(t)}e^{i
S_{\rm tot}(t)/\hbar},
\label{ge2}
\end{eqnarray}
where $\rho$ is the pseudo rest-mass density
and $S_{\rm tot}=(1/2)i\hbar\ln(\varphi^*/\varphi)$  is the total 
action of the SF. The total energy  of the SF (including its rest
mass  energy $mc^2$) is
\begin{eqnarray}
E_{\rm tot}(t)= -\frac{d S_{\rm tot}}{d t}.
\label{ge3}
\end{eqnarray}

Substituting Eq. (\ref{ge2}) into the KG equation (\ref{hsf1}) and taking the
imaginary part, we obtain the conservation equation \cite{abrilphas}
\begin{eqnarray}
\frac{d}{dt}\left (\rho E_{\rm tot} a^3\right )=0.
\label{ge3a}
\end{eqnarray}
It expresses the conservation of the
charge of the SF.\footnote{The density
of charge is proportional to $\rho E_{\rm tot}$ (see \cite{chavmatos} and footnote 4 of \cite{abrilphas}).} It can be
integrated into 
\begin{eqnarray}
\frac{E_{\rm tot}}{mc^2}=\frac{Qm}{\rho a^3},
\label{ge4}
\end{eqnarray}
where $Q$ is the charge of the SF.
These equations are equivalent to Eqs. (\ref{ext6}) and
(\ref{ext7}).\footnote{To make the
link between the SF variables and the hydrodynamical variables, we use
$|\varphi|=(\hbar/m)\sqrt{\rho}$,
$\theta=S_{\rm tot}/\hbar$ and $\omega=E_{\rm tot}/\hbar$.} 
Next, substituting Eq. (\ref{ge2}) into the KG equation
(\ref{hsf1}), taking the real part, and making the TF approximation
$\hbar\rightarrow 0$, we obtain the Hamilton-Jacobi (or Bernoulli) equation
\cite{abrilphas}
\begin{eqnarray}
E_{\rm tot}^2=m^2c^4+2m^2c^2V'(\rho).
\label{ge3b}
\end{eqnarray}
This equation is equivalent to Eq. (\ref{ext5}). It can be
rewritten as
\begin{eqnarray}
E_{\rm tot}=mc^2\sqrt{1+\frac{2}{c^2} V'(\rho)}. 
\label{ge5}
\end{eqnarray}
Note that Eq. (\ref{ge3b}) requires that
\begin{eqnarray}
1+\frac{2}{c^2} V'(\rho)>0,
\label{ge5b}
\end{eqnarray}
corresponding to $V'_{\rm tot}(\rho)>0$.
Combining Eqs. (\ref{ge4})  and (\ref{ge5}), we obtain
\begin{eqnarray}
\rho \sqrt{1+\frac{2}{c^2}V'(\rho)}=\frac{Qm}{a^3},
\label{ge6}
\end{eqnarray}
which is equivalent to Eq. (\ref{suna2}).  Finally, writing Eqs. (\ref{hsf2})
and 
(\ref{hsf3}) in terms of hydrodynamic variables, making the TF approximation
$\hbar\rightarrow 0$, and using the the Hamilton-Jacobi (or Bernoulli) equation
(\ref{ge3b}), we get
\cite{abrilphas}
\begin{eqnarray}
\epsilon=\rho c^2+V(\rho)+\rho V'(\rho), 
\label{ge7}
\end{eqnarray}
\begin{eqnarray}
P=\rho
V'(\rho)-V(\rho),
\label{ge8}
\end{eqnarray}
which are equivalent to Eqs. (\ref{ext12}) and (\ref{ext13}). Eq.
(\ref{ge8})
determine the equation of state $P(\rho)$ for a given potential
$V(\rho)$.\footnote{We can add a 
term of the form $A\rho$ in the potential without changing the pressure. This
adds a term $2A\rho$ in the energy density. If we add a constant term $C$
(cosmological constant) in the potential, this adds a term $C$ in the energy
density and a term $-C$ in the pressure.} Inversely, for a given equation of
state, the potential is given by
\begin{eqnarray}
V(\rho)=\rho\int \frac{P(\rho)}{\rho^2}\, d\rho.
\label{ge9b}
\end{eqnarray}

The 
correspondances with the results of the previous sections
show that the fast oscillation regime ($\omega\gg H$) is equivalent to the TF
or semiclassical approximation ($\hbar\rightarrow 0$). We note that we cannot
directly take $\hbar=0$ in the KG equation (this is why we have to average
over the oscillations) while we can take 
$\hbar=0$ in the hydrodynamic equations (see Refs.
\cite{abrilphas,action,cspoly} for
more
details). This is an interest of the hydrodynamic representation of the SF.
It can be shown  \cite{btv,action,cspoly} that Eqs.
(\ref{ge7}) and (\ref{ge8}) remain valid for a spatially inhomogeneous
SF in the TF approximation.\footnote{Equation (\ref{ge8}) is also valid
for a nonrelativistic SF
in the general case, i.e., for a possibly spatially
inhomogeneous SF beyond the TF approximation
\cite{chavtotal,action}.} They determine
the equation of state
$P=P(\epsilon)$ of the SF
in parametric form.  The equation of state
parameter can be written
as
\begin{eqnarray}
w=\frac{P}{\epsilon}=\frac{\rho
V'(\rho)-V(\rho)}{\rho c^2+V(\rho)+\rho V'(\rho)},
\label{ge8b}
\end{eqnarray}
which is equivalent to Eq.
(\ref{ext14}). We note that the condition from Eq. (\ref{ge5b}) implies $w>-1$
so that a complex SF in the fast oscillation regime has never a phantom
behavior. 
The pseudo squared speed of sound is
\begin{eqnarray}
c_s^2=P'(\rho)=\rho V''(\rho),
\label{ge9p}
\end{eqnarray}
while the
true  squared speed of sound is
\begin{eqnarray}
c_s^2=P'(\epsilon)=\frac{\rho V''(\rho)}{c^2+2V'(\rho)+\rho V''(\rho)}.
\label{ge9}
\end{eqnarray}

{\it Remark:} We note that Eq. (\ref{ge6}) can be obtained 
directly from the energy equation (\ref{hsf4}) with Eqs. (\ref{ge7}) and
(\ref{ge8}) \cite{abrilph}. Indeed, combining these equations we obtain
\begin{equation}
\left\lbrack c^2+2V'(\rho)+\rho V''(\rho)\right\rbrack
\frac{d\rho}{dt}=-3H \left\lbrack\rho c^2+2\rho V'(\rho)\right\rbrack,
\label{hj24}
\end{equation}
leading to
\begin{eqnarray}
\int \frac{c^2+2V'(\rho)+\rho V''(\rho)}{\rho c^2+2\rho 
V'(\rho)}\, d\rho=-3\ln a.
\label{hj25}
\end{eqnarray}
Eq. (\ref{hj25}) integrates to give Eq. (\ref{ge6}).

\subsection{Rest-mass density and internal energy}
\label{sec_rmd}

The rest-mass density $\rho_m=n m$ (proportional to the charge density) of a
spatially homogeneous SF is given by
\cite{action,cspoly}
\begin{eqnarray}
\rho_m=\rho\frac{E_{\rm
tot}}{mc^2}=\rho\frac{\hbar\omega}{mc^2}=-\rho\frac{\dot S_{\rm tot}}{mc^2}. 
\label{rmd1}
\end{eqnarray}
It is equal to $\rho_m=J_0/c$, where $J_0=-\rho \partial_0S_{\rm tot}/m$
is the time component of the current of charge. This formula is general
for a homogeneous SF, being valid beyond the TF approximation. According to
Eq.
(\ref{ge4}), we have 
\begin{eqnarray}
\rho_m=\frac{Qm}{a^3}.
\label{rmd2}
\end{eqnarray}
The rest-mass density (or the charge density) decreases as
$a^{-3}$. This  expresses the conservation
of the charge of the SF or, equivalently, the conservation of the boson number
(provided that anti-bosons are counted negatively).\footnote{Inversely,
Eq. (\ref{rmd1}) can be directly obtained from Eq.  (\ref{ge4}) by using Eq.
(\ref{rmd2}).} In the TF approximation,
using the Hamilton-Jacobi (or Bernoulli) equation (\ref{ge5}), we find that the
relation between the rest-mass density $\rho_m$ and the pseudo rest-mass density
$\rho$ is 
\begin{eqnarray}
\rho_m=\rho\sqrt{1+\frac{2}{c^2}V'(\rho)}.
\label{rmd3}
\end{eqnarray}
From the knowledge of $P(\rho)$ we can then
obtain $P=P(\rho_m)$. It can be shown \cite{action,cspoly} that Eq. (\ref{rmd3})
remains valid for an inhomogeneous SF in the TF approximation.

The energy density can be written as
\begin{eqnarray}
\epsilon=\rho_m c^2+u(\rho_m),
\label{rmd4}
\end{eqnarray}
where the first term is the rest-mass energy and the second term is the
internal energy.  The internal energy is related to the equation of state 
$P(\rho_m)$, expressed in terms of the rest-mass density, by\footnote{This
relation can be obtained by integrating the first law of thermodynamics at $T=0$
yielding
$d(\epsilon/\rho_m)=-Pd(1/\rho_m)$ \cite{epjp}. Combining the first law of
thermodynamics at $T=0$ written as $d\epsilon/d\rho_m=(P+\epsilon)/\rho_m$  with
the energy conservation equation $d\epsilon/dt+3H(\epsilon+P)=0$, we
get $d\rho_m/dt+3H\rho_m=0$, which integrates to give $\rho_m\propto
a^{-3}$ \cite{epjp}.}
\begin{eqnarray}
u(\rho_m)=\rho_m\int \frac{P(\rho_m)}{\rho_m^2}\, d\rho_m.
\label{rmd5}
\end{eqnarray}
It is argued in
\cite{epjp} that the rest-mass density $\rho_m$ represents DM and
that the internal energy $u(\rho_m)$ represents DE. This provides an
interesting interpretation of these two components. From Eqs.
(\ref{ge7}), (\ref{rmd3}) and (\ref{rmd4}), we obtain
\begin{eqnarray}
u=\rho c^2+V(\rho)+\rho V'(\rho)-\rho c^2\sqrt{1+\frac{2}{c^2}V'(\rho)}.
\label{rmd6}
\end{eqnarray}
Therefore,
the rest-mass density (DM) is determined by Eq. (\ref{rmd3}) and the
internal energy density (DE) is determined by Eq. (\ref{rmd6}). We can then
obtain $u=u(\rho_m)$. Equation (\ref{rmd6}) remains
valid for an
inhomogeneous SF in the TF approximation. 

{\it Remark:} Owing to our interpretation of DM and DE, we can
write
\begin{eqnarray}
\rho_m c^2=\frac{\Omega_{\rm m,0}\epsilon_0}{a^3}
\label{mtd11}
\end{eqnarray}
and
\begin{eqnarray}
\label{mtd2b}
\epsilon=\frac{\Omega_{\rm m,0}\epsilon_0}{a^3}+u\left (\frac{\Omega_{\rm
m,0}\epsilon_0}{c^2a^3} \right ),
\end{eqnarray}
where $\epsilon_0$ is the present energy density of the universe and
$\Omega_{\rm m,0}$ is the present proportion of DM. We can then solve
the Friedmann equation (\ref{hsf5}) with Eq. (\ref{mtd2b})
to obtain the temporal evolution of the scale factor $a(t)$. We note
that, in this interpretation, the constant $Qmc^2$ (proportional to the charge
of the SF) is equal to the
present energy density of DM $\epsilon_{{\rm m},0}=\Omega_{{\rm m},0}\epsilon_0$
[compare Eqs. (\ref{rmd2}) and (\ref{mtd11})].

\subsection{Two-fluid model}
\label{sec_twofluids}

In our model, we have a single SF (or a single dark fluid). Still, the
energy density (\ref{rmd4}) is the sum of two terms, a rest-mass density
term
$\rho_m$ which mimics DM and an internal energy term $u(\rho_m)$ which mimics
DE. It is interesting to consider a two-fluid model which leads to the
same results as the single dark fluid model, at least for what concerns the
evolution of the homogeneous background. In this two-fluid model, one fluid
corresponds to pressureless DM with an equation of state $P_{\rm m}=0$ and a
density $\rho_m c^2=\Omega_{\rm m,0}\epsilon_0/a^3$ determined by the energy
conservation equation for DM, and the other fluid corresponds to DE with an
equation of state $P_{\rm de}(\epsilon_{\rm de})$ and an energy density
$\epsilon_{\rm de}(a)$ determined by the energy
conservation equation for DE. We can obtain the equation of state of DE yielding
the same results as the one-fluid model by taking
\begin{eqnarray}
P_{\rm de}=P(\rho_m),\qquad \epsilon_{\rm de}=u(\rho_m).
\end{eqnarray}
In other words, the  equation of state $P_{\rm de}(\epsilon_{\rm de})$ of DE
 in the two-fluid model corresponds to the
relation $P(u)$ in the single fluid model. Explicit examples of the
correspondance between the one-fluid model and the two-fluid model are given in
\cite{action,cspoly} and
in Sec. \ref{sec_ie}. We note that
although
the one and two-fluid models are equivalent for the evolution  of the
homogeneous background, they may differ for what concerns the formation of the
large-scale structures of the Universe and for inhomogeneous systems in
general.

\section{Logarithmic potential and logotropic equation of state}
\label{sec_log}

The previous equations are general. We now apply them to a specific model of
Universe called the logotropic model. We assume that DM and DE are the
manifestation of a single substance and that
this substance can be described by a complex  SF (or an exotic dark fluid)
governed by a KG equation with a logarithmic potential of the form
\begin{equation}
V(|\varphi|^2)=-A\ln \left
(\frac{m^2|\varphi|^2}{\hbar^2\rho_P}\right )-A.
\label{log0}
\end{equation}
Using the hydrodynamic
variables introduced previously, the SF potential can be written as
\begin{eqnarray}
V(\rho)=-A\ln\left (\frac{\rho}{\rho_P}\right )-A,
\label{log1}
\end{eqnarray}
where $A/c^2$ and $\rho_P$ are two positive constants
with the dimensions of a mass density. We will give the physical meaning and the value of
these constants in Sec.
\ref{sec_valf}. In the fast oscillation regime, using
Eq. (\ref{ge8}), we find that the pressure is given
by\footnote{Inversely, we
could
start from the equation of state  (\ref{log2}) and integrate Eq. (\ref{ge9b}) to
obtain the
potential $V(\rho)$.}  
\begin{eqnarray}
P=A\ln\left (\frac{\rho}{\rho_P}\right ).
\label{log2}
\end{eqnarray}
This equation is similar to the logotropic equation of state [see Eq.
(\ref{intro1})] introduced in our
previous papers \cite{epjp,lettre,jcap,pdu}. However, as we shall see, the
present model is
substantially different from the model of Refs. \cite{epjp,lettre,jcap,pdu}. In
particular, we stress
that $\rho$ represents here the pseudo rest mass density defined by Eq.
(\ref{ge1}), not
the true rest mass density $\rho_m=nm$ used in Refs.
\cite{epjp,lettre,jcap,pdu}.
It is only in the nonrelativistic limit that they coincide. The relation between the different logotropic models is
discussed in Appendix \ref{sec_lmt} (see also \cite{action}).

For the logarithmic potential (\ref{log1}) the equations of the
problem valid in the fast oscillation regime [see Eqs. (\ref{ge5})-(\ref{ge7})]
are
\begin{eqnarray}
\rho \sqrt{1-\frac{2A}{\rho c^2}}=\frac{Qm}{a^3},
\label{log3}
\end{eqnarray}
\begin{eqnarray}
\epsilon=\rho c^2-A\ln\left (\frac{\rho}{\rho_P}\right )-2A,
\label{log4}
\end{eqnarray} 
\begin{eqnarray}
\frac{E_{\rm tot}}{mc^2}=\sqrt{1-\frac{2A}{\rho c^2}}. 
\label{log5}
\end{eqnarray}
They depend on three parameters $A$, $Qm$ and $\rho_P$. The
first equation can be solved explicitly giving
\begin{eqnarray}
\rho c^2=A+\sqrt{A^2+\frac{(Qmc^2)^2}{a^6}}.
\label{log3b}
\end{eqnarray}
On the other hand, eliminating $\rho$
between Eqs. (\ref{log2}) and (\ref{log4}), we find that the equation of
state $P(\epsilon)$ is given under the inverse form $\epsilon(P)$ by
\begin{eqnarray}
\epsilon=\rho_P c^2 e^{P/A}-P-2A.
\label{log6}
\end{eqnarray} 
Finally, the equation of state parameter is given by 
\begin{equation}
w=\frac{P}{\epsilon}=\frac{A\ln\left (\frac{\rho}{\rho_P}\right )}{\rho c^2-A
\ln\left
(\frac{\rho}{\rho_P}\right )-2A}.
\label{wd2}
\end{equation} 
We note from Eq. (\ref{log2}) that $P>0$ when $\rho>\rho_P$ and  $P<0$ when
$\rho<\rho_P$. The pressure vanishes ($P=w=0$) when $\rho=\rho_P$. We will see
that the logotropic model is valid for $\rho\ll\rho_P$. Therefore, in practice,
the pressure of the LDF is always negative.

\section{Rest-mass density and internal energy}
\label{sec_ie}

According to Eqs. (\ref{rmd3}) and (\ref{log1}), the rest-mass density $\rho_m$
of the LDF is related to its pseudo rest-mass density $\rho$ by
\begin{equation}
\rho_m=\rho\sqrt{1-\frac{2A}{\rho c^2}}.
\label{ie1}
\end{equation} 
This equation can be inverted to give
\begin{equation}
\rho=\frac{A}{c^2}+\sqrt{\frac{A^2}{c^4}+\rho_m^2},
\label{ie2}
\end{equation} 
where $\rho_m$ is given by Eq. (\ref{rmd2}). Using Eqs. (\ref{rmd6}),
(\ref{log1}) and (\ref{ie2}), we find that the internal energy is given by
\begin{eqnarray}
u=&-&A+\sqrt{A^2+\rho_m^2c^4}-\rho_m c^2\nonumber\\
&-&A\ln\left\lbrack
\frac{A}{\rho_Pc^2}+\sqrt{\frac{A^2}{\rho_P^2c^4}+\frac{\rho_m^2}{\rho_P^2}}
\right\rbrack.
\label{ie3}
\end{eqnarray}
Finally, according to Eqs. (\ref{log2}) and (\ref{ie2}), we obtain the equation
of state of the SF in terms of the rest-mass density as
\begin{eqnarray}
P=A\ln\left\lbrack
\frac{A}{\rho_Pc^2}+\sqrt{\frac{A^2}{\rho_P^2c^4}+\frac{\rho_m^2}{\rho_P^2}}
\right\rbrack.
\label{ie4}
\end{eqnarray} 
As we have recalled in Sec. \ref{sec_rmd}, the rest mass density $\rho_m$ of the
SF
represents DM and the internal energy $u$ of the SF represents DE \cite{epjp}.

{\it Remark:} In the two-fluid model associated with the logotropic model (see
Sec. \ref{sec_twofluids}), the DE
has an equation of state $P_{\rm de}(\epsilon_{\rm
de})$ which is obtained by eliminating $\rho_m$ between Eqs. (\ref{ie3}) and
(\ref{ie4}), and by identifying $P(u)$ with $P_{\rm de}(\epsilon_{\rm
de})$. It can be written in inverse form as\footnote{This relation can be
obtained simply by solving Eq. (\ref{ie4}) to get $\rho_m(P)$
and by using Eqs. (\ref{rmd4}) and (\ref{log6}).}
\begin{eqnarray}
\epsilon_{\rm de}=\rho_P c^2 e^{P_{\rm de}/A}-P_{\rm de}-2A\nonumber\\
-\rho_P
c^2 \sqrt{e^{2P_{\rm de}/A}-\frac{2A}{\rho_P c^2}e^{P_{\rm de}/A}}.
\label{log6ju}
\end{eqnarray}

\section{The evolution of the parameters with the scale factor}
\label{sec_evo}

In our model, strictly speaking, there is no DM and no DE. There is just a
single SF (or a
single DF). This is an example of unified models of
DM and DE that are refered to as unified dark matter (UDM) models or
``quartessence''
models \cite{makler}. The
logotropic model is therefore 
fundamentally different from the $\Lambda$CDM model in which DM and DE are
interpreted  as two distinct entities (see
Appendices \ref{sec_dml} and \ref{sec_dmde}).\footnote{Actually, the
$\Lambda$CDM model can also be
regarded as a UDM model as discussed in Appendix \ref{sec_df}.} Nevertheless,
since the
$\Lambda$CDM model works remarkably well
in describing the large scale structure of the Universe, it is important to make
a connection between the logotropic model and the  $\Lambda$CDM model. This
connection will allow us to determine in which limit the $\Lambda$CDM model
is valid from the viewpoint of our more general model and to obtain the
parameters  of the LDF  by using the values of the parameters that have
been obtained  from cosmological observations interpreted in the framework of
the $\Lambda$CDM model.

\subsection{Early Universe: DM-like regime}
\label{sec_evoe}

In the early Universe ($a\rightarrow 0$), the general equations of Sec.
\ref{sec_log} reduce to
\begin{eqnarray}
\rho \sim \frac{Qm}{a^3},
\label{evo1}
\end{eqnarray}
\begin{eqnarray}
\epsilon\sim \rho c^2\sim \frac{Qmc^2}{a^3},
\label{evo2}
\end{eqnarray} 
\begin{eqnarray}
\frac{E_{\rm tot}}{mc^2}\rightarrow 1,
\label{evo5}
\end{eqnarray}
\begin{eqnarray}
P\sim A\ln \left (\frac{Qm}{\rho_P a^3}\right ),
\label{evo3}
\end{eqnarray} 
\begin{eqnarray}
P\sim A\ln \left (\frac{\epsilon}{\rho_Pc^2}\right ),
\label{evo4}
\end{eqnarray}
\begin{equation}
w\sim \frac{A}{\rho c^2} \ln\left (\frac{\rho}{\rho_P}\right ),
\label{wd3}
\end{equation} 
\begin{equation}
w\sim \frac{Aa^3}{Qm c^2} \ln\left (\frac{Qm}{\rho_Pa^3}\right ),
\label{wd4}
\end{equation} 
\begin{equation}
w\sim \frac{A}{\epsilon} \ln\left (\frac{\epsilon}{\rho_Pc^2}\right ).
\label{wd5}
\end{equation}
Since $\epsilon\propto a^{-3}$ and $w\simeq 0$ we see that the LDF behaves at
early times similarly to DM. If we impose that the LDF matches the $\Lambda$CDM
model for
$a\ll 1$ (see
Appendix \ref{sec_dmde}) we obtain 
\begin{eqnarray}
Qmc^2=\Omega_{\rm m,0}\epsilon_0.
\label{evo7}
\end{eqnarray}
Therefore, the quantity $Qmc^2$ which is proportional to the charge
of the SF corresponds to the present
energy density of DM ($\epsilon_{\rm m,0}=\Omega_{\rm
m,0}\epsilon_0$)  in the
$\Lambda$CDM model.\footnote{To simplify the presentation, 
we ignore the presence of baryons and take $\epsilon_{m}\simeq \epsilon_{\rm
dm}$.} Using the values of Appendix \ref{sec_nvnew}, we get
\begin{eqnarray}
Qm=2.66\times 10^{-24}\, {\rm g}\, {\rm m}^{-3}.
\label{evo8}
\end{eqnarray}

{\it Remark:} in the DM-like era, the energy and the pulsation of the SF are
given by  $E_{\rm tot}\sim mc^2$ and $\omega\sim mc^2/\hbar$ like for a free SF.
They are constant. For a boson mass $m\sim 10^{-22}\, {\rm eV/c^2}$ (see
Appendix \ref{sec_fdmw}), we 
get $\omega\sim 10^{-7}\, {\rm s}^{-1}$. On the other hand, in
the DM-like era, the
pseudo rest-mass density $\rho$ coincides with the true rest-mass density 
$\rho_m$ (see Sec. \ref{sec_ie}).

\subsection{Late Universe: DE-like  regime}
\label{sec_evol}

In the late Universe $a\rightarrow +\infty$, the general equations of
Sec.
\ref{sec_log} reduce to
\begin{eqnarray}
\rho \rightarrow  \rho_{\rm min}=\frac{2A}{c^2},
\label{evo9}
\end{eqnarray}
\begin{eqnarray}
\epsilon\rightarrow \epsilon_{\rm min}=A\ln\left (\frac{\rho_Pc^2}{2A}\right ),
\label{evo10}
\end{eqnarray} 
\begin{eqnarray}
\frac{E_{\rm tot}}{mc^2}\rightarrow 0,
\label{evo11}
\end{eqnarray}
\begin{eqnarray}
P\rightarrow P_{\rm min}=-\epsilon_{\rm min},
\label{evo12}
\end{eqnarray}
\begin{equation}
w\rightarrow w_{\rm min}=\frac{P_{\rm min}}{\epsilon_{\rm min}}=-1.
\label{wd6}
\end{equation} 
Since the energy density $\epsilon$ tends to a constant $\epsilon_{\rm min}$
and since the equation of state parameter
$w\rightarrow -1$ we see that
the LDF behaves at late times similarly to DE. We shall come back to the value
of $\epsilon_{\rm min}$ in Sec. \ref{sec_diff}. We can check that the equation
of
state
parameter $w$ is always strictly larger than $-1$ so the
Universe does not
become phantom.\footnote{This is a general result for a complex
SF. It is shown
after Eq. (\ref{ge8b}) that a complex SF in the fast oscillation regime can
never have a
phantom behavior whatever the form of the self-interaction potential. This is
because we have considered a SF with a Lagrangian
$L=\frac{1}{2c^2}|\dot\varphi|^2-V_{\rm tot}(|\varphi|^2)$ involving  a
{\it positive}
kinetic term. A complex SF has either a normal behavior (if it has a positive
kinetic term)
or a phantom behavior (if it has a negative kinetic term)  but it cannot pass
from a normal
to a phantom regime.} It
asymptotically tends to a de Sitter-like solution. This is an
important difference with our previous logotropic
model \cite{epjp,lettre,jcap,pdu}, based on a different equation of state [see
Eq. (\ref{intro1})], which
displays a phantom
behavior at late times (in that case, the scale factor has a super-de Sitter
behavior).

{\it Remark:} the asymptotic value $\rho_{\rm min}$ of the pseudo-rest mass
density corresponds to the case where the rest mass term $m^2c^2/\hbar^2$ in the
KG equation (\ref{hsf1}) is compensated by the self-interaction term $2
dV/d|\varphi|^2$.
In that
case, $\omega\simeq 0$ according to Eq. (\ref{ext5}) and  the fast oscillation
regime ceases to be valid (see Sec. \ref{sec_v}).

\subsection{Intermediate regime: stiff matter}

Considering the subleading terms in Eqs. (\ref{log2}), (\ref{log4})
and (\ref{log3b}) for large
values of $a$, one obtains the following expressions for the pseudo rest-mass
density, energy density and pressure:
\begin{equation}
\rho=\rho_{\rm min}+\frac{(Qmc^2)^2}{2Aa^6}+...
\label{ir1}
\end{equation} 
\begin{equation}
\epsilon=\epsilon_{\rm min}+\frac{(Qmc^2)^2}{4Aa^6}+...
\label{ir2}
\end{equation} 
\begin{equation}
P=P_{\rm min}+\frac{(Qmc^2)^2}{4Aa^6}+...
\label{ir3}
\end{equation} 
These equations describe the mixture of a cosmological constant
$\epsilon_{\rm min}$ (first terms in Eqs. (\ref{ir2}) and (\ref{ir3})) with a
form of ``stiff'' matter described
by the equation of state $P=\epsilon$ (second terms in Eqs. (\ref{ir2}) and
(\ref{ir3})) in which the speed of sound is equal to
the speed of light.\footnote{This stiff matter era is
completely different from the one considered in Sec. \ref{sec_eosso}.}
Therefore,
the logotropic model interpolates between different
phases of the Universe. Initially, the Universe behaves as if it were dominated
by a pressureless (dust) fluid. Ultimately, the density becomes asymptotically
constant implying a de Sitter evolution. There is
also an intermediate phase which can be described by a cosmological constant
mixed with a stiff matter fluid. The interesting point is that such an
evolution is accounted for by a single fluid. This is similar to the Chaplygin
gas model \cite{kmp}.

\section{The value of the fundamental constant of our model}
\label{sec_valf}

\subsection{An important identity obtained in the present Universe}
\label{sec_impid}

Applying the general equations (\ref{log4}) and (\ref{log3b}) at the
present
time ($a=1$) we get
\begin{eqnarray}
\epsilon_0=\rho_0 c^2-A\ln\left (\frac{\rho_0}{\rho_P}\right )-2A
\label{impid2}
\end{eqnarray} 
and
\begin{eqnarray}
\rho_0c^2=A+\sqrt{A^2+(Qmc^2)^2}.
\label{impid3}
\end{eqnarray}
Substituting Eq. (\ref{impid3}) into Eq. (\ref{impid2}) we obtain
\begin{equation}
\epsilon_0=\sqrt{A^2+(Qmc^2)^2}-A\ln\left \lbrack
\frac{A+\sqrt{A^2+(Qmc^2)^2}}{\rho_Pc^2}\right \rbrack-A.
\label{impid4}
\end{equation} 
Using Eq.
(\ref{evo7}), this relation can be rewritten as
\begin{eqnarray}
\epsilon_0=-A+\sqrt{A^2+(\Omega_{\rm m,0}\epsilon_0)^2}\nonumber\\
-A\ln\left
\lbrack \frac{A+\sqrt{A^2+(\Omega_{\rm
m,0}\epsilon_0)^2}}{\rho_Pc^2}\right\rbrack.
\label{impid4b}
\end{eqnarray} 
Assuming that  $A$ and $\rho_P$ are universal constants, this equation gives a
relation between $\Omega_{\rm m,0}$ and $\epsilon_0$. Inversely, we can use
Eq. (\ref{impid4b}) and the measured values of $\epsilon_0$ and $\Omega_{\rm
m,0}$ to determine the constants of our model.

As in our previous papers \cite{epjp,lettre,jcap,pdu}, it is convenient
to write
\begin{eqnarray}
A=B\epsilon_\Lambda,
\label{impid5}
\end{eqnarray}
where $B$ is a dimensionless constant and 
\begin{eqnarray}
\epsilon_\Lambda=\rho_{\Lambda}c^2=(1-\Omega_{\rm
m,0})\epsilon_0
\label{impid6}
\end{eqnarray}
is the present density of DE. Numerically,
\begin{eqnarray}
\rho_{\Lambda}=5.96\times 10^{-24} {\rm g}\, {\rm m}^{-3}.
\label{impid6b}
\end{eqnarray}
In the $\Lambda$CDM model (see Appendix
\ref{sec_lcdm}), $\rho_{\Lambda}$ represents the cosmological density
which is
related to the Einstein cosmological constant $\Lambda$ by
\begin{eqnarray}
\label{defrl}
\rho_{\Lambda}=\frac{\Lambda}{8\pi G}
\end{eqnarray} 
with $\Lambda=1.00\times 10^{-35}\, {\rm s}^{-2}$. For given $\rho_{\Lambda}$,
Eq. (\ref{impid5})
is just a change of notation.
In the following, we shall work with $B$ instead of $A$. In that case, Eqs.
(\ref{impid2}) and (\ref{impid3}) can be rewritten as 
\begin{eqnarray}
\frac{1}{1-\Omega_{\rm
m,0}}=\frac{\rho_0}{\rho_{\Lambda}}-B\ln\left(\frac{\rho_0}{\rho_{\Lambda}}\frac
{\rho_{\Lambda}}{\rho_P}\right )-2B
\label{impid7}
\end{eqnarray}
and
\begin{eqnarray}
\frac{\rho_0}{\rho_{\Lambda}}=B+\sqrt{B^2+\left
(\frac{Qm}{\rho_{\Lambda}}\right )^2}.
\label{impid8}
\end{eqnarray}
Using Eqs. (\ref{evo7}) and (\ref{impid6}),  we also have
\begin{eqnarray}
\frac{\rho_0}{\rho_{\Lambda}}=B+\sqrt{B^2+\left
(\frac{\Omega_{\rm
m,0}}{1-\Omega_{\rm
m,0}}\right )^2}.
\label{impid9}
\end{eqnarray}
Substituting Eq. (\ref{impid9}) into Eq. (\ref{impid7}) we obtain the exact
identity
\begin{eqnarray}
\frac{1}{1-\Omega_{\rm
m,0}}=-B+\sqrt{B^2+\left
(\frac{\Omega_{\rm
m,0}}{1-\Omega_{\rm
m,0}}\right )^2}\nonumber\\
-B\ln\left\lbrack B+\sqrt{B^2+\left
(\frac{\Omega_{\rm
m,0}}{1-\Omega_{\rm
m,0}}\right )^2}\right \rbrack
+B\ln\left(\frac{\rho_P}{\rho_{\Lambda}}\right ),
\label{impid10}
\end{eqnarray}
which is equivalent to Eq. (\ref{impid4b}). We can also rewrite Eq. (\ref{evo7})
as
\begin{eqnarray}
Qmc^2=\frac{\Omega_{\rm m,0}}{1-\Omega_{\rm m,0}}\rho_{\Lambda}c^2.
\label{evo7b}
\end{eqnarray}

\subsection{The value of $B$}
\label{sec_b}

Eq. (\ref{impid10}) determines the relation between $B$ and $\rho_P$ from the
measured
values of $\Omega_{m,0}$ and
$\rho_{\Lambda}=(1-\Omega_{\rm
m,0})\epsilon_0/c^2$. We will find that $B\ll
1$ so we can make the approximation
\begin{eqnarray}
B=\frac{1}{\ln\left(\frac{\rho_P}{\rho_{\Lambda}}\right )-1-\ln\left
(\frac{\Omega_{\rm
m,0}}{1-\Omega_{\rm
m,0}}\right )}.
\label{b1}
\end{eqnarray}
We will also find that $1+\ln\left \lbrack {\Omega_{\rm m,0}}/({1-\Omega_{\rm
m,0}})\right \rbrack$ is much smaller
than $\ln\left({\rho_P}/{\rho_{\Lambda}}\right )$ so we can
make the additional approximation
\begin{eqnarray}
B=\frac{1}{\ln\left(\frac{\rho_P}{\rho_{\Lambda}}\right )}.
\label{b2}
\end{eqnarray}
This is the same result as in our previous papers \cite{epjp,lettre,jcap,pdu}.
Eq. (\ref{b2}) can
be rewritten as 
\begin{eqnarray}
\frac{\rho_P}{\rho_\Lambda}=e^{1/B}.
\label{b3}
\end{eqnarray} 
Now the crucial remark is to observe that Eq. (\ref{b3}) is
analogous to the
fundamental identity
\begin{eqnarray}
\frac{\rho_P}{\rho_\Lambda}=10^{123}
\label{b4}
\end{eqnarray} 
expressing the fact  that the Planck density 
\begin{eqnarray}
\rho_P=\frac{c^5}{\hbar G^2}=5.16\times 10^{99}\, {\rm g/m^3}
\label{a1}
\end{eqnarray}
and the cosmological density
\begin{eqnarray}
\rho_{\Lambda}=\frac{\Lambda}{8\pi G}=5.96\times 10^{-24} {\rm g}\, {\rm
m}^{-3}
\end{eqnarray} 
differ by $123$ orders of magnitude. Following our previous
works \cite{epjp,lettre,jcap,pdu}, this analogy prompts us to identify $\rho_P$
with the
Planck density.\footnote{Actually, the density $\rho_*$ that appears in the 
logotropic equation of state $P=A\ln(\rho/\rho_*)$ \cite{epjp} could be smaller
than the Planck density $\rho_P$, being equal for example to the characteristic
scale $\rho_{\rm GUT}\sim 10^{-3}\rho_P$ of a generic grand unified theory
(GUT). However, for definiteness, we shall take $\rho_*=\rho_P$.} In that
case,
$B$ is fully determined by Eq. (\ref{b2}). Its numerical value is
\begin{eqnarray}
B=3.53\times 10^{-3}.
\label{b5}
\end{eqnarray}
We note that $B\simeq 1/[123\, \ln(10)]$, so that  $B$ is essentially the
inverse of the famous number $123$ (up to a conversion factor from neperian to
decimal logarithm). We note that $B$ has a small but {\it nonzero} value. This
is because $B$
depends on the Planck constant $\hbar$ through the Planck density  $\rho_P$
in Eq. (\ref{b2}) and because $\hbar$ has a small but
nonzero value. In the semiclassical limit
$\hbar\rightarrow 0$, we find that $\rho_P\rightarrow +\infty$ and $B\rightarrow
0$. In that case, we recover the $\Lambda$CDM model (see Sec. \ref{sec_rec}).
The fact that
$B$ is nonzero means that quantum effects ($\hbar\neq 0$) play
a fundamental role in the logotropic model. Indeed, $\rho_P$ explicitly appears
in 
the logarithmic potential from Eq. (\ref{log1}). Since the effects of
$B$ manifest themselves in the late Universe (see below), this
implies -- surprisingly -- that quantum mechanics affects the
late acceleration of the Universe. As we shall see in Sec. \ref{sec_diff},
quantum mechanics provides  (in the framework of our model)  a small correction
to the Einstein cosmological constant.

\subsection{The value of $A$}
\label{sec_a}

The logarithmic potential from Eq. (\ref{log1}) involves two
constants $A$ and $\rho_P$. We have seen that $\rho_P$ is the Planck
density. On the other hand, in line with our previous works
\cite{epjp,lettre,jcap,pdu}, we interpret the logotropic constant $A$ as
a new
fundamental constant of physics which  supersedes (in the
framework of our model) the Einstein cosmological
constant $\Lambda$ or the Einstein cosmological density
$\rho_{\Lambda}=\Lambda/8\pi G$. Indeed, the logotopic
constant $A$ is responsible for the late acceleration of the
Universe. According to Eqs. (\ref{impid5}) and (\ref{b2}) we have
\begin{eqnarray}
A=\frac{\rho_{\Lambda}c^2}{\ln\left(\frac{\rho_P}{\rho_{\Lambda}}\right )}.
\label{a2b}
\end{eqnarray}
Its numerical value is
\begin{eqnarray}
A/c^2=2.10\times 10^{-26}\, {\rm g}\, {\rm m}^{-3}.
\label{a2}
\end{eqnarray}
We note that $A/c^2$ is equal to the Einstein cosmological
density $\rho_{\Lambda}$ divided by $123$ (up to a logarithmic conversion
factor). More precisely, the logotropic constant $A$ is related to the  Einstein
cosmological constant $\Lambda$ by
\begin{eqnarray}
A=B\frac{\Lambda c^2}{8\pi G}
\label{a2bh}
\end{eqnarray}
with $B=1/\ln(\rho_P/\rho_\Lambda)=3.53\times 10^{-3}$. We stress, however,
that, in the logotropic model, the DE density is not constant (see
Sec. \ref{sec_ie}).

{\it Remark:} Using Eq. (\ref{a2b}), the logotropic equation of state
(\ref{log2}) can be rewritten as
\begin{eqnarray}
P=-\frac{\rho_{\Lambda}c^2}{\ln\left(\frac{\rho_P}{\rho_{\Lambda}}\right
)}\ln\left (\frac{\rho_P}{\rho}\right).
\label{a2bc}
\end{eqnarray}
We note that $P=-\rho_{\Lambda}c^2$ at
$\rho=\rho_{\Lambda}$, i.e., when the pseudo rest-mass density
is equal to the present DE density.

\subsection{Validity of our approximations and a curious result}
\label{sec_c}

We can now check the validity of our approximations. Since $B=3.53\times
10^{-3}\ll 1$, the
approximation leading from Eq. (\ref{impid10}) to Eq. (\ref{b1}) is valid. We
also observe that $1+\ln\left \lbrack
{\Omega_{\rm m,0}}/({1-\Omega_{\rm
m,0}})\right \rbrack=0.195$ is much smaller
than $\ln\left({\rho_P}/{\rho_{\Lambda}}\right )=283$ so we can make
the additional approximation leading from Eq. (\ref{b1}) to Eq. (\ref{b2}).

As an interesting (and intriguing) remark, we note the following. If we
assume that $B$ is given {\it exactly} by Eq. (\ref{b2}), then, according to
Eq. (\ref{b1}), we get
\begin{eqnarray}
1+\ln\left (\frac{\Omega_{\rm m,0}}{1-\Omega_{\rm
m,0}}\right )=0.
\label{b13}
\end{eqnarray}
This equation determines the value of $\Omega_{\rm m,0}$ which, in
the $\Lambda$CDM model, represents the present proportion of DM.\footnote{In
the framework of our model where there is no DM and no DE (just a single DF),
$\Omega_{\rm m,0}$ represents the coefficient that appears in the asymptotic behavior $\epsilon/\epsilon_0\sim
\Omega_{\rm m,0}/a^3$ of the energy density  when $a\ll 1$ [see Eq. (\ref{evo2})
with Eq. (\ref{evo7})]. This coefficient, which is related to
the charge $Qmc^2$ of the SF, is expected to be universal.}
We get 
\begin{eqnarray}
\Omega_{\rm m,0}^{\rm th}=\frac{1}{1+e}=0.269.
\label{b13w}
\end{eqnarray}
Remarkably, this value is reasonably close to the measured
value $\Omega_{\rm m,0}=0.3089$. This result was
previously obtained in \cite{pdu} in the framework of the original logotropic
model.

{\it Remark:} For the simplicity of the presentation, we have ignored the
presence of baryonic matter. If we take into account the presence 
of baryons (with a proportion $\Omega_{\rm b,0}$)
and redo the preceding analysis, we obtain the proportion of DM and DE:
\begin{eqnarray}
\Omega_{\rm dm,0}^{\rm
th}=\frac{1}{1+e}(1-\Omega_{\rm b,0}),
\label{f11a}
\end{eqnarray}
\begin{eqnarray}
\Omega_{\rm de,0}^{\rm th}=\frac{e}{1+e}(1-\Omega_{\rm b,0}).
\label{f11b}
\end{eqnarray}
If we neglect baryonic matter $\Omega_{\rm b,0}=0$ we obtain the pure numbers
$\Omega_{\rm de,0}^{\rm th}=\frac{e}{1+e}=0.731059...$ and
$\Omega_{\rm dm,0}^{\rm
th}=\frac{1}{1+e}=0.268941...$ which give the correct
proportions $70\%$ and $25\%$ of DE and DM \cite{pdu}. If we take 
baryonic matter into account and use the measured value of
$\Omega_{\rm b,0}=0.0486\pm 0.0010$, we get
$\Omega_{\rm de,0}^{\rm th}=0.6955\pm 0.0007$ and $\Omega_{\rm dm,0}^{\rm
th}=0.2559\pm
0.0003$ which are very close to the
observed values $\Omega_{\rm de,0}=0.6911\pm 0.0062$  and
$\Omega_{\rm dm,0}=0.2589\pm
0.0057$ within the error bars. We note that the ratio $\Omega_{\rm de,0}^{\rm
th}/\Omega_{\rm dm,0}^{\rm
th}=e=2.71828...$ is independent of $\Omega_{\rm b,0}$ and close to $\Omega_{\rm
de,0}/\Omega_{\rm dm,0}=2.66937\pm 0.08$. Finally, combining the foregoing
formulae, we find that the charge $Qmc^2=\Omega_{\rm dm,0}\epsilon_0$ of the SF
[see Eq. (\ref{evo7})] can be written as
$Qmc^2=\rho_{\Lambda}/e$. The postulate from Eq.
(\ref{b13}) means that the fundamental constant $A$ is equal to 
$\rho_{\Lambda}c^2/\ln ({\rho_P}/{\rho_{\Lambda}})$ where $\rho_{\Lambda}$ is
the  {\it
present}
DE density. This can be viewed as
a strong cosmic coincidence \cite{pdu} giving to our epoch a central
place in the history of the universe. The same results are obtained with
the original logotropic model. These important results will be developed in a
specific paper \cite{oufsuite}.

\subsection{Validity of the nonrelativistic regime}
\label{sec_vnr}

According to the results of Sec. \ref{sec_evo}, the nonrelativistic regime is
valid provided that\footnote{More generally, the nonrelativistic regime is
valid when
the rest-mass energy density $\rho_m c^2$ (DM) is much larger than the internal
energy $u$ (DE).}
\begin{eqnarray}
\rho c^2 \gg \epsilon_{\Lambda},\qquad \frac{Qmc^2}{a^3}\gg \epsilon_{\Lambda}.
\label{vnr1}
\end{eqnarray} 
Using Eqs. (\ref{evo7}) and (\ref{impid6}), these conditions can be rewritten as
\begin{eqnarray}
\rho \gg \rho_{\Lambda},\qquad a\ll a_t=\left (\frac{\Omega_{m,0}}{1-\Omega_{m,0}}\right )^{1/3},
\label{vnr2}
\end{eqnarray} 
like for the $\Lambda$CDM model (see Appendix \ref{sec_lcdm}). The scale factor
$a_t=0.765$ determines the transition between the DM and DE eras. 
In the nonrelativistic regime, we have $\epsilon\sim\rho c^2$, $E_{\rm tot}\sim
mc^2$, $\rho\sim
\Omega_{m,0}(\epsilon_0/c^2)/a^3$ and $w\ll 1$. The SF behaves at large
(cosmological) scales as pressureless DM. Note, however, that the logotropic
pressure manifests itself at small (galactic) scales even in the
nonrelativistic regime and can solve the problems
of the CDM model such as the core-cusp problem and the missing 
satellite problem (see Sec. \ref{sec_diff}).

\section{Dimensionless equations}
\label{sec_dime}

\subsection{General equations}
\label{sec_dg}

\begin{figure}[!h]
\begin{center}
\includegraphics[clip,scale=0.3]{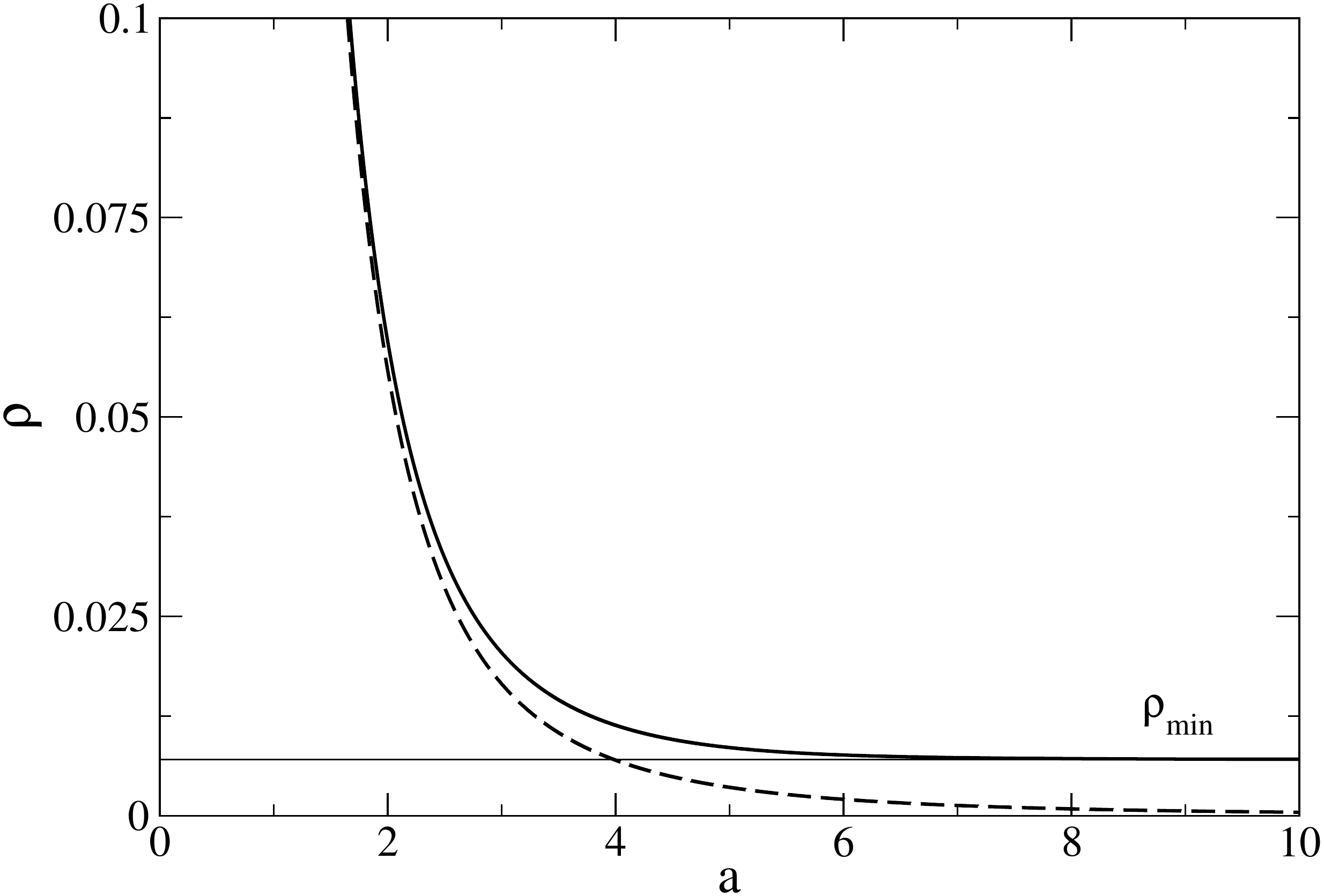}
\caption{Pseudo rest-mass density as a function of the
scale factor. Here and in the following figures, the dashed line corresponds to
the $\Lambda$CDM model ($B=0$).}
\label{arho}
\end{center}
\end{figure}

\begin{figure}[!h]
\begin{center}
\includegraphics[clip,scale=0.3]{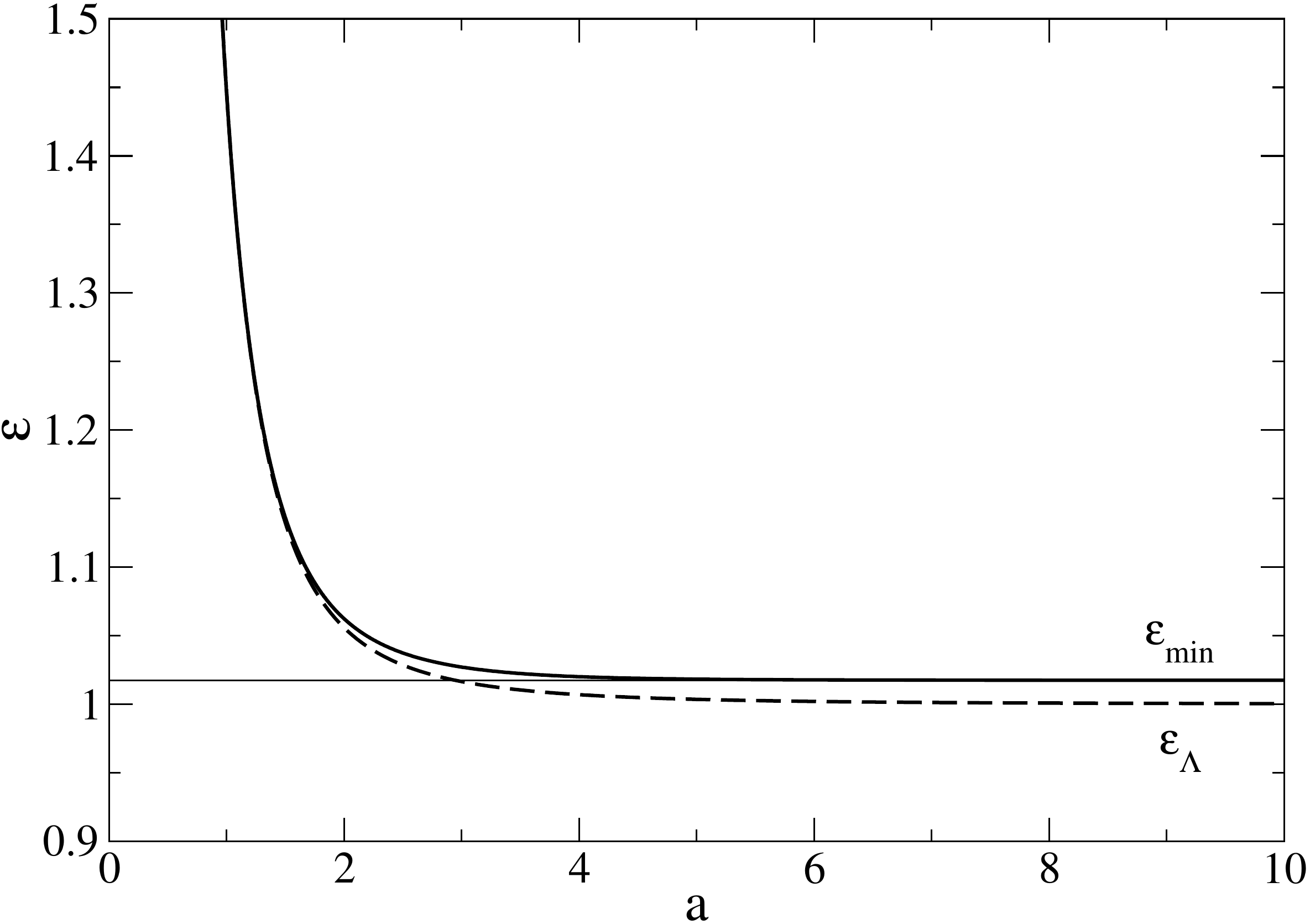}
\caption{Energy density as a function of the
scale factor.}
\label{aeps}
\end{center}
\end{figure}

\begin{figure}[!h]
\begin{center}
\includegraphics[clip,scale=0.3]{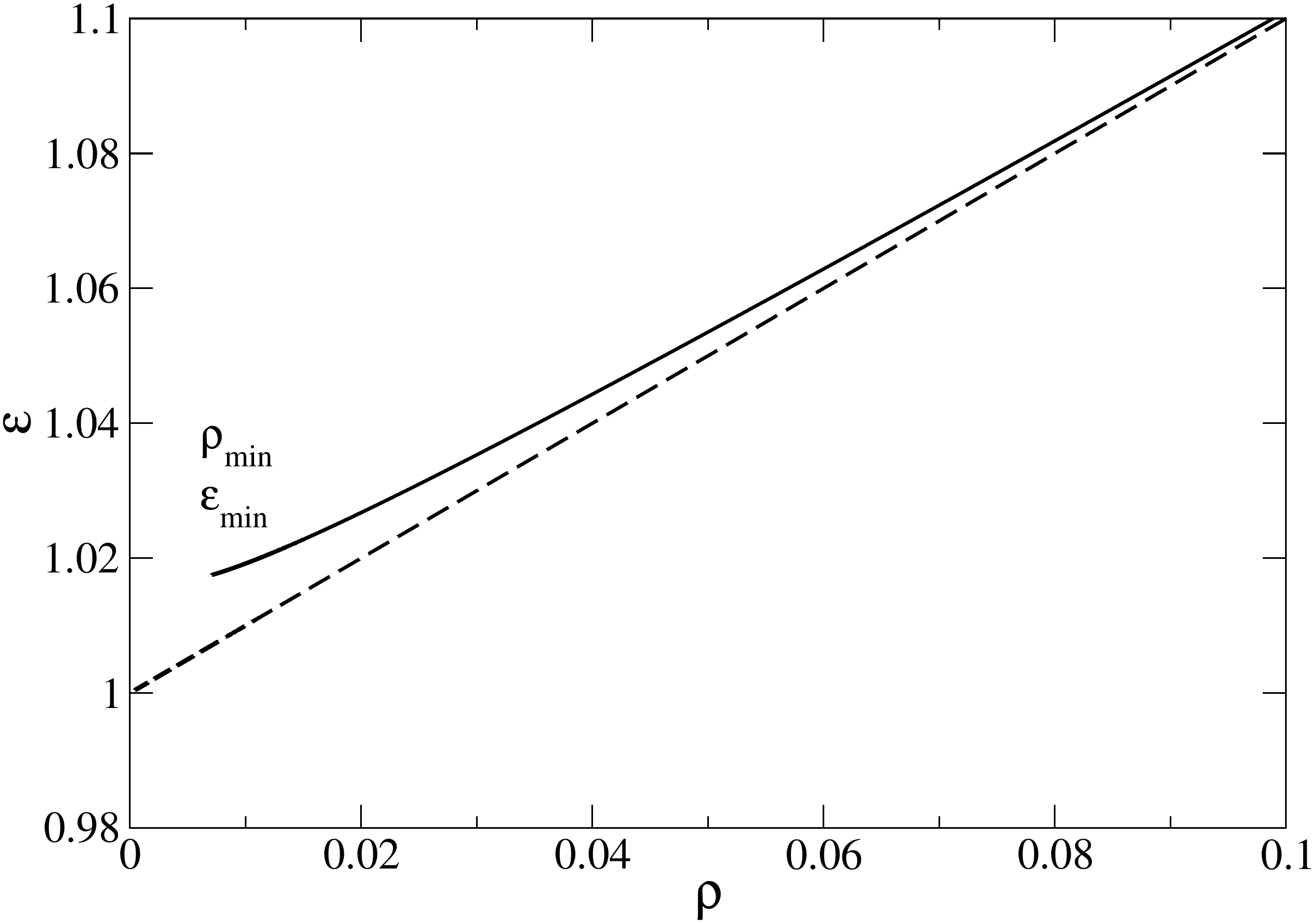}
\caption{Relation between the energy density and the pseudo rest-mass density.}
\label{rhoeps}
\end{center}
\end{figure}

\begin{figure}[!h]
\begin{center}
\includegraphics[clip,scale=0.3]{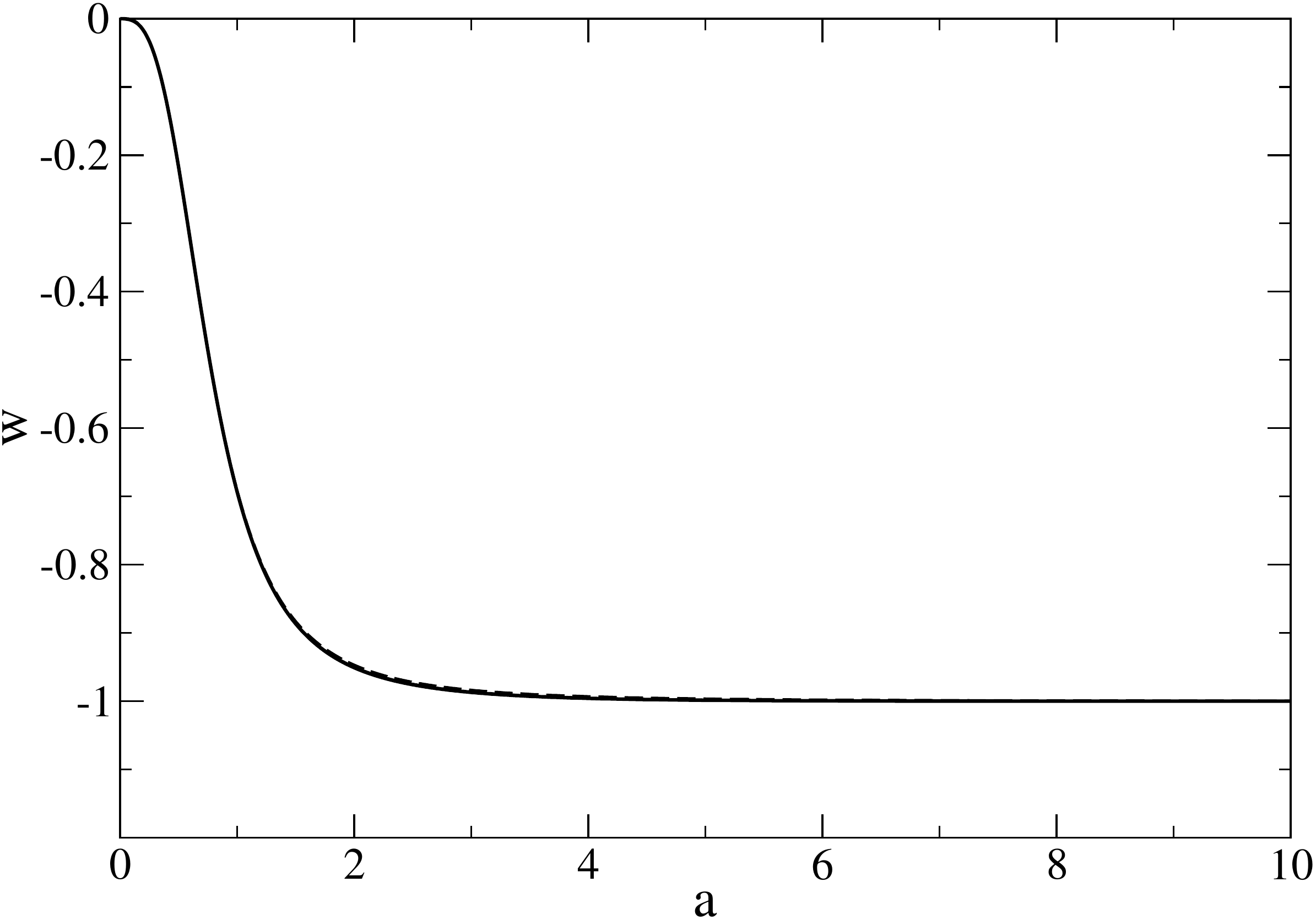}
\caption{Equation of state parameter as a function of the scale factor.}
\label{aw}
\end{center}
\end{figure}

\begin{figure}[!h]
\begin{center}
\includegraphics[clip,scale=0.3]{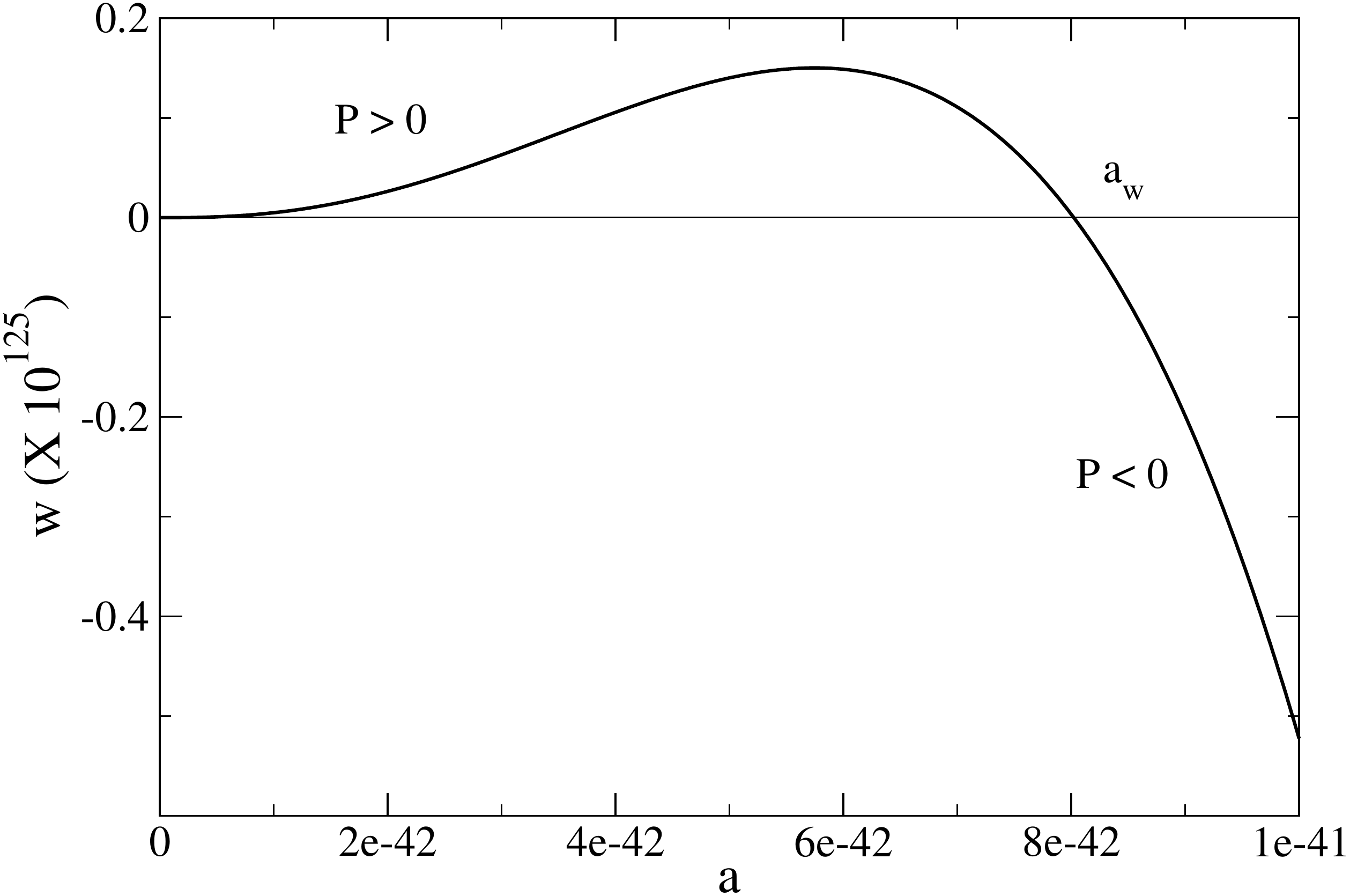}
\caption{Zoom of Fig. \ref{aw} at very small values of the scale factor where
the
pressure passes from positive to negative values.}
\label{awzoom}
\end{center}
\end{figure}

\begin{figure}[!h]
\begin{center}
\includegraphics[clip,scale=0.3]{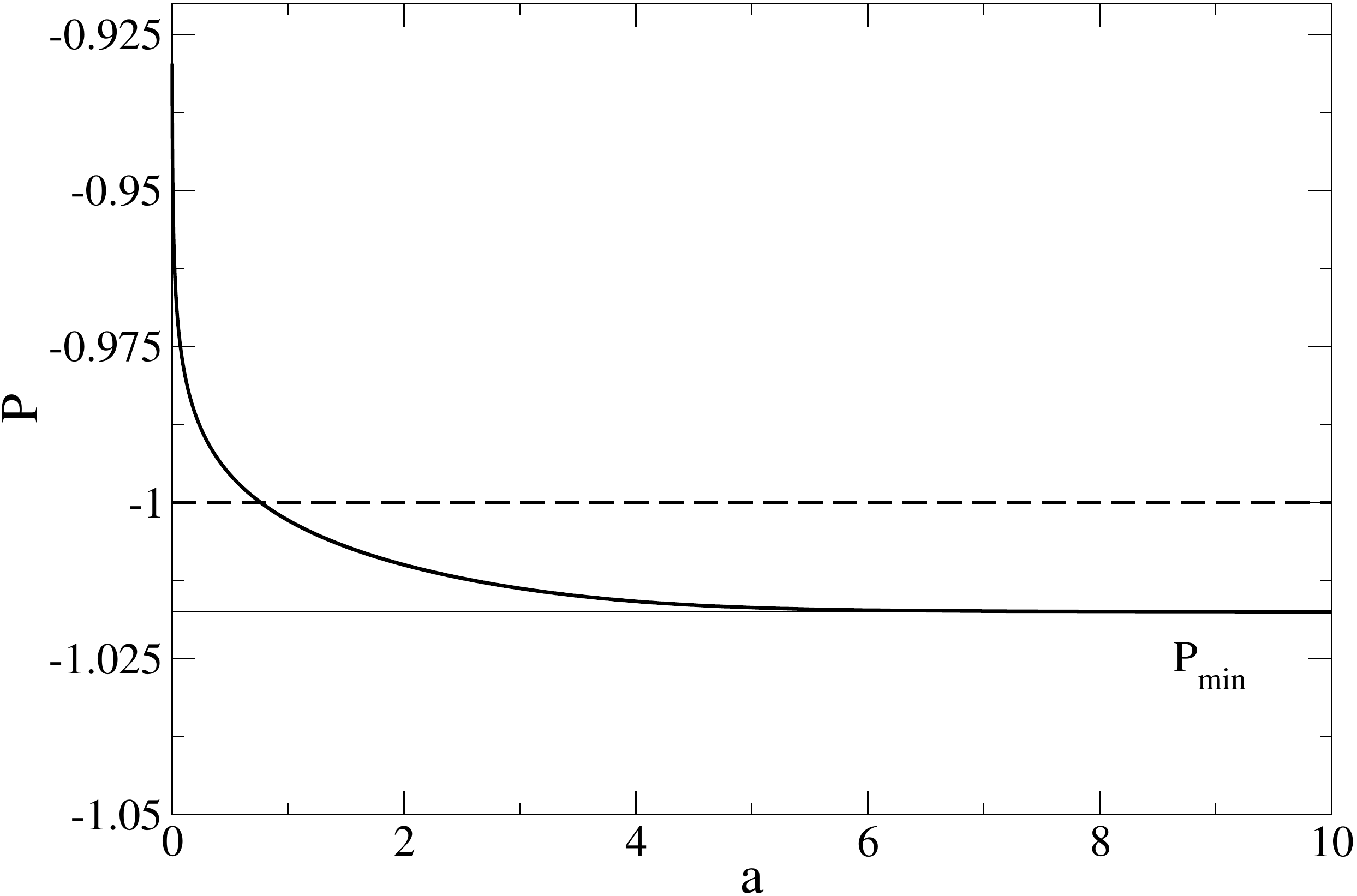}
\caption{Pressure as a function of the scale factor (${\tilde P}=-1$ for
$a=0.765$). }
\label{ap}
\end{center}
\end{figure}

\begin{figure}[!h]
\begin{center}
\includegraphics[clip,scale=0.3]{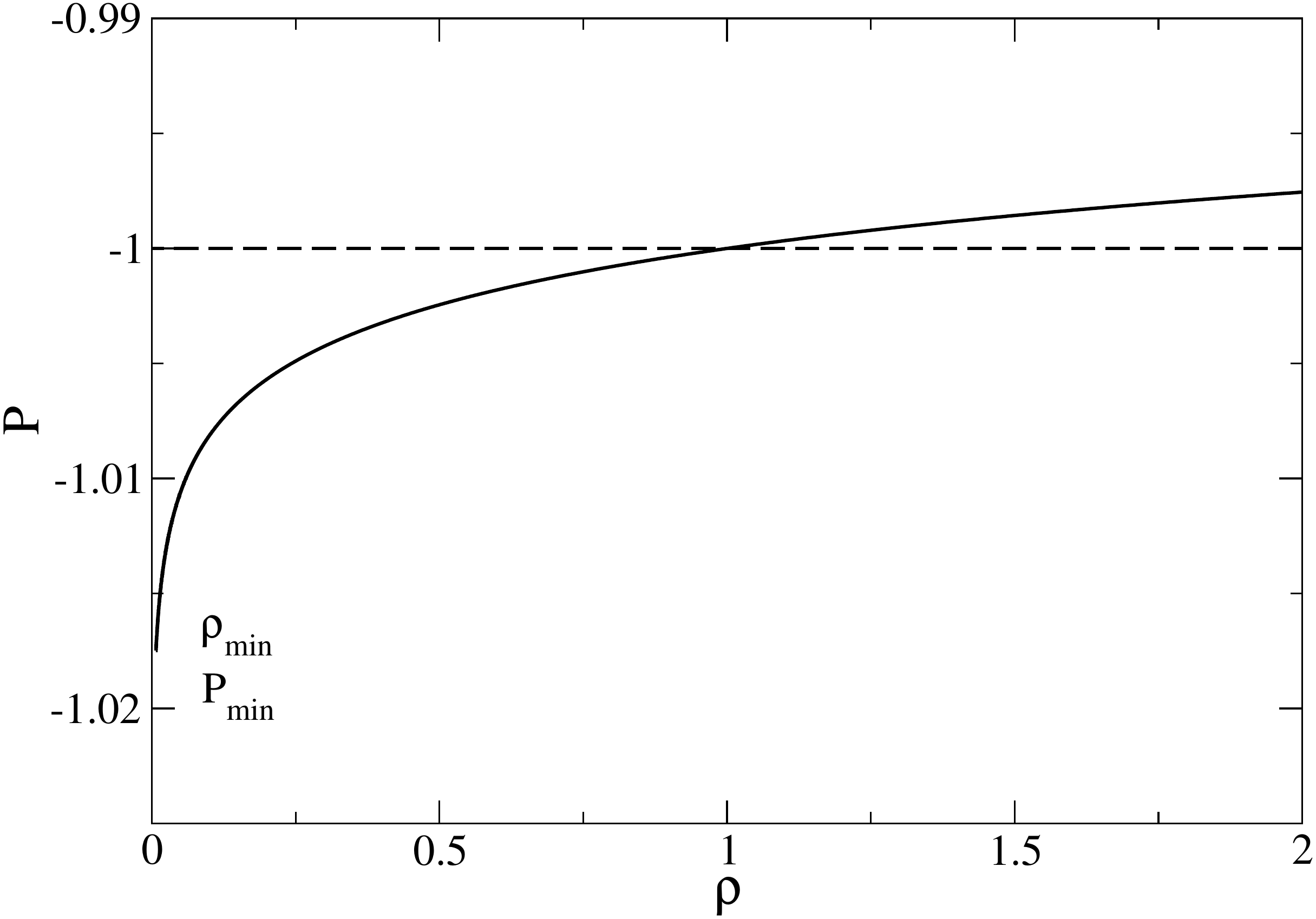}
\caption{Pressure as a function of the pseudo rest-mass
density (${\tilde P}=-1$ for ${\tilde \rho}=1$).}
\label{rhop}
\end{center}
\end{figure}

\begin{figure}[!h]
\begin{center}
\includegraphics[clip,scale=0.3]{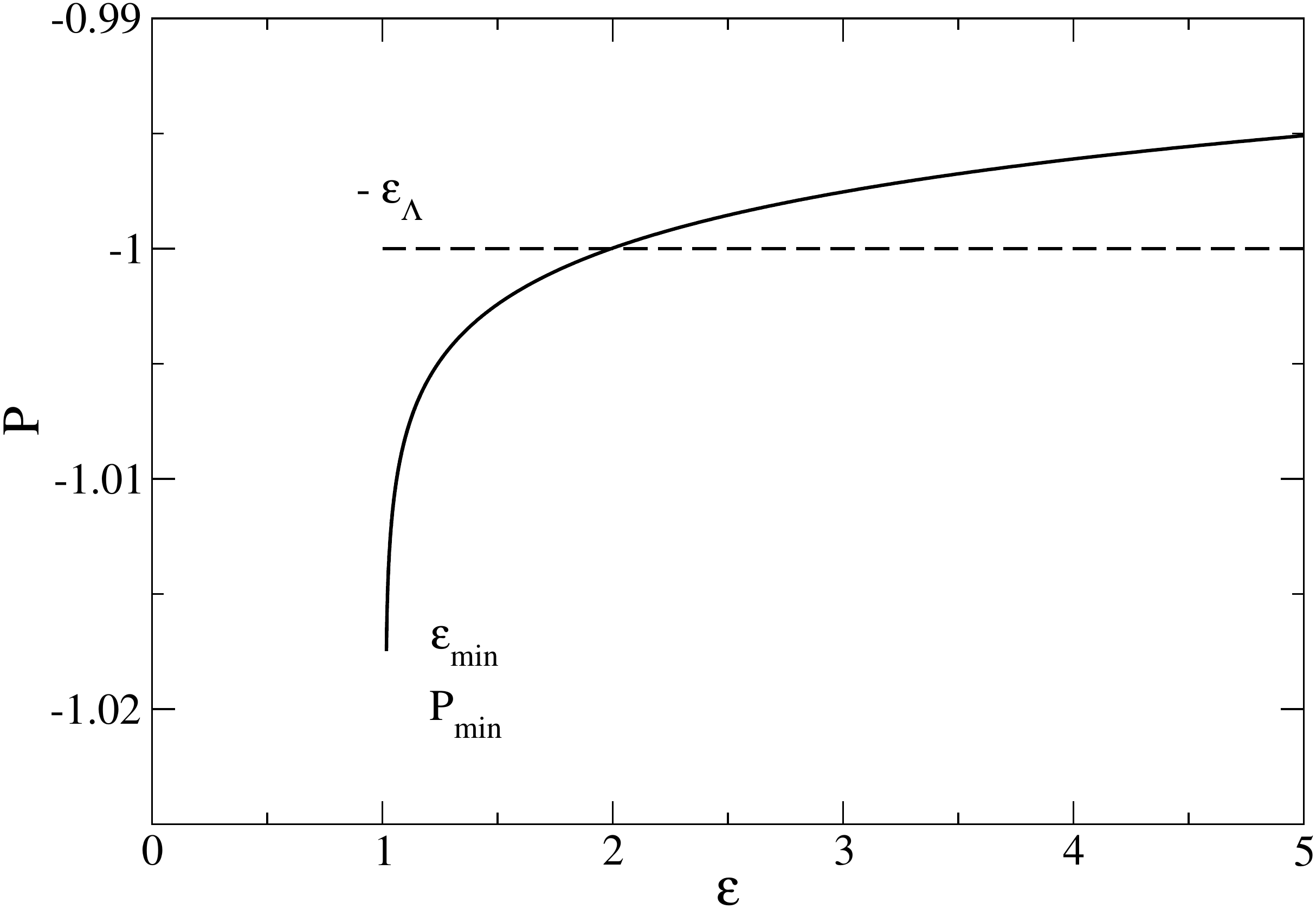}
\caption{Pressure as a function of the energy density (${\tilde P}=-1$ for
${\tilde \epsilon}=1.99$).}
\label{epsp}
\end{center}
\end{figure}

It is convenient to write the equations of the problem in terms of dimensionless
variables. Introducing $\tilde\rho=\rho/\rho_\Lambda$,
$\tilde\epsilon=\epsilon/\rho_\Lambda c^2$,
$\tilde P=P/\rho_\Lambda c^2$ and ${\tilde E}_{\rm tot}=E_{\rm tot}/mc^2$, we
obtain
\begin{eqnarray}
\tilde\rho\sqrt{1-\frac{2B}{\tilde\rho}}
=\frac{\Omega_{\rm m,0}}{1-\Omega_{\rm m,0}}\frac{1}{a^3},
\label{dg1}
\end{eqnarray} 
\begin{eqnarray}
\tilde\epsilon=\tilde\rho-B\ln\tilde\rho+1-2B,
\label{dg2}
\end{eqnarray} 
\begin{eqnarray}
{\tilde E}_{\rm tot}=\sqrt{1-\frac{2B}{\tilde\rho}},
\label{dg5}
\end{eqnarray} 
\begin{eqnarray}
{\tilde P}=B\ln {\tilde\rho}-1,
\label{dg3}
\end{eqnarray} 
\begin{eqnarray}
{\tilde \epsilon}=e^{1/B}e^{\tilde P/B}-{\tilde P}-2B,
\label{dg4}
\end{eqnarray} 
\begin{eqnarray}
w=\frac{\tilde P}{\tilde\epsilon}=\frac{B\ln
{\tilde\rho}-1}{\tilde\rho-B\ln\tilde\rho+1-2B}.
\label{dg6}
\end{eqnarray} 
We can easily solve the first equation to express the pseudo rest-mass density
in terms
of the scale factor as
\begin{eqnarray}
\tilde\rho=B+\sqrt{B^2+\left
(\frac{\Omega_{\rm m,0}}{1-\Omega_{\rm m,0}}\right )^2\frac{1}{a^6}}.
\label{dg7}
\end{eqnarray} 
We can then inject this relation into the other equations to obtain the
evolution of the different dimensionless variables as a function
of $a$.
Their evolution is
represented in solid lines in Figs. \ref{arho}-\ref{epsp}. The dashed lines in
these figures correspond to the $\Lambda$CDM model which is recovered from the
logotropic model when $B=0$ (see Sec. \ref{sec_rec}).

The pressure vanishes ($\tilde P=w=0$) when
\begin{equation}
{\tilde\rho}_w=e^{1/B}=8.65\times 10^{122},
\label{wn3kj}
\end{equation} 
corresponding to $\rho=\rho_P$. Using Eqs. (\ref{dg1}) and (\ref{dg2}), this
corresponds to
a scale factor
\begin{equation}
a_w\simeq \left (\frac{\Omega_{\rm m,0}}{1-\Omega_{\rm m,0}}\right
)^{1/3} e^{-1/(3B)}=8.02\times 10^{-42}
\label{wn3a}
\end{equation} 
and an energy density
\begin{equation}
{\tilde\epsilon}_w\simeq e^{1/B}=8.65\times 10^{122}.
\label{wn3ab}
\end{equation}
The pressure is positive ($P>0$) when $a<a_w$ and negative ($P<0$) when
$a>a_w$.\footnote{Since $\rho_P$ has been identified with the Planck density,
and
since the logotropic model is expected to unify DM and DE but not the early
inflation where the density is of the order of the Planck scale, we conclude
that the logotropic model is valid only for $\rho\ll\rho_P$. In that
regime, the pressure is always negative.} We note that the equation of state
parameter $w$
reaches a maximum value $w_{\rm
max}=1.50\times 10^{-126}$ at $a_*=5.75\times 10^{-42}$.

The pressure is equal to $\tilde P=-1$ (i.e.
$P=-\rho_{\Lambda}c^2$) when
\begin{equation}
{\tilde\rho}=1,
\label{own3kj}
\end{equation} 
corresponding to $\rho=\rho_\Lambda$. Using Eqs. (\ref{dg1}) and (\ref{dg2}),
this
corresponds to
a scale factor
\begin{equation}
a=\left (\frac{\Omega_{\rm m,0}}{1-\Omega_{\rm m,0}}\right
)^{1/3}\frac{1}{(1-2B)^{1/6}}=0.765
\label{kown3a}
\end{equation} 
and an energy density
\begin{equation}
{\tilde\epsilon}=2(1-B)=1.99.
\label{own3a}
\end{equation}
At that point $w=-1/[2(1-B)]=-0.502$. This corresponds typically to the time of
equality between DM and DE in the $\Lambda$CDM model (see Appendix
\ref{sec_lcdm}).

\subsection{Early Universe}
\label{sec_dge}

In the early Universe ($a\rightarrow 0$), we get
\begin{eqnarray}
\tilde\rho\sim\tilde\epsilon\sim\frac{\Omega_{\rm m,0}}{1-\Omega_{\rm
m,0}}\frac{1}{a^3},
\label{dg8}
\end{eqnarray} 
\begin{eqnarray}
\tilde E_{\rm tot}\rightarrow 1,
\label{dg9}
\end{eqnarray}
\begin{eqnarray}
\tilde P\simeq B\ln \left (\frac{\Omega_{\rm m,0}}{1-\Omega_{\rm
m,0}}\frac{1}{a^3}\right )-1,
\label{dg9b}
\end{eqnarray}
\begin{eqnarray}
\tilde P\simeq B\ln\tilde\epsilon-1,
\label{dg10}
\end{eqnarray}
\begin{equation}
w\sim \frac{B\ln\tilde\rho-1}{\tilde\rho},
\label{wn4}
\end{equation} 
\begin{equation}
w\sim \frac{1-\Omega_{\rm m,0}}{\Omega_{\rm m,0}}a^3\left\lbrack
B\ln\left (\frac{\Omega_{\rm m,0}}{1-\Omega_{\rm
m,0}}\frac{1}{a^3}\right )-1\right\rbrack,
\label{wn5}
\end{equation} 
\begin{equation}
w\sim \frac{B\ln\tilde\epsilon-1}{\tilde\epsilon}.
\label{wn6}
\end{equation}

\subsection{Late Universe}
\label{sec_dgl}

In the late Universe ($a\rightarrow +\infty$), we get
\begin{eqnarray}
\tilde\rho\rightarrow \tilde\rho_{\rm min}=2B=7.065\times 10^{-3},
\label{dg11}
\end{eqnarray} 
\begin{eqnarray}
\tilde\epsilon\rightarrow \tilde\epsilon_{\rm min}=1-B\ln(2B)=1.02,
\label{dg12}
\end{eqnarray} 
\begin{eqnarray}
\tilde E_{\rm tot}\rightarrow 0,
\label{dg9bh}
\end{eqnarray}
\begin{eqnarray}
\tilde P\rightarrow \tilde P_{\rm min}=-\tilde\epsilon_{\rm min},
\label{dg13}
\end{eqnarray}
\begin{equation}
w\rightarrow w_{\rm min}=\frac{{\tilde P}_{\rm min}}{{\tilde\epsilon}_{\rm
min}}=-1.
\label{wn7}
\end{equation}

\subsection{Recovery of the $\Lambda$CDM model when $B=0$}
\label{sec_rec}

When $B=0$, the general equations of Sec. \ref{sec_dg} reduce to
\begin{eqnarray}
\tilde\rho=\frac{\Omega_{\rm m,0}}{1-\Omega_{\rm
m,0}}\frac{1}{a^3},
\label{dg14}
\end{eqnarray}
\begin{eqnarray}
\tilde\epsilon=\tilde\rho+1,
\label{dg15}
\end{eqnarray}
\begin{eqnarray}
\tilde\epsilon=\frac{\Omega_{\rm m,0}}{1-\Omega_{\rm
m,0}}\frac{1}{a^3}+1,
\label{dg16}
\end{eqnarray}
\begin{eqnarray}
\tilde P=-1,
\label{dg17}
\end{eqnarray}
\begin{eqnarray}
{\tilde E}_{\rm tot}=1,
\label{dg17b}
\end{eqnarray}
\begin{equation}
w=-\frac{1}{\tilde\rho+1}=-\frac{1}{\tilde\epsilon},
\label{wn8}
\end{equation}
\begin{equation}
w=-\frac{1}{\frac{\Omega_{\rm
m,0}}{1-\Omega_{\rm
m,0}}\frac{1}{a^3}+1}.
\label{wn8b}
\end{equation} 
Therefore, for $B=0$, we recover the equations of the  $\Lambda$CDM
model (see Appendix \ref{sec_lcdm}). Since the $\Lambda$CDM
model works very well at large scales, the logotropic model should work well too
provided that $B$ is small enough. We recall that $B$ is not a free
parameter of our model that could be tuned in order to
fit the data. It is actually determined by the
theory (see Sec. \ref{sec_b}). Indeed, if we
identify $\rho_P$ with the Planck density, this automatically fixes $B$ through
Eq.
(\ref{b2}).
Therefore, our model is fully predictive. As we have seen in Sec. \ref{sec_b},
the limit
$B\rightarrow 0$ corresponds to $\rho_P\rightarrow +\infty$ or $\hbar\rightarrow
0$ (quantum effects negligible). {\it Therefore, the
$\Lambda$CDM model
corresponds to the semiclassical limit $\hbar\rightarrow 0$ of the logotropic
model.} However, because of the fundamentally nonzero value of $\hbar$, the
logotropic model with a nonzero
value of $B=3.53\times 10^{-3}$ should be priviledged over the
$\Lambda$CDM model (corresponding to $B=0$).

{\it Remark:} It is instructive to establish the
connection between the LDF and
the $\Lambda$CDM model directly from the dimensional equation of state
(\ref{log2}). This equation can be rewritten as
\begin{eqnarray}
P=A\ln\left (\frac{\rho}{\rho_\Lambda}\right )-A\ln\left
(\frac{\rho_P}{\rho_\Lambda}\right ).
\label{cui1}
\end{eqnarray}
Taking the limit $A\rightarrow 0$ and $\rho_P\rightarrow +\infty$ with
$A\ln(\rho_P/\rho_\Lambda)=\rho_{\Lambda}c^2$ fixed [see Eq.
(\ref{a2b})], we obtain 
\begin{eqnarray}
P=\frac{\rho_{\Lambda}c^2}{\ln\left (\frac{\rho_P}{\rho_\Lambda}\right
)}\ln\left (\frac{\rho}{\rho_\Lambda}\right )-\rho_{\Lambda}c^2\simeq
-\rho_{\Lambda}c^2.
\label{cui2}
\end{eqnarray}
This returns the constant equation of state of the $\Lambda$CDM model in its
UDM interpretation (see
Appendix \ref{sec_df}). 

\section{Effective DM and DE}
\label{sec_effdmde}

\begin{figure}[!h]
\begin{center}
\includegraphics[clip,scale=0.3]{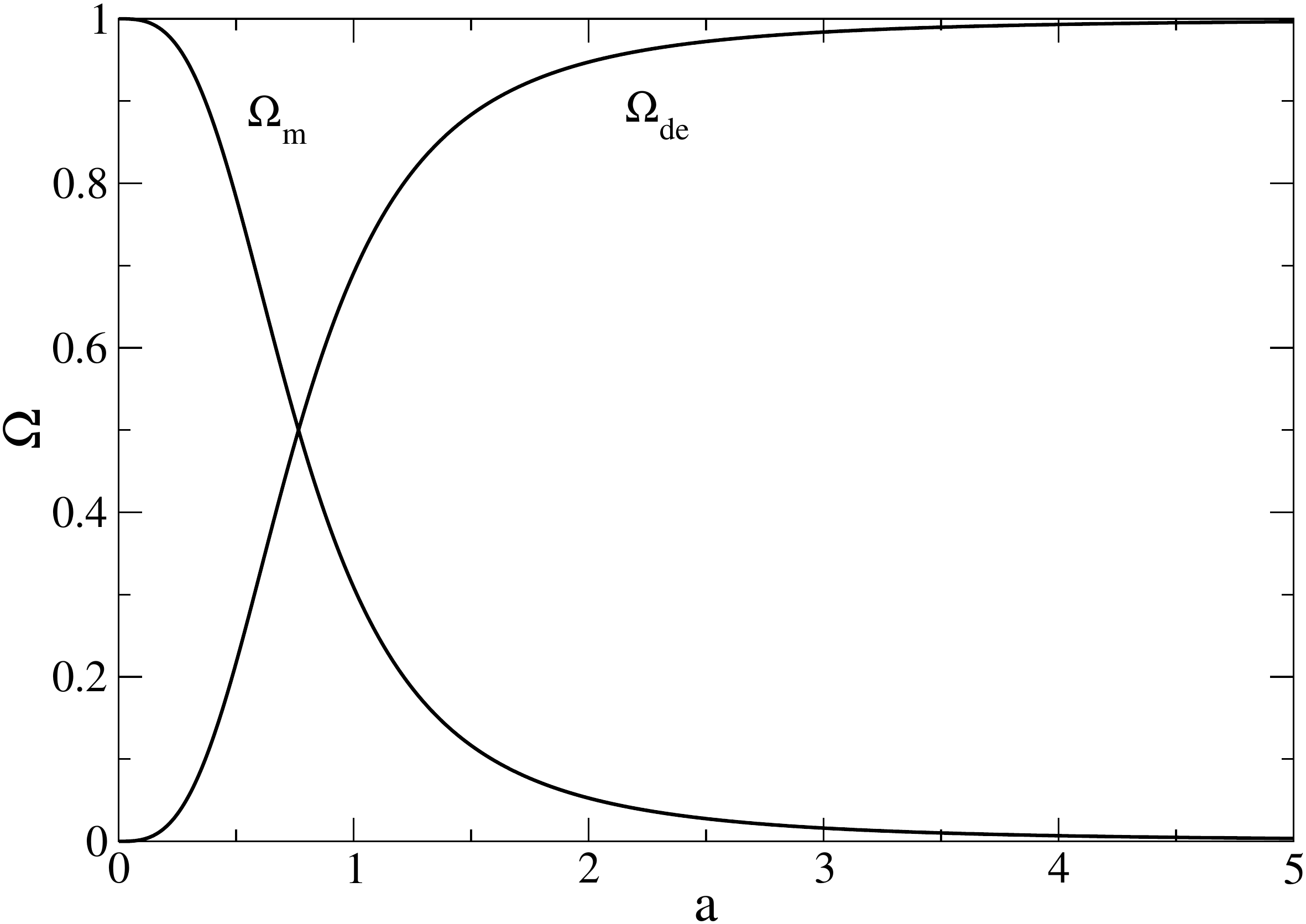}
\caption{Proportion of DM (rest-mass) and DE (internal energy) as a function of
the scale factor ($\Omega_m=\rho_m c^2/\epsilon$ and $\Omega_{\rm
de}=u/\epsilon$).}
\label{fraction}
\end{center}
\end{figure}

In terms of dimensionless variables, the rest-mass density is given by (see
Sec. \ref{sec_ie})
\begin{eqnarray}
\tilde\rho_m=\tilde\rho\sqrt{1-\frac{2B}{\tilde\rho}}
\label{dmde1}
\end{eqnarray} 
and the internal energy $u=\tilde\epsilon-\tilde\rho_m$ is given by
\begin{eqnarray}
\tilde
u=\tilde\rho-B\ln\tilde\rho+1-2B-\tilde\rho\sqrt{
1-\frac{2B}{\tilde\rho}}.
\label{dmde2}
\end{eqnarray} 
Using Eq. (\ref{dg7}), they evolve with the
scale factor $a$ as
\begin{eqnarray}
\tilde\rho_m=\frac{\Omega_{\rm m,0}}{1-\Omega_{\rm m,0}}\frac{1}{a^3}
\label{dmde3}
\end{eqnarray} 
and
\begin{eqnarray}
\tilde
u=1-B+\sqrt{B^2+\left
(\frac{\Omega_{\rm m,0}}{1-\Omega_{\rm m,0}}\right
)^2\frac{1}{a^6}}\nonumber\\
-B\ln\left\lbrack  B+\sqrt{B^2+\left
(\frac{\Omega_{\rm m,0}}{1-\Omega_{\rm m,0}}\right )^2\frac{1}{a^6}} 
\right\rbrack
-\frac{\Omega_{\rm m,0}}{1-\Omega_{\rm
m,0}}\frac{1}{a^3}.\nonumber\\
\label{dmde4}
\end{eqnarray} 
As indicated previously, {\it the rest-mass density $\rho_m$ can be interpreted
as
DM and the internal energy density $u$ can be interpreted as DE \cite{epjp}.}
The proportion of DM and DE as a function of the scale factor
$a$ is plotted in Fig. \ref{fraction}. At early times, the universe is
dominated by DM
($\rho_m c^2\gg u$) and at late times, the universe is dominated by DE ($\rho_m
c^2\ll u$). The DE density increases monotonically from
${\tilde u}\sim 3B\ln a\rightarrow -\infty$ when $a\rightarrow 0$  to
${\tilde u}\rightarrow 1-B\ln(2B)$ when $a\rightarrow +\infty$. Since
the DE density corresponds to the internal energy density $u$ of the LDF, it can
very well be negative as long as the total energy density
$\epsilon$ is positive. In the regime of interest
($\rho_{\rm m}\ll\rho_P$) where the logotropic model is valid, the DE density
$\epsilon_{\rm de}$ is positive.

{\it Remark:} For the $\Lambda$CDM model ($B=0$), we find that
$\rho=\rho_m\propto a^{-3}$ and $u=-\rho_{\Lambda}c^2$ (see also Appendices
\ref{sec_lcdm} and \ref{sec_gfdm}).

\section{Deceleration parameter}
\label{sec_q}

\begin{figure}[!h]
\begin{center}
\includegraphics[clip,scale=0.3]{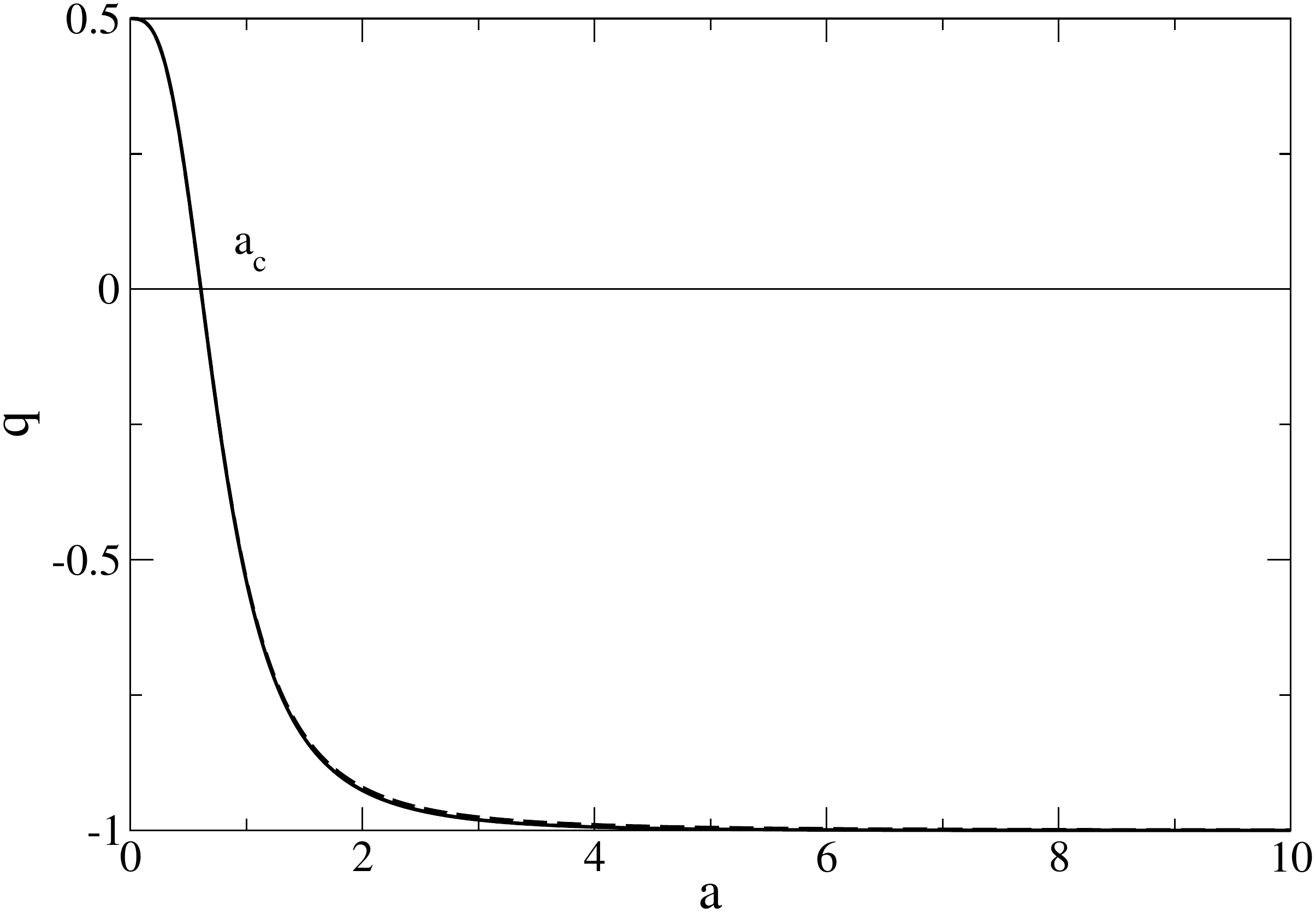}
\caption{Deceleration parameter as a function of the scale factor. The value
$a_c=0.607$ corresponds to the moment at which the Universe starts
accelerating.}
\label{aq}
\end{center}
\end{figure}

In a flat Universe without cosmological constant ($k=\Lambda=0$), the
deceleration parameter $q=-{\ddot a} a/{\dot a}^2$ is related to the equation
of state parameter $w$
by (see, e.g., \cite{cosmopoly1})  
\begin{equation}
q=\frac{1+3w}{2}.
\label{q1}
\end{equation} 
Therefore, we can easily deduce the evolution of $q$ from the evolution of $w$
obtained in Sec. \ref{sec_dime}. The function $q(a)$ is represented in Fig.
\ref{aq}.

The Universe starts accelerating when $q=0$
corresponding to $w_c=-1/3$. At that point ${\tilde\rho}_c\simeq 2$,
${\tilde\epsilon}_c\simeq 3$ and $a_c\simeq 0.607$ like for the $\Lambda$CDM
model (the difference is less than $1\%$).

The present value of the deceleration parameter is
\begin{equation}
q_0=\frac{1+3w_0}{2},
\label{q2}
\end{equation} 
where
\begin{equation}
w_0=\frac{B\ln\tilde\rho_0-1}{\tilde\rho_0-B\ln\tilde\rho_0+1-2B}
\label{q3}
\end{equation} 
with 
\begin{eqnarray}
\tilde\rho_0=B+\sqrt{B^2+\left
(\frac{\Omega_{\rm m,0}}{1-\Omega_{\rm m,0}}\right )^2}.
\label{q4}
\end{eqnarray} 
We get $\tilde\rho_0=0.4505$, $\tilde\epsilon_0=1.45$, ${\tilde E}_{\rm
tot}=0.992$, ${\tilde P}=-1.00$, $w_0=-0.693$ and $q_0=-0.540$. For the
$\Lambda$CDM model,
we obtain $\tilde\rho_0=0.447$, $\tilde\epsilon_0=1.45$, ${\tilde E}_{\rm
tot}=1$, ${\tilde P}=-1$, $w_0=-0.691$ and $q_0=-0.537$. The values of the
two models are very
close to each other differing by less than $1\%$.

\section{Speed of sound}
\label{sec_cs}

\subsection{Dimensional variables}
\label{sec_csd}

\begin{figure}[!h]
\begin{center}
\includegraphics[clip,scale=0.3]{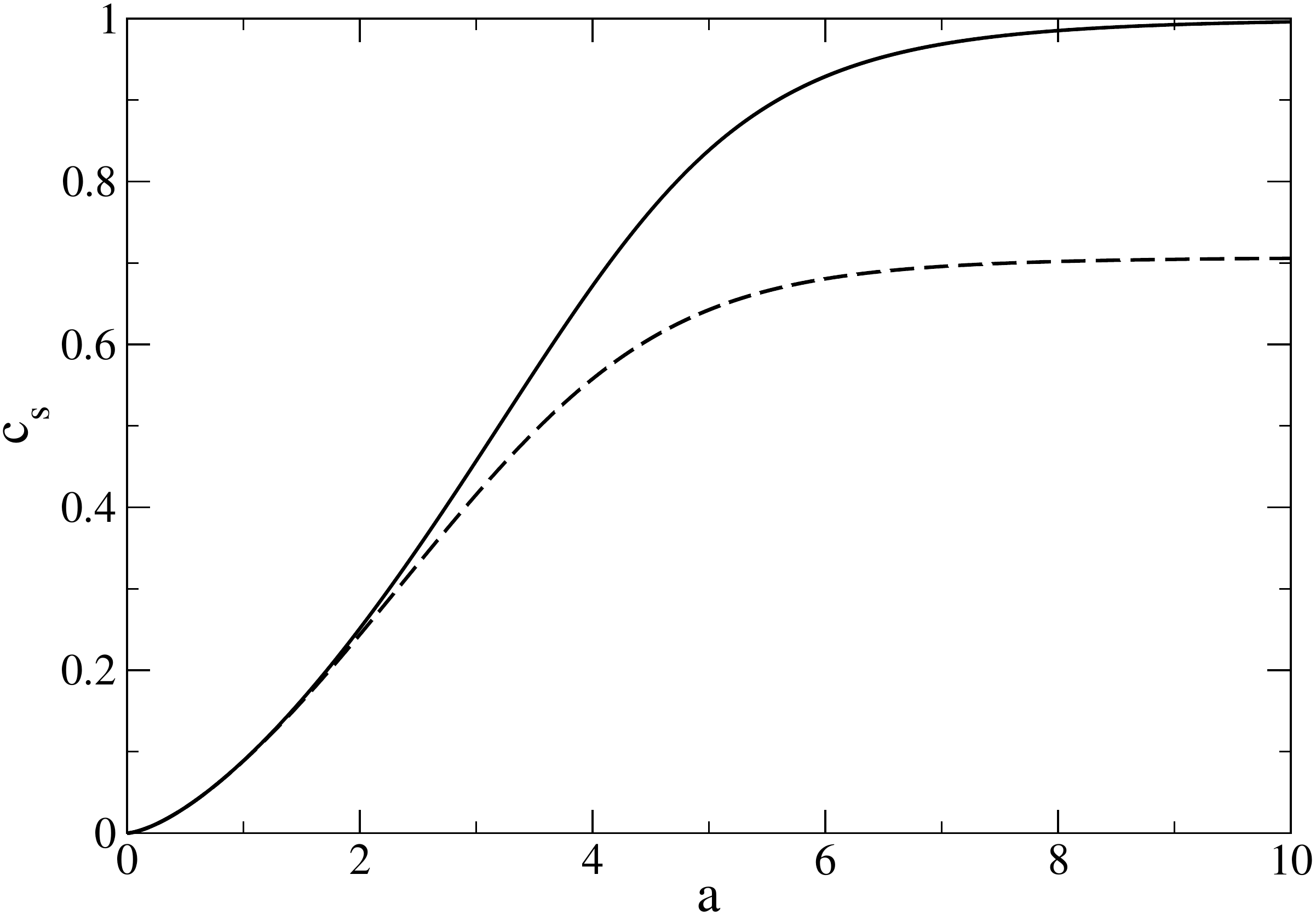}
\caption{Speed of sound as a function of the scale factor (the dashed line
corresponds to the pseudo speed of sound). We
note that $c_s=0$ in the $\Lambda$CDM model.}
\label{acs}
\end{center}
\end{figure}

The speed of sound $c_s$ is defined by
\begin{equation}
c_s^2=P'(\epsilon)c^2.
\label{csd1}
\end{equation} 
Differentiating Eq. (\ref{log6}) with respect to $\epsilon$ and using  Eq.
(\ref{log2})  we obtain
\begin{equation}
\frac{c_s^2}{c^2}=\frac{1}{\frac{\rho c^2}{A}-1}.
\label{csd2}
\end{equation} 
Since $\rho\ge\rho_{\rm min}=2A/c^2$ we find that $c_s^2\ge 0$ and
$c_s<c$.  The
speed of sound tends to zero ($c_s\rightarrow 0$) when $\rho\rightarrow
+\infty$ and  to the speed of light  ($c_s\rightarrow c$)
when $\rho\rightarrow \rho_{\rm min}$ (see Fig. \ref{acs}).\footnote{We note
that the speed of sound in the LDF is positive in
spite of the fact that
its pressure is negative. This is a very important property because, in many
cases, fluids with negative pressure obeying a barotropic equation of state
suffer from  hydrodynamic or  tachyonic  instabilities at small
scales due to an imaginary speed of sound. This does not occur in the present
model. In addition, the speed of
sound is always less than the speed of light.
By contrast, in the original logotropic model \cite{epjp}, the speed of
sound diverges as we enter the phantom era, before becoming imaginary.}

In the early Universe:
\begin{equation}
\frac{c_s^2}{c^2}\sim \frac{A}{\rho c^2},
\label{csd3}
\end{equation} 
\begin{equation}
\frac{c_s^2}{c^2}\sim \frac{A}{Qm c^2} a^3,
\label{csd4}
\end{equation} 
\begin{equation}
\frac{c_s^2}{c^2} \sim \frac{A}{\epsilon}.
\label{csd5}
\end{equation} 
In the late Universe:
\begin{equation}
c_s\rightarrow c.
\label{csd6}
\end{equation}

{\it Remark:} The pseudo speed of sound $c_s^*$ defined by
$(c_s^*)^2=P'(\rho)$ is given by
\begin{equation}
(c_s^*)^2=\frac{A}{\rho}.
\label{csn1we}
\end{equation} 
It coincides with the true speed of sound  $c_s$ in the nonrelativistic regime
$\rho\gg A/c^2$ (early universe). On the other
hand, in the late universe, we get
$c_s^*=c/\sqrt{2}$ when $\rho=\rho_{\rm min}=2A/c^2$.

\subsection{Dimensionless variables}
\label{sec_csn}

Introducing the dimensionless speed of sound ${\tilde
c}_s={c_s}/{c}$ and using the dimensionless variables defined previously, we get
\begin{equation}
{\tilde c_s}^2=\frac{1}{\frac{\tilde\rho}{B}-1}.
\label{csn1}
\end{equation} 
In the early Universe:
\begin{equation}
{\tilde c_s}^2\sim \frac{B}{\tilde\rho},
\label{csn2}
\end{equation} 
\begin{equation}
{\tilde c_s}^2\sim B \frac{1-\Omega_{\rm m,0}}{\Omega_{\rm m,0}}a^3,
\label{csn3}
\end{equation} 
\begin{equation}
{\tilde c_s}^2 \sim \frac{B}{\tilde\epsilon}.
\label{csn4}
\end{equation} 
In the late Universe:
\begin{equation}
{\tilde c_s}\rightarrow 1. 
\label{csn5}
\end{equation} 
The present value of the squared speed of sound is
\begin{equation}
({\tilde c_s}^2)_0=\frac{1}{\frac{\tilde\rho_0}{B}-1}=7.90\times 10^{-3},
\label{csn1web}
\end{equation} 
showing that the present Universe is strongly special relativistic [$(c_s)_0\sim
0.1 c$]. The pseudo squared speed of sound
is $({\tilde c_s^*})^2=B/\tilde\rho$ and its 
present value is $({\tilde
c_s}^*)^2_0={B}/{\tilde\rho_0}=7.84\times
10^{-3}$. As discussed in Sec. \ref{sec_fw} and in Appendix \ref{sec_lin} the
present value
of the squared speed of sound $c_s^2/c^2\sim 10^{-2}$ is too large to
enable the formation of clusters of galaxies 
and to
account for the observations of the power spectrum. This is a serious problem of
the
logotropic model.

\subsection{$\Lambda$CDM model ($B=0$)}
\label{sec_cscdm}

For $B=0$, corresponding to $\rho_P\rightarrow +\infty$ (no quantum
effects), we find that
\begin{equation}
c_s=0.
\label{cscdm}
\end{equation} 
The speed of sound vanishes in the $\Lambda$CDM model
since  the pressure $P=-\rho_{\Lambda}c^2$ is constant (see Appendix
\ref{sec_df}). The vanishing of the speed of sound
in the $\Lambda$CDM model (implying the absence of pressure gradient to
balance the gravitational attraction in DM halos) is at the origin of the
small scale crisis of the CDM model. The fact that the speed of
sound is nonzero in the logotropic model ($B=3.53\times 10^{-3}$) while it
vanishes in the $\Lambda$CDM model ($B=0$) is an important difference
between the two models. Indeed, a nonzero speed of sound may solve the
CDM small scale crisis. However, the fact that the
speed of
sound increases with the scale factor  in the logotropic model (see Fig. \ref{acs}) poses new problems regarding the formation of structures as
discussed in Sec. \ref{sec_fw}.

\section{Evolution of the scale factor}
\label{sec_eu}

\begin{figure}[!h]
\begin{center}
\includegraphics[clip,scale=0.3]{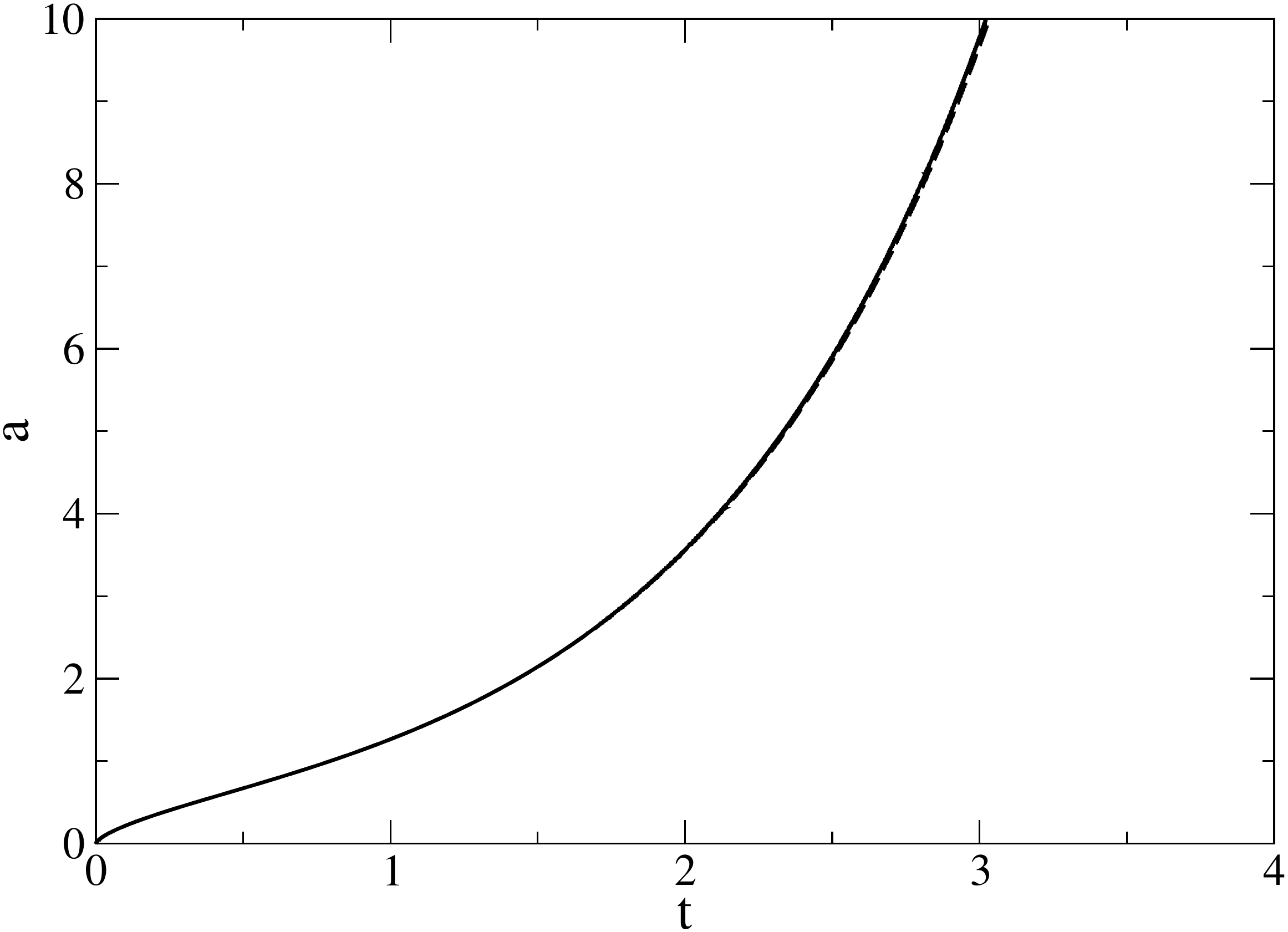}
\caption{Scale factor as a function of time. The logotropic model starts to
deviate from the $\Lambda$CDM model for $a\gtrsim 2$ but the difference between
the two models  is hardly perceptible on this 
representation.
}
\label{at}
\end{center}
\end{figure}

The evolution of the scale factor of the Universe is determined by the
Friedmann equation (\ref{hsf5}) combined with the relation
$\epsilon(a)$ between the energy density and the scale factor. This yields an
equation of the form
\begin{equation}
H=\frac{\dot a}{a}=\left (\frac{8\pi G}{3c^2}\right )^{1/2}\epsilon^{1/2}(a).
\label{eu1}
\end{equation}
Introducing the dimensionless energy
$\tilde\epsilon=\epsilon/\rho_{\Lambda}c^2$ and the dimensionless time $\tilde
t=(8\pi G\rho_{\Lambda}/3)^{1/2}t$, this equation can be rewritten as
\begin{equation}
\frac{\dot a}{a}={\tilde\epsilon}^{1/2}(a).
\label{eu2}
\end{equation}
It can be integrated into
\begin{equation}
{\tilde t}=\int_0^a\frac{dx}{x
{\tilde\epsilon}^{1/2}(x)}\equiv {\tilde t}(a),
\label{eu3}
\end{equation}
which gives $a(\tilde t)$ in reversed form. In the logotropic model, the
relation  $\tilde\epsilon(a)$ between the dimensionless energy and the scale
factor is
determined  by Eqs. (\ref{dg2}) and (\ref{dg7}). One can then solve Eq.
(\ref{eu3}) numerically. The function $a(\tilde t)$ is
plotted in Fig. \ref{at}.

In the early Universe $t\rightarrow 0$, using Eq. (\ref{dg8}), we get
\begin{equation}
a=\left\lbrack \frac{3}{2}\left (\frac{8\pi G\rho_{\Lambda}}{3}\right
)^{1/2} \left (\frac{\Omega_{\rm m,0}}{1-\Omega_{\rm m,0}}\right
)^{1/2}t\right\rbrack^{2/3},
\label{eu4}
\end{equation}
which can be rewritten as
\begin{equation}
a=\left (\frac{3}{2}\sqrt{\Omega_{m,0}}H_0 t\right )^{2/3},
\label{eu5}
\end{equation}
where $H_0$ is the present value of the Hubble constant. This is the usual
Einstein-de Sitter (EdS)
solution.

In the late Universe $t\rightarrow +\infty$, using Eq. (\ref{dg12}), we get
\begin{equation}
a\propto e^{\left (\frac{8\pi G\rho_{\Lambda}}{3}\right
)^{1/2} \left \lbrack 1-B\ln(2B)\right\rbrack^{1/2}t}.
\label{eu6}
\end{equation}
This is the de Sitter solution with a $B$-modified cosmological constant. As
discussed in Secs. \ref{sec_valf} and \ref{sec_einstein}, this
modification has a quantum
origin.

The age of the Universe in the logotropic model is
\begin{equation}
t_0=\left (\frac{3}{8\pi G\rho_{\Lambda}}\right
)^{1/2}\int_0^1\frac{dx}{x {\tilde\epsilon}^{1/2}(x)}.
\label{eu7}
\end{equation}
We obtain ${\tilde t}_0=0.795$ giving $t_0=13.8\, {\rm Gyrs}$ like for the
$\Lambda$CDM model corresponding to $B=0$ (the difference is less than $1\%$).

\section{Total potential}
\label{sec_tp}

\subsection{Dimensional variables}
\label{sec_tpd}

\begin{figure}[!h]
\begin{center}
\includegraphics[clip,scale=0.3]{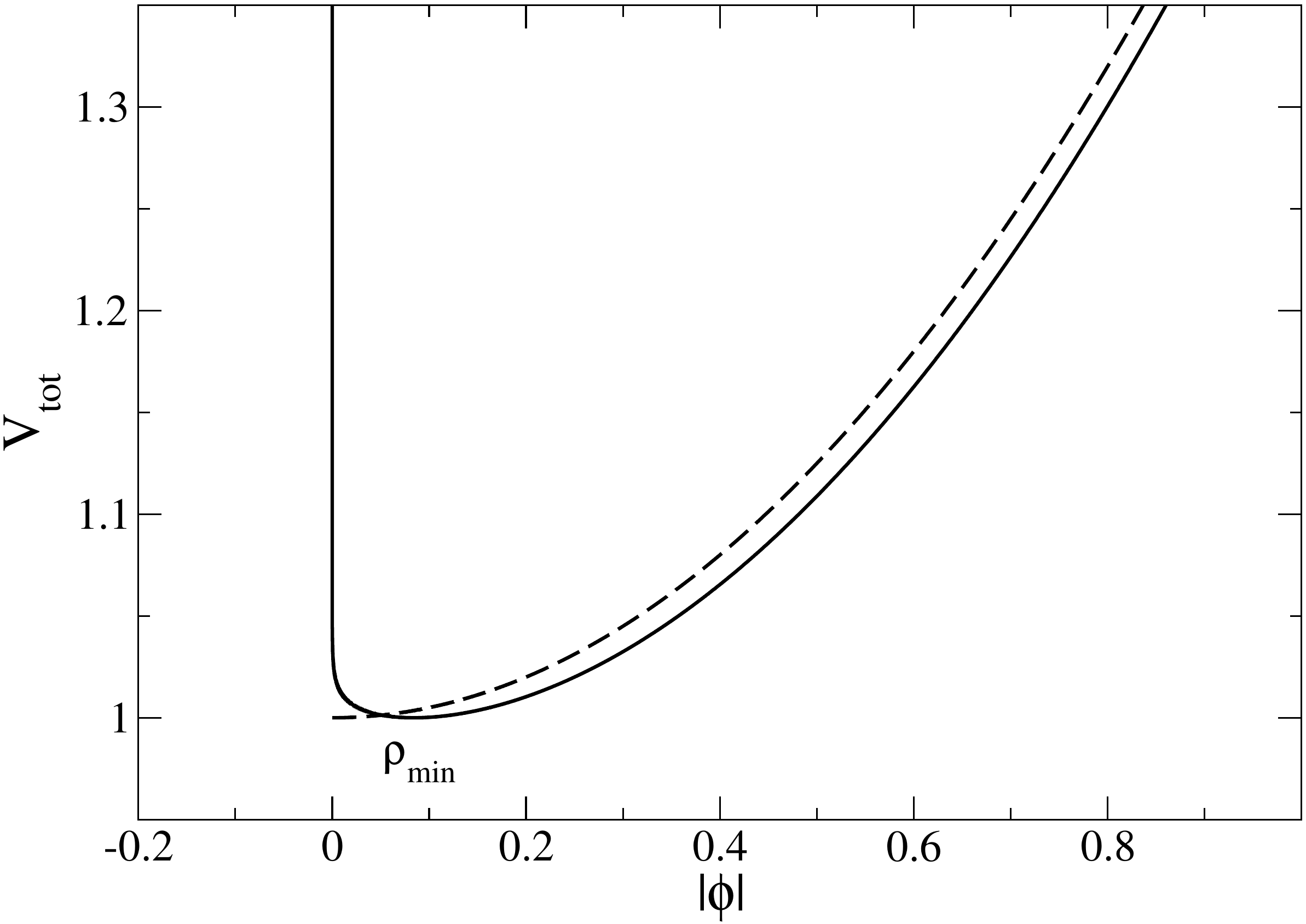}
\caption{Total potential of the logotropic SF. The SF descends the potential by
rapidly spinning around the vertical axis. Only the exterior branch
$\rho\ge\rho_{\rm min}$
is accessible. The dashed line  corresponds to the total potential 
of the $\Lambda$CDM model in its SF representation
(see Appendix
\ref{sec_gfdm}).}
\label{vtot}
\end{center}
\end{figure}

In the logotropic model, the total potential of the SF including the rest-mass
term and the logarithmic term 
[see Eqs. (\ref{hsf1b}) and (\ref{log0})] is
\begin{equation}
V_{\rm tot}(|\varphi|^2)=\frac{m^2c^2}{2\hbar^2}|\varphi|^2-A\ln \left
(\frac{m^2|\varphi|^2}{\hbar^2\rho_P}\right )-A.
\label{tp4}
\end{equation}
Introducing the pseudo rest-mass density defined by Eq. (\ref{ge1}) it can be
rewritten as
\begin{equation}
V_{\rm tot}=\frac{1}{2}\rho c^2-A\ln \left
(\frac{\rho}{\rho_P}\right )-A.
\label{tp1}
\end{equation}
It is represented in Fig. \ref{vtot}. It behaves as $V_{\rm tot}\sim -A\ln\rho$
for $\rho\rightarrow 0$ and as $V_{\rm tot}\sim (1/2)\rho c^2$ for
$\rho\rightarrow +\infty$.  It has a
minimum at
\begin{equation}
\rho_{\rm min}=\frac{2A}{c^2},\qquad V_{\rm min}=\epsilon_{\rm min}=A\ln \left
(\frac{\rho_Pc^2}{2A}\right ).
\label{tp2}
\end{equation}
We note that $\rho_{\rm min}$ corresponds to the asymptotic value of the pseudo
rest-mass density for $a\rightarrow +\infty$ (see
Sec. \ref{sec_evol}). Since $\rho\ge\rho_{\rm min}$, only the exterior
branch of
the potential is accessible. For a complex SF, the potential is symmetric with
respect to the origin $|\varphi|=0$ and, by rotation around the vertical axis,
the exterior
branch defines a surface similar to the surface of a ``bowl'' (there is also
a central ``wall'' corresponding to the
interior branch). The
SF slowly descends the
potential on the surface of the bowl by rapidly spinning around the vertical
axis. We note that the SF does not reach the origin $|\varphi|=\rho=0$ because
of the presence of the
central wall. This is a particularity of the logotropic model. In the SF
representation of the 
$\Lambda$CDM model, there is no
central wall. In that case,
$\rho_{\rm min}=0$ and  the SF can reach the origin (see Appendix
\ref{sec_fcdm}).
We
also note that the modulus $|\varphi|$ of a complex SF does {\it not} oscillate,
contrary to
the case of a real SF. Only its phase $\theta$ oscillates. This corresponds to
the spintessence phenomenon described in Sec. \ref{sec_spin}. In this sense,
the evolution of a
complex SF is very different from the evolution of a real SF.

\subsection{Dimensionless variables}
\label{sec_tpn}

Introducing the dimensionless variables ${\tilde V}_{\rm tot}=V_{\rm
tot}/\rho_{\Lambda}c^2$ and
$\tilde\varphi=(m/\hbar)\varphi/\sqrt{\rho_{\Lambda}}$ in addition to those
defined previously, we can rewrite the total SF potential of the logotropic
model under the form
\begin{equation}
{\tilde V}_{\rm tot}=\frac{1}{2}|\tilde\varphi|^2-B\ln |\tilde\varphi|^2 
-B+1,
\label{tp8}
\end{equation}
or
\begin{equation}
{\tilde V}_{\rm tot}=\frac{1}{2}\tilde\rho-B\ln\tilde\rho-B+1,
\label{tp7}
\end{equation}
where
\begin{equation}
\tilde\rho=|\tilde\varphi|^2.
\label{tp6}
\end{equation}

\subsection{$\Lambda$CDM model ($B=0$)}
\label{sec_tpcdm}

For $B=0$, the foregoing equations reduce to 
\begin{equation}
{\tilde V}_{\rm tot}=\frac{1}{2}|\tilde\varphi|^2+1=\frac{1}{2}\tilde\rho+1.
\label{tp9}
\end{equation}
Coming back to the original variables, or taking the limit $A\rightarrow 0$
and $\rho_P\rightarrow +\infty$ with
$A\ln(\rho_P/\rho_\Lambda)\rightarrow \rho_{\Lambda}c^2$ fixed [see Eq.
(\ref{a2b})] in Eqs.
(\ref{tp4}) and (\ref{tp1}), we obtain 
\begin{equation}
V_{\rm
tot}(|\varphi|^2)=\frac{m^2c^2}{2\hbar^2}|\varphi|^2+\rho_{\Lambda}c^2=\frac{1}{
2}\rho c^2+\rho_{\Lambda}c^2.
\label{tp4k}
\end{equation}
We recover the  constant 
potential $V=\epsilon_{\Lambda}=\rho_{\Lambda}c^2$ of the complex SF associated
with the $\Lambda$CDM model that we call the $\Lambda$FDM model (see Appendix
\ref{sec_gfdm}).

\section{Validity of the fast oscillation regime (TF approximation) in cosmology}
\label{sec_v}

The previous results are valid in the fast oscillation regime of the complex
SF. We have seen that it corresponds to the TF approximation. Let us
determine the domain of validity of this approximation. The fast oscillation
regime is valid provided that $\omega\gg H$, where
$\omega=\dot\theta={\dot S}_{\rm tot}/\hbar=-E_{\rm tot}/\hbar$ is the
pulsation of the SF and $H=\dot a/a$ is the Hubble constant which is related to
the
energy density by the Friedmann equation $H^2=(8\pi
G/3c^2)\epsilon$. In terms of the
dimensionless variables introduced previously, the fast oscillation regime is
valid provided that
\begin{eqnarray}
\frac{\tilde\epsilon}{{\tilde E}_{\rm tot}^2}\ll \sigma, 
\label{v1}
\end{eqnarray} 
where
\begin{eqnarray}
\sigma=\frac{3m^2c^4}{8\pi G\hbar^2\rho_\Lambda}
\label{v2}
\end{eqnarray}
is a dimensionless parameter. It can be written as 
\begin{eqnarray}
\sigma=\left (\frac{m}{m_\Lambda}\right )^2,
\label{v3}
\end{eqnarray} 
where
\begin{equation}
m_\Lambda=\frac{\hbar}{c^2}\left (\frac{8\pi
G\rho_\Lambda}{3}\right
)^{1/2}=\frac{\hbar}{c^2}\sqrt{\frac{\Lambda}{3}}=1.20\times
10^{-33}\, {\rm
eV/c^2}
\label{v4}
\end{equation} 
is the cosmon mass.\footnote{This mass scale
is often interpreted as the smallest mass of the elementary particles predicted
by string theory \cite{axiverse} or as the upper bound on the mass
of the graviton \cite{graviton}. The mass $m_\Lambda$ also represents the
quantum of mass in
theories of extended supergravity \cite{tsujikawa}. The mass
scale $m_\Lambda$ is
simply obtained by equating the Compton wavelength of the particle
$\lambda_C=\hbar/mc$ with the Hubble radius $R_\Lambda=c/H_0$ (the typical size
of the visible Universe) giving $m_{\Lambda}=\hbar H_0/c^2\sim
\hbar\sqrt{\Lambda}/c^2$ (since $H_0^2\sim G\rho_\Lambda\sim\Lambda$). The mass
$m_\Lambda$ corresponds to Wesson's \cite{wesson} minimum mass interpreted as a
quantum of DE (Wesson's maximum mass $M_{\Lambda}=({4}/{3})\pi
({\epsilon_0}/{c^2})
R_{\Lambda}^3={c^3}/{2GH_0}=9.20\times 10^{55}\,
{\rm g}$ is of the order of the mass
of the Universe). These mass scales were also introduced in \cite{pdu}. B\"ohmer
and Harko \cite{bhcosmon} proposed to call the
elementary
particle of DE having the mass $m_\Lambda$ the ``cosmon''.  Cosmons were
originally introduced by Peccei {\it et al.} \cite{psw} to name SFs that could
dynamically adjust the cosmological constant to zero (see also
\cite{wcos,solacos1,solacos2}). The name
cosmon was also used in a different context \cite{ssdilaton} to designate a very
light scalar
particle (dilaton) of mass $\sim 10^{-3}\, {\rm eV/c^2}$ which could mediate new
macroscopic forces in the submillimeter range. } The fast oscillation regime
will be valid over a large period of
time provided that $\sigma\gg 1$, i.e.,  
\begin{equation}
m\gg m_\Lambda.
\label{v4b}
\end{equation} 
Therefore, the mass of the SF has to be much larger than the cosmon
mass.\footnote{In particular,
the validity of our approach requires that the mass of the SF is nonzero.}  The
mass of the boson required in the FDM
model to explain DM halos -- one of the smallest
particle mass quoted in the literature -- is of the order of $m_{22}=10^{-22}\,
{\rm eV/c^2}$ (see Appendix \ref{sec_gfdm}). For this value, we get
$\sigma_{22}=6.93\times 10^{21}\gg 1$ implying that the fast oscillation regime 
is valid over a large period of time. For future comparison, we note that the
criterion (\ref{v4b}) determining the validity of the fast oscillation regime
(or TF
approximation) in cosmology can also be written as
\begin{equation}
m\gg \frac{B\sqrt{8\pi G\hbar^2\rho_{\Lambda}^3}}{A},
\label{v4bb}
\end{equation} 
where we have used Eq. (\ref{impid5}).

\begin{figure}[!h]
\begin{center}
\includegraphics[clip,scale=0.3]{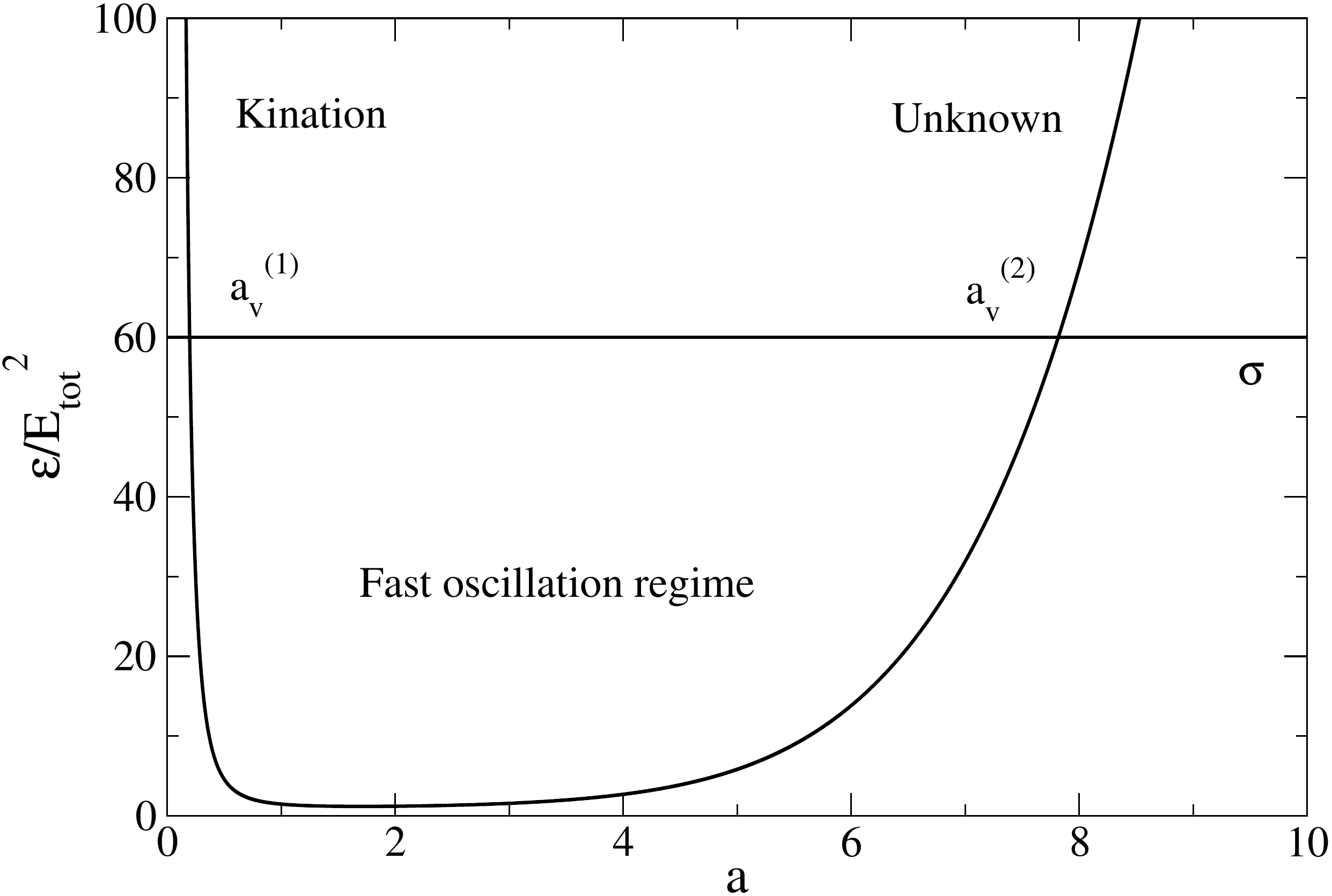}
\caption{Graphical construction determining the range of validity of the fast
oscillation regime in the logotropic model. }
\label{validity}
\end{center}
\end{figure}

In the logotropic model, the quantities $\tilde\epsilon$ and ${\tilde E}_{\rm
tot}$ are given as a function of the scale factor $a$ by Eqs. (\ref{dg2}),
(\ref{dg5}) and (\ref{dg7}). The curve
${\tilde\epsilon}/{\tilde E}_{\rm tot}^2(a)$ is plotted in Fig. \ref{validity}.
It presents a minimum value $({\tilde\epsilon}/{\tilde
E}_{\rm tot}^2)_{\rm min}=1.18$ at $a=1.71$. The condition
${\tilde\epsilon}/{\tilde E}_{\rm tot}^2<\sigma$ can be fulfilled provided that
$\sigma\ge \sigma_{\rm
min}=1.18$,
i.e., $m\ge 1.09\, m_{\Lambda}=1.30\times 10^{-33}\, {\rm
eV/c^2}$. When this condition is satisfied,
we find that
the fast oscillation regime is valid for
$a_v^{(1)}\ll a\ll a_v^{(2)}$, where
$a_v^{(1)}$ and $a_v^{(2)}$ are given by
\begin{eqnarray}
\frac{a_v}{a_t}=f\left
(\frac{3m^2c^4}{8\pi G\hbar^2\rho_\Lambda}\right )
\label{v5}
\end{eqnarray} 
with
\begin{eqnarray}
f(\sigma)=\frac{1}{r^{1/3}(1-2B/r)^{1/6}}
\label{v6}
\end{eqnarray} 
and
\begin{eqnarray}
\sigma=\frac{r-B\ln r+1-2B}{1-{2B}/{r}}.
\label{v7}
\end{eqnarray} 
We have introduced the transition scale factor $a_t$ from Eq. (\ref{vnr2}).
Equations (\ref{v6}) and (\ref{v7}) define the two-valued function $f(\sigma)$
in parametric form. When $\sigma\gg 1$, we
find that
\begin{eqnarray}
\frac{a_v^{(1)}}{a_t}\sim \frac{1}{\sigma^{1/3}}
\label{v8}
\end{eqnarray} 
and 
\begin{equation}
\frac{a_v^{(2)}}{a_t}\sim \frac{\sigma^{1/6}}{(2B)^{1/3}\left\lbrack
1-B\ln(2B)\right\rbrack^{1/6}}.
\label{v9}
\end{equation} 
For a SF of mass $m_{22}=10^{-22}\, {\rm eV/c^2}$, corresponding to
$\sigma_{22}=6.93\times 10^{21}$, we obtain $a_v^{(1)}=4.01\times 10^{-8}$ and
$a_v^{(2)}=1.73\times 10^4$. Therefore, the range of validity of the fast
oscillation regime is large.  For a larger mass $m$ of the SF, the range
of validity of the fast oscillation regime is even larger.

\begin{figure}[!h]
\begin{center}
\includegraphics[clip,scale=0.3]{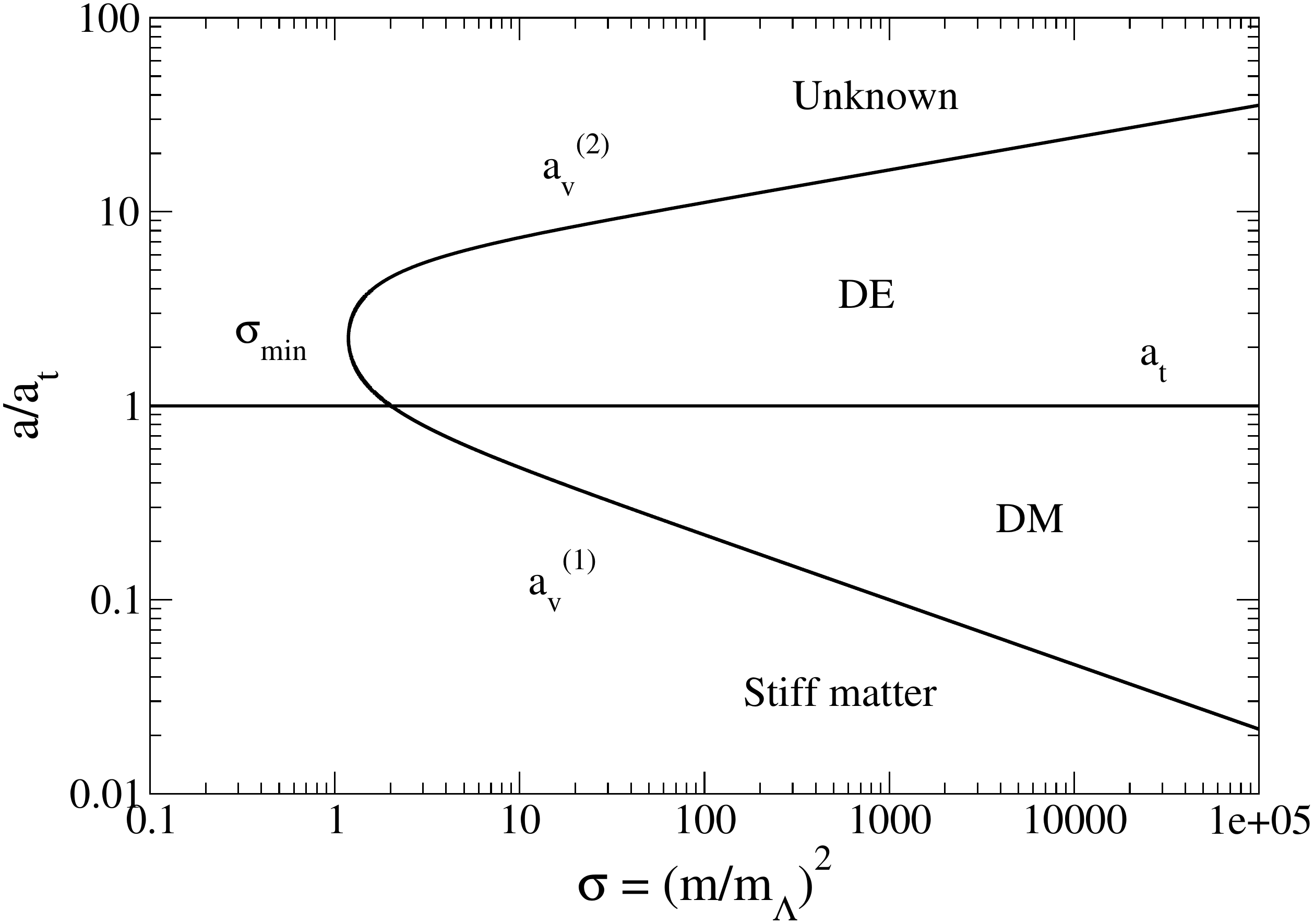}
\caption{Dynamical phase diagram of the logotropic model showing
the different eras experienced by the SF during the evolution of the Universe as
a function
of its mass $m$ (this figure also determines the validity
of the fast oscillation regime). We see how the fundamental cosmon mass
$m_{\Lambda}$ comes into play in the problem.}
\label{phasediag}
\end{center}
\end{figure}

According to the previous discussion, the fast
oscillation regime is valid for $m\gg m_{\Lambda}$ on the period $a_v^{(1)}\ll a\ll a_v^{(2)}$.
 During this period, we have seen that the SF behaves successively as
DM and DE. The transition
between the DM-like era and the DE-like era  corresponds to a scale
factor (see Sec. \ref{sec_vnr})\footnote{To define the transition
between the DM-like era and the DE-like era, we could have alternatively used
the
value
$a_c=0.607$ at which
the Universe starts accelerating.} 
\begin{equation}
a_t=0.765.
\label{v9b}
\end{equation} 
The fast oscillation regime is not valid at very early
times (i.e. $a<a_v^{(1)}$). In that case, the SF is in a slow oscillation
regime of kination (see  Sec.
\ref{sec_eosso}). As discussed in \cite{shapiro,abrilphas}, this gives rise to a stiff matter era. The stiff matter era usually takes place in the very
early Universe.  Therefore, $a_v^{(1)}$ marks the end of the stiff matter era
and the begining of the DM era.\footnote{If the SF has an additional
$|\varphi|^4$ self-interaction (see the Remark
below), a radiationlike era may be present between
the stiff matter era and the DM era \cite{shapiro,abrilphas}.} The logotropic SF
successively experiences a stiff matter era, a DM era and a DE era (this is also
the case for the $\Lambda$FDM model discussed in Appendix
\ref{sec_gfdm}). More surprisingly, the fast
oscillation regime ceases to be valid at very  late times (i.e. $a>a_v^{(2)}$).
This shows that quantum mechanics becomes important in the very late Universe.
In that case, we have to come back to the full set
of KGF equations, or their hydrodynamic representation  
\cite{abrilphas}, and take the terms in $\hbar$ into account (i.e., we have to
go beyond the TF approximation). Quantum mechanics will change the
results  derived on the basis of the fast oscillation (or TF)
approximation. Therefore, in the logotropic model, the very late Universe will
not remain in a de Sitter stage. It may
experience a stiff matter era again, or
another
(unknown) era, passing from a phase of
acceleration to a phase of deceleration. It should return
to a de Sitter stage ultimately as it falls in the bottom of the potential.
Note, by contrast, that the
fast oscillation regime is always valid at late times in the $\Lambda$FDM model
(see Appendix \ref{sec_gfdm}).

We can represent the previous results on a dynamical phase diagram
(see Fig. \ref{phasediag}) where we plot the transition scales $a_v^{(1)}$ and
$a_v^{(2)}$ as a function of the mass $m$ of the SF. For
$m>1.09\, m_\Lambda$, the logotropic complex SF undergoes four successive
eras: a stiff matter era for $a<a_v^{(1)}$, a DM era for $a_v^{(1)}<a<a_t$,
a DE era for $a_t<a<a_v^{(2)}$, and another (unknown) era for $a>a_v^{(2)}$.

\begin{figure}[!h]
\begin{center}
\includegraphics[clip,scale=0.3]{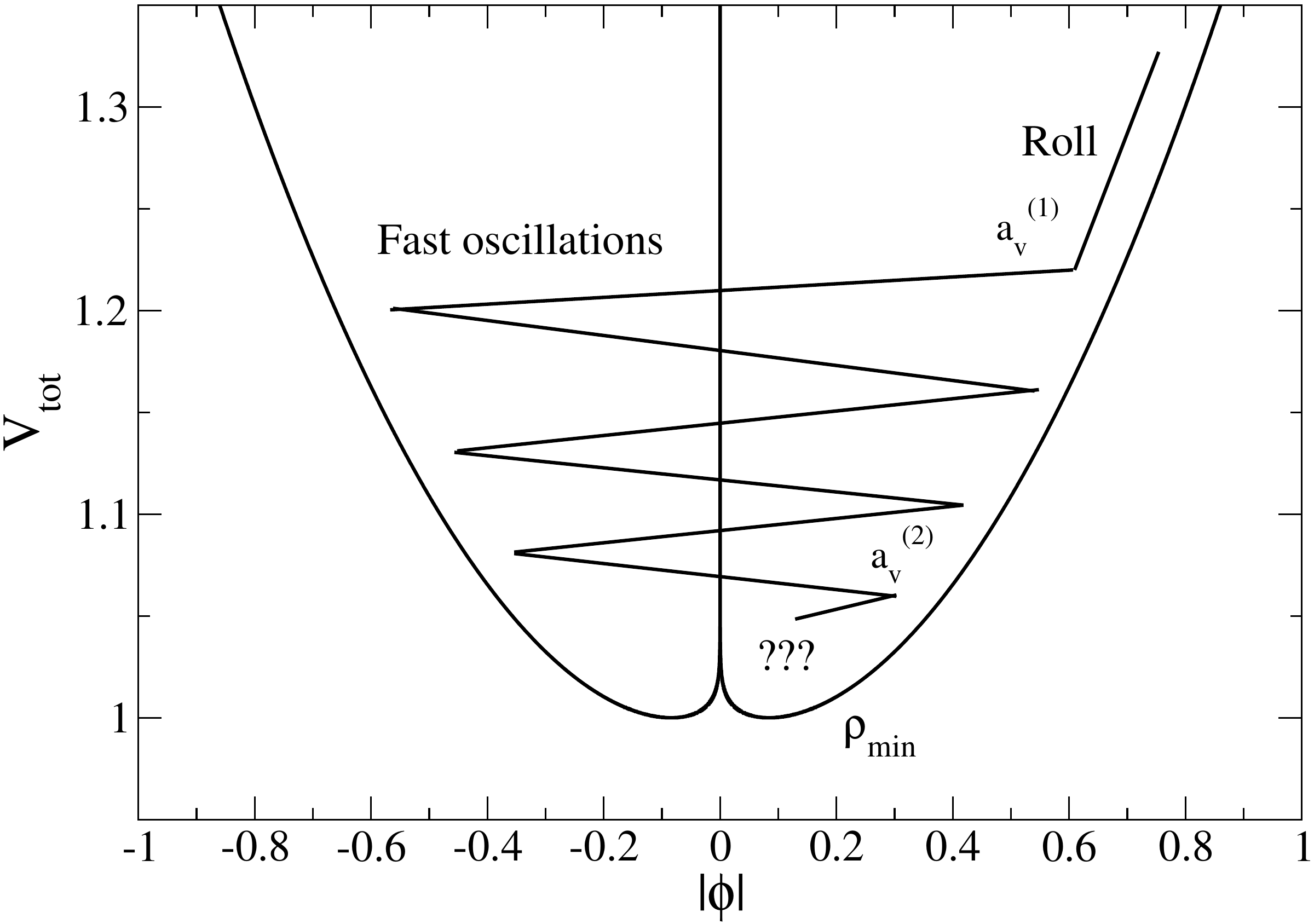}
\caption{Schematic evolution of the logotropic complex SF in the
total potential $V_{\rm tot}(|\varphi|^2)$ (the scales are not respected). For 
$a<a_v^{(1)}$, it rolls down the potential well without oscillating (stiff
matter
era); for $a_v^{(1)}<a<a_v^{(2)}$, it oscillates rapidly (DM and DE eras); for
$a>a_v^{(2)}$, it stops oscillating rapidly (its evolution remains to be
characterized in detail).}
\label{vtotzigzag}
\end{center}
\end{figure}

It is interesting, in parallel, to discuss how the complex SF
evolves in the potential $V_{\rm tot}(|\varphi|^2)$ during these different
periods. During the stiff matter era ($a<a_v^{(1)}$), corresponding to a slow
oscillation regime, the SF rolls down the potential well without
oscillating. Then,
for $a>a_v^{(1)}$, the SF enters in the fast oscillation regime and
descends the potential by oscillating rapidly about the vertical axis as
explained in Sec. \ref{sec_tp}. This covers the DM and DE eras. Finally, for
$a>a_v^{(2)}$, the SF stops oscillating rapidly again. Its detailed behaviour, which
corresponds to an evolution different from an exponential  (de Sitter)
expansion, is unknown.
This evolution -- roll versus oscillations -- is represented schematically in
Fig.
\ref{vtotzigzag}.

{\it Remark:} Combining the present results with those of
\cite{shapiro,abrilphas}, we can propose a more general complex SF model based
on
a potential of the form 
\begin{equation}
V_{\rm tot}(|\varphi|^2)=\frac{m^2c^2}{2\hbar^2}|\varphi|^2+\frac{2\pi a_s
m}{\hbar^2}|\varphi|^4-A\ln \left
(\frac{m^2|\varphi|^2}{\hbar^2\rho_P}\right )-A.
\label{tp4bh}
\end{equation}
This potential applies to a self-interacting relativistic Bose-Einstein
condensate (BEC) in the framework of
the logotropic model. Indeed, it includes a
$|\varphi|^4$ self-interaction potential, proportional to
the scattering length $a_s$ of the bosons, in addition to
the logarithmic potential. When $a_s>0$ (repulsive
self-interaction), the  $|\varphi|^4$ potential produces a radiationlike era in
the fast oscillation regime 
preceding the DM era (see  \cite{shapiro,abrilphas} for details). Therefore, 
a complex SF evolving in the potential defined by Eq. (\ref{tp4bh}) experiences
successively
a stiff matter
era, a (dark) radiationlike era, a DM era and a DE era (the case of an
attractive
self-interaction is more complicated \cite{abrilphas}). For $A\rightarrow
0$ and $\rho_P\rightarrow +\infty$ with
$A\ln(\rho_P/\rho_\Lambda)=\rho_{\Lambda}c^2$ fixed [see Eq.
(\ref{a2b})], the DE era is equivalent to a cosmological
constant (see
Appendix \ref{sec_gfdm}) and the potential from Eq. (\ref{tp4bh}) reduces to 
\begin{equation}
V_{\rm tot}(|\varphi|^2)=\frac{m^2c^2}{2\hbar^2}|\varphi|^2+\frac{2\pi a_s
m}{\hbar^2}|\varphi|^4+\epsilon_{\Lambda}.
\label{tp4bhj}
\end{equation}
It applies to a self-interacting  relativistic self-interacting Bose-Einstein
condensate (BEC) in the presence of a
cosmological constant.

\section{Analogies and differences between the logotropic model and the
$\Lambda$CDM model}
\label{sec_diff}

In this section, we compare the predictions of the logotropic and $\Lambda$CDM
models.

\subsection{Minimum energy density}
\label{sec_einstein}

Since the $\Lambda$CDM model corresponds to $B=0$ and since the predicted value
of  $B=3.53\times 10^{-3}$ is relatively small, we expect that the logotropic
model will not
differ substantially from the $\Lambda$CDM model regarding the
description of the
large scale structure of the Universe. This is a pre-requisit to any
viable cosmological model since the $\Lambda$CDM model works well at large
scales. Actually, the two models are almost
indistinguishable for what concerns the evolution of the
cosmological background up to the present epoch, and they will only slightly
differ in the far future. The two
models both tend to a constant energy density, ultimately leading to a de Sitter
era, but the values of this minimum energy density slightly differ.

In the  $\Lambda$CDM model,  the energy density tends, for $a\rightarrow
+\infty$, to
the Einstein cosmological density
\begin{eqnarray}
\epsilon_{\rm min}^{\Lambda{\rm CDM}}=\rho_{\Lambda}c^2,
\label{diff1}
\end{eqnarray} 
which is the constant density of DE. In the logotropic model, the energy density tends to the value
\begin{eqnarray}
\epsilon_{\rm min}^{\rm LDF}=\rho_{\Lambda}c^2\left\lbrack
1-B\ln(2B)\right\rbrack.
\label{diff2}
\end{eqnarray} 
Their ratio is
\begin{eqnarray}
\frac{\epsilon_{\rm min}^{\rm LDF}}{\epsilon_{\rm min}^{\Lambda{\rm CDM}}}=
1-B\ln(2B)=1.02.
\label{diff3}
\end{eqnarray} 
They differ by $2\%$.\footnote{Such a difference may be accessible
to the precision of modern cosmology. It would be interesting to
carefully compare the logotropic
model with the observations to see if it can relieve some tensions
experienced by the
$\Lambda$CDM model or, on the
contrary, if it increases them.} The
difference,
which is due to the nonzero value of $B$, may be interpreted as a quantum
correction to the Einstein cosmological constant (since
$B$ depends on $\rho_P$). 
 It is interesting to find a logarithmic
correction. Similar logarithmic corrections due to quantum effects arise in
particle physics and in the context of black hole thermodynamics.

{\it Remark:} We note that the present results  substantially differ from those
obtained in the framework of the original logotropic model developed in
\cite{epjp,lettre,jcap,pdu}. In these former works, we found that the
logotropic model is
indistinguishable from the $\Lambda$CDM model up to the present epoch but, at
later times, the energy density in the logotropic model increases logarithmically with the scale factor
(implying a phantom era) while in the $\Lambda$CDM model the energy density
always decreases and tends to
a constant. This leads to a super de Sitter behavior  instead of a standard de
Sitter behavior. It was shown in
\cite{epjp,lettre,jcap,pdu} that the two models would substantially differ in
about $25\,
{\rm Gyrs}$ when the logotropic Universe becomes phantom. The present
logotropic model does
not display a phantom behavior. It rather evolves towards a de Sitter era, like
the $\Lambda$CDM model, but with
a quantum modified cosmological constant. This may be an advantage of the new
logotropic model over the original one because phantom models are known to lead
to pathologies. By contrast, a model that tends 
to a de Sitter era is well-behaved. On the
other hand, in the original logotropic model \cite{epjp,lettre,jcap,pdu}, the
speed of sound  becomes
larger than the speed of light as we approach the phantom regime, then
becomes imaginary. We do not have
such anomalies in the present
model since the speed of sound is always real and smaller than the speed of
light ($0\le c_s\le c$). As discussed in Sec. \ref{sec_cs}, it increases from
$0$ to $c$ as the Universe expands. In comparison, the speed of sound is always
equal to zero in the $\Lambda$CDM model.

\subsection{Logotropic DM halos}
\label{sec_ldm}

As discussed in our previous papers \cite{epjp,lettre,jcap,pdu} (see
also Appendix \ref{sec_pldm}), the
main interest of the
logotropic model
with respect to the $\Lambda$CDM model becomes manifest when this model is
applied to DM
halos. When treating DM halos, one can use
Newtonian gravity. Furthermore, in
this section, we shall make the Thomas-Fermi approximation which amounts to
neglecting the quantum potential.\footnote{The domain of validity of the TF
approximation is discussed in Sec. \ref{sec_lwe}.} In
that case, the
equilibrium state of a logotropic DM halo results from the balance between
the gravitational attraction and the repulsion due to the pressure
force. It is described by the classical
equation of
hydrostatic equilibrium
\begin{eqnarray}
\nabla P+\rho\nabla\Phi={\bf 0}
\label{diff4}
\end{eqnarray}
coupled to the Poisson equation
\begin{eqnarray}
\Delta\Phi=4\pi G\rho.
\label{diff5}
\end{eqnarray} 
These equations  can be combined into a single differential equation
\begin{eqnarray}
-\nabla\cdot \left(\frac{\nabla
P}{\rho}\right )=4\pi G\rho,
\label{diff5b}
\end{eqnarray}
which determines, together with the equation of state (\ref{log2}), the density
profile of a logotropic DM halo.

In the framework of the $\Lambda$CDM model (see Appendix \ref{sec_lcdm}), the
pressure is zero ($P=0$) or
constant ($P=-\epsilon_{\Lambda}$) so there is no pressure gradient to balance
the gravitational attraction. This leads to cuspy density profiles. This
also leads to the formation of structures at all scales since the Jeans length
vanishes  owing to the fact that the
speed of sound is zero: $\lambda_J/2\pi=c_s/\sqrt{4\pi G\rho}=0$. These two
predictions of
the  $\Lambda$CDM model are in contradiction
with the observations which reveal that DM halos have a core instead of a
cusp
(core-cusp problem) and that there is no DM halo below a certain scale
of order $M\sim 10^8\, M_{\odot}$ and $R\sim 1\, {\rm kpc}$ (missing
satellite problem). The fact that the pressure, or pressure  gradient, vanishes
in the
$\Lambda$CDM model is the basic reason of the so-called CDM small-scale crisis.

In the framework of the logotropic model, the equation of state is given
by Eq. (\ref{log2}) where $\rho$ can be assimilated, in the nonrelativistic
regime, to the mass density. Since
the pressure is nonzero (and nonconstant),
the pressure gradient can balance the gravitational
attraction leading to cores instead of cusps.\footnote{We
introduced the logotropic model in \cite{epjp} by looking for the equation of
state that is the closest to a constant in order to have cored density profiles
at small (galactic) scales while producing the smallest deviation from the
$\Lambda$CDM model at large (cosmological) scales (see Appendix \ref{sec_mot}).}
The
structure of the logotropic
DM halos is studied in detail in Sec. 5 of Ref. \cite{epjp} (see
also Appendix \ref{sec_pldm}). Their density profile can
be obtained by numerically solving the Lane-Emden equation of index $n=-1$. It
presents a core for $r\rightarrow 0$ and decreases as $\rho\sim (A/8\pi
G)^{1/2}r^{-1}$ for
$r\rightarrow +\infty$. In addition,
the Jeans
length in the
logotropic model is nonvanishing and can account for the absence of
structures below a certain scale as discussed in  Sec. 6
of Ref. \cite{epjp} (see also Sec. \ref{sec_fw}). These results remain valid
in the present logotropic model because, in the nonrelativistic regime, the
equation of state (\ref{log2}) coincides with the equation of state studied in
our former
works \cite{epjp,lettre,jcap,pdu}. 

A remarkable result of the
logotropic model is to predict that all the DM halos (of any size)
have the same surface density $\Sigma_0=\rho_0 r_h$, where $\rho_0$ is the
central density and $r_h$ is the halo radius at which  the central density is
divided by $4$. Furthermore, the logotropic model predicts that this
universal surface density is given by 
\begin{eqnarray}
\Sigma_0^{\rm th}=\left (\frac{A}{4\pi G}\right )^{1/2}\xi_h=133\,
M_{\odot}/{\rm pc}^2,
\label{diff6}
\end{eqnarray} 
where $A$ is the fundamental constant of Eq. (\ref{a2}) and $\xi_h=5.8458...$ is
the dimensionless halo radius obtained by solving
the Lane-Emden equation of index $n=-1$ numerically (see Ref. \cite{epjp} and
Appendix \ref{sec_pldm}). It turns out
that the
theoretical value (\ref{diff6}) is in very good agreement with the value
$\Sigma_0^{\rm
obs}=\rho_0 r_h=141_{-52}^{+83}\, M_{\odot}/{\rm pc}^2$
obtained   from the observations \cite{donato}. This
is
remarkable because there is no
free (or
ajustable) parameter in our model. As discussed in Sec. \ref{sec_valf}, the
value of the logotropic constant $A$ is
determined by cosmological  considerations (large scales) while the result from
Eq. (\ref{diff6}) applies to DM halos (small scales). This suggests that there
is a
connection between the acceleration of the Universe and the universality of the
surface density of DM halos. They are both due to the logotropic constant $A$.
Indeed, the logarithmic
potential from Eq. (\ref{log0}) or the logotropic equation of state from Eq.
(\ref{log2})
accounts both
for the acceleration of the
Universe
and for the universality of the
surface density of DM halos.

{\it Remark:} We can write the universal surface density of DM halos given by
Eq. (\ref{diff6}) in terms of the Einstein cosmological constant $\Lambda$. Using
$A=B\rho_{\Lambda}c^2$ and
$\rho_{\Lambda}=\Lambda/(8\pi G)$, we get\footnote{Recalling that
$\rho_{\Lambda}$ represents 
the present density of DE, it may be more relevant to express $\Sigma_0^{\rm
th}$ in terms of the present value of the Hubble constant $H_0$. Using
$\Lambda=3(1-\Omega_{m,0})H_0^2$ obtained from Eqs. (\ref{hsf5}),
(\ref{impid6}) and (\ref{defrl}), we get
\begin{eqnarray}
\Sigma_0^{\rm th}=0.02815\frac{H_0
c}{G}.
\label{diff7b}
\end{eqnarray}
}
\begin{eqnarray}
\Sigma_0^{\rm th}=\left (\frac{B}{32}\right
)^{1/2}\frac{\xi_h}{\pi}\frac{c\sqrt{\Lambda}}{G}=0.01955\frac{c\sqrt{
\Lambda}}{G},
\label{diff7}
\end{eqnarray}
where we have used the numerical value of $B$ from Eq. (\ref{b5}). Recalling
that $B$ is given by Eq. (\ref{b2}) with $\rho_P/\rho_{\Lambda}=8\pi
c^5/\hbar G\Lambda$, we also have
\begin{eqnarray}
\Sigma_0^{\rm th}= 0.329  
\frac{\frac{c\sqrt{\Lambda}}{G}}{\sqrt{\ln\left(\frac{8\pi c^5}{\hbar
G\Lambda}\right )}}.
\label{diff8}
\end{eqnarray}
These identities express the universal surface density of DM halos in terms of
the fundamental constants of physics $G$, $c$, $\Lambda$, and $\hbar$. We
stress that the prefactors are also determined by our model. We note that the 
identities from Eqs. (\ref{diff7b})-(\ref{diff8}), which can be checked by a
direct numerical application, are interesting in themselves 
even in the case where the logotropic model would
turn out to be wrong. Furthermore, as observed in \cite{pdu}, the surface
density of DM halos is of the same order of magnitude as the
surface density of the electron. As a result, the identities from Eqs.
(\ref{diff7b})-(\ref{diff8}) allow us to express the mass of the electron in
terms of the cosmological constant and of the other fundamental constants of
physics as \cite{pdu}
\begin{eqnarray}
m_e\sim \left (\frac{\Lambda\hbar^4}{G^2c^2}\right )^{1/6}\quad {\rm
or}\quad m_e\sim \left (\frac{H_0\hbar^2}{Gc}\right )^{1/3},
\label{edd}
\end{eqnarray}
returning the Eddington-Weinberg relation \cite{eddington,weinbergbook}. This
provides a curious connection between microphysics
and macrophysics \cite{oufsuite}.

\section{Logotropic wave equations}
\label{sec_lwe}

The logotropic model developed in this paper is based on a complex SF theory
relying on the KG equation taking into account quantum effects ($\hbar\neq 0$).
In the
previous sections, we have neglected
quantum effects by making the TF approximation ($\hbar\rightarrow 0$). In this
section, we present more
general equations that are valid beyond this approximation.

\subsection{Logotropic  KG equation}
\label{sec_lkge}

For a spatially inhomogeneous complex SF, the  KGE equations read (see, e.g.,
\cite{playa})
\begin{equation}
\Box\varphi+\frac{m^2c^2}{\hbar^2}\varphi+2\frac{dV}{d|\varphi|^2}\varphi=0,
\label{lwe1}
\end{equation}
\begin{equation}
R_{\mu\nu}-\frac{1}{2}g_{\mu\nu}R=\frac{8\pi G}{c^4}T_{\mu\nu},
\label{lwe2}
\end{equation}
where $\Box$ is the d'Alembertian operator, $R_{\mu\nu}$ is the Ricci tensor,
$R$ is the Ricci scalar and 
\begin{eqnarray}
\label{lwe2b}
T_{\mu\nu}&=&\frac{1}{2}(\partial_{\mu}\varphi^*\partial_{\nu}\varphi+\partial_{
\nu}\varphi^*\partial_{\mu}\varphi)
\nonumber\\
&-&g_{\mu\nu}\left\lbrack
\frac{1}{2}g^{\rho\sigma}\partial_{\rho}\varphi^*\partial_{\sigma}
\varphi-V_{\rm tot}(|\varphi|^2)\right\rbrack
\end{eqnarray}
is the energy-momentum tensor of the
SF. For the logarithmic potential (\ref{log0}),
the wave equation (\ref{lwe1}) becomes
\begin{equation}
\Box\varphi+\frac{m^2c^2}{\hbar^2}\varphi-\frac{2A}{|\varphi|^2}\varphi=0.
\label{lwe3}
\end{equation}
This is the logotropic KG equation \cite{epjp}. 
This equation involves a nonlinear term, measured by the logotropic constant
$A$, which is
responsible for the late
acceleration of the Universe. In
Sec. \ref{sec_valf} we have interpreted $A$ as a
fundamental constant of physics superseding the Einstein cosmological constant.
Therefore, instead of introducing a cosmological constant
$\Lambda$ in the geometric part of the  equations of general
relativity, i.e. on the left hand side of Eq. (\ref{lwe3}), as Einstein does,
we
introduce  a new
fundamental constant $A$ directly in the wave equation (\ref{lwe3}). This is a
radically different point of view. We have seen in Sec.
\ref{sec_ldm} that this term accounts not only for the present acceleration of
the Universe but also for the universal surface density of
the DM halos. We cannot obtain this last result with the $\Lambda$CDM model.
Therefore,
our approach is substantially different from the $\Lambda$CDM model.

{\it Remark:} If we include a $|\varphi|^4$ self-interaction potential in
addition to the logarithmic potential in the complex SF potential [see Eq.
(\ref{tp4bh})], we obtain the generalized KG equation
\begin{equation}
\Box\varphi+\frac{m^2c^2}{\hbar^2}\varphi+\frac{8\pi
a_s m}{\hbar^2}|\varphi|^2\varphi-\frac{2A}{|\varphi|^2} \varphi=0.
\label{lwe3rad}
\end{equation}

\subsection{Logotropic GP equation} 
\label{sec_fws}

In the nonrelativistic limit $c\rightarrow +\infty$, using the Klein
transformation,
\begin{eqnarray}
\label{klein}
\varphi({\bf r},t)=\frac{\hbar}{m}e^{-imc^2t/\hbar}\psi({\bf r},t),
\end{eqnarray}
the KGE equations
(\ref{lwe1}) and (\ref{lwe2}) reduce
to the GPP equations\footnote{We consider here a static
background ($a=1$) since we will discuss these equations in the context of DM
halos where the expansion of the Universe can be neglected.} (see, e.g.,
\cite{playa})
\begin{eqnarray}
i\hbar\frac{\partial\psi}{\partial
t}=-\frac{\hbar^2}{2 m }\Delta\psi+m\Phi \psi
+m\frac{dV}{d|\psi|^2}\psi,
\label{lwe4}
\end{eqnarray}
\begin{eqnarray}
\Delta\Phi=4\pi G|\psi|^2, 
\label{lwe5}
\end{eqnarray}
where $\psi$ is the wavefunction such that $\rho=|\psi|^2$ represents the mass
density. For the logarithmic potential (\ref{log1}), the
nonrelativistic wave equation (\ref{lwe4}) becomes
\begin{eqnarray}
i\hbar \frac{\partial\psi}{\partial
t}=-\frac{\hbar^2}{2m}\Delta\psi+m\Phi\psi-\frac{Am}{|\psi|^2}\psi.
\label{lwe6}
\end{eqnarray} 
This is the logotropic GP equation \cite{epjp}. For $A=0$ we recover the
Schr\"odinger-Poisson equations which correspond to the FDM model (see Appendix
\ref{sec_fdmw}).

{\it Remark:} If we include a $|\psi|^4$ self-interaction potential in
addition to the logarithmic potential in the complex SF potential [see Eq.
(\ref{tp4bh})], we obtain the generalized GP equation
\begin{equation}
i\hbar \frac{\partial\psi}{\partial
t}=-\frac{\hbar^2}{2m}\Delta\psi+m\Phi\psi
+\frac{4\pi a_s\hbar^2}{m^2}|\psi|^2\psi
-\frac{Am}{|\psi|^2}\psi.
\label{lwe6red}
\end{equation}

\subsection{Madelung transformation} 
\label{sec_mad}

Writing the wave function as
\begin{equation}
\label{mad1}
\psi({\bf r},t)=\sqrt{{\rho({\bf r},t)}} e^{iS({\bf r},t)/\hbar},
\end{equation}
where $S({\bf r},t)$ is the action,
and making the Madelung \cite{madelung} transformation
\begin{equation}
\label{mad2}
{\bf u}=\frac{\nabla S}{m},
\end{equation}
where ${\bf u}({\bf r},t)$ is the velocity field, the GPP
equations (\ref{lwe4})-(\ref{lwe6}) can be written under the form of
hydrodynamic equations
\begin{equation}
\label{mad3}
\frac{\partial\rho}{\partial t}+\nabla\cdot (\rho {\bf u})=0,
\end{equation}
\begin{equation}
\label{mad4}
\frac{\partial {\bf u}}{\partial t}+({\bf u}\cdot
\nabla){\bf
u}=-\frac{1}{m}\nabla
Q_B-\frac{1}{\rho}\nabla P-\nabla\Phi,
\end{equation}
\begin{equation}
\label{mad5}
\Delta\Phi=4\pi G \rho,
\end{equation}
where
\begin{equation}
\label{mad6}
Q_B=-\frac{\hbar^2}{2m}\frac{\Delta
\sqrt{\rho}}{\sqrt{\rho}}=-\frac{\hbar^2}{4m}\left\lbrack
\frac{\Delta\rho}{\rho}-\frac{1}{2}\frac{(\nabla\rho)^2}{\rho^2}\right\rbrack
\end{equation}
is the Bohm quantum potential taking into account the Heisenberg uncertainty
principle and $P(\rho)$ is the pressure determined by Eq. (\ref{ge8}). For the
logarithmic potential (\ref{log1}), we obtain the logotropic equation
of state (\ref{log2}).

{\it Remark:} If we include a $|\varphi|^4$ self-interaction potential in
addition to the logarithmic potential in the complex SF potential [see Eq.
(\ref{tp4bh})], we need to account for an additional pressure term
\begin{equation}
P=\frac{2\pi a_s\hbar^2}{m^3}\rho^2
\label{??????}
\end{equation}
in the quantum Euler equation (\ref{mad4}).

\subsection{Condition of quantum  hydrostatic
equilibrium} 
\label{sec_madeq}

The condition of quantum  hydrostatic
equilibrium is expressed by the equation
\begin{eqnarray}
\label{lwe8}
\frac{\rho}{m}\nabla
Q_B+\nabla P+\rho\nabla\Phi={\bf 0}
\end{eqnarray}
coupled to the Poisson equation
\begin{eqnarray}
\Delta\Phi=4\pi G\rho.
\label{lwe9}
\end{eqnarray}
These equations describe the balance between the repulsion due to the quantum
potential, the repulsion due to the logotropic pressure, and the gravitational
attraction. 
In the TF approximation where the quantum
potential can be neglected, we recover the classical condition of hydrostatic
equilibrium (\ref{diff4}). This leads to classical
logotropic DM
halos such as those studied in Sec. 5 of \cite{epjp} and in Appendix
\ref{sec_pldm}. However, in the general
case ($Q_B\neq 0$), Eq.
(\ref{lwe8}) implies that
logotropic DM halos have, like in the FDM model (see
Appendix \ref{sec_fdmw}), a quantum core (soliton) in which the pressure is
provided
by the Heisenberg uncertainty principle. This quantum
core is surrounded 
by a logotropic envelope where the density decreases as $r^{-1}$.

\subsection{Generalized Lane-Emden equation}
\label{sec_glee}

Combining Eqs. (\ref{lwe8}) and (\ref{lwe9}), and using Eq.  (\ref{mad6}), we
obtain the fundamental
differential equation of quantum hydrostatic equilibrium 
\begin{eqnarray}
\frac{\hbar^2}{2m^2}\Delta\left (\frac{\Delta\sqrt{\rho}}{\sqrt{\rho}}\right
)-\nabla\cdot \left(\frac{\nabla P}{\rho}\right )=4\pi G\rho.
\label{glm1}
\end{eqnarray}
This equation determines the density profile of
BECDM halos
described by the GPP
equations.\footnote{More precisely, Eq. (\ref{glm1}) determines
the ground state of a self-gravitating BEC. This solution describes ultracompact DM
halos -- dwarf spheroidals (dSphs) like Fornax -- or the quantum core (soliton)
of large DM halos. In large DM halos, the soliton is surrounded by an extended
envelope 
which arises from the quantum interferences of excited states \cite{wignerPH}.
On a
coarse-grained scale, this envelope has a
structure similar to the  NFW profile (see Appendix \ref{sec_fdmw}).}
For the logotropic equation of state
(\ref{log2}), it
becomes
\begin{eqnarray}
\frac{\hbar^2}{2m^2}\Delta\left (\frac{\Delta\sqrt{\rho}}{\sqrt{\rho}}\right
)+A\Delta\left(\frac{1}{\rho}\right )=4\pi G\rho.
\label{glm2}
\end{eqnarray}
If we define
\begin{equation}
\label{glm3}
\theta=\frac{\rho_0}{\rho},\qquad \xi=\left (\frac{4\pi
G\rho_0^2}{A}\right )^{1/2}r,
\end{equation}
where $\rho_0$ is the central density, we find that Eq. (\ref{glm2}) takes
the form of a
generalized Lane-Emden equation
\begin{equation}
\label{glm4}
\chi\Delta\left
(\frac{\Delta\theta^{-1/2}}{\theta^{-1/2}}\right
)+\Delta\theta=\frac{1}{\theta}
\end{equation}
with a quantum coefficient
\begin{equation}
\label{glm5}
\chi=\frac{2\pi G\hbar^2\rho_0^3}{m^2A^2}.
\end{equation}
In the TF approximation $\chi\ll 1$, Eq. (\ref{glm4}) reduces to the usual
Lane-Emden equation of index $n=-1$ (see Ref. \cite{epjp} and Eq.
(\ref{lel3})).

\subsection{Validity of the TF approximation for DM halos}
\label{sec_tfdm}

The TF
approximation for DM halos is valid when $\chi\ll 1$, i.e., when
\begin{equation}
\label{glm7}
m\gg m_0\equiv \frac{\sqrt{2\pi G\hbar^2\rho_0^3}}{A}.
\end{equation}
If we consider an ultracompact DM halo of typical density $\rho_0\sim  10^8\,
M_{\odot}/{\rm
kpc}^3$ (Fornax), we find that $m_0=3.57\times 10^{-22}\, {\rm eV/c^2}$.
Remarkably, this mass scale is precisely of the same order of magnitude as the mass $m\sim 10^{-22}\, {\rm eV/c^2}$ of the
ultralight boson that
occurs in the
FDM model (see Appendix \ref{sec_gfdm}).\footnote{We note that
the criterion
$m\gg m_0=3.57\times 10^{-22}\, {\rm eV/c^2}$ determining the validity of the TF
approximation at the scale of DM halos differs by $11$ orders of magnitude  from
the 
criterion $m\gg m_{\Lambda}=1.20\times
10^{-33}\, {\rm
eV/c^2}$ determining the validity of the TF
approximation at the cosmological level (see Sec. \ref{sec_v}).} The mass $m$ of
the SF
determines the importance of
the quantum core (soliton) relative to the logotropic envelope in a DM halo. When $m\ll
m_0$, the DM halo is dominated by the quantum core, like in the FDM model, and
the logotropic envelope is negligible. Inversely, in the
TF approximation $m\gg m_0$, there is no quantum core. In that case, the DM halo
is
purely logotropic and the mass of the SF disappears from the equations (see
Sec. \ref{sec_ldm} and Appendix \ref{sec_pldm}). When
$m\sim m_0$, we have to take into account both the presence of the quantum
core (soliton) and the
logotropic envelope. This is the case in particular for the ultralight boson  of mass  $m\sim 10^{-22}\, {\rm eV/c^2}$  that
occurs in the
FDM model.

{\it Remark:} Using the fact that $\rho_0=k\rho_{\Lambda}$ with $k\sim 
10^6$ and $A=B\rho_{\Lambda}c^2$, we find that
\begin{equation}
\label{glm7g}
m_0=\frac{\sqrt{3}\, k^{3/2}}{2B}m_\Lambda,
\end{equation}
where $m_\Lambda=1.20\times 10^{-33}\, {\rm
eV/c^2}$ is the cosmon mass [see Eq. (\ref{v4})]. We get
$m_0\sim 3\times 10^{11}
m_\Lambda$. Therefore, the mass scale $m_0$ is equal to the cosmon mass
multiplied by a large prefactor.

\subsection{Interpretation of the logotropic term}

There are two manners to interpret the logotropic term in
Eqs. (\ref{lwe3}) and (\ref{lwe6}). Naively, we could  interpret this term as
a property of the SF measuring,
for example, the strength of its self-interaction. However, since $A$ is a
fundamental constant of physics rather than being a property of the SF like its mass $m$ or
its scattering length $a_s$,
it is more relevant to interpret
this term as an {\it intrinsic} term, independent of the SF, that is always
present in
the wave equation. In many situations,
this term is negligible and we recover the standard KG and Schr\"odinger
equations. However, when considering galactic or cosmological scales, this term
becomes important and is responsible for
the
accelerating expansion of the Universe (DE) and for the
universal surface density of the DM halos.\footnote{We note that the logotropic
term $A/(|\psi|^2 c^2)\sim \rho_{\Lambda}/\rho$ becomes important at very low
densities, typically when $\rho$ becomes
comparable to the cosmological density $\rho_{\Lambda}=5.96\times 10^{-24}
{\rm g}\, {\rm m}^{-3}$ which is the absolute minimum density in the universe.
At higher
densities, the logotropic term is negligible because the value of $A/c^2\sim
\rho_{\Lambda}$ is extremely small. This forces us to properly define what
we call ``vacuum''. For example, a density $\rho_{\rm lab}$ may look small at
the laboratory scale although it is much larger than $\rho_{\Lambda}$.
Therefore, we should not take $\rho_{\rm lab}=0$  in Eqs. (\ref{lwe3}) and
(\ref{lwe6}) because
that would make the  logotropic term $A/(\rho_{\rm lab}c^2)$ diverge while in
reality this term is negligible. If we
interpret $\rho_{\Lambda}$ as typically representing the
smallest possible value of the density in the Universe [in line with Eq.
(\ref{evo10})], then the logotropic term is
always less than unity.} It can therefore account for the
effects of DM and DE in a unified manner.  We suggest therefore that Eqs.
(\ref{lwe3}) and (\ref{lwe6}) could be fundamental equations of physics  superseding the
standard KG and Schr\"odinger equations. We note that 
these wave equations are nonlinear.  In
this point of view, the standard KG and Schr\"odinger equations appear as
{\it approximations} of the more general nonlinear wave equations (\ref{lwe3})
and
(\ref{lwe6}).

\subsection{Analogies and differences between the logotropic model and the $\Lambda$FDM model}
\label{sec_kg}

The  $\Lambda$FDM model (see
Appendix  \ref{sec_gfdm}) is based on a complex SF with a constant
potential
$V(|\varphi|^2)=\epsilon_{\Lambda}$ equal to the cosmological density. In
that case, the relativistic wave equation (\ref{lwe1}) reduces to the standard
KG equation (\ref{hlwe1})
and the nonrelativistic wave equation (\ref{lwe4}) reduces to the standard
Schr\"odinger
equation (\ref{hlwe4}) like in the FDM model. Therefore, the constant SF
potential $V(|\varphi|^2)=\epsilon_{\Lambda}$  does not explicitly appear in the fundamental wave equations of quantum
mechanics since only the derivative of $V$ matters. However, the constant potential  $V(|\varphi|^2)=\epsilon_{\Lambda}$  appears in the energy density and in the pressure of
the SF [see Eqs. (\ref{hsf2}) and (\ref{hsf3})]. In the fast oscillation
regime, a homogeneous complex SF with a constant potential
$V(|\varphi|^2)=\epsilon_{\Lambda}$ behaves as a gas with a constant pressure (see Appendix \ref{sec_vcdm})
\begin{equation}
P=-\epsilon_{\Lambda}.
\label{pco}
\end{equation}
As a result, it is equivalent to the
$\Lambda$CDM model and can therefore account for the accelerating expansion of
the Universe and to the clustering of DM.\footnote{A complex SF with a
constant potential
$V(|\varphi|^2)=\epsilon_{\Lambda}$, corresponding to the $\Lambda$FDM
model, provides a simple unification of DM and DE. By contrast, a complex SF
with a vanishing potential
$V(|\varphi|^2)=0$, corresponding to the FDM model, has a vanishing pressure ($P=0$) in the fast
oscillation regime and behaves only as DM.} If we apply this model to DM halos
and ignore quantum effects (TF approximation),  we
recover the small-scale problems of the CDM model. Indeed, since the pressure is
uniform [see Eq. (\ref{pco})], there is no pressure gradient to balance the
gravitational attraction. This leads to cuspy density profiles. However, if
we take quantum effects into account (see Appendix \ref{sec_fdmw}) the quantum potential can stabilize the system against
gravitational collapse and produce a core instead of a cusp. At the level of DM halos, a
complex SF
with a constant potential is equivalent to the FDM
model which can possibly solve the small-scale crisis of CDM.

There remains, however, an important problem with this model. Indeed, the FDM
model, unlike the logotropic
model, does not account for the universal surface density of DM halos. In the
FDM model, the core mass-radius relation scales as $M\sim \hbar^2/(G m^2 R)$
(see
Appendix \ref{sec_fdmw}) and, consequently, the surface density $\Sigma\sim
M/R^2$ of DM halos scales with the radius as
\begin{eqnarray}
\Sigma\propto \frac{\hbar^2}{G m^2 R^3}.
\label{lwe9cc}
\end{eqnarray}
Therefore, the surface density of FDM halos decreases as the size of the DM
halos increases instead of being constant.
Correspondingly, the mass of the
FDM halos decreases as their radius increases. This is in sharp contrast with
the observations of DM halos which reveal that their mass increases with their
radius as $M\propto R^2$ in agreement with a constant surface
density ($\Sigma\sim 1$) \cite{donato}.

The problems of the FDM model were mentioned by the author at
several occasions (see, e.g.,
Appendix F of Ref. \cite{clm2}, the
Introduction of Ref. \cite{chavtotal} and
Appendix L of \cite{modeldm}) and they have
been recently emphasized by
Burkert \cite{burkertfdm} and Deng {\it et al.} \cite{deng}. These are
serious drawbacks of
the FDM model.\footnote{The fermionic DM model and the BECDM
model with
a repulsive self-interaction experience the same problems (see
Appendix L of \cite{modeldm}).} It has been advocated
that these
problems could be
solved by taking into account
the effect of an isothermal halo and distinguishing between the quantum core
radius $R_c$ and the isothermal core radius $r_0$ (see Ref. \cite{modeldm}
and Appendix \ref{sec_cqmdm} for more details).
Alternatively,
we note that the logotropic model based on the nonlinear Schr\"odinger
equation (\ref{lwe6}) does not suffer from the problems of
the FDM model based on the usual Schr\"odinger equation (\ref{hlwe4}) since it
leads, 
in the TF approximation, to a constant surface density $\Sigma\sim 1$ and a
$M\propto
R^2$ mass-radius relation in agreement with the observations (see Sec.
\ref{sec_ldm}),
unlike the
FDM model.\footnote{This remark suggests that the DM halos should be in the TF
regime so that they are dominated by the logotropic profile, not by the
solitonic profile. According to the criterion
from Eq. (\ref{glm7}), this implies that $m\gg m_0$
with $m_0 \sim 10^{-22}\, {\rm eV/c^2}$.}

Finally, we expect that the logotropic GPP equations (\ref{lwe5}) and
(\ref{lwe6}), similarly to the Schr\"odinger-Poisson equations (\ref{hlwe4})
and (\ref{hlwe9}) of the FDM model, undergo a process of
violent relaxation and gravitational cooling (see Appendix \ref{sec_fdmw}). 
This should lead, in the general case, to DM halos possessing a  quantum core
(soliton) $+$ an inner logotropic envelope whose density
decreases as $r^{-1}$  (yielding a universal surface density) $+$ an outer
envelope with a
density profile decreasing as
$r^{-3}$ (consistent with the NFW profile). In the TF regime valid when $m\gg
m_0$, the quantum core should be replaced by a classical logotropic core. The resulting
structure made of a logotropic core $+$ a NFW halo turns out to be in agreement with the observed structure of DM halos.

\section{Jeans instability in a logotropic Universe}
\label{sec_fw}

In this section, we study the Jeans instability of a spatially homogeneous
self-gravitating logotropic gas in the  
expanding Universe. We use a nonrelativistic
approach\footnote{In principle, the nonrelativistic approximation is valid for
$a\ll a_t=0.765$ (see Sec. \ref{sec_vnr}). Since our discussion is essentially
qualitative,
we shall extrapolate our nonrelativistic results up to 
the present Universe ($a=1$).} and make
the TF approximation which amounts to neglecting quantum effects.\footnote{The
validity of the TF approximation for the Jeans problem
is discussed in Sec. \ref{sec_vtf}. A more
general study going beyond the TF approximation will be reported in a
forthcoming paper \cite{prep}.}  This approximate treatment will be sufficient
to
point out important problems encountered by the logotropic model regarding
the formation of the  large-scale  structures of the Universe.

\subsection{The Jeans scales}
\label{sec_jsl}

We first study how the Jeans length $\lambda_J$ 
and the Jeans mass $M_J$ of the logotropic gas depend on the density of
the Universe $\rho$.
In the nonrelativistic regime (DM-like era) the
density evolves with time as \cite{suarezchavanisprd3}
\begin{equation}
\label{jsl1}
\frac{\rho}{\rm g/m^3}=2.25\times 10^{-24}\, a^{-3},
\end{equation}
where $a$ is the scale factor. The beginning of
the nonrelativistic regime  which can be identified with the epoch
of matter-radiation equality (i.e. the transition between the radiation era and
the DM era) occurs at $a_{\rm
eq}=2.95\times 10^{-4}$ (corresponding to a redshift $z_{\rm
eq}=1/a_{\rm eq}-1=3390$). At
that moment, the density of the universe 
is
$\rho_{\rm eq}=8.77\times 10^{-14}\, {\rm g}/{\rm m}^3$. The present density of
the universe  is
$\rho_0=2.25\times
10^{-24}\, {\rm g/m^3}$.

In the nonrelativistic $+$ TF approximation, the Jeans wavenumber
$k_J$ is given by \cite{prd1}
\begin{eqnarray}
k_J=\left (\frac{4\pi G\rho}{c_s^2}\right )^{1/2},
\label{jsl2}
\end{eqnarray}
where $c_s^2=P'(\rho)$ is the squared speed of sound. The Jeans length
is
$\lambda_J=2\pi/k_J$ and the
comoving Jeans length is $\lambda^c_J=\lambda_J/a$.  The Jeans radius and the Jeans mass are defined
by
\begin{eqnarray}
R_J=\frac{\lambda_J}{2},\qquad M_J=\frac{4}{3}\pi\rho R_J^3.
\label{jsl3}
\end{eqnarray}
They represent the
minimum radius and the minimum mass of a fluctuation
that can collapse at a given
epoch. They are therefore expected to provide an order of magnitude 
of the minimum size and minimum mass of DM halos.

For the logotropic equation of state
\begin{eqnarray}
P=A\ln\left (\frac{\rho}{\rho_P}\right ),
\label{jsl4}
\end{eqnarray}
the squared speed of sound reads
\begin{eqnarray}
c_s^2=P'(\rho)=\frac{A}{\rho}.
\label{jsl4b}
\end{eqnarray}
The speed of sound increases as the density decreases. The Jeans length and the Jeans mass are given by
\begin{eqnarray}
\lambda_J=2\pi \left (\frac{A}{4\pi G}\right )^{1/2}\frac{1}{\rho},
\label{jsl5}
\end{eqnarray}
\begin{eqnarray}
M_J=\frac{4}{3}\pi^4 \left (\frac{A}{4\pi G}\right )^{3/2}\frac{1}{\rho^2}.
\label{jsl6}
\end{eqnarray}
They can be written as
\begin{eqnarray}
\frac{\lambda_J}{\rm pc}=9.67\times 10^{-15}\, \frac{\rm
g/m^3}{\rho},
\label{jsl7}
\end{eqnarray}
\begin{eqnarray}
\frac{M_J}{M_{\odot}}= 6.99\times 10^{-27}\, \left
(\frac{\rm g/m^3}{\rho}\right )^{2}.
\label{jsl8}
\end{eqnarray}
Using Eq. (\ref{jsl1}), we find that during the expansion of the Universe the
Jeans length
increases as
$a^{3}$ and the
Jeans mass increases as $a^{6}$ (the comoving
Jeans length increases as
$a^{2}$). Eliminating the
density between  Eqs.
(\ref{jsl5}) and (\ref{jsl6}), we obtain 
\begin{eqnarray}
M_J=\frac{\pi^2}{3} \left (\frac{A}{4\pi G}\right )^{1/2}\lambda_J^2.
\label{jsl9}
\end{eqnarray}
This relation  is similar to the mass-radius
relation $M_h(r_h)$  of logotropic DM halos (see Appendix \ref{sec_pldm}).

At the epoch of matter-radiation equality, we find
$\lambda_J=0.110\, {\rm pc}$ and
$M_J=0.910\, M_{\odot}$ (the comoving Jeans length
is $\lambda_J^c=\lambda_J/a=374\, {\rm pc}$).\footnote{The Jeans mass computed
at the
epoch of matter-radiation equality where structures start to
form gives a lower
bound on the mass of the DM halos observed today. Indeed, the Jeans instability
leads to clumps of mass $M_J$ and size $\lambda_J$. These clumps can merge to
form bigger structures but, in general, their mass cannot decrease.}
 In the case of CDM where $c_s=0$, the Jeans length and the Jeans
mass vanish. Therefore, structures can form at all scales. This is in
contradiction with the observations which reveal that DM halos exist only above
a
minimum size $R\sim 1\, {\rm kpc}$ and above a minimum mass $M\sim 10^8\,
M_{\odot}$
corresponding to typical dSphs. In the framework of the logotropic model, DM
halos can form only above $\lambda_J=0.110\, {\rm pc}$ and
$M_J=0.910\, M_{\odot}$.  The logotropic model implies the existence of a
``minimum halo''  but the size and  mass of this minimum halo are much too small
to
solve the missing satellite problem. We shall
come back to this problem in Sec. \ref{sec_jeansfdm}.

At the present epoch, we find
$\lambda_J=4.30\times 10^3\, {\rm Mpc}$ and
$M_J=1.38\times 10^{21}\, M_{\odot}$.  These values are of the order of the size
and mass of the Universe (see below). Therefore,
the
Jeans instability is inhibited in the present Universe even at very large
scales, i.e., at the scale of the clusters of galaxies. We shall come
back to this problem in Sec. \ref{sec_pom}.

{\it Remark:} We can rewrite the Jeans length (\ref{jsl5})
and the Jeans mass (\ref{jsl6}) as 
\begin{eqnarray}
\lambda_J=2\pi\left\lbrack \frac{2B(1-\Omega_{\rm m,0})}{3\Omega_{\rm
m,0}^2}\right\rbrack^{1/2} R_{\Lambda}a^3,
\label{klj1}
\end{eqnarray}
\begin{eqnarray}
M_J=\pi^3\left\lbrack \frac{2B(1-\Omega_{\rm m,0})}{3\Omega_{\rm
m,0}^2}\right\rbrack^{3/2} \Omega_{\rm m,0} M_{\Lambda}a^6,
\label{klj2}
\end{eqnarray}
where $R_{\Lambda}=c/H_0=4.44\times 10^3\, {\rm Mpc}$ is the size of the
visible Universe and $M_{\Lambda}=(4/3)\pi
(\epsilon_0/c^2)R_{\Lambda}^3=c^3/2GH_0=4.62\times
10^{22}\, M_{\odot}$ is its
mass. To obtain Eqs. (\ref{klj1}) and (\ref{klj2}), we have used
$\rho=\Omega_{\rm
m,0}(\epsilon_0/c^2)a^{-3}$, $H_0^2=(8\pi
G/3c^2)\epsilon_0$ and Eqs. (\ref{impid5}) and (\ref{impid6}). We see more
clearly on these expressions that the present values of the Jeans length and
Jeans mass are of the order of the size and mass of the Universe. This is due to
the fact 
that the speed of sound approaches the speed of light ($c_s\sim c$) when
$\rho\rightarrow \rho_\Lambda$. As a result, the Jeans length
$\lambda_J\sim c_s/\sqrt{G\rho_0}$ with $H_0^2=8\pi
G\rho_0/3$ becomes comparable to the Hubble length $\lambda_H=c/H$ (horizon)
and this
prevents the formation of structures (see below).\footnote{For the same reason,
structure formation is impossible during the radiation era where
$c_s=c/\sqrt{3}$.}

\subsection{Theory of perturbations in the linear regime}
\label{sec_edc}

In the nonrelativistic $+$ TF approximation, the equation determining the
evolution of the density contrast $\delta_k=\delta\rho_{k}/\rho$ in the linear regime of structure formation is
given by \cite{aacosmo}
\begin{eqnarray}
\frac{d^2\delta_k}{da^2}+\frac{3}{2a}\frac{d\delta_k}{da}+\frac{3}{2a^2}
\left (\frac{c_s^2k^2}{4\pi G\rho
a^2}
-1\right )\delta_k=0,
\label{edc2}
\end{eqnarray}
where  $c_s^2=P'(\rho)$ is the squared speed of sound from  Eq. (\ref{jsl4b}). For the logotropic
equation of state, the comoving Jeans wavenumber
is
\begin{equation}
k_J^c=\left (\frac{4\pi G\rho
a^2}{c_s^2}\right )^{1/2}=\left (\frac{4\pi G\rho^2a^2}{A}\right )^{1/2}.
\label{edc5}
\end{equation}
Recalling that $\rho\propto a^{-3}$ it can be written as $k_J^c=\kappa_J/a^2$
where $\kappa_J=\left(4\pi G\rho^2a^6/A\right)^{1/2}$ is a constant independent
of
time (it is equal to the present Jeans wavenumber). In terms of
this
parameter, Eq. (\ref{edc2})  can
be
rewritten as 
\begin{eqnarray}
\frac{d^2\delta_k}{da^2}+\frac{3}{2a}\frac{d\delta_k}{da}+\frac{3}{2a^2}
\left (\frac{k^2a^4}{\kappa_J^2}
- 1\right )\delta_k=0.
\label{edc6}
\end{eqnarray}

The CDM model is recovered by taking $\kappa_J\rightarrow +\infty$ in Eq. 
(\ref{edc6})
yielding
\begin{eqnarray}
\frac{d^2\delta_{\rm
CDM}}{da^2}+\frac{3}{2a}\frac{d\delta_{\rm CDM}}{da}-\frac{3}{ 2a^2 }
\delta_{\rm CDM}=0.
\label{edc7}
\end{eqnarray}
The growing solution is $\delta_{\rm CDM}\propto a$ (there is also a decaying
solution proportional to $a^{-3/2}$).  It is usually
considered
 that $\delta_i\sim 10^{-5}$ at the initial
time $a_i\sim 10^{-4}$ of matter-radiation equality
\cite{dodelson}.
Therefore, the
growing evolution of the density contrast in the
CDM model can be written as
\begin{equation}
\delta_{\rm CDM}(a)=\frac{\delta_i}{a_i} a.
\label{edc8}
\end{equation}
We will take this CDM result as a reference and compare it  with the prediction
of the logotropic model. We note that Eq. (\ref{edc6})  for the density contrast
of the logotropic gas reduces to Eq. (\ref{edc7})  when $k\rightarrow
0$  and when 
$a\rightarrow 0$ because  the logotropic term
${k^2}a^4/{\kappa_J^2}$ becomes negligible in these two limits. Therefore, the
logotropic gas
is expected to behave similarly to CDM at large scales and at early times as
specified below.

\subsection{Evolution of the density contrast}
\label{sec_evolution}

In this section,  we study the evolution of the density contrast
$\delta_k(a)$ 
in the linear regime of structure formation. It turns out that Eq.
(\ref{edc6}) can be solved
analytically
\cite{aacosmo}. The growing solution is given by
\begin{equation}
\delta_k(a)=\frac{{\cal
A}(k)}{a^{1/4}}J_{\frac{5}{8}}\left(\frac{\sqrt{6}}{4}\frac
{ k } {
\kappa_J}a^2\right),
\label{si2}
\end{equation}
where $J_{5/8}(x)$ is the Bessel function of order
$5/8$ (there is also a decaying solution proportional to  $J_{-5/8}(x)$).
The amplitude ${\cal A}(k)$ is determined by requiring that the asymptotic
behavior
of
Eq. (\ref{si2}) for $a\rightarrow 0$ exactly matches the solution
(\ref{edc8}) of the CDM
model. This gives 
\begin{equation}
{\cal A}(k)=\Gamma\left (\frac{13}{8}\right )\frac{8^{5/8}}{6^{5/16}}
\left(\frac{\kappa_J}{k}\right)^{5/8}\frac{\delta_i}{a_i}.
\label{si3mm}
\end{equation}
Eqs. (\ref{si2}) and (\ref{si3mm}) determine the evolution of
the density contrast  $\delta_k(a)$  in the logotropic gas. We can identify two
regimes:

(i) {\it Early times/large wavelengths:} We first consider the case
$ka^2/\kappa_J\ll
1$. For a given wavenumber $k$, this corresponds to a scale factor
$a\ll(\kappa_J/k)^{1/2}$. Alternatively, for a given scale factor $a$,
this corresponds
to a wavelength $\lambda\gg \lambda_J^c(a)$. Since the
wavelength of the perturbation is larger that the comoving Jeans
length, the density contrast $\delta_k(a)$ increases. Using the asymptotic
expansion of the Bessel function 
for large arguments, we
find that  
\begin{eqnarray}
\delta_k(a)\sim \frac{\delta_i}{a_i} a,
\label{si4}
\end{eqnarray}
independently of $k$. This solution is valid for $a\ll
(\kappa_J/k)^{1/2}$.  In that case, the perturbation grows like in the CDM model [see
Eq. (\ref{edc8})].

(ii) {\it Late times/small wavelengths:} We now consider the case
$ka^2/\kappa_J\gg
1$. For a given wavenumber $k$, this corresponds to a scale factor $a\gg
(\kappa_J/k)^{1/2}$. Alternatively, for a given scale factor $a$, this
corresponds
to a wavelength $\lambda\ll \lambda_J^c(a)$. Since the
wavelength of the perturbation is smaller that the comoving Jeans
length, the density contrast $\delta_k(a)$ displays damped oscillations
similar to acoustic oscillations (with Hubble damping).
Using the asymptotic expansion of the Bessel function for small arguments, we
find that  
\begin{eqnarray}
\delta_k(a)\sim
\Gamma\left (\frac{13}{8}\right )\frac{8^{9/8}}{6^{9/16}}\frac{1}{\sqrt{\pi}}
\frac{\delta_i}{a_i}\frac{1}{a^{5/4}}\left (\frac{\kappa_J}{k}\right
)^{9/8}\nonumber\\
\times\cos\left(\frac{\sqrt{6}}{4}\frac{k}{\kappa_J}a^2-\frac{9\pi}{16}\right
).
\label{si4b}
\end{eqnarray}
This solution is valid for $a\gg
(\kappa_J/k)^{1/2}$. We see that the amplitude of
the oscillations decreases like $a^{-5/4}$ as the Universe expands.

In conclusion,  when  $a\ll
(\kappa_J/k)^{1/2}$ or 
$\lambda\gg\lambda_J^c(a)$, the perturbation grows linearly with
the scale factor
like in the CDM model; when  $a\gg
(\kappa_J/k)^{1/2}$ or  $\lambda\ll\lambda_J^c(a)$, the perturbation
oscillates with
a decreasing amplitude scaling as $a^{-5/4}$

\begin{figure}[!h]
\begin{center}
\includegraphics[clip,scale=0.3]{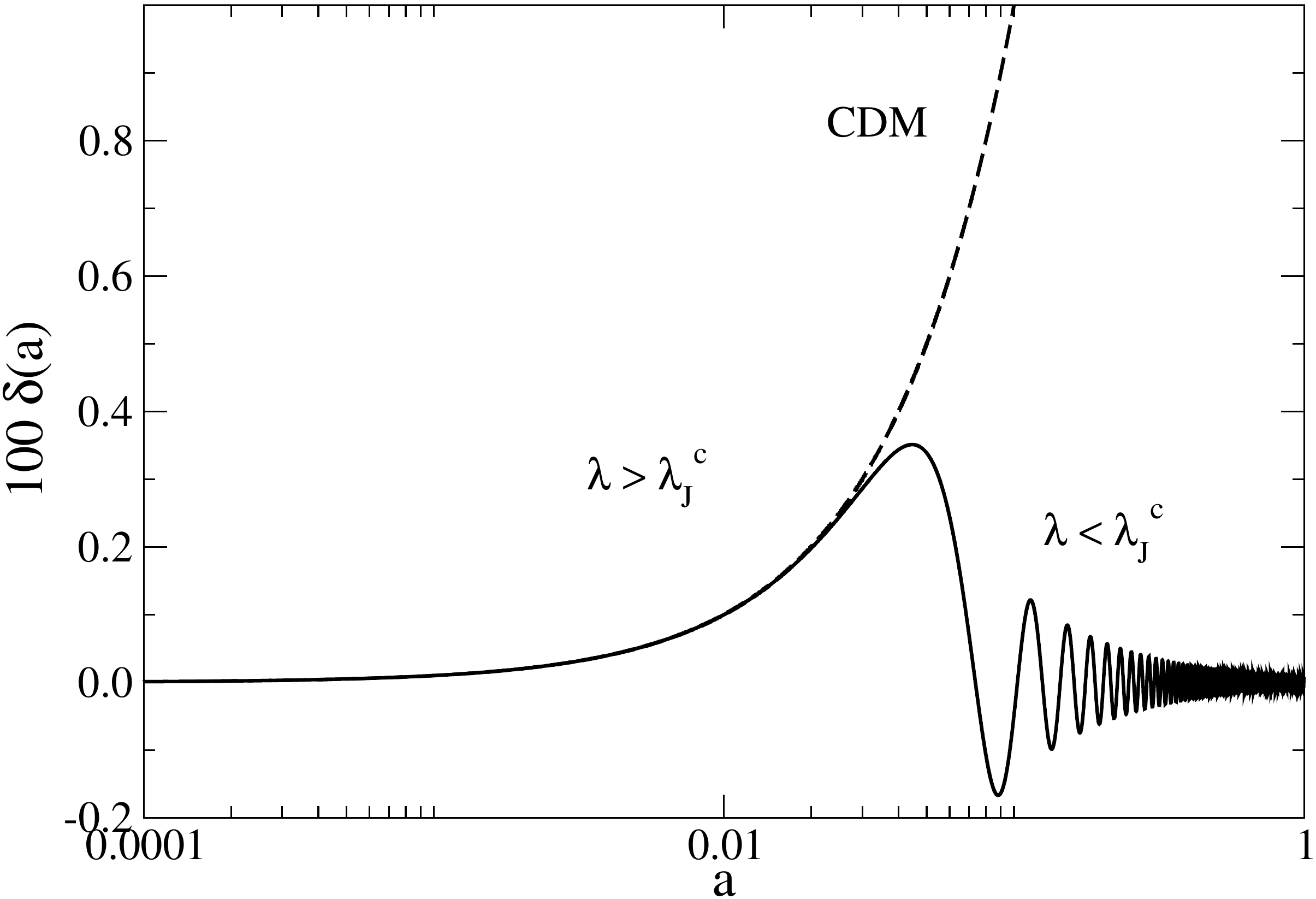}
\caption{Evolution of
the density contrast $\delta_k(a)$ in the logotropic model for $k/\kappa_J=1000$
(semi-log plot). The comoving Jeans length $\lambda_J^c(a)$ increases as the
Universe expands. As a result, the perturbation grows at early times like in the
CDM model ($\lambda>\lambda_J^c(a)$)  and undergoes damped oscillations
at late times ($\lambda<\lambda_J^c(a)$).}
\label{deltalogotropeK1000}
\end{center}
\end{figure}

A typical example of evolution of the density contrast is represented in Fig.
\ref{deltalogotropeK1000}. We assume that the matter era starts at
$a_i=10^{-4}$ and
we study the evolution of the density contrast up to the present time ($a_0=1$).
We consider a perturbation with a wavelength
$\lambda>\lambda_J^c(a_i)$. This
perturbation first starts to grow like in the
CDM model. However, at late times, the perturbation
decays and undergoes damped oscillations.
This behavior can be understood as follows. Initially, for small $a$, the
LDF behaves as pressureless CDM and all relevant
scales are gravitationally unstable ($\lambda>\lambda_J^c(a)$).  Therefore, the
LDF exhibits
growing modes and clusters like ordinary matter. Thus, the density constrast
increases as $\delta\propto a$. However, as the Universe expands, the comoving
Jeans length increases significantly until
there are no relevant gravitationally unstable scales ($\lambda<\lambda_J^c(a)$).
The
perturbation $\delta_k(a)$ stops growing
and begins to oscillate and decrease to zero when we enter the
DE era,
becoming a smooth component of the Universe.\footnote{Similarly, it is
well-known that the density perturbation in a Universe
dominated by the cosmological constant is zero  (i.e. $\delta_{\rm cc}=0$).}
Therefore, because of the increase of the comoving Jeans length
with $a$ (which is due to the increase of the speed of sound), the formation of
structures is blocked as the Universe expands.

\begin{figure}[!h]
\begin{center}
\includegraphics[clip,scale=0.3]{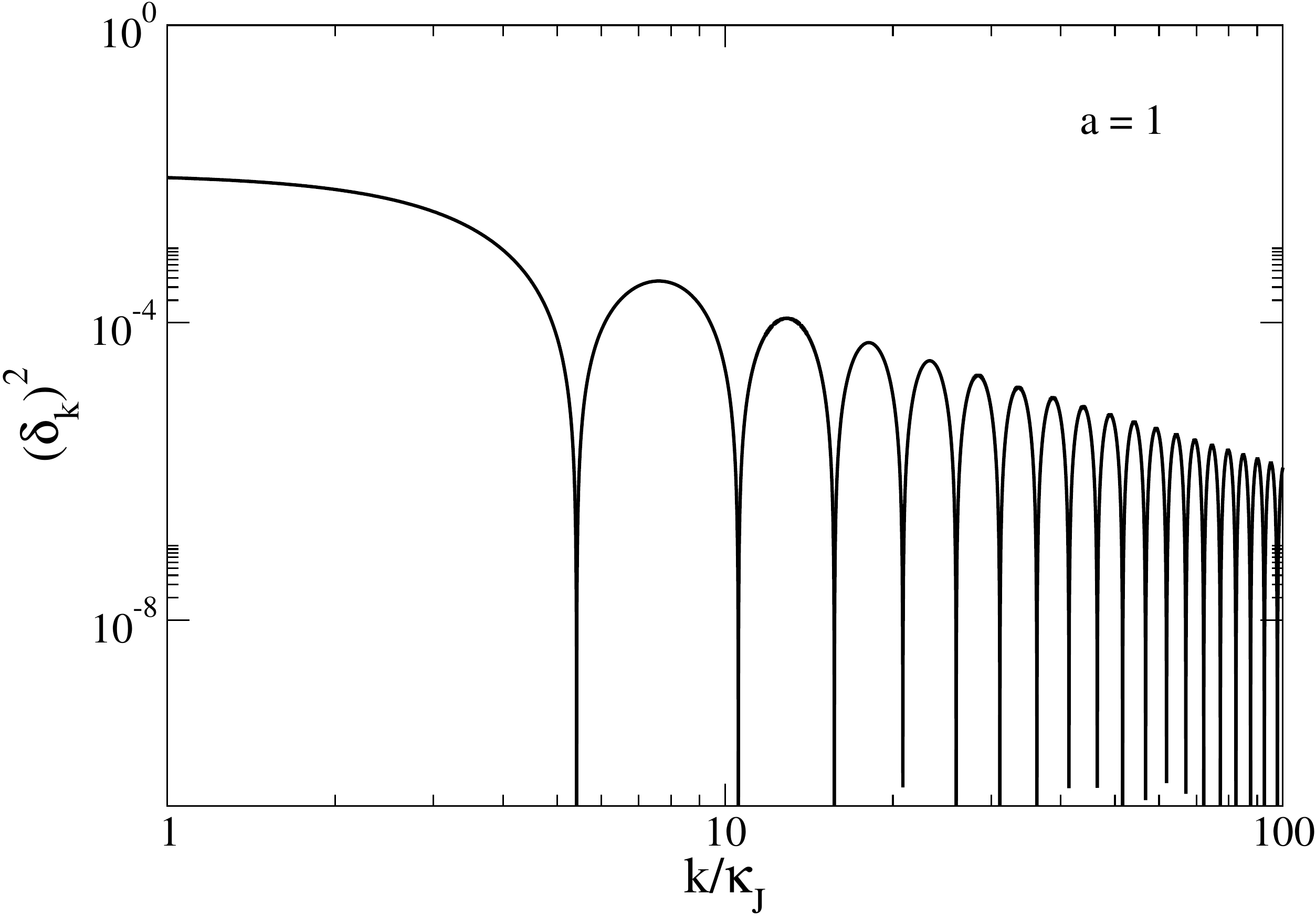}
\caption{Squared density contrast $(\delta_k)^2$ at the present time ($a=1$) as
a function of the wavenumber $k$ of the perturbation. The matter power spectrum has the same structure, displaying oscillations at large $k$.}
\label{spectre}
\end{center}
\end{figure}

The transition between the growing regime and the
oscillating regime occurs when $k\sim k_J^c(a)=\kappa_J/a^2$, i.e., when
$a=a_*(k)$ with 
\begin{eqnarray}
a_*(k)=\left (\frac{\kappa_J}{k}\right )^{1/2}.
\label{po1}
\end{eqnarray}
Therefore, the typical value of the maximum density contrast  achieved by a
perturbation of wavelength $k$ is $(\delta_k)_{\rm max}\sim \delta_k[a_*(k)]$,
i.e.,
\begin{equation}
(\delta_k)_{\rm max}=\Gamma\left (\frac{13}{8}\right
)\frac{8^{5/8}}{6^{5/16}}\frac{\delta_i}{a_i}
J_{\frac{5}{8}}\left(\frac{\sqrt{6}}{4}
\right)\left(\frac{\kappa_J}{k}\right)^{1/2}.
\label{si2h}
\end{equation}
Since $(\delta_k)_{\rm max}\sim a_*(k)\sim \left ({\kappa_J}/{k}\right
)^{1/2}$ and $\kappa_J\sim 1/R_{\Lambda}$, we see that a perturbation with a
wavelength $\lambda$ smaller than
the horizon $R_\Lambda$ cannot achieve a large  density contrast during its
evolution. Therefore, it cannot trigger the nonlinear regime
leading to the formation of the large-scale structures of the universe that we
observe today. This illustrates the blocking effect of the logotropic gas.

In Fig. \ref{spectre}, we plot the squared density
contrast $(\delta_k)^2$ at the present epoch ($a=1$) as a function of the
wavenumber $k$ of the perturbation. For $k\rightarrow 0$, it tends to
$(\delta_i/a_i)^2$. For $k\rightarrow +\infty$, it  decreases as
$(k/\kappa_J)^{-9/4}$ by oscillating. This function gives an idea of the
behaviour of the matter power spectrum in the logotropic model that is
discussed in Sec. \ref{sec_pom}.

\subsection{Comparison between the logotropic model and the FDM model}
\label{sec_jeansfdm}

In the previous sections we have made the TF approximation in the logotropic
model which amounts to
neglecting quantum effects. For comparison, it is
interesting to consider the
opposite limit where we take   quantum effects into account but neglect the
logotropic pressure. In that case we are led back to the FDM model (see
Appendix \ref{sec_gfdm}). The Jeans instability of the FDM model has been
studied in detail in our previous papers
\cite{prd1,aacosmo,abrilph,suarezchavanisprd3,jeansunivers,jeansMR}. We recall
below
the main results of these studies.

In the FDM model, the quantum Jeans wavenumber is given by \cite{prd1}
\begin{eqnarray}
k_J=\left (\frac{16\pi G\rho m^2}{\hbar^2}\right )^{1/4}.
\label{qwn}
\end{eqnarray}
During the expansion of the Universe, the
Jeans length
increases as $\lambda_J\propto a^{3/4}$ and the
Jeans mass decreases as $M_J\propto a^{-3/4}$  (the comoving
Jeans length decreases as $\lambda_J^c\propto a^{-1/4}$). As a result, the Jeans
mass-radius relation $M_J(\lambda_J)$ decreases, similarly to the core
mass-radius relation $M_c(R_c)$  of FDM halos (see Eq. (\ref{hni2}) and
\cite{jeansMR}).
Let us consider a boson mass $m=2.92\times
10^{-22}\, {\rm eV}/c^2$ representative of the FDM model \cite{jeansMR}. At the
epoch of matter-radiation equality, we find $\lambda_J=124\, {\rm pc}$
and
$M_J=1.31\times 10^9\, M_{\odot}$ (the comoving Jeans length
is $\lambda_J^c=\lambda_J/a=0.420\, {\rm Mpc}$). At the present epoch, we find
$\lambda_J=55.3\, {\rm kpc}$ and
$M_J=2.94\times 10^6\, M_{\odot}$.

In the nonrelativistic regime of the  FDM model, the equation
determining the
evolution of the density contrast in the linear regime of structure formation is
given by \cite{aacosmo}
\begin{equation}
\frac{d^2\delta_k}{da^2}+\frac{3}{2a}\frac{d\delta_k}{da}+\frac{3}{2a^2}
\left (\frac{\hbar^2k^4}{16\pi G\rho m^2 a^4}-1\right )\delta_k=0.
\label{edc2fdm}
\end{equation}
This
equation, which is based on the Schr\"odinger-Poisson equations, takes
quantum effects into account. It has been studied in detail
in \cite{aacosmo,abrilph}. It is found  (see Fig. 4 of \cite{abrilph}) that
$\delta_k(a)$ first oscillates
for small $a$ (quantum regime) then grows
like in the CDM model for large $a$ (classical regime).

This behavior can be understood as follows. Initially, for small $a$, most
scales are stable ($\lambda<\lambda_J^c(a)$)  and the perturbation oscillate. 
However, as the Universe expands, the comoving Jeans length decreases
significantly and the
relevant scales become gravitationally unstable ($\lambda>\lambda_J^c(a)$). In that case, FDM behaves as
pressureless CDM. It exhibits
growing modes and clusters like ordinary matter. Thus, the density constrast
increases as $\delta\propto a$. Therefore, because
of the decrease of the comoving Jeans length
with $a$, the formation of structures is facilitated as the Universe expands.

These results are reversed as compared to those obtained in the logotropic model. Indeed, as
the Universe expands, the
Jeans mass $M_J$ and the comoving Jeans length $\lambda_J^c$ decrease in
the FDM model while they increase in the logotropic model.\footnote{The Jeans length
$\lambda_J$ increases in the two models. Consequently, the Jeans mass-radius
relation $M_J(\lambda_J)$ decreases in the FDM model and increases in the logotropic model.} As a result, in the FDM model, the density contrast initially
oscillates then grows like CDM while, in the logotropic model, it  first
grows like CDM then undergoes damped oscillations. On the other hand, the
value of the Jeans mass $M_J=1.31\times 10^9\, M_{\odot}$
at the epoch of matter-radiation equality computed in the
framework of the FDM model is much larger
than the Jeans mass $M_J=0.910\, M_{\odot}$ computed in the framework
of the logotropic model. The Jeans mass $M_J=1.31\times 10^9\, M_{\odot}$  is of the order of  the mass of the smallest DM halos (dSphs) observed at present. Therefore, the FDM model is
consistent with the
observations and can solve the missing satellite problem (unlike the
classical logotropic
model). The main drawback of the FDM model is that (i) it does not
account for the universal surface density of DM halos and (ii) it does not
account for the present acceleration of the Universe (without adding an
additional DE component like a cosmological constant). By contrast, the
logotropic model can account for these two features simultaneously.

These results suggest that, regarding the formation of the large scale
structures of the Universe (Jeans
problem), it is important to take into account quantum effects
in the logotropic model, i.e., to go beyond the TF approximation. The
general expression of the Jeans wavenumber of a complex SF
(including quantum effects and self-interaction) is given by 
\cite{prd1}
\begin{eqnarray}
k_J^2=\frac{2m^2}{\hbar^2}\left (-c_s^2+\sqrt{c_s^4+\frac{4\pi G\rho\hbar^2}{m^2}}\right ).
\label{qwn2}
\end{eqnarray}
On the other hand, the general equation determining the
evolution of the density contrast of a nonrelativistic complex SF in the linear regime of
structure formation is given by  \cite{aacosmo}
\begin{equation}
\frac{d^2\delta_k}{da^2}+\frac{3}{2a}\frac{d\delta_k}{da}+\frac{3}{2a^2}
\left (\frac{c_s^2k^2}{4\pi G\rho
a^2}+\frac{\hbar^2k^4}{16\pi G\rho m^2 a^4}-1\right )\delta_k=0.
\label{edc2fdm2}
\end{equation}
This
equation, which is based on the GPP equations, takes quantum effects into
account in addition to a nonzero speed of sound like in Eq. (\ref{jsl4b}) for
the logotropic gas. This equation will be studied in a specific paper
\cite{prep}
for the logotropic equation of state\footnote{It has been studied in
\cite{abrilph}
for a quadratique equation of state corresponding to self-interacting BECs.} but we
can already mention its main properties. At early times, we can neglect
the logotropic pressure and we recover the results of the FDM model. Quantum
effects prevent the formation of structure below a 
minimum mass  ($\lesssim 1.31\times 10^9\, M_{\odot}$) and solve the missing
satellite problem as we have just seen. At late times,
we can neglect
the quantum potential and we recover the results of the logotropic
model in the TF approximation (see Sec. \ref{sec_evolution}). Generically, the
perturbation $\delta_k(a)$
first oscillates like in the FDM model, then grows like in the CDM model, and
finally undergoes damped oscillations like in the classical
logotropic model \cite{prep}. 

\subsection{Validity of the TF approximation in the Jeans instability analysis}
\label{sec_vtf}

Considering the order of magnitude of the quantum and logotropic terms 
in Eq. (\ref{qwn2}) [using Eq. (\ref{jsl4b})] or comparing Eqs. (\ref{jsl2}) and
(\ref{qwn}), we find that the TF approximation is valid when
\begin{equation}
c_s^2\gg\left (\frac{G\rho\hbar^2}{m^2}\right )^{1/2}\quad {\rm i.e.}\quad \rho\ll \rho_t\sim \left (\frac{m^2A^2}{G\hbar^2}\right )^{1/3}.
\label{qwn3}
\end{equation}
For a boson mass $m\sim 10^{-22}\, {\rm eV/c^2}$ we obtain
$\rho_t=5.34\times 10^{-18}\, {\rm g/m^3}$ (corresponding to $a_t=7.50\times
10^{-3}$ and $z_t=132$). Quantum effects are
important for $\rho\gg \rho_t$ while
they can be neglected (TF approximation) 
for $\rho\ll \rho_t$. In
particular, we must take into account quantum effects at the beginning of the
epoch of structure formation corresponding to
$\rho_{\rm eq}=8.77\times 10^{-14}\, {\rm g}/{\rm m}^3$. Quantum
effects could be neglected at this epoch
(TF approximation) provided that 
\begin{equation}
m\gg m_t=\frac{\sqrt{G\hbar^2\rho_{\rm eq}^3}}{A}.
\label{qwn4}
\end{equation}
We find $m_t=2.10\times 10^{-16}\, {\rm
eV/c^2}$.\footnote{We note that the criterion
$m\gg m_t=2.10\times 10^{-16}\, {\rm
eV/c^2}$ determining the validity of the TF approximation for the Jeans problem
at the epoch of matter-radiation equality differs by $6$ orders of magnitude 
from the criterion $m\gg m_0=3.57\times 10^{-22}\, {\rm eV/c^2}$ determining the
validity of the TF approximation at the scale of ultracompact DM halos (see Sec.
\ref{sec_tfdm}), and by $17$ orders of magnitude  from the criterion
$m\gg m_{\Lambda}=1.20\times
10^{-33}\, {\rm eV/c^2}$ determining the validity of the TF approximation at the
cosmological level (see Sec. \ref{sec_v}). In particular, for a particle
mass $m\sim 10^{-22}\, {\rm eV/c^2}$, the TF approximation is not valid for
the Jeans problem at the epoch of matter-radiation equality while it is valid 
at the cosmological level to describe the evolution of the background and
marginally valid at the scale of ultracompact DM halos to determine their
structure.} Since $m\ll m_t$ in general (see Appendix \ref{sec_gfdm}),
the TF
approximation is not valid at the
epoch of matter-radiation equality. On the contrary, the quantum pressure is
more
important than the logotropic pressure.  At
early times, the system is equivalent to the
FDM model 
and the results of Sec. \ref{sec_jeansfdm} apply. Therefore, the {\it quantum}
logotropic model can solve the missing satellite problem.

{\it Remark:} At the present epoch, the TF approximation is
valid if
\begin{equation}
m\gg m'_0=\frac{\sqrt{G\hbar^2\rho_{0}^3}}{A},
\end{equation}
where $\rho_0$ denotes here the present density of the universe (not the
central density of DM halos). We find $m'_0=2.73\times 10^{-32}\, {\rm
eV/c^2}$, which is of the order of the cosmon mass $m_{\Lambda}$. Since $m\gg
m_\Lambda$ in general (see Appendix \ref{sec_gfdm}), the TF approximation is
always valid at the present
epoch.

\subsection{The problem of the oscillations in the matter power spectrum}
\label{sec_pom}

In the logotropic model, the speed of sound increases as the density of the
Universe decreases. At early time, the  speed of sound is small and the LDF
clusters identically to CDM. In this regime, the Jeans length is small
so that
most
fluctuations are gravitationally unstable and grow. As one approaches the
present time, when
the LDF starts behaving like DE, the 
speed of sound increases. Correspondingly, the Jeans scale becomes large. This
prevents gravitational collapse and clustering from happening, even at large
scales. Fluctuations with wavelength below the comoving Jeans scale $\lambda_J^c$ are
pressure-supported (the pressure effectively opposes gravity) and oscillate
rather than grow. Therefore, the large speed of sound produces 
oscillations in the matter power spectrum (see Fig. \ref{spectre} for a schematic
view).

These oscillations in the matter power spectrum are not seen in observed data.
To be a successful model for UDM, the LDF should mimic the
inhomogeneous Universe as in the $\Lambda$CDM model. For this, it is necessary
that the LDF clusters similarly to CDM at all observable scales. Accordingly,
the agreement
with the observations will be obtained provided that
$B$ is small
enough since, for $B=0$, the logotropic model becomes equivalent to the
$\Lambda$CDM model which has $c_s=0$. Developing this argument,
Ferreira and Avelino \cite{fa} showed that $B$ must be smaller than $B_{\rm
max}\sim 6 \times
10^{-7}$. Unfortunately, this upper bound is smaller than our theoretical
prediction $B=3.53\times 10^{-3}$. This is an important problem of the
logotropic model.\footnote{This constraint can be understood
as follows. The matter power spectrum of the logotropic model displays
oscillations when
$\lambda<\lambda_J^c(a)$ or, equivalently, when
$k>k_J^c(a)$. Since these oscillations are not observed, we need
$\lambda_J^c(a=1)<R$ where $R\sim 15\, {\rm Mpc}$ is the typical size of
the clusters of galaxies. The present value of the Jeans
length must be smaller than the size of the clusters of galaxies  $R\sim 15\,
{\rm Mpc}$ so that the
linear growth of
cosmic structures on comoving scales larger than $R$ is not
significantly affected with respect to the standard $\Lambda {\rm CDM}$ result.
From Eq. (\ref{klj1}), we have $\lambda_J^c(a=1)\sim 10\sqrt{B}R_{\Lambda}$.
Therefore, we need  $100B<(R/R_{\Lambda})^2\sim 10^{-5}$, i.e.,
$B<10^{-7}$. We note that the constraint
$\lambda_J^c(a=1)<R$
is satisfied in the FDM model with $m=2.92\times
10^{-22}\, {\rm eV}/c^2$ since $\lambda_J^c(a=1)=55.3\, {\rm
kpc}$.}

These problems were first encountered in the context of the
GCG model
\cite{fabris2002,fabris2002GR,sandvik,cf,afbc},\footnote{Fig.
\ref{deltalogotropeK1000} can be compared to Fig. 2 of
\cite{fabris2002GR} and to Fig. 2 of \cite{cf}. Fig. \ref{spectre} can be
compared to Fig. 1 of \cite{sandvik}.} based on an equation of
state of the form
\begin{equation}
P=-\frac{A}{(\epsilon/c^2)^\alpha}
\label{gcg}
\end{equation}
with $A>0$ and $0\le\alpha\le 1$, and they actually arise
in any UDM model. In particular, Sandvik {\it et al.}
\cite{sandvik} ruled out a
broad class of UDM models by showing
that they produce oscillations (or exponential blowups) of the DM
power spectrum inconsistent with observations. For the GCG model, they showed
that $99.999\%$ of the parameter space is excluded. In order to
obtain the mass power spectra that we observe today, one needs
$|\alpha|<10^{-5}$ rendering the GCG 
indistinguishable from the standard $\Lambda$CDM  model
corresponding to $\alpha=0$ (see Appendix \ref{sec_df}). Similar conclusions
were reached
by Carturan and Finelli \cite{cf} and Amendola {\it et al.} \cite{afbc} who
studied  the effect of the GCG on density perturbations and on cosmic
microwave background (CMB) anisotropies and found that GCG 
strongly increases the amount of integrated Sachs-Wolfe effect.

More generally, these results apply to any UDM model where $P$
is a unique function of $\epsilon$. Such models are ruled out if the speed of sound
is large, i.e., if the function $P(\epsilon)$ departs substantially from a
constant over the range where pressure is important. Quantitatively, we must
have $|d\ln P/d\ln\epsilon|<10^{-5}$ (see footnote 32) leaving essentially only
the standard $\Lambda$CDM model.\footnote{This criterion is
not valid for a linear equation of state. 
The corresponding criterion is given in Appendix \ref{sec_lin}.} In
other words, a viable UDM model must have
negligible pressure gradient, i.e., the pressure must be essentially spatially
constant like a $\Lambda$ term.

In conclusion, UDM or quartessence models
can often
correctly explain the evolution of the homogeneous background (zeroth
order cosmology) but they fail at explaining the growth of linear
perturbations (first order cosmology) because they produce unphysical
features in the matter power spectrum in the form of huge oscillations or
exponential blow-ups which are not seen in the observed matter power spectrum.
If a solution to these problems cannot be provided, this would appear as an
evidence for an independent origin of DM and DE (i.e. they are two distinct
substances) and the demise of UDM models \cite{sandvik}.

\subsection{Possible solutions to the problems of the logotropic model}

Some solutions to the problems mentioned above have been proposed in the
context of the GCG. Since the LDF
experiences the same
problems as the GCG, these solutions could also be invoked for the LDF. 
We review these different solutions below.

{\it Two-fluid models:} In the begining of the matter era the GCG agglomerates
in the same way as CDM.
Later, it behaves as DE and becomes a
smooth component of the total matter  existing in the Universe. It does not
cluster anymore and produces decaying oscillations (or exponential blow up) in
the matter power spectrum. As we have seen, this is a problem of
any UDM model.\footnote{Quintessence models
have no such problems. Although they have high speeds of sound, this does
not prevent DM from clustering since it is a
separate component. Quintessence models would fail if they were
tightly coupled to DM and this is effectively what happens with UDM models
since DM and DE are one and the same substance.} 
Therefore, the GCG model
needs 
additional CDM in order to explain the dynamics of
the clusters of galaxies since a fraction of the total DM must remain clustered
until today. Consequently, a more realistic model is a two-fluid model where,
besides the GCG, normal fluid must be present.  Therefore, some authors
\cite{fabris2002,fabris2002GR,cf,afbc}  (see also 
\cite{avelinotwo,daj,ajd,multa,bd}) have proposed that GCG
describes only DE and that it must be
mixed with CDM. In this viewpoint, the GCG simply plays the role of DE
like in
quintessence models. This ``Chaplygin quintessence'' scenario would solve the
above mentioned problems but the original interest of the GCG as a
UDM (quartessence) model has been lost.\footnote{
We have seen in Sec. \ref{sec_twofluids} that a single fluid model
like the Chaplygin gas or the LDF can be viewed as a two-fluid
model made of effective DM and DE (the effective equation of state of DE in the
Chaplygin
gas model and in the original logotropic model has been determined in Appendix
D.3 of
\cite{action}). The single-fluid model and
the two-fluid models are equivalent at the level of the homogeneous
background but they
differ from each other for what concerns the formation of structures. The
two-fluid model
does not present the problems of the single fluid model reported above.}

{\it Baryons:} Some authors \cite{avelinoB,cf,afbc} proposed to  include
baryons in analyses of UDM scenarios. Indeed, while pressure effects prevent
the Chaplygin gas from collapsing, the baryon fluctuations can still 
keep growing since this is an independent
component with a low speed of sound. Therefore, baryons keep on clustering at
all times after
decoupling, even after the end of the Jeans instability for the GCG component.
Amendola {\it et al.} \cite{afbc} showed that the inclusion of baryons affects
the total linear matter power spectrum,
smoothing out the oscillations of the GCG component 
and
improving the agreement with observations. As a result, the inclusion of baryons
in the
analysis leads to less stringent bounds on
the GCG parameter $\alpha$. However, this parameter remains tightly
constrained by cosmological observations. Therefore, including baryons in UDM
models may not be sufficient to save the model.

{\it Nonlinear effects:} The importance of nonlinear
effects in UDM models was first mentioned by \cite{cf,afbc}. In the context of
the Chaplygin gas,  Bilic {\it et al.}
\cite{bilicNL} proposed to
take  
nonlinear effects into account in the growth of inhomogeneities by generalizing
the Zeldovich approximation and the spherical model so as to  include sonic
horizon effects. They showed that if the initial perturbation is above a certain
threshold then the perturbation always grows like in the $\Lambda$CDM model (in
contrast to linear theory where the speed of sound eventually stops $\delta(a)$
from growing irrespective of the initial value of the perturbation). If the
initial perturbation is  below the critical threshold, the perturbation does not
grow even in the nonlinear regime. Therefore, a fraction of the Chaplygin gas
condensates (i.e., collapses in gravitationally bound
structures) and never reaches a stage where its properties change from
DM to DE. Unfortunately, the detailed calculations of Bilic {\it et al.}
\cite{bilicNL} show that the collapse fraction (the fraction of Chaplygin gas that
goes into condensate) is not sufficient to solve the problems reported
above. Nonlinear condensate, while present, is
insufficient to save the Chaplygin gas model.\footnote{In
a later work, Bilic {\it et al.} \cite{bilicNL3,bilicNL2} repeated their study
for a
tachyon condensate model (a k-essence model corresponding to the
string-inspired tachyon Lagrangian that extends the Born-Infeld
Lagrangian of the original Chaplygin gas model) in full general relativity and
obtained, this time, gravitational condensates in significant quantities. This
is
because this model reduces the Jeans length by several orders of magnitude.}
The importance of nonlinear effects was also pointed out by Avelino {\it et
al.} \cite{avelinoNL} using simple considerations. They argued that nonlinear
effects severely complicate the analysis and render linear results invalid even
on large cosmological scales. However, in the case of the Chaplygin gas,
similarly to Bilic {\it et al.} \cite{bilicNL}, they argued that nonlinear
effects are too small to significantly affect the linear results.
In a more recent work, Avelino {\it et al.}  \cite{avelinoNL2} relaxed
earlier simplifying assumptions and showed that if clustering is strong
enough, the linear theory results no longer hold and the backreaction of
the small scale structures on the large scale evolution of the Universe
render the Chaplygin gas model virtually indistinguishable from the $\Lambda$CDM
model for all possible values of the GCG parameter $\alpha$. They concluded
that the GCG may be consistent with observational constraints over a wide region
of parameter space, provided there is a high level of nonlinear clustering  of
the UDM component on small scales. A detailed analysis of non-linear effects
would nevertheless require solving the full Einstein field equations for the
evolution of realistic cosmological fluctuations, which is a formidable
task. [Note: While this paper was in course of redaction, we
came accross the very interesting paper of Abdullah {\it et al.} \cite{zant} who
argue that a cosmological scenario based on the Chaplygin gas may
not be ruled out from the viewpoint of structure formation as usually claimed.
Indeed, a nonlinear analysis may predict collapse rather than a re-expansion of
small-scale perturbations so that nonlinear clustering may occur in the
Chaplygin gas. This is because pressure forces in UDM fluids decrease with
increasing density so that systems that are stable against self-gravitating
collapse in the linear regime may become unstable in the nonlinear regime. As a
result, the problem of acoustic oscillations in the linear power spectrum of
UDM models may not be as serious as usually assumed provided the hierarchical
structure formation process is adequately taken into account. These
arguments also apply to the logotropic model.]

{\it Nonadiabatic perturbations:} A possible solution to the problem of
oscillations would be to allow for
nonadiabatic perturbations in the
Jeans stability analysis  to make the effective speed of sound vanish, even in
the nonperturbative regime.\footnote{In the adiabatic
case, the effective speed of sound and the adiabatic speed of sound are equal.
However, this may not be true anymore if entropy perturbations are present
\cite{huseul}.} Indeed, the
isentropic perfect
fluid approximation might break down at sufficiently large densities or small
scales.
Reis {\it et al.} \cite{reis1,reis2} have shown that if nonadiabatic
perturbations are allowed, the  quartessence GCG models may be
compatible with observations. Indeed, entropy perturbations
eliminate instabilites and oscillations in the mass power spectrum of these
models.

{\it Braneworld models:} Another possible solution, proposed by  Bilic {\it et
al.}
\cite{bilicNL}, would
be to exploit the
braneworld
connection of the Lagrangian associated with the Chaplygin gas. In braneworld
models \cite{maartens}, the Einstein equations are modified, e.g., by dark
radiation. Similar changes are also brought about
by the radion mode \cite{ktv} which yields a scalar-tensor gravity.

{\it Higher order derivatives:} The GCG can be obtained from a field theory
based on a k-essence Lagrangian.
For the original Chaplygin gas ($\alpha=1$), this yields the
Born-Infeld Lagrangian for $d$-brane in a ($d+1,1$) space
time \cite{jackiw}. Creminelli {\it et al.}
\cite{creminelli1} have shown
that, for a k-essence Lagrangian, one can add a specific  higher derivative
operator in
the original action  that  does not change the background evolution for the
field or
its energy density and pressure.  But
for the perturbations, this extra higher
derivative operator leads to a vanishing
speed of sound
($c_{s}^2=0$). In such a scenario, the pressure
perturbation
vanishes and the k-essence clusters at all scales like the nonrelativistic
matter. These are called ``clustering quintessence'' models. Given the fact
that GCG as a UDM model fails because of the large speed of sound through the
fluid during the DE domination, Kumar and Sen \cite{ks} proposed to apply
this idea to the ``clustering GCG'' model and explored its consequences.  In
that
case, they showed that the
matter power spectrum for the  parameter values of $0\le \alpha\le 0.043$ are
well behaved without
any unphysical features (note that the original Chaplygin gas $\alpha=1$ is
ruled out). Therefore, by properly modifying the  k-essence Lagrangian, we
can ensure that the GCG clusters at all scales similarly to the CDM model
leaving, at the same time, the background evolution of the Universe unaltered
(i.e., the GCG behaves like CDM in the early time and like DE in the
late time). This added clustering property makes the GCG a suitable candidate
for UDM models. Thus, the study of Kumar and Sen \cite{ks} renewed interest in the GCG as a viable option
for
UDM models. It would be interesting to redo their analysis in the framework of
the logotropic model in order to obtain an enlarged range of allowed
values for the parameter  $B$ and see if the theoretical value
$B=3.53\times 10^{-3}$ is included in that range.

{\it Scale dependence:} Another
way to try to avoid these problems could be by introducing
some sort of scale dependence into the equations. For example, Padmanabhan 
and Choudhury \cite{pc} discussed a model based on a tachyonic SF that exhibits
different equations of state at different scales. The field behaves like
pressureless DM on small scales and like smoothly distributed DE on large
scales.

The solutions introduced in the context of the GCG model could be applied to
the logotropic model as well. It must be recognized that none of them brings an
undisputable answer. Therefore, the problems essentially remain. In spite of
these difficulties, we think that the GCG and logotropic models deserve further
investigation. It is possible that these models are incomplete rather than being
ruled out. On the other hand, a thorough investigation of the
nonlinear regime
of the growth of inhomogeneities through extensive numerical simulations is
needed for a definite
conclusion concerning the compatibility of the GCG and logotropic cosmologies
with the observable large-scale structure of the Universe.

{\it Remark:} When applied to DM halos, the
logotropic equation of state cannot be valid everywhere because it yields halos
with an infinite mass (see Appendix \ref{sec_pldm}). Indeed,
only the core of DM halos is expected to be logotropic (its density decays as
$r^{-1}$). In practice, the logotropic core is
surrounded by an envelope where the density decreases more rapidly as $r^{-2}$
or $r^{-3}$ (see Appendix \ref{sec_pldm}). This suggests that the
logotropic
equation of state is valid only at large scales in an ``average'' sense, which allows us to correctly describe the evolution
of the cosmological background. However, it may cease to be
valid everywhere at small scales when considering the more complicated problem
of structure formation. In particular, one has to be careful when treating
strongly inhomogeneous structures such as DM halos in the nonlinear regime. A
full numerical solution of the nonlinear problem (accounting for relativistic
and quantum effects) may lead to a matter power spectrum different from the one
obtained in the linear regime where it is assumed that the logotropic
equation of state holds everywhere.

\section{Conclusion}
\label{sec_con}

In this paper, we have proposed a unification of DM and DE based on a complex SF
described by the KGE equations (\ref{lwe1}) and (\ref{lwe2}) with
a potential of the form
\begin{equation}
V_{\rm
tot}(|\varphi|^2)=\frac{m^2c^2}{2\hbar^2}|\varphi|^2-A\ln
\left
(\frac{m^2|\varphi|^2}{\hbar^2\rho_P}\right )-A,
\label{co1}
\end{equation}
which is the sum of a rest-mass term and a logarithmic term. This model is
associated with a logotropic equation of state 
\begin{eqnarray}
P=A\ln\left (\frac{\rho}{\rho_P}\right ),
\label{co2}
\end{eqnarray}
where $\rho=(m^2/\hbar^2)|\varphi|^2$ is the pseudo rest-mass density.

The logotropic model is able to account for the present accelerating expansion
of the Universe  while solving at the same time
the small-scale crisis of the $\Lambda$CDM model. Indeed, at cosmological
scales, the logotropic model is almost indistinguishable from the $\Lambda$CDM
model  up to the present time and even far in the future. However, at galactic
scales, it leads to DM halos presenting a central core instead of a cusp.
Furthermore,  it
predicts their universal surface density $\Sigma_0^{\rm th}=133\,
M_{\odot}/{\rm pc}^2$ (in agreement with the observations giving $\Sigma_0^{\rm
obs}=141_{-52}^{+83}\, M_{\odot}/{\rm
pc}^2$)
without adjustable parameter.

The new logotropic model introduced in the present paper is different from
the original one \cite{epjp,lettre,jcap,pdu} which is characterized
by the equation of state
(\ref{lmt3}) where $\rho_m$ represents the true rest-mass density.
The interest of the new logotropic model is that (i) it is based on a complex SF
theory;
(ii) it avoids the pathologies of the original logotropic model such as a
phantom
behavior violating the dominant-energy condition and leading to a Little Rip,
and a superluminal or imaginary speed of sound;  (iii) it
asymptotically approaches a well-behaved de Sitter era at late times.

At the cosmological level, and for the evolution of the homogeneous
background, we have shown that the TF approximation is equivalent to the fast
oscillation regime where the complex SF rapidly spins. In this spintessence
regime, the SF is described by the logotropic equation of state
(\ref{co2}). It behaves as DM in the early universe ($a\ll a_t=0.765$) and as DE
in
the late universe ($a\gg a_t=0.765$). At the cosmological level, the TF
approximation is valid for a large
period
of time when $m\gg m_{\Lambda}$, where $m_{\Lambda}=1.20\times 10^{-33}\, {\rm
eV/c^2}$ is the cosmon mass.  For a boson mass $m\sim 10^{-22}\, {\rm eV/c^2}$,
the TF approximation is
valid from $a_v^{(1)}=4.01\times 10^{-8}$ to $a_v^{(2)}=1.73\times 10^4$. In
the very early universe ($a<a_v^{(1)}=4.01\times 10^{-8}$), the fast oscillation
regime is
not valid anymore and the SF experiences a kination regime where it behaves as
stiff matter. Therefore,
the homogeneous SF successively experiences a stiff matter era, a DM-like era
and a DE-like
era.  The logotropic model has an intrinsically quantum
nature (even in the TF regime) because the equation of state (\ref{co2})
involves $\rho_P$, and it returns the $\Lambda$CDM model in the
semiclassical limit
$\hbar\rightarrow 0$.

At the level of DM halos, the logotropic model differs from the $\Lambda$CDM
model because it generates a pressure which is either of quantum origin (as in
the FDM model) or due to the logarithmic potential. Following a process of
violent
relaxation \cite{lb} and gravitational cooling \cite{seidel94}, the logotropic
DM halos acquire a ``core-halo'' structure with a quantum or logotropic core
surrounded by a classical NFW  (or quasi-isothermal) atmosphere resulting from
quantum interferences of excited states \cite{wignerPH}. The pressure
effects  can
solve the core-cusp problem of the $\Lambda$CDM model. In the TF
approximation, the core is purely logotropic. The logotropic equation of state
implies a constant surface density $\Sigma_0^{\rm th}=5.85\,\left ({A}/{4\pi
G}\right )^{1/2}=133\, M_{\odot}/{\rm pc}^2$ which is in agreement with the
observations. Therefore, the logotropic model avoids the problems of the FDM
model reported by the author \cite{clm2,chavtotal,modeldm} and by 
\cite{burkertfdm,deng}. At the level of DM halos, the TF approximation is
valid for $m\gg m_0=3.57\times 10^{-22}\, {\rm eV/c^2}$. For a boson
mass $m\sim 10^{-22}\, {\rm eV/c^2}$ we are just at the limit of validity of
the TF pproximation so we
have to
take into account a quantum
core + a logotropic inner halo + a NFW (or isothermal) outer halo.

We have also discussed the formation of structures (Jeans problem) within the
logotropic model. In that case, there are two difficulties: (i) If we naively
make the TF approximation, we find that the Jeans mass
$M_J=0.910\, M_{\odot}$ at the epoch of
matter-radiation equality is much too  small to solve the
missing satellite problem. However, the
TF approximation is valid at this period only if $m\gg m_t=2.10\times
10^{-16}\, {\rm eV/c^2}$. For  a boson mass $m\sim 10^{-22}\, {\rm eV/c^2}$, we
are in the opposite limit where quantum effects are more important than the
logotropic
pressure. In that case, the logotropic model reduces to the FDM
model. Quantum effects yield a much larger Jeans mass $M_J=1.31\times 10^9\,
M_{\odot}$
that is able to solve the missing satellite problem. At later
times ($a\gg 7.50\times 10^{-3}$) the TF approximation becomes valid. (ii)
In the logotropic model, the
density contrast $\delta(a)$ first grows like in the $\Lambda$CDM model
(after the FDM era mentioned above) then undergoes decaying
oscillations (see Fig. \ref{deltalogotropeK1000}). This is because the squared
speed of sound increases as the density decreases. As a result, the
comoving Jeans length becomes very high and prevents the formation of
structures. This gives rise to oscillations in the matter power
spectrum.  These features (large Jeans length and oscillations) are in severe
disagreement with the observations.
The Chaplygin gas model, and more generally most UDM models, share
the same problems \cite{sandvik}. We have reviewed several possible
solutions proposed in the literature but none of these solutions has gained
complete acceptance so far. This remains an important weakness of the
logotropic and Chaplygin gas models. The recent paper of
Abdullah {\it et al.} \cite{zant} suggests, however, that these problems may not
be as insurmountable as previously thought provided that an adequate
nonlinear analysis of structure formation is developed.

We note that the criteria (\ref{v4b}), (\ref{glm7}) and (\ref{qwn4}) determining
the validity of the TF approximation at the cosmological level, at the level of
ultracompact DM
halos, and for the Jeans problem at the epoch of matter-radiation
equality involve different densities ($\rho_\Lambda$,
$\rho_0$ and $\rho_{\rm
eq}$) yielding different critical particle masses $m_{\Lambda}=1.20\times
10^{-33}\, {\rm
eV/c^2}$, $m_0=3.57\times 10^{-22}\, {\rm eV/c^2}$ and $m_t=2.10\times
10^{-16}\, {\rm eV/c^2}$. As a result, for a boson mass $m\sim 10^{-22}\, {\rm
eV/c^2}$, the TF approximation is valid during a long period of time for what
concerns the
evolution of the homogeneous background (quantum terms can be neglected) while
it is marginally valid to describe ultracompact DM halos (both quantum and
logotropic terms have to be taken into account), and not valid at all to
describe the formation of structures at the begining of the matter era
(logotropic terms can be neglected). 

We have argued that the logarithmic term in 
Eq. (\ref{co1}) is a fundamental term that is always
present in the KG equation. It is not a particular attribute of the SF
(such as its mass or self-interaction constant) but rather an intrinsic property
of spacetime. In other words, the ordinary (linear) KG equation is an
approximation of the more fundamental wave equation (\ref{lwe3}). This equation
involves a new  fundamental constant of physics $A$
superseeding the Einstein cosmological
constant $\Lambda$. This term accounts simultaneously for the accelerating
expansion
of the universe and for the universal surface density of DM halos. The
logarithmic potential manifests itself only at extremely low densities and this
is why the
ordinary (linear) KG and Schr\"odinger equations are so successful at the
laboratory scale where $\rho\gg\rho_{\Lambda}$. However,  the logarithmic
potential becomes important at
astrophysical and cosmological scales and leads to a logotropic dark fluid which
unifies DM and DE. If the logarithmic term in Eq.
(\ref{co1}) is replaced by a
constant $V=\rho_\Lambda c^2$ mimicking a cosmological constant, we obtain the
$\Lambda$FDM model which is associated with a constant equation of state
$P=-\rho_\Lambda c^2$ (see Appendix \ref{sec_gfdm}). In the TF approximation, it
reduces to the $\Lambda$CDM model (see
Appendix \ref{sec_lcdm}).\footnote{For $V=0$ we get the FDM model which reduces
to the CDM
model in the TF approximation.} The
$\Lambda$FDM model accounts for the accelerating
expansion of the universe and solves the core-cusp problem and the missing
satellite problem due to quantum effects. However, it does not account for
the universal surface density of DM halos, contrary to the logotropic model.
This is an important advantage of the  logotropic model.

As discussed above, the KG equation with the potential from Eq.
(\ref{co1}) describes a {\it noninteracting} SF. Indeed, we have argued that the
logarithmic term in Eq. (\ref{co1}) is a fundamental  term which is rooted
in the KG equation. We can now consider more general models, where the bosons
have a self-interaction, by including additional terms in the SF potential.

At the end of Secs. \ref{sec_v}, \ref{sec_lkge} and \ref{sec_fws} we have
briefly considered the case of a relativistic BEC with a
repulsive $|\varphi|^4$ self-interaction 
[see Eqs. (\ref{tp4bh}),
(\ref{lwe3rad}) and (\ref{lwe6red})]. At a cosmological level, the 
$|\varphi|^4$ self-interaction is responsible, in the fast oscillation (or TF)
regime, for an additional radiationlike era before the mattelike era. Therefore,
the homogeneous SF successively experiences a stiff matter era, a radiationlike
era, a DM-like era and a DE-like
era. On the other hand, logotropic DM halos with a repulsive $|\varphi|^4$
self-interaction possess an additional hydrodynamic core
stabilized by the self-interaction (see, e.g., \cite{prd1,modeldm}) in
addition
to
the quantum core (soliton) due to the Heisenberg uncertainty principle, the 
logotropic core due to the logarithmic potential and the NFW halo resulting
from quantum interferences of excited states.

More generally, we can consider a relativistic  BEC with a potential of the form
\begin{eqnarray}
V_{\rm tot}(|\varphi|^2)=\frac{m^2c^2}{2\hbar^2}|\varphi|^2+\frac{2\pi a_s
m}{\hbar^2}|\varphi|^4+\frac{32\pi^4a_s^2}{9c^2\hbar^2}|\varphi|^6\nonumber\\
+\frac{m k_B T}{\hbar^2}|\varphi|^2\left\lbrack \ln \left
(\frac{m^2|\varphi|^2}{\hbar^2\rho_*}\right )-1\right\rbrack-A\ln \left
(\frac{m^2|\varphi|^2}{\hbar^2\rho_P}\right )-A.\nonumber\\
\label{co3}
\end{eqnarray}
This potential includes a $|\varphi|^2$ rest-mass term, a
$|\varphi|^4$
self-interaction which can be repulsive ($a_s>0$) or attractive   ($a_s<0$), a
repulsive $|\varphi|^6$ self-interaction of relativistic origin that can
stabilize the system when $a_s<0$, a $|\varphi|^2
\ln |\varphi|^2 $ self-interaction which arises from effective or real
thermal effects, and the intrinsic  logarithmic $\ln
|\varphi|^2 $ self-interaction discussed
above.\footnote{Instead of the logarithmic
term we can consider a constant term $V_0=\epsilon_\Lambda$ mimicking a
cosmological constant like in Appendix \ref{sec_gfdm}. It is associated with a
constant
equation of state $P=-\epsilon_\Lambda$.} A power-law potential (see
Appendix C of \cite{cspoly})
\begin{eqnarray}
V(|\varphi|^2)=\frac{K}{\gamma-1}\left (\frac{m}{\hbar}\right
)^{2\gamma}|\varphi|^{2\gamma}
\label{co4}
\end{eqnarray}
is associated with a polytropic equation of state
\begin{eqnarray}
P=K\rho^{\gamma}.
\label{co5}
\end{eqnarray}
In particular, for the $|\varphi|^4$ and $|\varphi|^6$ self-interaction, we have
\begin{eqnarray}
V(|\varphi|^2)=\frac{2\pi a_s
m}{\hbar^2}|\varphi|^4\quad \Rightarrow\quad P=\frac{2\pi
a_s\hbar^2}{m^3}\rho^2,
\label{co6}
\end{eqnarray}
\begin{eqnarray}
V(|\varphi|^2)=\frac{32\pi^4a_s^2}{9c^2\hbar^2}|\varphi|^6\quad \Rightarrow\quad
P=\frac{64\pi^4 a_s^2\hbar^4}{9m^6c^2}\rho^3.
\label{co7}
\end{eqnarray}
On the other hand, the  $|\varphi|^2
\ln |\varphi|^2 $ self-interaction is associated with an isothermal equation of
state
\begin{eqnarray}
P=\rho\frac{k_B T}{m}.
\label{co8}
\end{eqnarray}
When $T>0$, it can take into account the finite temperature of DM halos. The KG
equation
associated
with the potential (\ref{co3}) is [see Eq. (\ref{lwe1})] 
\begin{eqnarray}
\square\varphi+\frac{m^2c^2}{\hbar^2}\varphi+\frac{8\pi a_s
m}{\hbar^2}|\varphi|^2\varphi+\frac{64\pi^4a_s^2}{3c^2\hbar^2}
|\varphi|^4\varphi\nonumber\\
+\frac{2 m k_B T}{\hbar^2}\ln \left
(\frac{m^2|\varphi|^2}{\rho_*\hbar^2}\right )\varphi-\frac{2A}{|\varphi|^2}
\varphi=0,
\label{co9}
\end{eqnarray}
and the corresponding GP equation, valid in the nonrelativistic regime, is [see
Eq. (\ref{lwe4})]
\begin{eqnarray}
i\hbar \frac{\partial\psi}{\partial
t}=-\frac{\hbar^2}{2m}\Delta\psi+m\Phi\psi
+\frac{4\pi a_s\hbar^2}{m^2}|\psi|^2\psi\nonumber\\
+\frac{32\pi^4 a_s^2\hbar^4}{3m^5c^2}|\psi|^4\psi
+k_B T\ln \left (\frac{|\psi|^2}{\rho_*}\right )\psi
-\frac{Am}{|\psi|^2}\psi.
\label{co10}
\end{eqnarray}
At the
cosmological level, the rest-mass term is responsible for a DM-like era
($\epsilon\sim a^{-3}$), the repulsive $|\varphi|^4$ self-interaction is
responsible for
a radiationlike era ($\epsilon\sim a^{-4}$),  the $|\varphi|^6$
self-interaction is
responsible for a new primordial era ($\epsilon\sim a^{-9/2}$), the $|\varphi|^2
\ln
|\varphi|^2 $ self-interaction is responsible for a DM-like era with
logarithmic corrections, and the logarithmic $\ln |\varphi|^2 $
self-interaction is responsible for a DE-like era. At the level of DM halos,
the rest-mass term produces a quantum core and a NFW envelope, the $|\varphi|^4$
and
$|\varphi|^6$ potentials produce a hydrodynamic core, the 
$|\varphi|^2 \ln |\varphi|^2 $ potential produces an isothermal envelope
(when $T>0$) and the
logarithmic $\ln
|\varphi|^2 $ self-interaction produces a logotropic envelope. 
The $|\varphi|^4$ self-interaction has been studied in
\cite{prd1,shapiro,abrilphas,carvente}, the $|\varphi|^6$
self-interaction has been studied in \cite{phi6}, the $|\varphi|^{2\gamma}$
self-interaction has been studied in \cite{sfpoly}, the  $|\varphi|^2
\ln |\varphi|^2 $ self-interaction has been studied in \cite{modeldm,sfpoly},
and the $\ln |\varphi|^2 $ self-interaction has been studied in the present
paper.

\appendix

 \section{Motivation of the logotropic model}
\label{sec_mot}

In this Appendix, we recall the arguments that led us to introduce the
logotropic model in Ref. \cite{epjp}. In short,
we assumed that DM and DE are the manifestation of a single DF and we tried to
construct a UDM model with a nonconstant pressure that is as 
close as possible to the standard $\Lambda$CDM
model.\footnote{As
discussed in Appendix \ref{sec_lmt}, we can introduce different types
of logotropic models. The following arguments apply to all of them.}

Let us consider a DF described by the polytropic equation of state
\begin{eqnarray}
P=K\rho^{\gamma},
\label{mot1}
\end{eqnarray}
where $K$ is the polytropic constant and  $\gamma=1+1/n$ is the polytropic
index. Up to a slight change of notations, this corresponds to the equation
of state of the GCG. As shown in Appendix \ref{sec_df}, the
$\Lambda$CDM
model (interpreted as a UDM model)
is equivalent to a single DF with a constant pressure
\begin{eqnarray}
P=-\rho_{\Lambda}c^2,
\label{mot2}
\end{eqnarray}
where $\rho_{\Lambda}$ is the cosmological density.
Equation (\ref{mot2})  can be viewed as a particular  polytropic equation of
state with index  $\gamma=0$ and negative polytropic constant
$K=-\rho_{\Lambda}c^2$. In this sense, the $\Lambda$CDM model is the simplest  
UDM model that one can imagine. Since the $\Lambda$CDM
model works well at large scales, a viable
model must necessarily be close to the $\Lambda$CDM model. However, it should
not coincide with it otherwise it would  not be able to solve the CDM small
scale crisis such as the core-cusp problem and the
missing satellite problem. Therefore, we need a model with a nonzero pressure
gradient which can balance the
gravitational attraction
in DM halos and avoid singularities. In addition, a successful model should
account for the constant surface density of DM halos $\Sigma_0^{\rm
obs}=141_{-52}^{+83}\, M_{\odot}/{\rm
pc}^2$,
something that the 
$\Lambda$CDM model does not do. Following  \cite{epjp}, we look for the
simplest 
extension of the standard $\Lambda$CDM model viewed as a UDM model. 

A first possibility would be to consider the polytropic equation of state
(\ref{mot1}) with an index $\gamma$ very close to zero (but nonzero).
Such a model can be  as successful as the $\Lambda$CDM model at
large scales. However, it seems hard to explain theoretically why a polytropic
index like, e.g.,
$\gamma=-0.0123$ should be selected by nature. Furthermore, if we let
$\gamma\rightarrow
0$ with $K$ fixed we recover the $\Lambda$CDM model so we have not gained
anything (in particular the small scale crisis remains).

Alternatively, in \cite{epjp} we considered
the limit $\gamma\rightarrow 0$ and $K\rightarrow \infty$ in such a way that
$A=K\gamma$ is finite. Interestingly, this leads to a model close to, but
different from, the $\Lambda$CDM model. This is how we justified the logotropic
model in \cite{epjp}. We recall below how the logotropic equation of state can 
be obtained from the polytropic equation of state in that limit
\cite{logo,epjp}. 

To that purpose we consider a nonrelativistic DM halo described
by the condition of hydrostatic
equilibrium 
\begin{eqnarray}
\nabla P+\rho\nabla\Phi={\bf 0}.
\label{diff4bis}
\end{eqnarray}
For the polytropic equation of state (\ref{mot1}),
this condition can be written as 
\begin{eqnarray}
K\gamma\rho^{\gamma-1}\nabla\rho+\rho\nabla\Phi={\bf 0}.
\label{mot3}
\end{eqnarray}
Taking the limit $\gamma\rightarrow 0$ and $K\rightarrow\infty$ with
$A=K\gamma$ finite, we obtain
\begin{eqnarray}
\frac{A}{\rho}\nabla\rho+\rho\nabla\Phi={\bf 0}.
\label{mot4}
\end{eqnarray}
Comparing this equation with Eq. (\ref{diff4}), we see that the pressure
involved in this expression corresponds to the logotropic equation of
state\footnote{Of course, we can obtain the logotropic equation of state
(\ref{mot5}) directly
from Eq. (\ref{mot1}) by
writing
\begin{eqnarray}
P=Ke^{\gamma\ln \rho}
\label{mot6}
\end{eqnarray}
and expanding the right hand side  for $\gamma\rightarrow 0$, yielding
\begin{eqnarray}
P=K(1+\gamma\ln\rho+...).
\label{mot6a}
\end{eqnarray}
In the limit $\gamma\rightarrow 0$ and $K\rightarrow \infty$ with
$A=K\gamma$ finite, we get
\begin{eqnarray}
P=A\ln\rho+K.
\label{mot6b}
\end{eqnarray}
The drawback with this calculation is that it yields an infinite constant
($K\rightarrow +\infty$) in addition to the logotropic equation of state so that
the procedure is not well-justified
mathematically. By contrast, the calculation based on Eq. (\ref{mot3})  avoids 
dealing explicitly with infinite constants since they disappear in the
gradients.
}
\begin{eqnarray}
P=A\ln\left (\frac{\rho}{\rho_*}\right ),
\label{mot5}
\end{eqnarray}
where $\rho_*$ is a constant of integration. It is interesting to note that the
logotropic equation of state, when coupled to gravity, yields the
Lane-Emden equation of index $n=-1$ (see Appendix \ref{sec_pldm}). Therefore, a
logotrope is closely related to a polytrope of index $\gamma=0$ (or
$n=-1$).\footnote{Note that the logotropic model differs from a
pure polytrope of index $\gamma=0$
(or $n=-1$) and fixed $K$ which has a constant pressure $P=K$. For this constant
pressure model, equivalent to the $\Lambda$CDM
model, the condition of hydrostatic
equilibrium (\ref{mot3}) has no solution (there is no equilibrium state) since
there is no pressure gradient.
As a result, in the framework of the
polytropic equation of state \cite{chandra,mcmh}, the Lane-Emden equation of
index $n=-1$
is ill-defined (the scale radius $r_0$ defined by Eq. (A7) of \cite{mcmh}
vanishes). Therefore, the limit
 $\gamma\rightarrow 0$ and $K\rightarrow \infty$ with $A=K\gamma$ finite leading
to the logotropic equation of state is very peculiar. The logotropic model
allows us to give a physical meaning to the Lane-Emden equation of index $n=-1$
which is
excluded by the usual polytropic model. In this sense, the logotropic model
naturally
completes the polytropic model.} In this sense, the logotropic model 
may
be viewed as the simplest extension  of the $\Lambda$CDM model (corresponding to $\gamma=0$) in the framework
of UDM models \cite{epjp}.

{\it Remark:} As explained above, the $\Lambda$CDM model is
equivalent to a fluid with a pressure that is independent of the density. On
the other hand, the logotropic equation of state depends on the density
very weakly (logarithmically). This is the argument that led us to introduce the
logotropic model \cite{epjp}. Interestingly, by developing this model,
we found
that the
constants $\rho_*$ and $A$ that appear in the logotropic equation of state
(\ref{mot5}) can
be determined by theoretical considerations and by observations (namely the
measured
values of $\Omega_{\rm m,0}$ and $H_0$). As a result, there is no adjustable
parameter in our model. Furthermore, this model can account for the
observed value of the surface density of DM halos $\Sigma_0^{\rm
obs}=141_{-52}^{+83}\, M_{\odot}/{\rm
pc}^2$. Following our paper
\cite{epjp}, some authors \cite{cal1,ootsm,cal2} have
introduced a simple extension of the logotropic model  by considering an
equation of state of the form
\begin{eqnarray}
P=A\left (\frac{\rho}{\rho_*}\right )^{-n}\ln\left (\frac{\rho}{\rho_*}\right ),
\label{mot7}
\end{eqnarray}
where $n$ is a free 
parameter. Interestingly, this equation
of state is similar to the Anton-Schmidt \cite{as} equation of state for
crystalline solids in the Debye approximation \cite{debye}. In that case, the
index $n$ can be written as $n=-1/6-\gamma_G$ where $\gamma_G$ is the so-called
Gr\"uneisen \cite{gruneisen} parameter.  The original logotropic model is
recovered for $n=0$. However, since  $n$ is a free parameter the generalized
logotropic model (\ref{mot7}) introduces some indetermination (or freedom) in
the analysis while the
original logotropic model \cite{epjp} is completely predictive. By
comparing  the results
of the generalized logotropic
model (\ref{mot7}) with cosmological observations, the authors
of \cite{cal1,ootsm,cal2} found that $B\simeq 3.54\times
10^{-3}$ and $n=-0.147_{-0.107}^{+0.113}$. This confirms the robustness of
the value of the fundamental constant $B= 3.53\times 10^{-3}$ introduced in
\cite{epjp,lettre}. On the
other hand, up to the error bars,  the value of $n$ is close to
$n=0$, corresponding to the
logotropic model (see also \cite{jcap}). This suggests that the logotropic model
tends to be selected
among more general families of models containing additional parameters
$\lbrace n\rbrace$.

\section{Logotropic models of type I, II and III}
\label{sec_lmt}

As explained in  \cite{action} we can introduce three types of barotropic equations of
state with the same functional
form depending on whether the pressure $P$ is expressed in terms of the energy
density $\epsilon$ (model I), the rest-mass density $\rho_m=n m$ (model II), or
the pseudo rest-mass density $\rho$ (model III). These models are equivalent in
the nonrelativistic limit but they differ from each other in the relativistic
regime. A detailed discussion of these models and their interrelations is given
in \cite{action} (see also \cite{epjp,partially,stiff,abrilphas}). In this
Appendix, we briefly discuss
these models in the framework of  the logotropic equation of state.

\begin{figure}[!h]
\begin{center}
\includegraphics[clip,scale=0.3]{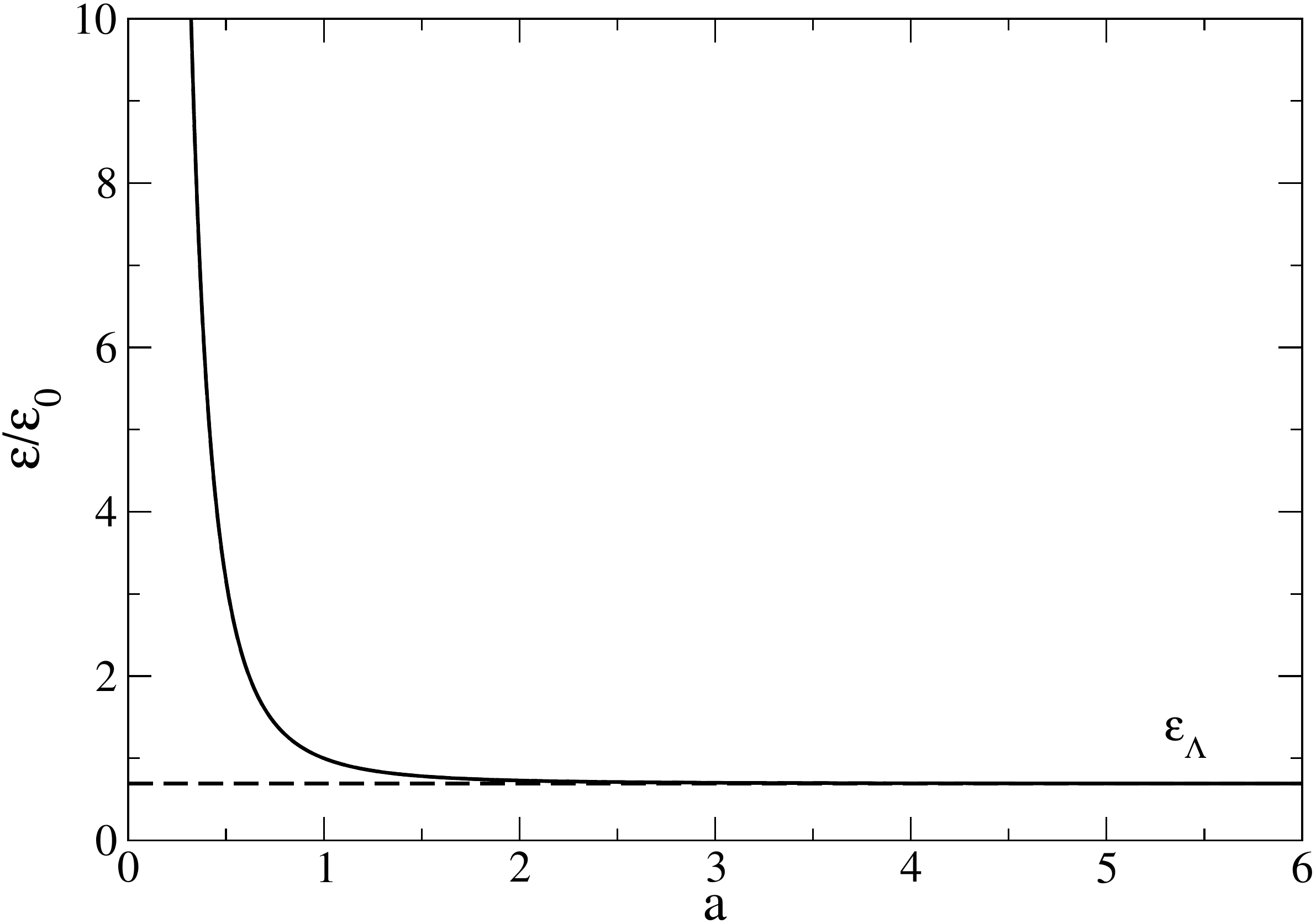}
\caption{Normalized energy density $\epsilon/\epsilon_0$ as a function of the
scale factor $a$ for the logotropic model of type I. It is compared with the
$\Lambda$CDM model. The two curves are indistinguishable on the figure.}
\label{aepsN}
\end{center}
\end{figure}

Barotropic models of type I correspond to an equation of state of
the form $P=P(\epsilon)$, where $\epsilon$ is the energy density. The logotropic
model of type I is therefore
\begin{eqnarray}
P=A\ln\left (\frac{\epsilon}{\rho_Pc^2}\right ).
\label{lmt1}
\end{eqnarray}
The energy conservation equation (\ref{hsf4}) combined with the
logotropic
equation of state (\ref{lmt1}) yields
\begin{equation}
\label{dim1}
\ln a=-\frac{1}{3}\int_{\epsilon_0}^{\epsilon}
\frac{d\epsilon'}{\epsilon'+A\ln\left (\frac{\epsilon'}{\rho_P c^2}\right )},
\end{equation}
where $\epsilon_0$ denotes the present energy density of the universe (when
$a=1$).  This equation determines the evolution of the energy density
$\epsilon(a)$ as a function of the scale factor. When $a\rightarrow 0$, we get
$\epsilon\propto a^{-3}$ similar to DM. When
$a\rightarrow +\infty$ we get $\epsilon\rightarrow \epsilon_{\rm min}$
similar to DE where
$\epsilon_{\rm min}$ is the solution of the equation $\epsilon_{\rm
min}+A\ln\left ({\epsilon_{\rm min}}/{\rho_P c^2}\right )=0$. This leads to
an exponential (de Sitter) expansion like in the $\Lambda$CDM
model. If we identify $\epsilon_{\rm min}$ with the DE density
$\rho_{\Lambda}c^2$ in the $\Lambda$CDM
model (which coincides with the asymptotic value of $\epsilon$), we get
\begin{eqnarray}
A=\frac{\rho_{\Lambda}c^2}{\ln\left(\frac{\rho_P}{\rho_{\Lambda}}\right )}.
\label{dim2}
\end{eqnarray}
This returns  the relation obtained in the logotropic model of
type II \cite{epjp} and in the logotropic model of type III [see Eq.
(\ref{a2b})]. This strengthens the validity of this relation
\cite{prep}. If we set $x=\epsilon'/\epsilon_0$,
$A=B\rho_{\Lambda}c^2$ and
$B=1/\ln(\rho_P/\rho_{\Lambda})$ with
$\rho_{\Lambda}=\Omega_{\rm de,0}\epsilon_0/c^2$, we can rewrite Eq.
(\ref{dim1}) as
\begin{equation}
\label{dim3}
\ln a=-\frac{1}{3}\int_{1}^{\epsilon/\epsilon_0}
\frac{dx}{x+B\Omega_{\rm de,0}\left (\ln
x-\ln\Omega_{\rm de,0}-\frac{1}{B}\right )}.
\end{equation}
The function $\epsilon/\epsilon_0(a)$ is plotted in Fig. \ref{aepsN}. We have
taken $\Omega_{\rm de,0}=0.6911$ and $B=3.53\times 10^{-3}$. The logotropic
model of type I behaves similarly to the $\Lambda$CDM
model. This model will be studied in more detail in a future work
\cite{prep}.

Barotropic
models of type II correspond to an equation of state of
the form  $P=P(\rho_m)$, where $\rho_m=n m$ is the rest-mass
density ($n$ is the particle number density). The logotropic model of type II
is therefore
\begin{eqnarray}
P=A\ln\left (\frac{\rho_m}{\rho_P}\right ).
\label{lmt3}
\end{eqnarray}
This is the original  logotropic model introduced in
\cite{epjp}.

Barotropic
models of type III correspond to an equation of state of
the form  $P=P(\rho)$, where $\rho$ is the pseudo rest-mass
density associated with a complex SF (see Sec. \ref{sec_csf}). The
logotropic model of type III
is therefore
\begin{eqnarray}
P=A\ln\left (\frac{\rho}{\rho_P}\right ).
\label{lmt4}
\end{eqnarray}
This is the logotropic model studied in the present paper. It
could be called the logotropic complex SF model and
refered to as logotropic CSF (or LCSF) model.

In the nonrelativistic limit, we have $\rho=\rho_m$ and $\epsilon\sim \rho c^2$
so the three models become equivalent. They correspond to an equation of state
of the form $P=P(\rho)$, where $\rho$ is the mass
density. The nonrelativistic  logotropic equation of state is
\begin{eqnarray}
P=A\ln\left (\frac{\rho}{\rho_P}\right ).
\label{lmt5}
\end{eqnarray}
The structure of logotropic DM halos described by the equation of state
(\ref{lmt5}) has been studied  in  \cite{epjp} (see also
Appendix \ref{sec_pldm}).

\section{DM with a linear equation of state}
\label{sec_lin}

In the CDM model it is assumed that DM is pressureless ($P=0$). In this
Appendix, we consider the possibility that CDM is described by a linear equation
of state $P=\alpha\epsilon$ with $\alpha\ge 0$, yielding a constant (nonzero)
 speed of sound $c_s=\sqrt{\alpha}c$. We determine the condition that $\alpha$
must satisfy in order to be consistent with the observations of the matter power
spectrum.

In the nonrelativistic regime where $\epsilon\sim \rho c^2$, we can rewrite the
equation of state as $P=\alpha\rho c^2$.  This linear equation of state can be
interpreted  as an isothermal equation of state of the form $P=\rho k_B T_{\rm
eff}/m$ with $\alpha=k_B T_{\rm eff}/mc^2$. Here, $T_{\rm eff}$ is a temperature
which may be identified with the effective temperature of DM halos. For a
typical DM halo of mass $M_h=10^{11}\, M_{\odot}$ (medium spiral), one has $(k_B
T_{\rm eff}/m)^{1/2}=108\, {\rm km/s}$ \cite{modeldm}. This gives $\alpha\sim
10^{-7}$. 

According to Eqs. (\ref{jsl2}) and (\ref{jsl3}) the Jeans length and the Jeans
mass are given by
\begin{eqnarray}
\lambda_J=2\pi \left (\frac{\alpha c^2}{4\pi G}\right
)^{1/2}\frac{1}{\rho^{1/2}},
\label{iso1}
\end{eqnarray}
\begin{eqnarray}
M_J=\frac{4}{3}\pi^4 \left (\frac{\alpha c^2}{4\pi G}\right
)^{3/2}\frac{1}{\rho^{1/2}}.
\label{iso2}
\end{eqnarray}
Using Eq. (\ref{jsl1}), we find that during the expansion of the Universe the
Jeans length and the Jeans mass both increases as
$a^{3/2}$ (the comoving
Jeans length increases as
$a^{1/2}$). Eliminating the
density between  Eqs.
(\ref{jsl5}) and (\ref{jsl6}), we obtain  
\begin{eqnarray}
M_J=\frac{\pi^2}{6}\frac{\alpha c^2}{G}\lambda_J.
\label{iso3}
\end{eqnarray}
This relation  is similar to the mass-radius
relation of an isothermal self-gravitating system confined
within a box \cite{aaiso}.  
Repeating the calculations made at the end of Sec. \ref{sec_jsl} we can
rewrite the Jeans length and the Jeans mass as
\begin{eqnarray}
\lambda_J=2\pi\left ( \frac{2\alpha}{3\Omega_{\rm
m,0}}\right )^{1/2} R_{\Lambda}a^{3/2},
\label{iso4}
\end{eqnarray}
\begin{eqnarray}
M_J=\pi^3\left ( \frac{2\alpha}{3\Omega_{\rm
m,0}}\right )^{3/2} \Omega_{\rm m,0} M_{\Lambda}a^{3/2},
\label{iso5}
\end{eqnarray}
where $R_{\Lambda}=4.44\times 10^3\, {\rm Mpc}$ and $M_{\Lambda}=4.62\times
10^{22}\, M_{\odot}$ represent the typical radius and mass
of
the visible Universe. As
we have seen in footnote 45, observational constraints from the
matter power spectrum require that $\lambda_J(a=1)<R$ with $R\sim 15\, {\rm
Mpc}$. Since $\lambda_J(a=1)\sim 10\sqrt{\alpha}R_{\Lambda}$, we need 
$100\alpha<(R/R_{\Lambda})^2\sim 10^{-5}$. This imposes $\alpha<10^{-7}$ in
agreement with the findings of \cite{muller}.  Interestingly, the
value $\alpha\sim 10^{-7}$ obtained above from the effective temperature of DM
halos
satisfies this constraint.

\section{$\Lambda$CDM model}
\label{sec_lcdm}

In this Appendix, we discuss different equivalent manners to introduce the
$\Lambda$CDM model.

\subsection{DM $+$ $\Lambda$}
\label{sec_dml}

The usual manner to introduce the $\Lambda$CDM model in cosmology is to
assume that the
Universe is filled with DM (in addition to baryonic
matter and radiation that we do not consider here for brevity) and that the
Einstein cosmological constant $\Lambda$ has a nonzero value. DM is
introduced
to explain the formation of the large scale structures of the Universe and the
flat rotation curves of the galaxies (see Appendix
\ref{sec_cdmh}). A positive cosmological
constant is introduced to explain the present acceleration of the Universe.  

DM is usually treated as a pressureless fluid with an equation of state 
\begin{eqnarray}
P_{\rm m}=0.
\label{ecdm1}
\end{eqnarray}
Solving the energy conservation equation
(\ref{hsf4}) with the equation of state (\ref{ecdm1}), we obtain 
\begin{eqnarray}
\epsilon_{\rm m}=\frac{\epsilon_{\rm
m,0}}{a^3},
\label{ecdm2}
\end{eqnarray}
where $\epsilon_{\rm m,0}$ is a constant of integration which can be identified
with the
present energy density of DM.

On the other hand, considering the Friedmann equation (\ref{fe1}), we
see
that the cosmological constant $\Lambda$ is equivalent to a constant energy density
\begin{eqnarray}
\epsilon_{\Lambda}= \rho_{\Lambda}c^2=\frac{\Lambda
c^2}{8\pi G}.
\label{ecdm3}
\end{eqnarray}
Substituting Eq. (\ref{ecdm2}) into the Friedmann equation
(\ref{fe1}), we get
\begin{eqnarray}
H^2=\frac{8\pi G \epsilon_{\rm
m,0}}{3c^2a^3}+\frac{\Lambda}{3},
\label{ecdm4}
\end{eqnarray}
where we have assumed $k=0$. Eq. (\ref{ecdm4}) is equivalent to Eq. (\ref{hsf5}) with a total energy density
\begin{eqnarray}
\epsilon=\frac{\epsilon_{\rm
m,0}}{a^3}+\epsilon_{\Lambda}.
\label{ecdm5}
\end{eqnarray}

\subsection{DM $+$ DE}
\label{sec_dmde}

A second manner to introduce the
$\Lambda$CDM model is to assume that the Universe is filled with DM and DE
interpreted as two noninteracting fluids (in that case we take $\Lambda=0$ in
Eq. (\ref{fe1})). DM is 
treated as a pressureless fluid with the equation of state 
(\ref{ecdm1}). Its energy density evolves with the scale factor according to Eq.
(\ref{ecdm2}).
DE is treated as a fluid with a negative pressure determined by the linear
equation of state
\begin{eqnarray}
P_{\rm de}=-\epsilon_{\rm de}.
\label{ecdm6}
\end{eqnarray}
Solving the energy conservation equation
(\ref{hsf4}) with the equation of state (\ref{ecdm6}), we obtain
\begin{eqnarray}
\epsilon_{\rm de}=\epsilon_{\Lambda},
\label{ecdm7}
\end{eqnarray}
where $\epsilon_{\Lambda}$ is a constant of integration that is identified with
the cosmological density.

The total energy density of
the Universe is the sum of DM and DE: $\epsilon=\epsilon_{\rm m}+\epsilon_{\rm
de}$. Summing Eqs.
(\ref{ecdm2}) and  (\ref{ecdm7}), we get
\begin{eqnarray}
\epsilon=\frac{\epsilon_{\rm
m,0}}{a^3}+\epsilon_{\Lambda},
\label{ecdm8}
\end{eqnarray}
which is equivalent to Eq. (\ref{ecdm5}). Introducing the present energy
density  of the Universe $\epsilon_0=3c^2H_0^2/8\pi G$ (where $H_0$ is the
present value of the Hubble constant) and the
present fraction of DM and DE given by $\Omega_{\rm m,0}=\epsilon_{\rm
m,0}/\epsilon_0$ and
$\Omega_{\rm de,0}=\epsilon_{\Lambda}/\epsilon_0=1-\Omega_{ \rm
m,0}$, we obtain
\begin{eqnarray}
\frac{\epsilon}{\epsilon_0}=\frac{\Omega_{\rm m,0}}{a^3}+1-\Omega_{\rm
m,0}.
\label{ecdm9}
\end{eqnarray}

The $\Lambda$CDM model involves two unknown parameters $\epsilon_0$ and
$\Omega_{\rm m,0}$ that must be determined by the observations. When
$a\rightarrow 0$, the Universe is dominated by DM and we have
\begin{eqnarray}
\epsilon\sim\frac{\Omega_{\rm m,0}\epsilon_0}{a^3},
\label{ecdm10}
\end{eqnarray}
leading to a decelerated expansion (Einstein-de Sitter era).
When $a\rightarrow
+\infty$, the Universe is dominated by DE and we have
\begin{eqnarray}
\epsilon\rightarrow \epsilon_\Lambda=(1-\Omega_{\rm
m,0})\epsilon_0.
\label{ecdm11}
\end{eqnarray}
The energy density tends to a constant, leading to an exponential expansion (de
Sitter era).

{\it Remark:} Introducing the dimensionless variables of Sec. \ref{sec_dime},
we can rewrite Eq. (\ref{ecdm9}) as
\begin{eqnarray}
\tilde\epsilon=\frac{\Omega_{\rm m,0}}{1-\Omega_{\rm m,0}}\frac{1}{a^3}+1.
\label{ecdm12}
\end{eqnarray}
The equality between DM and DE in the $\Lambda$CDM
model corresponds to a scale factor
\begin{equation}
a_t=\left (\frac{\Omega_{\rm m,0}}{1-\Omega_{\rm m,0}}\right
)^{1/3}=0.765,
\label{atrans}
\end{equation} 
an energy density ${\tilde\epsilon}_t=2$, and a value 
of the equation of state parameter $w_t=-1/2$.

\subsection{DF}
\label{sec_df}

A third manner to introduce the
$\Lambda$CDM model is to assume that the Universe is filled with a single DF 
(in that case we take $\Lambda=0$ in Eq. (\ref{fe1}))
with a constant equation of state
\begin{eqnarray}
P=-\epsilon_{\Lambda},
\label{ecdm14}
\end{eqnarray}
where $\epsilon_{\Lambda}$ is identified with
the
cosmological density. 
We stress that this constant equation of state is different from the
linear
equation of state (\ref{ecdm6}). Solving the
energy conservation equation
(\ref{hsf4}) with the equation of state (\ref{ecdm14}), we obtain 
\begin{eqnarray}
\epsilon=\frac{\epsilon_{\rm
m,0}}{a^3}+\epsilon_{\Lambda},
\label{ecdm15}
\end{eqnarray}
where $\epsilon_{\rm
m,0}$ is a constant of integration. This equation is equivalent to Eq.
(\ref{ecdm5}) or Eq. (\ref{ecdm8}). The equation of state parameter is 
\begin{eqnarray}
w=\frac{P}{\epsilon}=\frac{-\epsilon_\Lambda}{\frac{\epsilon_{\rm
m,0}}{a^3}+\epsilon_{\Lambda}}.
\label{ecdm15b}
\end{eqnarray}
The squared speed of sound
$c_s^2=P'(\epsilon)c^2$ is
equal to zero. This single DF model, based on the  
equation of state (\ref{ecdm14}), provides the simplest unification of
DM and DE that one can imagine and it coincides with the usual $\Lambda$CDM
model from Appendices \ref{sec_dml} and \ref{sec_dmde}.\footnote{It is shown in
Refs. \cite{sandvik,avelinoZ} that the
constant pressure model (\ref{ecdm14}) is equivalent to the $\Lambda$CDM not
only for the evolution of the background but to all orders in perturbation
theory, even in the nonlinear clustering regime (contrary to the initial claim
of \cite {fabris2004GR}). If we consider
the affine equation of state
$P=\alpha\epsilon-\epsilon_\Lambda$, which yields a constant squared speed of
sound $c_s^2=\alpha c^2$, we obtain \cite{cosmopoly1} 
\begin{eqnarray}
\epsilon=\frac{\epsilon_{\rm
m,0}}{a^{3(1+\alpha)}}+\epsilon_{\Lambda}.
\label{ecdm15bu}
\end{eqnarray}
This is equivalent to a two-fluid model with $P=\alpha\epsilon$ (DM) and
$P=-\epsilon$ (DE).} In
this connection, the first term in Eq. (\ref{ecdm15}) plays the role of DM and
the second term plays the role of DE. As shown in \cite{epjp} at a 
general level, the effective DM term corresponds to the rest-mass energy
$\rho_m c^2$ of the
DF and the effective DE term corresponds to its internal energy $u$ (for the
$\Lambda$CDM model, Eq. (\ref{ecdm15}) can be obtained from Eqs. (\ref{rmd4})
and (\ref{rmd5}) with the equation of state (\ref{ecdm14}) yielding
$u=\epsilon_{\Lambda}$).

{\it Remark:} The relation (\ref{ecdm15}) between the energy
density and the scale factor can be rewritten as
\begin{equation}
\epsilon=\rho_\Lambda c^2 \left\lbrack \left (\frac{a_t}{a}\right
)^3+1\right\rbrack,
\label{ecdm15x}
\end{equation}
where $a_t$ is the transition scale factor defined by Eq. (\ref{atrans}).
Solving the Friedmann equation (\ref{hsf5}) with the energy density given by Eq.
(\ref{ecdm15x}), we find that the
temporal evolution of the scale factor is then given by
\begin{equation}
\frac{a}{a_t}=\sinh^{2/3}\left (\sqrt{6\pi G\rho_{\Lambda}}t\right ).
\label{atlcdm}
\end{equation}

\subsection{CDM halos}
\label{sec_cdmh}

Classical numerical simulations of CDM lead to DM halos with a universal density
profile that is well-fitted by the function
\begin{equation}
\label{comp1}
\rho(r)\propto \frac{1}{\frac{r}{r_s}\left
(1+\frac{r}{r_s}\right )^2},
\end{equation}
where $r_s$ is a scale radius that varies from halo to halo. This is the
so-called NFW profile \cite{nfw}. Such halos results from a process
of violent collisionless relaxation. The density decreases as $r^{-3}$
for $r\rightarrow +\infty$ and diverges as $r^{-1}$ for $r\rightarrow 0$. The
divergence of the density at short distances is related to the fact that
classical CDM halos are pressureless ($P=0$) so there is no pressure gradient
to balance the gravitational attraction. This divergence is not
consistent with observations that reveal that DM halos possess a core, not
a cusp. Observed DM halos are better fitted by the function 
\begin{equation}
\label{comp2}
\rho(r)=\frac{\rho_0}{\left (1+\frac{r}{r_h}\right )\left
(1+\frac{r^2}{r_h^2}\right )},
\end{equation}
where $\rho_0$ is the central density and $r_h$ is the halo radius defined as
the distance at which the
central density  $\rho_0$  is divided by $4$. This is the so-called Burkert
profile
\cite{burkert}. The density decreases as $r^{-3}$ for
$r\rightarrow
+\infty$,
similarly to the NFW profile, but displays a flat core for $r\rightarrow
0$ instead of a cusp. It is important to recall, however, that the Burkert
profile is purely empirical and has no fundamental justification.

\subsection{SF}
\label{sec_rep}

The previous models are purely classical (non quantum) since $\hbar$ does not
explicitly appear in the equations. However, it is possible to introduce SF
models that reproduce, in certain limits, the $\Lambda$CDM model. These models
are more general than the $\Lambda$CDM model since a SF, being governed by the
KG equation, has a quantum origin.

Let us first consider a spatially homogeneous real SF evolving
according to the KG equation\footnote{Here $V(\varphi)$ denotes the {\it total}
potential of the SF including the rest-mass term. In addition,
the time variable stands here for $ct$.}
\begin{equation}
\label{mtu4}
\ddot \varphi+3H\dot\varphi+\frac{dV}{d\varphi}=0
\end{equation}
coupled to the Friedmann equation (\ref{hsf5}). The SF tends
to run down the potential towards lower energies and is submitted to  an
Hubble friction. The density and the pressure of
the  SF are given by
\begin{equation}
\label{mtu5}
\epsilon=\frac{1}{2}\dot\varphi^2+V(\varphi),
\end{equation}
\begin{equation}
\label{mtu6}
P=\frac{1}{2}\dot\varphi^2-V(\varphi).
\end{equation}
We can easily check that these equations imply the energy conservation
equation (\ref{hsf4}) \cite{action}. For a
general equation of
state
$P(\epsilon)$, using standard techniques \cite{em,paddytachyon,cst,bamba} we can
obtain the SF potential as follows \cite{cosmopoly2}. From
Eqs. (\ref{mtu5}) and (\ref{mtu6}), we get
\begin{eqnarray}
\label{mtu8}
\dot\varphi^2=(w+1)\epsilon,
\end{eqnarray}
where we have defined $w=P/\epsilon$. Using
$\dot\varphi=(d\varphi/da)Ha$ and the
Friedmann equation (\ref{hsf5}), we find that the relation between
the SF and the
scale factor is given by\footnote{We assume a non-phantom Universe
$w>-1$.}
\begin{eqnarray}
\label{mtu9}
\frac{d\varphi}{da}=\left (\frac{3c^4}{8\pi G}\right
)^{1/2}\frac{\sqrt{1+w}}{a}.
\end{eqnarray}
On the other hand, according to Eqs. (\ref{mtu5}) and (\ref{mtu6}), the
potential of the SF is given by
\begin{eqnarray}
\label{mtu10}
V=\frac{1}{2}(1-w)\epsilon.
\end{eqnarray}
Therefore, the potential of the SF is determined in parametric form by the
equations 
\begin{equation}
\label{mtu11}
\varphi(a)=\left (\frac{3c^4}{8\pi G}\right )^{1/2}\int
\sqrt{1+w(a)}\, \frac{da}{a},
\end{equation}
\begin{equation}
\label{mtu12}
V(a)=\frac{1}{2}\left\lbrack 1-w(a)\right\rbrack
\epsilon(a).
\end{equation}
For the constant equation of state (\ref{ecdm14}) corresponding to the
$\Lambda$CDM model in its UDM interpretation, 
Eq. (\ref{mtu11}) with Eq. (\ref{ecdm15b})  is readily integrated leading to the
hyperbolic potential \cite{gkmp,cosmopoly2}
\begin{eqnarray}
V(\psi)=\frac{1}{2}\rho_\Lambda c^2 (\cosh^2\psi+1),
\label{sfr8}
\end{eqnarray}
where
\begin{eqnarray}
\psi=-\left (\frac{8\pi G}{3c^4}\right )^{1/2}\frac{3}{2}\varphi.
\label{sfr4}
\end{eqnarray}
The SF
is related to the scale factor by
\begin{eqnarray}
(a/a_t)^{-3/2}=\sinh\psi,
\label{sfr4b}
\end{eqnarray}
where $a_t$ is the transition scale factor defined
by Eq. (\ref{atrans}) and $\psi\ge 0$. We
note that this solution is exact in the sense that it does not rely on
any approximation. However, it corresponds to a very particular initial
condition of the KGF equations \cite{gkmp}. We also note that the SF
does not oscillate.
According to Eqs. (\ref{atlcdm}) and (\ref{sfr4b}) it gently descends the
potential.\footnote{We can
also associate to the $\Lambda$CDM model a tachyonic SF with a potential (see
\cite{gkmp,cosmopoly2} for details) 
\begin{eqnarray}
V(\psi)=\frac{\rho_{\Lambda}c^2}{\cos\psi},
\end{eqnarray}
where $\psi=-\sqrt{6\pi
G\rho_{\Lambda}/c^2}\varphi$. The SF is related to the scale
factor by $(a/a_t)^{-3/2}=\tan\psi$ with $0\le\psi\le \pi/2$.}

The $\Lambda$CDM model can also be obtained from a real SF model
with a potential
\begin{equation}
{V}(\varphi)=\frac{m^2c^2}{2\hbar^2}\varphi^2+\epsilon_{\Lambda}.
\label{vcdm1ki}
\end{equation}
In the fast oscillation regime, the SF experiences slowly damped oscillations
and behaves as DM. When it reaches the bottom of the potential, the energy
density becomes constant ($\epsilon=\epsilon_\Lambda$) and the SF behaves as DE
(cosmological constant).

We can also consider a complex SF with  a
potential
\begin{equation}
{V}_{\rm
tot}(|\varphi|^2)=\frac{m^2c^2}{2\hbar^2}|\varphi|^2+\epsilon_{\Lambda}.
\label{vcdm1}
\end{equation}
This model, referred to as the $\Lambda$FDM model, is considered
in detail in
Appendix \ref{sec_gfdm}. Here, we just note that, in the fast oscillation regime
where quantum effects can be neglected (TF approximation), the SF undergoes a process of 
spintessence (it slowly descends the potential by rapidly spinning about the
vertical axis) and behaves like the $\Lambda$CDM model.

We note that the shifted quadratic potentials
from Eqs. (\ref{vcdm1ki}) and (\ref{vcdm1}) are very different from
the hyperbolic potential from Eq. (\ref{sfr8}). In addition, the SF oscillates
or
spins rapidly
in the potentials from Eqs. (\ref{vcdm1ki}) and (\ref{vcdm1}) while it just descends the potential from Eq. (\ref{sfr8}) without oscillating. These remarks  show that several SF models can
behave just like the $\Lambda$CDM model while being fundamentally different from each
others. 

{\it Remark:} If we expand
Eq. (\ref{sfr8}) for $\varphi\rightarrow 0$ we find that
\begin{eqnarray}
V(\varphi)=\rho_\Lambda c^2+\frac{9m_{\Lambda}^2c^2}{8\hbar^2}\varphi^2+...
\label{sfr8exp}
\end{eqnarray}
We see that the minimum of the potential is equal to the cosmological
density $V_0=\rho_\Lambda c^2$ and that the mass of the SF is
$m=(3/2)m_{\Lambda}$, where $m_{\Lambda}=1.20\times 10^{-33}\, {\rm eV/c^2}$ is
the cosmon mass [see Eq. (\ref{v4})]. Our approach provides
therefore a physical
interpretation to the cosmon mass as being the mass of the SF responsible for
the DE in the late universe. To the best of our knowledge, this
interpretation has not been given before. In comparison, the
mass of the SF in the
$\Lambda$FDM model (\ref{vcdm1}) is of order $m\sim 10^{-22}\, {\rm eV/c^2}$
(see Appendix \ref{sec_gfdm}). 

\section{$\Lambda$FDM model}
\label{sec_gfdm}

In this Appendix, we consider a complex SF model with a
constant potential
$V=\epsilon_\Lambda$ equal to the cosmological density. This
model generalizes the relativistic FDM model described by the
KGE equations (\ref{lwe1}) and (\ref{lwe2}) with $V=0$. In the fast
oscillation regime or in the TF approximation (where quantum effects can be
neglected), it
coincides with the $\Lambda$CDM model.  On the other hand, when quantum
effects are taken into account but relativistic effects are neglected (as in
the case of DM halos), it coincides with the nonrelativistic FDM model 
\cite{hu} described
by the GPP equations (\ref{lwe4}) and (\ref{lwe5}) with $V=0$
reducing to the
Schr\"odinger-Poisson equations. We shall call it the
$\Lambda$FDM model.

\subsection{Potential of the $\Lambda$FDM model}
\label{sec_vcdm}

We assume that DM and DE are described by a single complex SF with a constant
potential
\begin{equation}
V=\epsilon_{\Lambda},
\label{vcdm3}
\end{equation}
where $\epsilon_\Lambda$ is the cosmological density. 
In the fast oscillation regime, using Eq.
(\ref{ge8}), we find that the pressure is given by
\begin{equation}
P=-\epsilon_{\Lambda}.
\label{vcdm4}
\end{equation}
Therefore, the pressure is constant as in the $\Lambda$CDM model [see Eq. (\ref{ecdm14})]. On the other
hand, the equations governing the evolution of the homogeneous background in
the fast oscillation regime [Eqs. (\ref{ge5})-(\ref{ge9})] are
\begin{equation}
\rho=\frac{Qm}{a^3}, 
\label{vcdm6}
\end{equation}
\begin{equation}
\epsilon=\rho c^2+\epsilon_{\Lambda},
\label{vcdm5}
\end{equation}
\begin{equation}
E_{\rm tot}=mc^2, 
\label{vcdm7}
\end{equation}
\begin{equation}
w=\frac{P}{\epsilon}=-\frac{\epsilon_{\Lambda}}{\rho c^2+\epsilon_{\Lambda}},
\label{vcdm7b}
\end{equation}
\begin{equation}
c_s=0.
\label{vcdm7c}
\end{equation}
They return the equations of the $\Lambda$CDM model (see Appendix
\ref{sec_lcdm}). In particular, combining Eqs. (\ref{vcdm6}) and (\ref{vcdm5}),
we obtain
\begin{equation}
\epsilon=\frac{Qmc^2}{a^3}+\epsilon_{\Lambda},
\label{vcdm5b}
\end{equation}
which is equivalent to Eq. (\ref{ecdm15}) with the identification
$Qmc^2=\epsilon_{\rm m,0}$. The
constant $Qmc^2$ (charge of the SF) is equal to the
present energy density of DM $\epsilon_{{\rm m},0}=\Omega_{{\rm
m},0}\epsilon_0$. This result is valid for an arbitrary potential $V$ (see
Sec. \ref{sec_rmd}). However, for a constant potential,  the pseudo
rest-mass density $\rho$ coincides with the rest-mass density $\rho_m$ [see
Eq. (\ref{rmd3})] and
plays the role of DM ($\rho=\rho_m$). On the other hand, the internal energy is
constant ($u=\epsilon_\Lambda$) and plays the role of DE [see Eq.
(\ref{rmd6})].

The total potential of the SF including the rest-mass term  is 
\begin{equation}
{V}_{\rm
tot}(|\varphi|^2)=\frac{m^2c^2}{2\hbar^2}|\varphi|^2+\epsilon_{\Lambda}.
\label{vcdm1kk}
\end{equation}
This is a shifted quadratic potential.
Using Eq. (\ref{ge1}), it can be written a
\begin{equation}
{V}_{\rm tot}=\frac{1}{2}\rho c^2+\epsilon_{\Lambda}.
\label{vcdm2}
\end{equation}
The total potential of the SF is represented by a dashed line in Fig.
\ref{vtot}. The SF descends the
potential on the surface of the ``bowl''  up to the origin
$|\varphi|=0$ by rapidly
spinning around the vertical axis (see Sec. \ref{sec_tpd}).

{\it Remark:} The ordinary FDM model corresponds to a complex SF with a
vanishing potential $V=0$. In the fast oscillation regime, it just describes
pressureless DM ($P=0$). Therefore, it does not provide a unification of DM and
DE. DE has to be introduced in a different manner, either by introducing another species
(like quintessence)
or through a nonvanishing cosmological constant $\Lambda$. The ordinary FDM
model ($V=0$) $+$ a
cosmological constant  is the
complex SF
generalization of the $\Lambda$CDM model of Appendix \ref{sec_dml}. We shall
call it the
$\Lambda$FDM model. The FDM model with
a constant potential $V=\epsilon_\Lambda$ provides a simple unification of DM
and DE. This is the complex SF generalization of the  $\Lambda$CDM model viewed
as a DF or a UDM model (see Appendix \ref{sec_df}). We shall 
also call it the  $\Lambda$FDM model.

\subsection{Validity of the fast oscillation regime}
\label{sec_fcdm}

\begin{figure}[!h]
\begin{center}
\includegraphics[clip,scale=0.3]{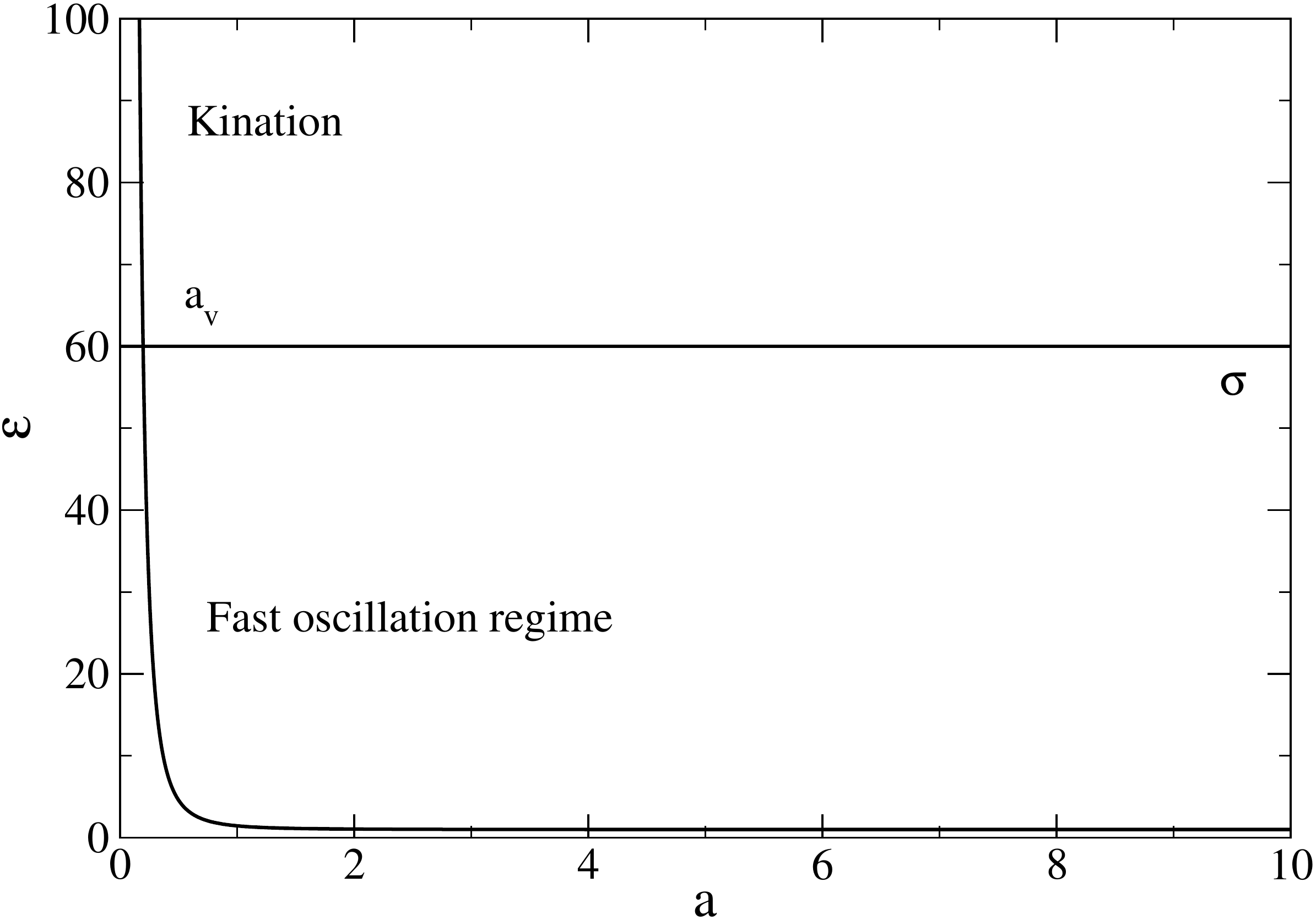}
\caption{Graphical construction determining the range of validity of the fast
oscillation regime in the $\Lambda$FDM model.}
\label{validityCDM}
\end{center}
\end{figure}

Introducing the dimensionless variables of Secs. \ref{sec_dime} and \ref{sec_v},
and using Eq. (\ref{ecdm12}), we find that the
fast oscillation regime of the $\Lambda$FDM model (where it is equivalent to
the
$\Lambda$CDM model) is valid for $\tilde\epsilon\ll\sigma$, where $\sigma$ is
defined by Eqs. (\ref{v2}) and (\ref{v3}). This criterion first requires that
$\sigma\gg 1$, i.e., 
$m\gg m_{\Lambda}=1.20\times
10^{-33}\, {\rm eV/c^2}$. Therefore, the mass of the SF must be much larger than
the cosmon mass. When this condition is fulfilled, the fast
oscillation regime
is valid for
$a\gg a_v$ (see Fig. \ref{validityCDM}) with 
\begin{eqnarray}
\frac{a_v}{a_t}=\left (
\frac{1}{\sigma-1}\right )^{1/3},
\label{fcdm1}
\end{eqnarray} 
where $a_t$ is the transition scale factor from
Eq. (\ref{atrans}). The fast oscillation regime is not valid for
$a<a_v$. In that case,
the SF is in a slow oscillation regime of kination. This gives rise to a stiff matter era 
as discussed in \cite{shapiro,abrilphas} and in Sec. \ref{sec_v}.

\begin{figure}[!h]
\begin{center}
\includegraphics[clip,scale=0.3]{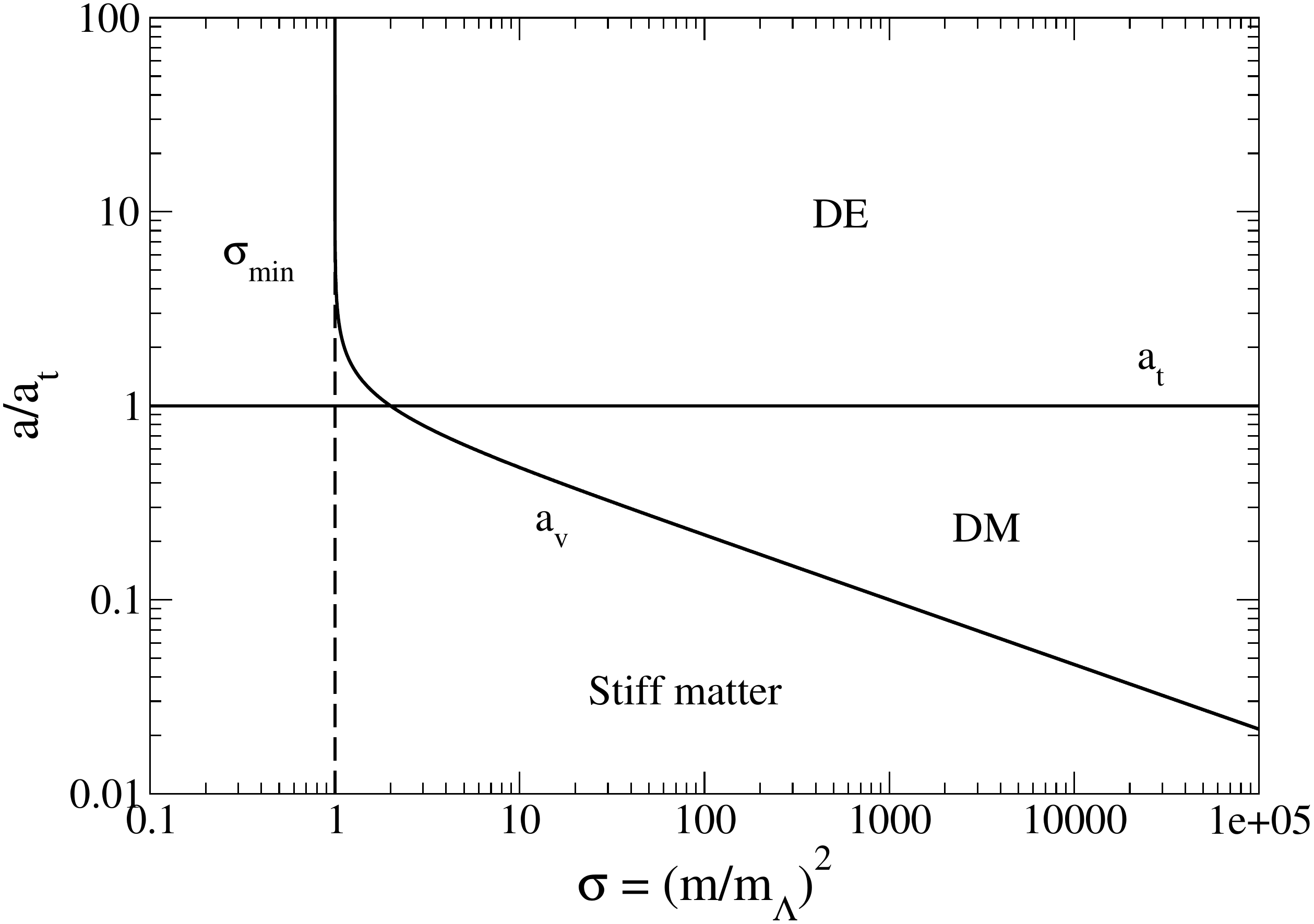}
\caption{Dynamical phase diagram of the $\Lambda$FDM model showing
the different eras experienced by  the SF during the evolution of the Universe
as a function
of its mass $m$ (this figure also determines the validity
of the fast oscillation regime).}
\label{phasediagLFDM}
\end{center}
\end{figure}

In the $\Lambda$FDM model the SF undergoes three successive eras: a stiff matter era for
$a<a_v$, a DM era for $a_v<a\ll a_t$, and a DE era for $a\gg a_t$ (we
recall that
$a_t=0.765$ corresponds to the transition between the DM and DE eras). If
the SF has a
$|\varphi|^4$ self-interaction, an additional radiationlike era occurs
between
the stiff matter era and the DM era (see the Remark at the end of Sec. 
\ref{sec_v}). These results are represented on the dynamical phase diagram of
Fig.
\ref{phasediagLFDM}, where we have plotted the transition scale $a_v$ as a
function of the mass $m$ of the SF.

\begin{figure}[!h]
\begin{center}
\includegraphics[clip,scale=0.3]{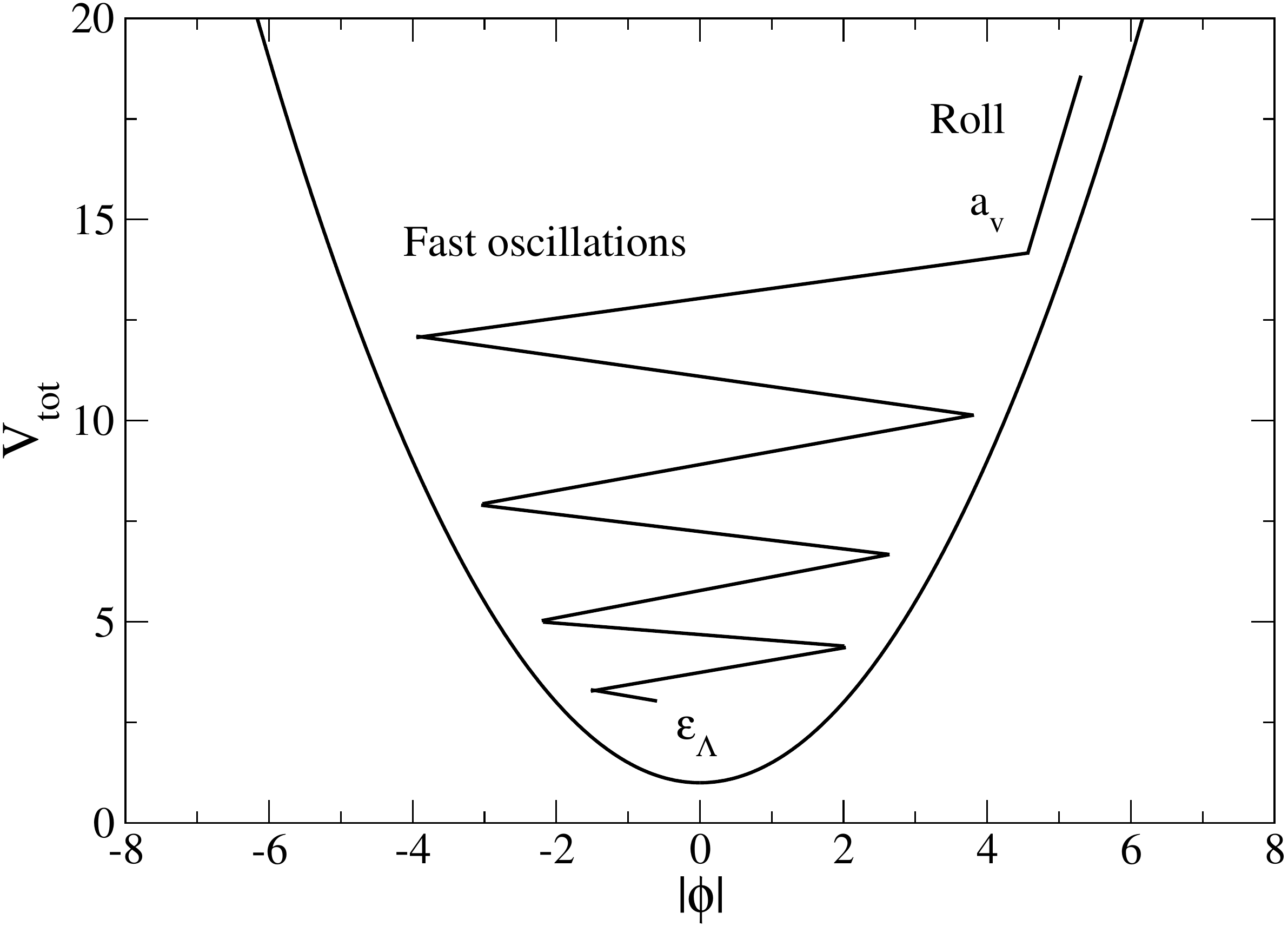}
\caption{Schematic evolution of the SF in the
total potential $V_{\rm tot}(|\varphi|^2)$ showing roll versus oscillations (the
scales are not respected). For 
$a<a_v$, the SF rolls down the potential well without oscillating (stiff matter
era); for $a>a_v$, it oscillates rapidly (DM and DE eras).}
\label{vtotzz}
\end{center}
\end{figure}

In Fig. \ref{vtotzz} we have represented the motion of the SF in the potential
$V_{\rm tot}(|\varphi|^2)$ during these different periods. During the stiff
matter era ($a<a_v^{(1)}$), corresponding to a slow
oscillation regime, the SF rolls down the potential well without
oscillating. Then,
for $a>a_v$, the SF enters in the fast oscillation regime and
descends the potential by oscillating rapidly about the vertical axis until it
falls at the bottom of the well $(V_{\rm tot})_{\rm min}=\epsilon_\Lambda$ and
achieves a constant energy density $\epsilon_\Lambda$. This
evolution successively describes  the DM
and DE eras.

\subsection{FDM halos}
\label{sec_fdmw}

Since the potential $V(|\varphi|^2)=\epsilon_\Lambda$ of the $\Lambda$FDM model is
constant, it disappears from the wave equations. As a result, the relativistic wave equation
(\ref{lwe1}) reduces to the standard KG equation
\begin{equation}
\Box\varphi+\frac{m^2c^2}{\hbar^2}\varphi=0,
\label{hlwe1}
\end{equation}
and the nonrelativistic (GP) wave equation (\ref{lwe4}) reduces to the standard
Schr\"odinger
equation
\begin{eqnarray}
i\hbar\frac{\partial\psi}{\partial
t}=-\frac{\hbar^2}{2 m }\Delta\psi+m\Phi \psi.
\label{hlwe4}
\end{eqnarray}
We thus recover the wave equations of the standard FDM model corresponding to
$V=0$.\footnote{If the SF has a
$|\varphi|^4$ self-interaction, it is described by the KGE or GPP equations
(\ref{lwe3rad}) and (\ref{lwe6red}) with $A=0$.}

When considering DM halos, we can make the nonrelativistic approximation. FDM
halos are therefore described by the Schr\"odinger-Poisson
equations. The
Schr\"odinger-Poisson equations are known to undergo a
process of gravitational cooling  and violent relaxation
\cite{seidel94,wignerPH}.
This leads to FDM halos
with a core-halo structure involving a quantum core (soliton) surrounded by an
atmosphere of scalar radiation  whose coarse-grained structure 
is consistent with the NFW density profile
of CDM halos at large
distances. This quantum core-halo structure is observed in numerical simulations
of
FDM halos \cite{ch2,ch3,
schwabe,mocz,moczSV,veltmaat,moczprl,moczmnras,veltmaat2}.

The quantum core (soliton) corresponds to the ground state of the
Schr\"odinger-Poisson
equations. In the Madelung hydrodynamical representation of the
Schr\"odinger-Poisson
equations (see Sec. \ref{sec_mad}), it is determined by the condition of
quantum 
hydrostatic equilibrium
\begin{eqnarray}
\label{hlwe8}
\frac{\rho}{m}\nabla
Q_B+\rho\nabla\Phi={\bf 0}
\end{eqnarray}
coupled to the Poisson equation
\begin{eqnarray}
\Delta\Phi=4\pi G\rho.
\label{hlwe9}
\end{eqnarray}
The solitonic core results from the balance between the gravitational
attraction and the quantum potentiel taking
into account the Heisenberg uncertainty principle. 
Its density profile can be
determined by solving numerically  the
differential equation 
\begin{eqnarray}
\label{hlwe8b}
\frac{\hbar^2}{2m^2}\Delta\left
(\frac{\Delta\sqrt{\rho}}{\sqrt{\rho}}\right )=4\pi G\rho,
\end{eqnarray}
obtained by combining Eqs. (\ref{hlwe8}) and (\ref{hlwe9}),
as done in \cite{membrado,prd2}. The
exact core mass-radius relation is given by 
\cite{membrado,prd2}
\begin{equation}
M =9.95\, \frac{\hbar^2}{G m^2 R_{99}},
\label{hni2}
\end{equation}
where $R_{99}$ is the radius enclosing $99\%$ of the mass. This mass-radius
relation is consistent with the characteristics of the smallest
-- ultracompact -- DM halos observed in the Universe (dSphs
like Fornax with $M\sim 10^8\, M_{\odot}$ and $R\sim 1\, {\rm kpc}$) provided
that the boson mass is of the order
of\footnote{Ultracompact DM halos (like dSphs) are assumed to correspond to a pure soliton 
without atmosphere, or a tiny one. This is the ground state of the
Schr\"odinger-Poisson equations. Large DM halos (like the Medium Spiral) have a solitonic core surrounded
by an extended envelope. The core mass -- halo mass relation $M_c(M_h)$ has been
obtained in different manners in
\cite{ch3,modeldm,mcmh,mcmhbh,veltmaat,mocz,egg,bbbs}}
\begin{equation}
m\sim 10^{-22}\, {\rm eV/c^2}.
\label{hni2b}
\end{equation}
The
exact density profile of the soliton is well-approximated by the Gaussian
\cite{prd1}
\begin{eqnarray}
\rho=\rho_0 e^{-r^2/R^2}
\label{hlwe9b}
\end{eqnarray}
with $M=5.57\rho_0R^3$ and $R_{99}=2.38R$. It can also be fitted by the function
\cite{ch2,ch3} 
\begin{eqnarray}
\rho=\frac{\rho_0}{\left\lbrack 1+\left ({r}/{R}\right )^2\right\rbrack^8}
\label{hlwe9c}
\end{eqnarray}
with $M=0.318\rho_0R^3$ and $R_{99}=1.151R$ (see
Fig 2 of \cite{modeldm} for a comparison between these two
profiles and the exact one). We note that the density presents a core,
not a cusp, when $r\rightarrow 0$. Quantum terms are important at ``small''
scales implying
that the soliton has a size comparable to the de Broglie length ($\lambda_{\rm
dB}\sim 1\, {\rm kpc}$).\footnote{The mass-radius relation of
the soliton scales as $M\sim h^2/Gm^2R$. Introducing a typical velocity
scale through
the virial relation $v^2\sim GM/R$, we obtain $R\sim h/mv=\lambda_{\rm dB}$.
Since $v\sim \sqrt{\alpha} c\sim 10^{-3}c$ (see Appendix \ref{sec_lin}), the de
Broglie
length is larger than the Compton length
$\lambda_C=\hbar/mc$ by about $3$ orders of magnitude.} Quantum mechanics
stabilizes the
system against
gravitational collapse and solves the core-cusp problem.

The halo of scalar radiation results from the quantum interferences of excited
states \cite{wignerPH}. It is made of uncondensed
bosons with an out-of-equilibrium DF. On the coarse-grained scale the
density of the halo is consistent with the
NFW density profile of CDM halos [see Eq. (\ref{comp1})] which decrease as
$r^{-3}$ at large
distances. It is also consistent with an isothermal profile with an effective
temperature $T_{\rm eff}$ as predicted by the statistical theory of
violent collisionless relaxation developed by Lynden-Bell \cite{lb}. Effective
thermal effects are important at ``large'' scales ($\ge 1\, {\rm kpc}$). An
approximately isothermal halo can account for
the flat rotation curves of the galaxies which have a constant circular velocity
(e.g., $v_{\infty}=(2k_B T_{\rm eff}/m)^{1/2}\sim 153\, {\rm km/s}$ for the
Medium Spiral). On the fine-grained scale, the halo has a granular structure
\cite{ch2,ch3}. It is made of ``quasiparticles'' \cite{hui} of the size of the
solitonic
core $\lambda_{\rm dB}\sim\hbar/m v\sim 1\, {\rm kpc}$ (de Broglie wavelength)
and with an effective mass $m_{\rm eff}\sim\rho\lambda_{\rm dB}^3\sim 10^7\,
M_{\odot}\gg m$. These quasiparticles can
induce a secular collisional evolution of the halo as discussed in
\cite{hui,bft,bft2,meff}.

In conclusion, in the FDM model, the quantum core (soliton) is able to solve the
core-cusp problem and the approximately isothermal halo accounts for the flat
rotation curves of the galaxies. This core-halo
structure is in
qualitative agreement with the observations. However, as discussed in Sec.
\ref{sec_kg}, the FDM model cannot account for the universality  of the surface density
of DM halos $\Sigma_0^{\rm
obs}=141_{-52}^{+83}\, M_{\odot}/{\rm
pc}^2$.
This suggests that the constant potential from Eq. (\ref{vcdm3}) should be
replaced by a more general potential such as the logarithmic potential of Eq.
(\ref{log0}) leading to
the logotropic model, which can account for the universality and the value of
$\Sigma_0^{\rm obs}$.

\section{The structure  of logotropic DM halos}
\label{sec_pldm}

In this Appendix, we describe in detail the structure of logotropic DM halos.
We use a nonrelativistic approach that is appropriate to DM halos.
This Appendix complements the discussion given in Sec. 5 of Ref. \cite{epjp} and
in Sec. \ref{sec_ldm} of the present paper. 

\subsection{Density profile}
\label{sec_dpro}

In the TF approximation, the differential equation of hydrostatic
equilibrium determining the density profile of a DM halo is given by Eq.
(\ref{diff5b}). For the logotropic equation of state (\ref{log2}), it becomes
\begin{eqnarray}
A\Delta\left(\frac{1}{\rho}\right )=4\pi G\rho.
\label{lel1}
\end{eqnarray}
If we define
\begin{equation}
\label{lel2}
\theta=\frac{\rho_0}{\rho},\qquad \xi=\left (\frac{4\pi
G\rho_0^2}{A}\right )^{1/2}r,
\end{equation}
where $\rho_0$ is the central density and $r_0=({A}/{4\pi
G\rho_0^2})^{1/2}$ is the logotropic core radius, we find that Eq.
(\ref{lel1})
reduces to the Lane-Emden equation of index $n=-1$
\cite{chandra}:
\begin{equation}
\label{lel3}
\Delta\theta=\frac{1}{\theta}
\end{equation}
with the boundary conditions $\theta=1$ and $\theta'=0$ at $\xi=0$.\footnote{As
expained in footnote 60 the Lane-Emden equation of index $n=-1$ cannot be
obtained from the
equation of state of a polytrope of index $\gamma=0$ (i.e. $n=-1$) which has a
vanishing pressure gradient. One has to consider the limit $\gamma\rightarrow 0$
and $K\rightarrow \infty$ with $A=K\gamma$ finite, leading to the logotropic
equation of state (\ref{log2}). In this sense, Eq. (\ref{lel3}) is a new
equation which completes the class of Lane-Emden equations for
standard polytrope.} This
equation has been studied in detail in  \cite{logo,epjp}. There exists an exact
analytical solution $\theta_s=\xi/\sqrt{2}$, corresponding to $\rho_s=(A/8\pi
G)^{1/2}r^{-1}$,
called
the singular logotropic sphere. The regular logotropic density profiles must be
computed numerically. The normalized density profile $\rho/\rho_0(r/r_0)$ is
universal.\footnote{This universality is related to the homology invariance of
the solutions of the Lane-Emden equation.} It is
plotted in Fig. 18  of \cite{epjp}. The density
profile of a logotropic DM halo has a core
($\rho\rightarrow {\rm cst}$ when $r\rightarrow 0$) and decreases at large distances as $\rho\sim
r^{-1}$. More precisely, for $r\rightarrow +\infty$, we have
\begin{eqnarray}
\rho\sim \left (\frac{A}{8\pi G}\right )^{1/2}\frac{1}{r},
\end{eqnarray}
like for the singular logotropic sphere. This profile has an infinite mass because the density does not decrease sufficiently rapidly with the distance. This implies
that, in the case of real DM
halos, the logotropic equation of state (\ref{log2}) or the logotropic profile determined by Eq. (\ref{lel1}) cannot be valid at
infinitely
large distances (corresponding to very low
densities).\footnote{This infinite mass problem does not rule out the logotropic
model. Actually, we
have the same problem with the isothermal sphere. The isothermal density profile
decreases at large distances as $\rho\sim 1/(2\pi G \beta m r^2)$, like the
singular isothermal sphere \cite{chandra}. It has an infinite mass. Despite
this problem, the isothermal density profile has often been used to model DM
halos because it provides a good fit of their central parts (up to a few halo
radii) and it can be
justified by Lynden-Bell's statistical theory of violent collisionless  relaxation \cite{lb,clm1,clm2,modeldm}. In reality, the density of DM halos
decreases more rapidly at large distances, typically as $r^{-3}$, like for the
Burkert \cite{burkert} and NFW \cite{nfw} profiles. This can be explained in
terms of incomplete relaxation (see, e.g., Appendix B of \cite{modeldm}). In
\cite{epjp,lettre} we have suggested that the logotropic model could be justified
by a notion of generalized thermodynamics. In this context, the constant $A$ in
the logotropic distribution plays the role of a generalized temperature which is
the counterpart of the temperature $T$ in the isothermal distribution. This generalized thermodynamical 
interpretation strengthens the analogy between the isothermal and logotropic
models.}  The logotropic profile is expected to be surrounded by an extended
envelope where the density decreases more rapidly like, e.g., $r^{-3}$ (see
footnote 70).  In practice, we shall consider the logotropic profile up to a
few
halo radii $r_h$ (see below).

We note that the density profiles of real DM halos obtained from observations display a
core ($\rho\sim r^0$) followed by a region where the density decreases as $r^{-1}$,
similarly to the logotropic density profile. This $r^{-1}$ decay can be seen in
Fig. 6 (right) of Oh {\it et al.} \cite{oh} and in Fig. 3 (plate U11583) of Robles and
Matos \cite{rm}. The fact that the
slope of the density profile of DM halos close to the core  radius $r_h$ is
approximately equal to $-1$ has
also been pointed out by Burkert \cite{burkert2} (see in particular the upper
right panel of his Fig. 1). These properties are in good agreement
with the logotropic model.\footnote{We note that the logotropic profile
$\rho\sim r^{-1}$ may be
wrongly interpreted in certain observations as a NFW cusp  $r^{-1}$ if the
logotropic core is not sufficiently well-resolved.  Indeed, in that case, we see
only the $r^{-1}$ tail of the logotropic distribution, not the core ($\rho\sim
r^0$). This may
lead to the illusion that certain DM halos are cuspy in agreement
with the NFW prediction while they are not \cite{blok}.} However, at
large distances,
the density of real DM
halos decreases more rapidly than $r^{-1}$, typically as $r^{-2}$ or $r^{-3}$,
consistently
with the asymptotic behavior of the isothermal sphere \cite{chandra} or with the
asymptotic behaviors of the Burkert \cite{burkert} and NFW \cite{nfw}
profiles. Now, we note that the
logotropic density profile defined by Eqs. (\ref{lel2}) and (\ref{lel3}) has
been
obtained by neglecting quantum (or wave) effects. If we consider the
logotropic GPP
equations (\ref{lwe5}) and (\ref{lwe6})  it is possible that, like in the
case of FDM (see
Appendix \ref{sec_gfdm}),
quantum interferences build up a halo whose average
density profile
decreases as $r^{-2}$ or $r^{-3}$ at large distances
\cite{wignerPH}.\footnote{This halo may 
also be obtained in a purely classical model based on the Euler-Poisson
equations with a logotropic equation of state. This corresponds to the TF
approximation
$\hbar\rightarrow 0$ of the logotropic GPP equations.} It would be interesting
to
investigate this idea numerically. If this idea is correct, the
``quantum'' logotropic halo would possess a core ($\rho\sim
r^0$) $+$ an intermediate logotropic profile ($\rho\sim r^{-1}$) $+$ an
extended isothermal ($\rho\sim r^{-2}$) or NFW ($\rho\sim
r^{-3}$) envelope, in agreement with the observations (see, e.g.,
\cite{oh,rm,burkert2}). This structure would be obtained  in the TF
approximation $m\gg m_0=3.57\times 10^{-22}\, {\rm eV/c^2}$ (see
Sec. \ref{sec_tfdm}). If we go beyond the TF approximation (which is not
satisfied for a boson mass $m\sim 10^{-22}\, {\rm eV/c^2}$), the DM halo should
also possess a quantum core (soliton) like in the FDM model (see
Appendix \ref{sec_gfdm}).

{\it Remark:} Using qualitative arguments, Ferreira and Avelino \cite{fa}  have
argued that logotropic DM halos
are dynamically unstable. However, the stability of logotropic spheres must be
considered carefully due to the fact that they have an infinite mass in an
unbounded domain. The stability of box-confined logotropic configurations has
been studied in detail in \cite{logo}. It is found that they are stable below a
critical
density contrast and unstable above it. These results are similar to those
obtained for box-confined self-gravitating isothermal spheres \cite{aaiso}. 
Isothermal spheres have been used in many models of DM halos despite the fact
that they have an infinite mass and that they are unstable in certain
conditions leading to core collapse. Similar properties are expected for
logotropic spheres.

\subsection{Halo mass}
\label{sec_hma}

The halo radius $r_h$ is defined as the distance at which the
central density  $\rho_0$  is divided by $4$. For logotropic DM halos, using Eq. (\ref{lel2}), it is
given by
\begin{eqnarray}
\label{lel4}
r_h=\left (\frac{A}{4\pi
G\rho_0^2}\right )^{1/2}\xi_h,
\end{eqnarray} 
where
$\xi_h$ is determined by the equation
\begin{eqnarray}
\label{lel5}
\theta(\xi_h)=4.
\end{eqnarray} 
The normalized density profile $\rho/\rho_0(r/r_h)$ of logotropic DM halos is
plotted in Fig. 19 of \cite{epjp}. The  halo mass
$M_h$,
which is the
mass $M_h=\int_0^{r_h} \rho(r') 4\pi {r'}^2\, dr'$ contained within the sphere
of radius $r_h$, is given by
\begin{eqnarray}
\label{lel6}
M_h=4\pi\frac{\theta'(\xi_h)}{{\xi_h}}\rho_0 r_h^3.
\end{eqnarray} 
Solving the Lane-Emden equation of index $n=-1$ [see Eq. (\ref{lel3})], we numerically find
\begin{eqnarray}
\xi_h=5.85,\qquad \theta'_h=0.693.
\label{lel7}
\end{eqnarray}
This yields
\begin{eqnarray}
r_h=5.85\, \left (\frac{A}{4\pi G}\right )^{1/2}\frac{1}{\rho_0}
\label{lel8}
\end{eqnarray}
and
\begin{eqnarray}
M_h=1.49\, \rho_0 r_h^3.
\label{lel9}
\end{eqnarray}

\subsection{Constant surface density}
\label{sec_cds}

Eliminating the central density between Eqs. (\ref{lel8}) and (\ref{lel9}), we
obtain the logotropic halo mass-radius relation
\begin{eqnarray}
M_h=8.71 \, \left (\frac{A}{4\pi G}\right
)^{1/2} r_h^2.
\label{lel11}
\end{eqnarray}
Since $M_h\propto r_h^2$ we see that the surface density $\Sigma_0$ is 
constant.\footnote{This is consistent with the fact that the density of a
logotropic DM halo decreases as $r^{-1}$ at large distances.} This is a very
important property of logotropic DM
halos \cite{epjp}. From Eq. (\ref{lel8}), we get 
\begin{eqnarray}
\Sigma_0=\rho_0 r_h=5.85 \, \left (\frac{A}{4\pi G}\right
)^{1/2}.
\label{lel12}
\end{eqnarray}
Therefore, all the logotropic DM halos have the same surface density, whatever
their size, provided that
$A$ is interpreted as a universal constant. With
the value of
$A/c^2=2.10\times
10^{-26}\, {\rm g}\, {\rm m}^{-3}$ obtained from cosmological
considerations (without free parameter) in Sec. \ref{sec_valf}  we obtain
$\Sigma_0^{\rm
th}=133\,
M_{\odot}/{\rm pc}^2$ in very good agreement with the value
$\Sigma_0^{\rm obs}=\rho_0 r_h=141_{-52}^{+83}\, M_{\odot}/{\rm
pc}^2$ obtained from the observations \cite{donato}.   On the other hand, Eq.
(\ref{lel9}) may
 be rewritten as 
\begin{eqnarray}
M_h=1.49\, \Sigma_0 r_h^2=1.49\, \frac{\Sigma_0^3}{\rho_0^2}.
\label{lel13}
\end{eqnarray}
We note that the ratio $M_h/(\Sigma_0 r_h^2)=1.49$ in Eq. (\ref{lel13}) is  in
good
agreement with the ratio $M_h/(\Sigma_0 r_h^2)=1.60$ obtained from the
observational Burkert profile (see Appendix D.4 of \cite{modeldm}). This is an
additional argument in favor of the logotropic model.

\subsection{The gravitational acceleration}
\label{sec_ga}

We can define an average DM halo surface density by the relation
\begin{eqnarray}
\label{lel14}
\langle\Sigma\rangle=\frac{M_h}{\pi r_h^2}.
\end{eqnarray}
For logotropic DM halos, we find
\begin{eqnarray}
\label{lel15}
\langle\Sigma\rangle_{\rm th}=\frac{M_h}{\pi
r_h^2}=\frac{1.49}{\pi}\Sigma_0^{\rm th}=63.1\,
M_{\odot}/{\rm pc}^2.
\end{eqnarray}
This theoretical value is in good agreement with the value
$\langle\Sigma\rangle_{\rm
obs}=72_{-27}^{+42}, M_{\odot}/{\rm pc}^2$ obtained from the observations \cite{gentile}.\footnote{This
measured value is based on the
observations and on a fit of the density profile of DM halos by
the Burkert profile \cite{gentile}.}

The gravitational acceleration at the halo radius is
\begin{eqnarray}
\label{lel16}
g=g(r_h)=\frac{GM_h}{r_h^2}=\pi G \langle\Sigma\rangle.
\end{eqnarray}
For logotropic DM
halos, we find
\begin{equation}
\label{lel17}
g_{\rm th}=\pi G \langle\Sigma\rangle_{\rm
th}=1.49 G \Sigma_0^{\rm
th}=2.76\times
10^{-11}\, {\rm m/s^2}.
\end{equation}
Again, this theoretical value is in good agreement with the measured value
$g_{\rm obs}=\pi G
\langle\Sigma\rangle_{\rm obs}=3.2_{-1.2}^{+1.8}\times 10^{-11}\, {\rm
m/s^2}$ of the gravitational acceleration \cite{gentile}.

The circular velocity at the halo radius  is 
\begin{equation}
\label{lel18}
v_h^2=\frac{GM_h}{r_h}.
\end{equation}
Using Eqs. (\ref{lel13})-(\ref{lel17}), we obtain the relation
\begin{equation}
\label{lel18bb}
v_h^4=GgM_h=\pi\langle\Sigma\rangle G^2M_h=1.49\Sigma_0G^2M_h,
\end{equation}
where $g$ and $\Sigma_0$ are universal constants. This relation is connected 
to the Tully-Fisher relation \cite{tf}  which involves the baryon mass
$M_b$
instead of the DM halo mass $M_h$ via the cosmic baryon fraction
$f_b=M_b/M_h\sim 0.17$. This yields $(M_{\rm
b}/v_h^4)^{\rm th}=46.4\,
M_{\odot}{\rm km}^{-4}{\rm s}^4$ which is close to the
observed value 
$(M_{\rm b}/v_h^4)^{\rm obs}=47\pm 6 \, M_{\odot}{\rm km}^{-4}{\rm s}^4$
\cite{mcgaugh}. The  Tully-Fisher  relation 
is also a prediction of the MOND
(modification of Newtonian dynamics) theory \cite{mond}. Using Eqs.
(\ref{lel17}) and (\ref{diff7b}),
we obtain
\begin{eqnarray}
g_{\rm th}=0.0291\sqrt{3(1-\Omega_{m,0})}H_0 c=0.0419\, H_0 c.
\label{diff7c}
\end{eqnarray}
This relation explains why the fundamental constant $a_0=g/f_b$ that appears in
the MOND theory is of order $H_0 c/4=1.65\times
10^{-10}\, {\rm m/s^2}$ (see the Remark in Sec. 3.3. of \cite{pdu} for a more
detailed discussion). Note, however, that our model is completely different from
the MOND
theory.

\subsection{Logarithmic density slope }
\label{sec_lds}

\begin{figure}[!h]
\begin{center}
\includegraphics[clip,scale=0.3]{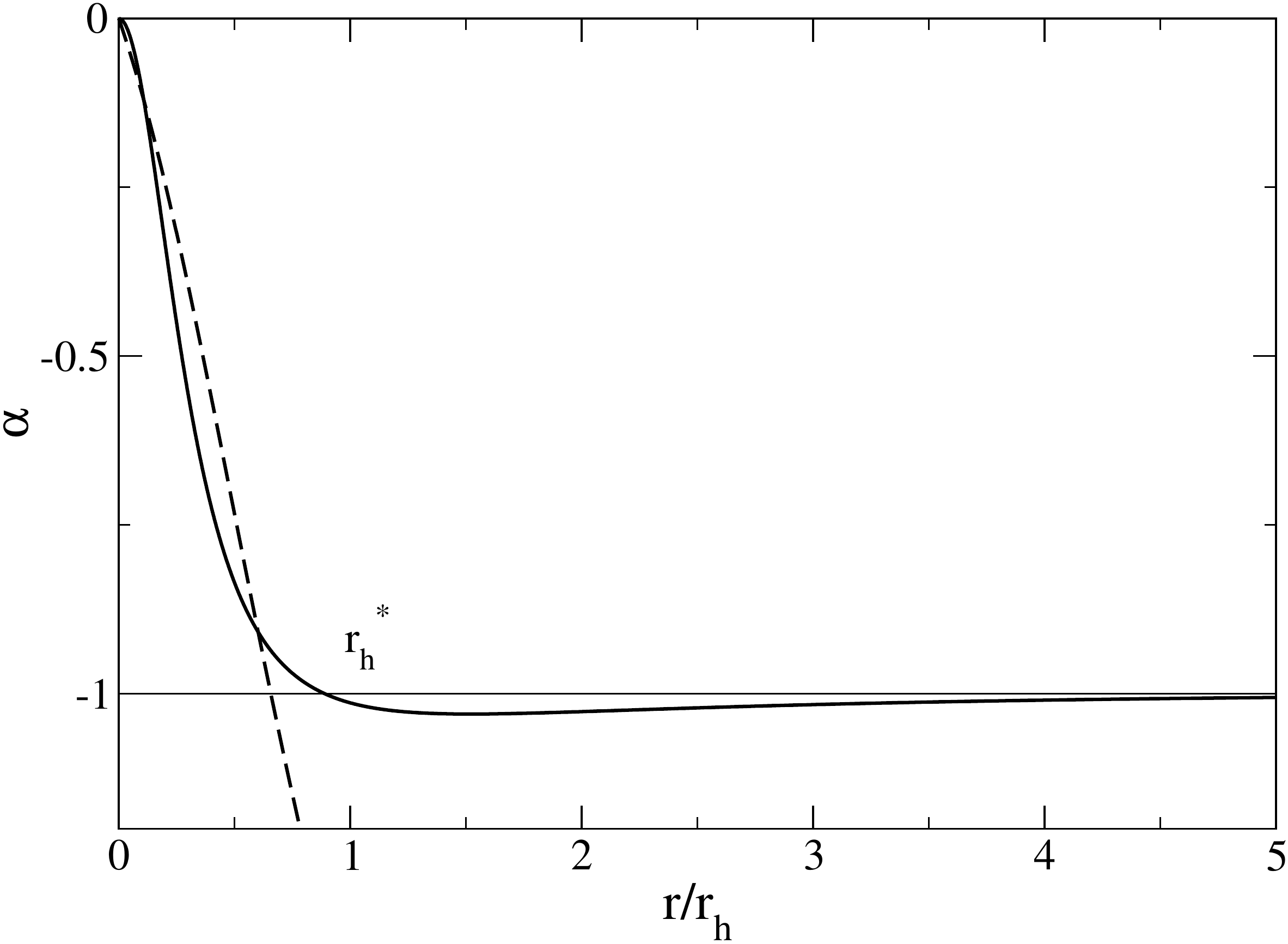}
\caption{Logarithmic density slope of the logotropic profile as a
function
of the radial distance normalized by the halo radius.
It is compared to the logarithmic density slope of
the Burkert profile (dashed line) from Eq.
(\ref{comp2}). The two profiles are relatively close to each other for
$r\le r_h$. At
$r\sim r_h$ the Burkert profile has an effective slope $\alpha\sim -1$ like the
asymptotic slope of the logotrope.}
\label{alpha}
\end{center}
\end{figure}

The logarithmic slope of the density profile of a DM halo  is defined by
\begin{eqnarray}
\label{lel18cc}
\alpha(r)=\frac{d\ln\rho}{d\ln r}.
\end{eqnarray}
For logotropic DM halos it can be expressed in terms
of the
Lane-Emden function $\theta$ by
\begin{eqnarray}
\label{lel18b}
\alpha(r)=-\frac{\xi\theta'}{\theta}=-v,
\end{eqnarray}
where $v=\xi\theta'/\theta$ is the Milne variable \cite{chandra}. The
logarithmic
density slope $\alpha(r)$ of a logotropic DM halo is plotted in Fig.
\ref{alpha}. It starts from
$\alpha=0$ at $r=0$ (core) and tends to $-1$ when $r\rightarrow +\infty$. It
reaches a  minimum value $\alpha_{\rm min}=-1.03$ at $r_*=1.52\, r_h$. We find
that $\alpha=-0.3$ (corresponding to the typical logarithmic inner
density slope of  real DM halos found by
\cite{blok,oh,rm}) at
$r=0.186\, r_h$. Robles and Matos \cite{rm} define the core radius $r_h^*$
 by the condition
$\alpha(r_h^*)=-1$. In the case of logotropic DM halos, $r_h^*$ is
consistent with our definition of the halo radius  $r_h$ since we find that
$r_h^*=0.890\, r_h\sim r_h$.

\subsection{Logarithmic circular velocity slope}
\label{sec_beta}

\begin{figure}[!h]
\begin{center}
\includegraphics[clip,scale=0.3]{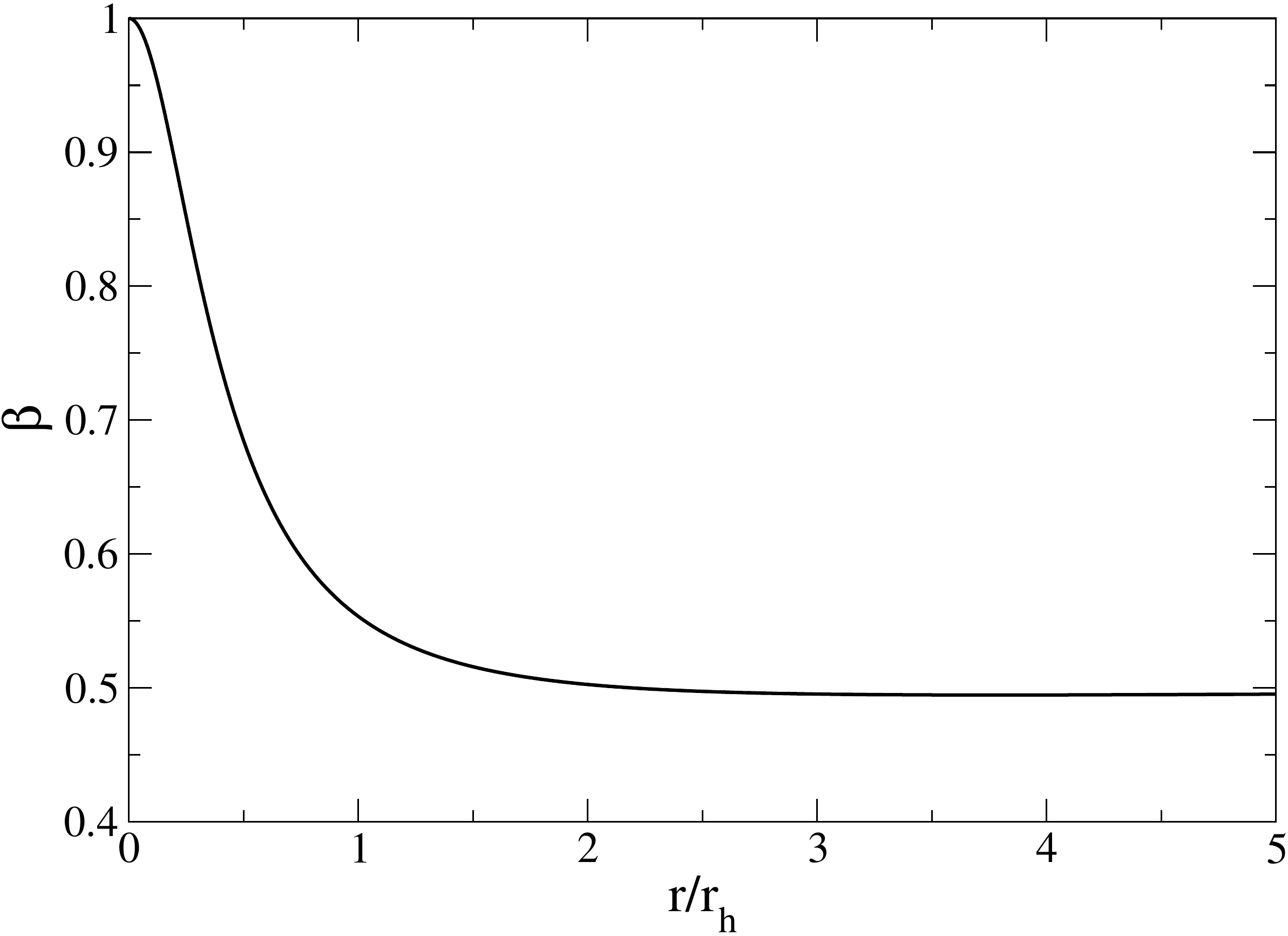}
\caption{Logarithmic circular velocity slope of the logotropic profile as a
function
of the radial distance normalized by the halo radius.}
\label{beta}
\end{center}
\end{figure}

The circular velocity of a DM halo is given by
\begin{eqnarray}
\label{circv}
v_c^2(r)=\frac{GM(r)}{r},
\end{eqnarray}
where $M(r)=\int_0^r\rho(r')4\pi {r'}^2\, dr'$ is the mass contained within the
sphere of radius $r$. The logarithmic slope of the circular velocity profile is
defined by
\begin{eqnarray}
\label{beta1}
\beta(r)=\frac{d\ln v_c}{d\ln r}.
\end{eqnarray}
Using Eq. (\ref{circv}) it can be written as
\begin{eqnarray}
\label{beta2}
\beta(r)=\frac{1}{2}\left (\frac{d\ln M(r)}{d\ln r}-1\right ).
\end{eqnarray}
For logotropic DM halos, using the Lane-Emden equation (\ref{lel3}), we can
establish that
\begin{eqnarray}
\label{beta3}
\frac{d\ln M(r)}{d\ln r}=\frac{\xi}{\theta\theta'}=u,
\end{eqnarray}
where $u=\xi/(\theta\theta')$ is the Milne variable \cite{chandra}. Therefore,
\begin{eqnarray}
\label{beta4}
\beta(r)=\frac{1}{2}(u-1).
\end{eqnarray}
The logarithmic  slope  $\beta(r)$ of the circular velocity of a logotropic DM halo is plotted in Fig.
\ref{beta}.

\subsection{Comparison between the logotropic model and the fermionic and
bosonic models of DM halos}
\label{sec_cqmdm}

It is an observational evidence that there is no DM halo below a certain mass and
below a certain size. The smallest and most compact DM halos observed in the
Universe are dSphs like Fornax. To fix
the ideas we shall consider that the smallest halo observed in the
Universe (what we call the ``minimum halo'') has a mass\footnote{We take this
value as a reference
in order to be consistent with our previous papers. If a possibly more relevant
minimum mass is adopted, our numerical applications should be reconsidered but our main conclusions should not be altered.} 
\begin{eqnarray}
\label{lel19}
(M_h)_{\rm min}=10^8\, M_{\odot} \qquad ({\rm Fornax}).
\end{eqnarray}
It is also  an observational fact that the surface density $\Sigma_0$  of DM
halos is constant and that it has the universal value \cite{donato}
\begin{eqnarray}
\Sigma_0^{\rm obs}=\rho_0 r_h=141_{-52}^{+83}\, M_{\odot}/{\rm
pc}^2.
\label{lel20}
\end{eqnarray}

In the case of fermionic or bosonic models of DM, the ``minimum
halo''  is expected to correspond to the ground state ($T=0$) of the
self-gravitating quantum system \cite{mcmh}. We can then combine the mass-radius
relation $M_h(r_h)$ of a fermionic or bosonic DM halo at $T=0$ (ground state)
with the universal surface density $\Sigma_0$ from Eq. (\ref{lel20}) in order to express
the mass $(M_h)_{\rm min}$, the radius $(r_h)_{\rm
min}$ and the central density $(\rho_0)_{\rm max}$ of the minimum halo as a
function of the characteristics (mass $m$ and scattering length $a_s$) of the DM
particle. For specified values of $m$ and $a_s$ we can thus obtain $(M_h)_{\rm
min}$,
$(r_h)_{\rm min}$ and $(\rho_0)_{\rm max}$. In practice, we proceed the
other way round. We use the observed value of $(M_h)_{\rm min}$ given by
Eq. (\ref{lel19}) to obtain the characteristics ($m$, $a_s$) of the DM particle.
Once these characteristics are known, we can obtain the radius $(r_h)_{\rm min}$
and the central density
$(\rho_0)_{\rm max}$ of the minimum halo. We can also plot the density profile
of the minimum halo for these different models. These calculations are developed
in detail in Sec. II of \cite{mcmh} (they are generalized in Sec. VII.C of
\cite{mcmh} to more general models by using a Gaussian ansatz). We can also obtain an estimate of the
minimum halo mass and minimum halo radius as a function of  $m$ and $a_s$  
from the Jeans instability theory (see, e.g., Refs.
\cite{suarezchavanisprd3,jeansMR} and Appendices H an I of \cite{mcmh}) but this
method is less accurate. 
We note that the fermionic or bosonic models of DM halos are not fully
predictive since (i) we have to
assume the value of $\Sigma_0$ from the observations [see Eq. (\ref{lel20})] and
(ii) we need to know
the value of $m$ and $a_s$ to determine $(M_h)_{\rm min}$, $(r_h)_{\rm min}$ and
$(\rho_0)_{\rm max}$ theoretically (alternatively, we have to know the value of
$(M_h)_{\rm min}$ to obtain $m$ and $a_s$, which then yield $(r_h)_{\rm min}$ and
$(\rho_0)_{\rm max}$).  

In the case of the logotropic model, there is no unknown parameter. The universal constant surface density
$\Sigma_0=133\,
M_{\odot}/{\rm pc}^2$ of the DM halos is predicted by this model (see
\cite{epjp} and Appendix \ref{sec_cds}). 
As a result, the characteristics of a
logotropic DM halo of mass  $M_h$  are fully characterized by Eq.
(\ref{lel13}). For
 $(M_h)_{\rm min}=10^8\,
M_{\odot}$, we obtain (without free parameter)
\begin{equation}
(r_h)_{\rm min}=710\, {\rm pc},\qquad  (\rho_0)_{\rm max}=0.187\, M_{\odot} {\rm
pc}^{-3}.
\label{lel21}
\end{equation}
We note, however, that the minimum halo mass $(M_h)_{\rm min}$ is not directly
determined by the logotropic model
since the mass-radius relation $M_h(r_h)$ of {\it all} the logotropic DM halos
satisfies the constraint from Eq. (\ref{lel20}).\footnote{In the fermionic or
bosonic DM
models, the value of the minimum halo mass $(M_h)_{\rm min}$ results from the
combination of the mass-radius relation of the ground state  $M_h(r_h)$ with the
constraint from Eq. (\ref{lel20}), for given values of $m$ and $a_s$
\cite{mcmh}. For
example, for noninteracting bosons of mass $m\sim 10^{-22}\, {\rm eV/c^2}$, we
get $(M_h)_{\rm min}\sim 10^8\, M_{\odot}$  and $(r_h)_{\rm min}\sim 1\, {\rm
kpc}$. However, for bigger halos, these purely quantum models cannot account for
the observed
constant surface density of DM halos (see Sec. \ref{sec_kg}). Note that if we
consider the quantum logotropic model (going beyond the TF approximation), the
core of
DM halos is both quantum and logotropic. It is possible that the quantum
core dominates in small halos (we have indeed seen that the TF approximation is
marginally valid in small DM halos of mass $(M_h)_{\rm min}\sim 10^8\,
M_{\odot}$ for $m\sim m_0\sim 10^{-22}\, {\rm eV/c^2}$) and that the logotropic
core
dominates in large DM halos.  In that case, the mass and
size $(M_h)_{\rm min}\sim 10^8\, M_{\odot}$  and $(r_h)_{\rm min}\sim 1\, {\rm
kpc}$ of the minimum halo could be determined by quantum effects (the boson
mass $m\sim 10^{-22}\, {\rm eV/c^2}$) like in the FDM model, in agreement  with
the Jeans study of Sec. \ref{sec_jeansfdm}, while the universal
surface density $\Sigma_0=141\, M_{\odot}/{\rm
pc}^2$ of bigger DM halos could be due to
logotropic effects (the fundamental constant $A/c^2=2.10\times
10^{-26}\, {\rm g}\, {\rm
m}^{-3}$ of our model). This important point is further discussed in Appendix
\ref{sec_mdm}.}

\begin{figure}[!h]
\begin{center}
\includegraphics[clip,scale=0.3]{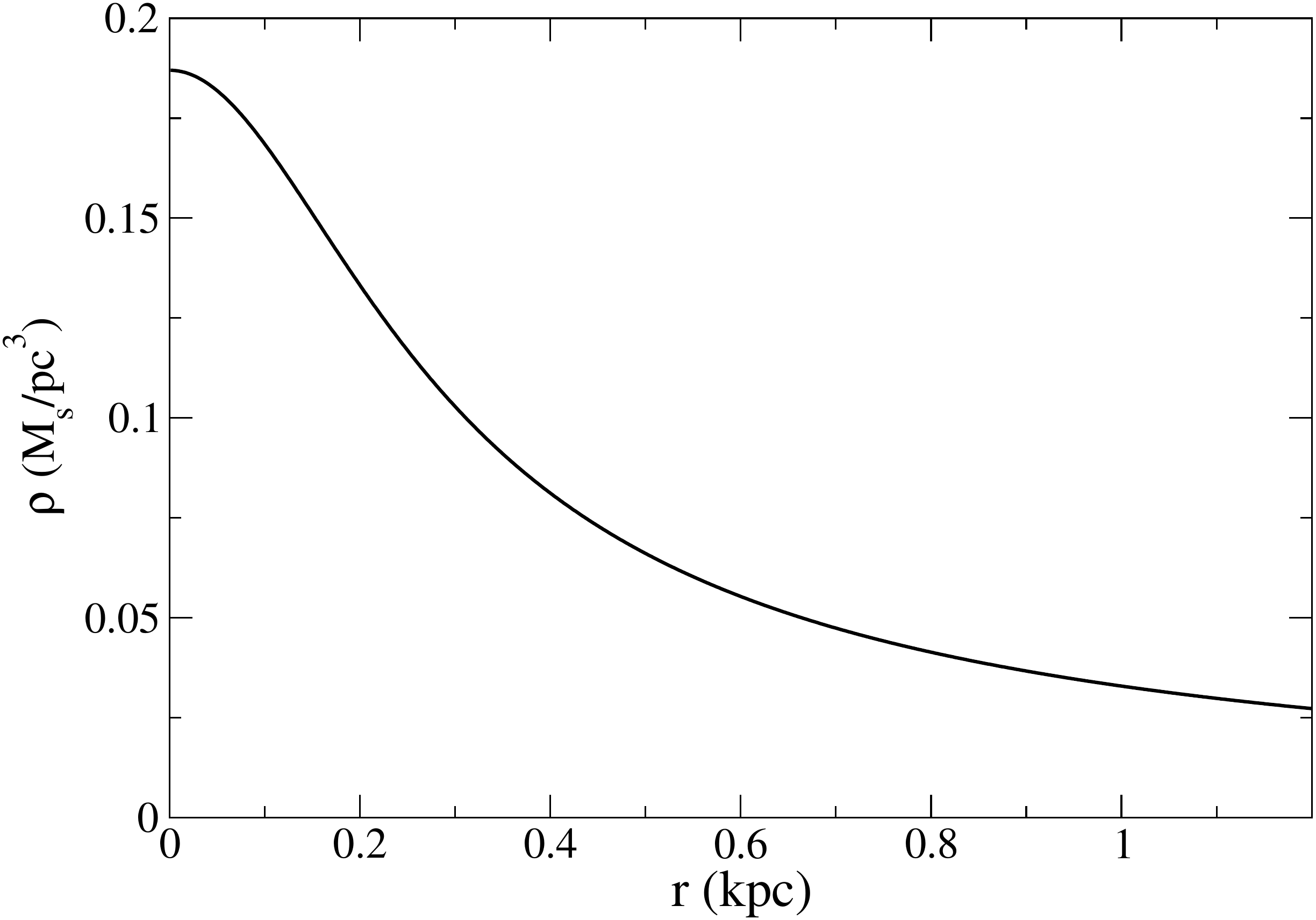}
\caption{Density profile of the ``minimum halo'' of mass  $(M_h)_{\rm
min}=10^8\, M_{\odot}$ in the logotropic model.}
\label{minhalo}
\end{center}
\end{figure}

The density profile of a logotropic DM halo  of mass $(M_h)_{\rm min}=10^8\,
M_{\odot}$ (minimum halo) is plotted in Fig. \ref{minhalo}. It can be
compared to the profiles obtained in Sec. II of \cite{mcmh} by
assuming
that DM is made of fermions (see Fig.
1 of \cite{mcmh}), noninteracting BECs (see Fig.
2 of \cite{mcmh}), or self-interacting BECs in the TF approximation (see Fig.
3 of \cite{mcmh}). We see
that the
density of a logotropic DM halo decreases less rapidly than the density
 of a fermionic or bosonic DM halo in its ground state. Indeed,  it
decays as 
$r^{-1}$ at large distances while the density profile of noninteracting BECs decreases exponentially rapidly
and the density profiles of fermions and BECs in the TF approximation  have a
compact support. As mentioned previously, the logotropic profile is not valid at
large
distances since it would have an infinite mass. In reality, the logotropic core is surrounded  by an outer envelope where the
density
decreases more rapidly than $r^{-1}$, presumably  as
$r^{-3}$.

{\it Remark:} As discussed in Appendix \ref{sec_dpro}, the $r^{-1}$
decay of the logotropic density profile is in agreement with the density profile
of real DM halos close to the halo radius. This $r^{-1}$
decay is responsible for the universal surface density of DM halos and
the fact that their mass-radius relation behaves as $M_h\propto r_h^2$ in
agreement with the observations \cite{donato}. By contrast, fermionic and
bosonic DM
halos do not present a region where the density decreases as $r^{-1}$.
As a result, they do not have a
constant surface density $\Sigma_0$ and their mass-radius relation is not in
agreement with the observations (see the discussion at the end of Sec. \ref{sec_kg}).
For large fermionic or bosonic DM halos, the constraint from Eq.
(\ref{lel20}) could be satisfied by taking into account the presence of an
isothermal halo surrounding the quantum core and assuming that the effective
``central'' density of the halo corresponds to the density at the contact
between the quantum core and the halo \cite{modeldm}. In that case, we have
to identify the
halo radius $r_h$ with the isothermal core radius $r_0$, not with the quantum
core radius $R_c$, and we have to allow the
temperature $T$ to change from halo to halo according to Eq. (168) of
\cite{modeldm} in order  to maintain a constant
surface density.
Alternatively, if the constraint (\ref{lel20}) cannot be satisfied in all  DM
halos, the (pure) fermionic and bosonic DM models are in trouble and the
logotropic
model may be an interesting substitute.

\section{The typical mass of the DM particle}
\label{sec_mdm}

Let us assume that DM halos are described by the quantum logotropic model and
that dSphs (with typical mass $10^8\, M_{\odot}$)
are just at the limit of validity of the TF approximation.
This means
that they can be marginally described both by the FDM model ($\hbar\neq 0$
and $A=0$) and by
the classical logotropic model ($A\neq 0$ and $\hbar=0$). 

The mass-radius relation of FDM halos is [see Eq. (\ref{hni2})]
\begin{equation}
M = 9.95\, \frac{\hbar^2}{G m^2 R}.
\label{ahni2}
\end{equation}
The mass-radius relation of classical logotropic DM halos is [see Eq.
(\ref{lel13})]
\begin{eqnarray}
M = 1.49\, \Sigma_0 R^2,
\label{alel13}
\end{eqnarray}
where $\Sigma_0=5.85\, (A/4\pi G)^{1/2}=133\, M_{\odot}/{\rm pc}^2$ [see Eq.
(\ref{lel12})] is the
universal surface density of DM halos. If we combine these
two relations, we get
\begin{eqnarray}
(M_h)_{\rm min}=5.28\, \left (\frac{\Sigma_0\hbar^4}{G^2m^4}\right )^{1/3}.
\label{alel13a}
\end{eqnarray}
This formula determines the mass $(M_h)_{\rm min}$ of the minimum halo as a
function of the
boson mass $m$. Inversely, knowing the minimum halo mass from the observations,
we can determine the boson mass. Taking $M=10^8\, M_{\odot}$ we find
$m=3.46\times 10^{-22}\, {\rm eV/c^2}$. This
explains why the mass $m_0=3.57\times 10^{-22}\, {\rm eV/c^2}$
determining the domain of validity of the TF approximation in DM halos similar
to dSphs
happens to
coincide with the boson mass $m\sim 10^{-22}\, {\rm eV/c^2}$ (see Sec.
\ref{sec_tfdm}). We
then obtain the minimum halo radius
\begin{eqnarray}
(r_h)_{\rm min}=1.88\, \left (\frac{\hbar^2}{Gm^2\Sigma_0}\right )^{1/3}=709\,
{\rm pc},
\end{eqnarray}
which is in good agreement with the typical size of dSphs.

On the other hand, the
Jeans mass
in the FDM
model is (see Appendix H of \cite{mcmh}) 
\begin{eqnarray}
M_J=\frac{\pi}{6}\left (\frac{\pi^3\hbar^2\rho_{\rm m,0}^{1/3}}{G m^2}\right
)^{3/4},
\label{alel13b}
\end{eqnarray}
where $\rho_{\rm m,0}$ is the present matter density in the homogeneous
background. Writing $(M_h)_{\rm min}= \chi M_J$ with $\chi\sim 10-100$ (see
Appendix I of 
\cite{mcmh}) and using Eqs. (\ref{alel13a}) and (\ref{alel13b}) we get
\begin{eqnarray}
m=4.89\, \chi^6\frac{\hbar\rho_{\rm m,0}^{3/2}}{\Sigma_0^2G^{1/2}}.
\label{alel13c}
\end{eqnarray}
Using $\rho_{\rm m,0}=0.0178\, \Lambda/G$ and $\Sigma_0=0.0207\,
c\sqrt{\Lambda}/G$ (see Eqs.
(155) and (156) of \cite{mcmh}), we finally obtain
\begin{eqnarray}
m=27.1\, \chi^6
\frac{\hbar\sqrt{\Lambda}}{c^2}.
\label{alel13cb}
\end{eqnarray}
Therefore, the DM boson mass $m$ is equal to the cosmon mass (\ref{v4})
multiplied by
a huge prefactor $\sim 10^{11}$. The cosmon mass gives the fundamental
mass scale of bosons \cite{mcmh,oufsuite}.

\section{Numerical values from observations}
\label{sec_nvnew}

We have taken the following values from the Planck Collaboration
\cite{planck2016}
\begin{equation}
H_0=2.195\, 10^{-18} {\rm s}^{-1},
\end{equation}
\begin{equation}
\epsilon_0=7.75\times 10^{-7} {\rm g}\, {\rm m}^{-1}\,{\rm s}^{-2},
\end{equation}
\begin{equation}
\epsilon_0/c^2=8.62\times 10^{-24} {\rm g}\, {\rm m}^{-3},
\end{equation}
\begin{equation}
\Omega_{\rm de,0}=0.6911,
\end{equation}
\begin{equation}
\Omega_{\rm dm,0}=0.2589,
\end{equation}
\begin{equation}
\Omega_{\rm b,0}=0.0486,
\end{equation}
\begin{equation}
\Omega_{\rm m,0}=0.3089.
\end{equation}


\begin{thebibliography}{99}


\bibitem{epjp}{\small  P.H. Chavanis, Eur. Phys. J. Plus {\bf 130}, 130 (2015)}
\bibitem{lettre}{\small  P.H. Chavanis, Phys. Lett. B {\bf 758}, 59 (2016)}
\bibitem{jcap}{\small  P.H. Chavanis, S. Kumar, J. Cosmol. Astropart. Phys.
{\bf 5}, 018 (2017)}
\bibitem{pdu}{\small  P.H. Chavanis, Phys. Dark Univ. {\bf 24}, 100271 (2019)}
\bibitem{fa}{\small V.M.C. Ferreira, P.P. Avelino, Phys. Lett. B {\bf 770}, 213
(2017)}
\bibitem{cal1}  {\small  S. Capozziello, R. D'Agostino, O. Luongo, Phys.
Dark Univ.  {\bf 20}, 1 (2018)}
\bibitem{ootsm}  {\small S.D. Odintsov, V.K. Oikonomou, A.V. Timoshkin, E.N.
Saridakis, R. Myrzakulov, Ann. Phys. {\bf 398}, 238 (2018)}
\bibitem{cal2}  {\small  S. Capozziello, R. D'Agostino, R. Giamb\`o, O. Luongo,
Phys. Rev. D {\bf 99}, 023532 (2019)}
\bibitem{bal}  {\small  K. Boshkayev, R. D'Agostino, O. Luongo,
Eur. Phys. J. C {\bf 79}, 332 (2019)}
\bibitem{mamon}  {\small  A. Al Mamon, S. Saha, Int. J. Mod. Phys. D {\bf 29},
2050097 (2020)}
\bibitem{bklmp}  {\small  K. Boshkayev, T.
Konysbayev, O. Luongo, M. Muccino, F. Pace, Phys. Rev. D {\bf 104}, 023520
(2021)}
\bibitem{logogen}{\small  H.B. Benaoum, P.H. Chavanis, H. Quevedo,
{\it Generalized Logotropic Models and their Cosmological Constraints}
[arXiv:2112.13318]}
\bibitem{caldwellprl}{\small R.R. Caldwell, M. Kamionkowski, N.N. Weinberg,
Phys. Rev. Lett. {\bf 91}, 071301 (2003)}
\bibitem{littlerip}{\small P.H. Frampton, K.J. Ludwick,
R.J. Scherrer, Phys.
Rev. D {\bf 84}, 063003 (2011)}
\bibitem{kormendy}{\small J. Kormendy, K.C. Freeman, in S.D. Ryder, D.J. Pisano,
M.A. Walker, K.C. Freeman, eds.,  Proc. IAU Symp. 220, Dark Matter in Galaxies.
Astron. Soc. Pac., San Francisco, p. 377 (2004)}
\bibitem{spano}{\small M. Spano, M. Marcelin, P. Amram, C. Carignan, B. Epinat,
O. Hernandez, Mon. Not. R. Astron. Soc. {\bf 383}, 297 (2008)}
\bibitem{donato}{\small F. Donato {\it et al.}, Mon. Not. R. Astron. Soc. {\bf
397}, 1169 (2009)}
\bibitem{strigari}{\small L.E. Strigari {\it et al.}, Nature {\bf 454}, 1096
(2008)}
\bibitem{mcgaugh}{\small S.S. McGaugh, Astron. J.
{\bf 143}, 40 (2012)}
\bibitem{abrilphas}{\small A. Su\'arez, P.H. Chavanis,  Phys. Rev. D {\bf 95},
063515 (2017)}
\bibitem{action}{\small P.H. Chavanis, {\it K-essence Lagrangians of polytropic
and logotropic unified dark matter and dark energy models}  [arXiv:2109.05963]}
\bibitem{cspoly}{\small P.H. Chavanis, {\it Cosmological models based on a
complex scalar field with a power-law potential associated with a polytropic
equation of state} [arXiv:2111.01828]}
\bibitem{clm2}{\small P.H. Chavanis, M. Lemou, F. M\'ehats,  Phys. Rev. D {\bf
92}, 123527 (2015)}
\bibitem{chavtotal}{\small P.H. Chavanis, Eur. Phys. J. Plus {\bf
132}, 248 (2017)}
\bibitem{modeldm}{\small P.H. Chavanis, Phys. Rev. D {\bf 100}, 083022  (2019)}
\bibitem{burkertfdm}{\small A. Burkert, Astrophys. J. {\bf 904}, 161
(2020)}
\bibitem{deng}{\small H. Deng {\it et al.}, Phys. Rev. D {\bf 98}, 023513
(2018)}
\bibitem{shapiro}{\small B. Li, T. Rindler-Daller, P.R. Shapiro, Phys. Rev. D
{\bf 89}, 083536 (2014)}
\bibitem{abrilph}{\small A. Su\'arez, P.H. Chavanis,  Phys. Rev. D {\bf 92},
023510 (2015)}
\bibitem{playa}{\small A. Su\'arez, P.H. Chavanis, J. Phys.: Conf. Series {\bf
654}, 012008 (2015)}
\bibitem{chavmatos}{\small P.H. Chavanis, T. Matos, Eur. Phys. J. Plus {\bf
132}, 30 (2017)}
\bibitem{guthinflation}{\small A.H. Guth, Phys. Rev. D  {\bf 23}, 347 (1981)}
\bibitem{planck2014}{\small Planck Collaboration, Astron.
Astrophys. {\bf 571}, 66 (2014)}
\bibitem{planck2016}{\small Planck Collaboration, Astron.
Astrophys. {\bf 594}, A13 (2016)}
\bibitem{weinbergbook}{\small S. Weinberg, Gravitation and Cosmology (John
Wiley, 2002)}
\bibitem{arbeycosmo}{\small A. Arbey, J. Lesgourgues, and P. Salati, Phys. Rev.
D {\bfseries 65}, 083514 (2002)}
\bibitem{gh}{\small J.-A. Gu and W.-Y.P. Hwang, Phys. Lett. B {\bf 517},
1 (2001)}
\bibitem{spintessence}{\small L.A. Boyle, R.R. Caldwell, and M. Kamionkowski,
Phys. Lett. B  {\bf 545}, 17 (2002)}
\bibitem{turner}{M.S. Turner, Phys. Rev. D {\bf 28},
1243 (1983)}
\bibitem{ford}{L.H. Ford, Phys. Rev. D {\bf 35},
2955 (1987)}
\bibitem{pv}{ P.J.E. Peebles, A. Vilenkin, Phys. Rev.
D {\bf 60}, 103506 (1999)}
\bibitem{mul}{T. Matos and L.A. Ure\~na-L\'opez, Phys. Rev.
D {\bf 63}, 063506 (2001)}
\bibitem{kination}{\small M. Joyce, Phys. Rev. D
{\bf 55}, 1875 (1997)}
\bibitem{btv}{\small N. Bilic, G.B. Tupper, R.D. Viollier, Phys. Lett. B {\bf
535}, 17 (2002)}
\bibitem{makler}{\small M. Makler, S.Q. Oliveira, I. Waga, Phys. Lett. B {\bf
555}, 1 (2003)}
\bibitem{kmp}{\small A. Kamenshchik, U. Moschella, V. Pasquier, Phys. Lett. B 
{\bf 511}, 265 (2001)}
\bibitem{oufsuite}{\small P.H. Chavanis, in preparation}
\bibitem{cosmopoly1}{\small P.H. Chavanis,  Eur. Phys. J. Plus  {\bf 129}, 38
(2014)}
\bibitem{axiverse}{\small A. Arvanitaki, S. Dimopoulos, S. Dubovsky, N. Kaloper,
J.  March-Russell,  Phys. Rev. D {\bf 81}, 123530 (2010)}
\bibitem{graviton}{\small A.S. Goldhaber, M.M. Nieto, Rev. Mod. Phys. {\bf 82},
939 (2010)}
\bibitem{tsujikawa}{\small S. Tsujikawa, Class. Quantum Grav. {\bf 30},
214003  (2013)}
\bibitem{wesson}{\small P.S. Wesson, Mod. Phys. Lett. A {\bf 19}, 1995  (2004)}
\bibitem{bhcosmon}{\small C.G. B\"ohmer, T. Harko, Found Phys. {\bf 38},
216 (2008)}
\bibitem{psw}{\small R.D. Peccei, J. Sola, C. Wetterich, Phys. Lett. B {\bf
195}, 183 (1987)}
\bibitem{wcos}{\small C. Wetterich, Nucl. Phys. B {\bf 302}, 668 (1988)}
\bibitem{solacos1}{\small  J. Sola,  Phys. Lett. B {\bf 228}, 317 (1989)}
\bibitem{solacos2}{\small  J. Sola, Int. J. Mod. Phys. A  {\bf 5}, 4225 (1990)}
\bibitem{ssdilaton}{\small  I.L. Shapiro, J. Sola,  Phys. Lett. B {\bf 475}, 236
(2000)}
\bibitem{eddington}{\small A.S. Eddington, Proc. Roy. Soc. A {\bf 133}, 605
(1931)}
\bibitem{madelung}{\small E. Madelung, Zeit. F. Phys. {\bf 40}, 322 (1927)}
\bibitem{wignerPH}{\small P.H. Chavanis, {\it A heuristic wave equation
parameterizing BEC dark matter halos with a quantum core and an isothermal
atmosphere} [arXiv:2104.09244]}
\bibitem{prep}{\small P.H. Chavanis, in preparation} 
\bibitem{suarezchavanisprd3}{\small A. Su\'arez, P.H. Chavanis, Phys. Rev. D
{\bf 98}, 083529 (2018)}
\bibitem{prd1}{\small P.H. Chavanis, Phys. Rev. D {\bf 84}, 043531 (2011)}
\bibitem{aacosmo}{\small P.H. Chavanis, Astron. Astrophys. {\bf 537}, A127
(2012)}
\bibitem{dodelson}{\small S. Dodelson, {\itshape Modern Cosmology} (Academic
Press, 2003)}
\bibitem{jeansunivers}{\small P.H. Chavanis, Universe
{\bf 6}, 226 (2020)}
\bibitem{jeansMR}{\small P.H. Chavanis, Phys. Rev. D {\bf 103}, 123551 (2021) }
\bibitem{fabris2002}{\small J.C. Fabris, S.V.B. Gon\c calves, P.E. de Souza,
Gen. Relat. Grav. {\bf 34}, 53 (2002)}
\bibitem{fabris2002GR}{\small J.C. Fabris, S.V.B. Gon\c calves, P.E. de Souza,
Gen. Relat. Grav. {\bf 34}, 2111 (2002)}
\bibitem{sandvik}{\small H.B. Sandvik, M. Tegmark, M. Zaldarriaga, I. Waga, 
Phys. Rev. D {\bf 69}, 123524 (2004)}
\bibitem{cf}{\small D. Carturan, F. Finelli,  Phys. Rev. D {\bf 68}, 103501
(2003)}
\bibitem{afbc}{\small L. Amendola, F. Finelli, C. Burigana, D. Carturan, J.
Cosmol. Astropart. Phys. {\bf 07}, 005
(2003)}
\bibitem{avelinotwo}{\small P.P. Avelino, L.M.G. Be\c ca, J.P.M. de Carvalho,
C.J.A.P. Martins, P. Pinto, Phys. Rev. D {\bf 67}, 023511 (2003)}
\bibitem{daj}{\small A. Dev, J.S. Alcaniz, D. Jain,  Phys. Rev. D {\bf 67},
023515 (2003)}
\bibitem{ajd}{\small  J.S. Alcaniz, D. Jain, A. Dev, Phys. Rev. D {\bf 67},
043514 (2003)}
\bibitem{multa}{\small T. Multam\"aki, M. Manera, E. Gazta\~naga, Phys. Rev. D
{\bf 68}, 023004 (2003)}
\bibitem{bd}{\small R. Bean, O. Dor\'e, Phys. Rev. D {\bf 68},
023515 (2003)}
\bibitem{avelinoB}{\small L.M.G. Be\c ca, P.P. Avelino, J.P.M. de
Carvalho, C.J.A.P. Martins, Phys. Rev. D {\bf 67}, 101301(R)
(2003)}
\bibitem{bilicNL}{\small N. Bilic, R.J. Lindebaum, G.B. Tupper, R.D. Viollier, 
J. Cosmol. Astropart. Phys. {\bf 11}, 008 (2004)}
\bibitem{bilicNL3}{\small N. Bilic, G.B. Tupper, R.D. Viollier, Phys. Rev. D
{\bf 80}, 023515 (2009)}
\bibitem{bilicNL2}{\small N. Bilic, G.B. Tupper, R.D. Viollier, in Physics
Beyond the Standard Models of Particles, Cosmology and Astrophysics. Edited by
H. V. Klapdor-Kleingrothaus, I. V. Krivosheina, and  R. Viollier (World
Scientific Publishing, 2011), pp. 503-510}
\bibitem{avelinoNL}{\small P.P. Avelino, L.M.G. Be\c ca, J.P.M. de Carvalho,
C.J.A.P. Martins, E.J. Copeland, Phys. Rev. D {\bf 69}, 041301(R) (2004)}
\bibitem{avelinoNL2}{\small P.P. Avelino, K. Bolejko, G.F. Lewis,
Phys. Rev. D {\bf 89}, 103004 (2014)}
\bibitem{zant}{\small A. Abdullah, A.A. El-Zant, A. Ellithi, {\it The growth of
fluctuations in Chaplygin gas cosmologies: A nonlinear Jeans scale for unified
dark matter} [arXiv:2108.03260]}
\bibitem{huseul}{\small W. Hu, Astrophys. J. {\bf 306}, 485 (1998)}
\bibitem{reis1}{\small R.R.R. Reis, I. Waga, M.O. Calv\~ao, S.E. Jo\'ras, 
Phys. Rev. D {\bf 68}, 061302 (2003)}
\bibitem{reis2}{\small R.R.R. Reis, M. Makler, I. Waga, 
Phys. Rev. D {\bf 69}, 0101301 (2004)}
\bibitem{maartens}{\small R. Maartens, Living Rev. Rel. {\bf 7}, 1 (2004)}
\bibitem{ktv}{\small J.E. Kim, G.B. Tupper, R.D. Viollier, Phys. Lett. B {\bf
593}, 209 (2004)}
\bibitem{jackiw}{\small R. Jackiw, {\it Lectures on Fluid Dynamics. 
A Particle Theorist's View of Supersymmetric, Non-Abelian, Noncommutative Fluid
Mechanics and $d$-branes} (New York, Springer, 2002) [arXiv:physics/0010042]}
\bibitem{creminelli1}{\small P. Creminelli, G. D'Amico, J. Norena, F.
Vernizzi, J. Cosmol. Astropart. Phys.  {\bf 02}, 018 (2009)}
\bibitem{ks}{\small S. Kumar, A.A. Sen, J. Cosmol. Astropart. Phys.  {\bf 10},
036 (2014)}
\bibitem{pc}{\small T. Padmanabhan, T. Roy Choudhury,
Phys. Rev. D {\bf 66}, 081301(R) (2002)}
\bibitem{lb}{\small D. Lynden-Bell, Mon. Not. R. Astron. Soc. {\bf 136}, 101
(1967)}
\bibitem{seidel94}{\small E. Seidel, W.M. Suen, Phys. Rev. Lett.
{\bf  72}, 2516 (1994)}
\bibitem{carvente}{\small B. Carvente, V. Jaramillo, C. Escamilla-Rivera, D.
N\'u\~nez, Mon. Not. R. astr. Soc. {\bf 503}, 4008 (2021)}
\bibitem{phi6}{\small P.H. Chavanis, Phys. Rev. D {\bf 98}, 023009
(2018)}
\bibitem{sfpoly}{\small P.H. Chavanis, in preparation} 
\bibitem{logo}{\small P.H. Chavanis, C. Sire, Physica A {\bf 375}, 140
(2007)}
\bibitem{chandra}{\small S. Chandrasekhar, An Introduction to the Study of
Stellar Structure (Dover, 1958)}
\bibitem{mcmh}{\small P.H. Chavanis, Phys. Rev. D {\bf 100}, 123506 (2019)}
\bibitem{as}  {\small  H. Anton, P.C. Schmidt, Intermetallics {\bf 5}, 449
(1997)}
\bibitem{debye}  {\small  P. Debye, Ann. Phys. {\bf 344}, 789
(1912)}
\bibitem{gruneisen}  {\small  E. Gr\"uneisen, Ann. Phys. {\bf 344}, 257
(1912)}
\bibitem{partially}{\small P.H. Chavanis,  Eur. Phys. J. Plus  {\bf 130}, 181
(2015)}
\bibitem{stiff}{\small P.H. Chavanis, Phys. Rev. D {\bf 92}, 103004
(2015)}
\bibitem{aaiso}{\small P.H. Chavanis, Astron. Astrophys. {\bf 381}, 340
(2002)}
\bibitem{muller}{\small C.M. M\"uller, Phys. Rev. D {\bf 71},
047302 (2005)}
\bibitem{avelinoZ}{\small P.P. Avelino, L.M.G. Beca,  J.P.M. de
Carvalho, C.J.A.P. Martins, J. Cosmol. Astropart. Phys. {\bf 09}, 002
(2003)}
\bibitem{fabris2004GR}{\small J.C. Fabris, S.V.B. Gon\c calves, R. de S\'a
Ribeiro, Gen. Relat. Grav. {\bf 36}, 211 (2004)}
\bibitem{nfw}{\small J.F. Navarro, C.S. Frenk, S.D.M. White, Mon. Not. R.
Astron.
Soc. {\bf 462}, 563 (1996)}
\bibitem{burkert}{\small A. Burkert, Astrophys. J. {\bf
447}, L25 (1995)}
\bibitem{em}{\small G.F.R. Ellis, M.S. Madsen,  Class. Quantum Grav. 
{\bf 8}, 667 (1991)}
\bibitem{paddytachyon}{\small T. Padmanabhan, Phys. Rev. D  {\bf 66}, 021301
(2002) }
\bibitem{cst}{\small E.J. Copeland, M. Sami, S. Tsujikawa, Int. J. Mod. Phys. D 
{\bf 15}, 1753 (2006)}
\bibitem{bamba}{\small K. Bamba, S. Capozziello, S. Nojiri, S.D.
Odintsov, Astrophys. Space Sci. {\bf 342}, 155 (2012)}
\bibitem{cosmopoly2}{\small P.H. Chavanis,  Eur. Phys. J. Plus  {\bf 129}, 222
(2014)}
\bibitem{gkmp}{\small V. Gorini, A. Kamenshchik, U. Moschella, V. Pasquier,
Phys. Rev. D {\bf 69}, 123512 (2004)}
\bibitem{hu}{\small W. Hu, R. Barkana, A. Gruzinov, Phys. Rev. Lett. {\bf  85},
1158 (2000)}
\bibitem{ch2}{\small H.Y. Schive, T. Chiueh, T. Broadhurst, Nature Physics {\bf
10}, 496 (2014)}
\bibitem{ch3}{\small H.Y. Schive {\it et al.}, Phys. Rev. Lett. {\bf 113},
261302 (2014)}
\bibitem{schwabe}{\small B. Schwabe, J. Niemeyer, J. Engels, Phys. Rev. D {\bf
94}, 043513 (2016)}
\bibitem{moczSV}{\small P. Mocz, L. Lancaster, A.Fialkov, F. Becerra, P.H.
Chavanis, Phys. Rev. D {\bf 97}, 083519 (2018)}
\bibitem{veltmaat}{\small J. Veltmaat, J.C. Niemeyer, B. Schwabe,
 Phys. Rev. D {\bf 98}, 043509 (2018)}
\bibitem{moczprl}{\small P. Mocz {\it et al.}, Phys. Rev. Lett.
{\bf 123}, 141301 (2019)}
\bibitem{moczmnras}{\small P. Mocz {\it et al.}, Mon. Not.
R. astr. Soc. {\bf 494}, 2027 (2020)}
\bibitem{veltmaat2}{\small J. Veltmaat,
B. Schwabe, J.C.
Niemeyer, Phys. Rev. D {\bf 101}, 083518 (2020)}
\bibitem{mocz}{\small P. Mocz {\it et al.}, Mon. Not. R. Astron. Soc.
{\bf 471}, 4559 (2017)}
\bibitem{membrado}{\small M. Membrado, A.F. Pacheco, J. Sanudo, Phys. Rev. A
{\bf  39}, 4207 (1989)}
\bibitem{prd2}{\small P.H. Chavanis, L. Delfini, Phys. Rev. D {\bf 84}, 043532
(2011)}
\bibitem{mcmhbh}{\small P.H. Chavanis, Phys. Rev. D {\bf 101}, 063532 (2020)}
\bibitem{egg}{\small B. Eggemeier, J.C. Niemeyer, Phys. Rev. D {\bf 100},
063528 (2019)}
\bibitem{bbbs}{\small N. Bar, D. Blas, K. Blum, S. Sibiryakov, Phys. Rev. D
{\bf 98}, 083027 (2018)}
\bibitem{hui}{\small L. Hui, J. Ostriker, S. Tremaine, E. Witten, Phys. Rev. D
{\bf 95}, 043541 (2017)}
\bibitem{bft}  {\small B. Bar-Or, J.B. Fouvry, S. Tremaine, Astrophys. J. {\bf
871}, 28 (2019)}
\bibitem{bft2}{\small B. Bar-Or, J.B. Fouvry, S. Tremaine, Astrophys. J. {\bf
915}, 27 (2021) }
\bibitem{meff}{\small P.H. Chavanis, Eur. Phys. J. Plus {\bf 136}, 703 (2021)}
\bibitem{clm1}{\small P.H. Chavanis, M. Lemou, F. M\'ehats, Phys. Rev. D {\bf
91}, 063531 (2015)}
\bibitem{oh}{\small S.-H. Oh {\it et al.}, Astrophys. J. {\bf 142}, 24 (2011)}
\bibitem{rm}{\small V.H. Robles, T. Matos, Mon. Not. R. Astron. Soc. {\bf
422}, 282 (2012)}
\bibitem{burkert2}{\small A. Burkert, Astrophys. J. {\bf 808}, 158
(2015)}
\bibitem{blok}{\small W.J.G. de Blok, S.S. McGaugh, A. Bosma, V.C. Rubin,
Astrophys. J. {\bf 552}, L23 (2001)}
\bibitem{gentile}{\small G. Gentile, B. Famaey, H. Zhao, P. Salucci, Nature {\bf
461}, 627 (2009)}
\bibitem{tf}{\small R.B. Tully, J.R. Fisher,  Astron. Astrophys.
{\bf 54}, 661 (1977)}
\bibitem{mond}{\small M. Milgrom, Astrophys. J. {\bf 270}, 365
(1983)}



\end{thebibliography}
\end{document}